\documentclass{article}

\usepackage{jheppub}

\makeatletter
\providecommand\@fpheader{} 
\providecommand\@preprint{} 
\makeatother

\usepackage{amsmath}
\usepackage{braket}
\usepackage[table]{xcolor}
\usepackage{mathrsfs}
\usepackage[english]{babel}
\usepackage{graphicx,multirow}
\usepackage{caption}
\usepackage{subcaption}
\usepackage{amssymb}
\usepackage{slashed}
\usepackage{makecell}
\usepackage{hhline}
\usepackage{tcolorbox}

\usepackage{upgreek}

\usepackage[makeroom]{cancel}

\def\nn{\nonumber}

\def\pd{\partial}

\def\cE{{\cal E}}

\def\exd{{\hbox{d}}}

\def\bea{\begin{eqnarray}}
\def\eea{\end{eqnarray}}
\def\ba{\begin{eqnarray}}
\def\ea{\end{eqnarray}}
\def\be{\begin{equation}}
\def\ee{\end{equation}}

\def \bal#1\eal  {\begin{align} #1 \end{align}}
\def\({\left(}
\def\){\right)}
\def\[{\left[}
\def\]{\right]}
\def\<{\langle}
\def\>{\rangle}
\def\d{\mathrm{d}}

\def\nn {\nonumber}
\def\d{\mathrm{d}}

\def\ssS{{\scriptscriptstyle S}}

\def\conn{{\mathrm{ conn } }}

\newcommand{\roughly}[1]{\mathrel{\raise.3ex\hbox{$#1$\kern-0.85em
\lower1ex\hbox{$\sim$}}}}

\newcommand{\stu}{St\"uckelberg }

\def\Tr{\mathrm{Tr}}
\def\tr{\mathrm{tr}}

\def\TrE{\underset{\mathcal{E}}{\mathrm{Tr}}}

\def\QFe{\mathscr{Q}_{\mathcal{F}}}
\def\QFeAST{\mathscr{Q}_{\mathcal{F}}^{\ast}}
\def\QWe{\mathscr{Q}_{\mathcal{W}}}

\def\Wvac{\mathcal{W}^{\mathrm{vac}}}

\def\Ivac{I^{\mathrm{vac}}}
\def\LFvac{L_{\mathcal{F}}^{\mathrm{vac}}}
\def\LWvac{L_{\mathcal{W}}^{\mathrm{vac}}}

\def\ti{t_{\mathrm{i}}}
\def\tf{t_{\mathrm{f}}}

\def\smath#1{\text{\scalebox{.85}{$#1$}}}
\def\sfrac#1#2{\smath{\frac{#1}{#2}}}

\newcommand*{\Scale}[2][4]{\scalebox{#1}{$#2$}}

\numberwithin{equation}{section}
\title{Gauging Open EFTs from the top down}

\author[a,b,c]{Greg Kaplanek,}
\author[d, e]{Maria Mylova,}
\author[a,f]{and Andrew J. Tolley}

\affiliation[a]{Abdus Salam Centre for Theoretical Physics, Imperial College, London, SW7 2AZ, UK}
\affiliation[b]{Department of Electrical Engineering and Computer Science, Syracuse University, NY 13210, USA}
\affiliation[c]{Institute for Quantum \& Information Sciences, Syracuse University, NY 13210, USA}
\affiliation[d]{Kavli Institute for the Physics and Mathematics of the Universe (WPI)
the University of Tokyo, Kashiwa, Chiba 277-8583, Japan}
\affiliation[e]{Perimeter Institute for Theoretical Physics, Waterloo, Ontario, N2L 2Y5, Canada}
\affiliation[f]{CERCA, Department of Physics, Case Western Reserve University, 10900 Euclid Ave, Cleveland,
OH 44106, USA}

\emailAdd{gkaplane@syr.edu, maria.mylova@ipmu.jp, a.tolley@imperial.ac.uk}

\date{today}

\begin{document}

\sloppy

\abstract{We present explicit top-down calculations of Open EFTs for gauged degrees of freedom with a focus on the effects of gauge fixing. Starting from the in-in contour with two copies of the action, we integrate out the charged matter in various $U(1)$ gauge theories to obtain the Feynman-Vernon influence functional for the photon, or, in the case of symmetry breaking, for the photon and St\"uckelberg fields. The influence functional is defined through a quantum path integral, which --- as is always the case when quantizing gauge degrees of freedom --- contains redundancies that must be eliminated via a gauge-fixing procedure. We implement the BRST formalism in this setting. The in-in boundary conditions break the two copies of BRST symmetry down to a single diagonal copy. Nevertheless the single diagonal BRST is sufficient to ensure that the influence functional is itself gauge invariant under two copies of gauge symmetries, retarded and advanced, regardless of the choice of state or symmetry-breaking phase. We clarify how this is consistent with the decoupling limit where the global advanced symmetry is broken by the in-in boundary conditions. We illustrate our results with several examples: a gauge field theory analogue of the Caldeira-Leggett model; spinor QED with fermions integrated out; scalar QED in a thermal state; the Abelian Higgs-Kibble model in the spontaneously broken state with the Higgs integrated out; and the Abelian Higgs-Kibble model coupled to a charged bath in a symmetry-broken phase. The latter two serve as examples of open systems for St\"uckelberg/Goldstone fields. Finally we show how to construct bottom-up EFTs in both the broken and unbroken phase.}

\maketitle


\section{Introduction}

Effective Field Theories (EFTs) are ubiquitous in modern physics, providing a systematic description of low-energy degrees of freedom after heavy or high-energy modes have been integrated out. In a broader sense, an EFT can be viewed as a special case of an open quantum system: part of a quantum field theory is tracked while the rest is unobserved and traced out, with or without the requirement that the unobserved sector be parametrically heavier than the retained degrees of freedom. This perspective has recently gained attention in high-energy and cosmological settings, where time-dependent backgrounds and the resulting non-equilibrium quantum states naturally give rise to dissipative effects, stochastic fluctuations, and decoherence --- phenomena not typically captured by ordinary EFTs. Motivated by this, recent work has pursued the construction of bottom-up ``Open EFTs'', which aim to capture the effective dynamics of open systems while remaining agnostic about the microscopic details of the environment. Initial focus was on the scalar sector \cite{LopezNacir:2011kk, Hongo:2018ant, Hongo:2019qhi, Salcedo:2024smn}, specifically in the context of inflation.  This program has been especially active lately in the context of gauge theories, including photons \cite{Salcedo:2024nex} and, of particular cosmological interest, gravitons \cite{Lau:2024mqm, Salcedo:2025ezu, Christodoulidis:2025ymc,Christodoulidis:2025vxz}.

Equally important, however, are explicit top-down examples in which the environment is integrated out of a known closed system or partial UV-completion. In this paper we consider examples of Abelian gauge theories such as QED/Abelian-Higgs model and extensions, where the charged matter or the symmetry breaking Higgs are integrated out. These simple examples are instructive because the resulting influence functional inherits the subtleties of gauge symmetry, forcing us to confront the role of gauge fixing in the open system setting.

The early development of the open quantum systems framework centered on oscillator models, in which both the system and its environment were treated as collections of scalar degrees of freedom. Building on foundational work by Schwinger \cite{Schwinger:1960qe} and Feynman-Vernon \cite{Feynman:1963fq}, Caldeira and Leggett \cite{caldeira1983path} provided a microscopic account of dissipation and noise by coupling a quantum system linearly to an oscillator bath, yielding a quantum Langevin/Fokker-Planck description. In the context of QED, subsequent analyses of dissipative phenomena used the Abraham-Lorentz-Dirac-Langevin framework, integrating out the photon field to describe radiation reaction and stochastic fluctuations \cite{ford1993relativistic, Johnson:2000qd, o2012radiation}.

Schwinger developed his closed time path (CTP) formalism \cite{Schwinger:1960qe} to show how the action formalism could be used to describe non-time-ordered correlation functions such as `in-in' expectation values and in doing so be able to describe open quantum systems. The formalism was first applied to field theory by Schwinger's students Mahanthappa and Bakshi \cite{Bakshi:1962dv,Bakshi:1963bn} and notably was applied to QED in \cite{Mahanthappa:1962ex,Bakshi:1963bn}. Subsequently, Keldysh developed an equivalent version of the formalism \cite{Keldysh:1964ud} making use of a convenient change of variables which significantly aids calculations. The resulting Schwinger-Keldysh/in-in/real-time/CTP formalism has wide application in condensed matter systems \cite{kamenev2023field}. An early application was to laser physics \cite{KORENMAN196672}.

Within high-energy physics, the Schwinger-Keldysh approach underlies modern studies of non-equilibrium dynamics in heavy-ion collisions. Numerical and functional methods, including lattice simulations, kinetic theory, and real-time effective approaches, are used to extract gauge-invariant quantities such as transport coefficients from correlation functions \cite{Asakawa:2000tr, Arnold:2002zm, caron2008heavy, meyer2009transport, ding2012charmonium, berges2014turbulent, brambilla2020lattice}. More recently, studies of heavy quarkonium have adopted Lindblad-type descriptions of the real-time evolution of the reduced density matrix, with gauge invariance maintained through Wilson-line dressing of correlators \cite{brambilla2017quarkonium, yao2021open, akamatsu2022quarkonium, scheihing2023real}.
Over the past decade, hydrodynamics itself has been reformulated as a Schwinger-Keldysh effective field theory, in which one writes the most general influence functional consistent with symmetries, conservation laws, and thermal (KMS) conditions \cite{Harder:2015nxa,Crossley:2015evo,Jensen:2017kzi,Zhou:2025mbq,Mo:2025nes}.

The real-time formalism has also found extensive application in cosmology, where the dynamics are inherently out of equilibrium. Open quantum systems tools provide a natural framework for studying time-dependent processes in the early universe including decoherence and quantum signatures \cite{Sakagami:1987mp,Brandenberger:1990bx,Matacz:1992mk,Lombardo:1995fg,Calzetta:1995ys,Polarski:1995jg,Kiefer:1998qe,Lombardo:2005iz,Burgess:2006jn,Prokopec:2006fc,Sharman:2007gi,Kiefer:2008ku,Kiefer:2010pb,Bachlechner:2012dg,Franco:2011fg,Burgess:2014eoa,Nelson:2016kjm,Hollowood:2017bil,boddy2017decoherence,Martin:2018zbe,Bao:2019ghe,Brahma:2020zpk,Martin:2021znx,Banerjee:2021lqu,Brahma:2022yxu,Espinosa-Portales:2022yok,Colas:2022hlq,Burgess:2022nwu,DaddiHammou:2022itk,Colas:2022kfu,Martin:2022kph,Sou:2022nsd,Boutivas:2023mfg,Tinwala:2024wod,Colas:2024xjy,Colas:2024ysu,deKruijf:2024ufs,Burgess:2024eng,Brahma:2024yor,Burgess:2025dwm,Cespedes:2025zqp,deKruijf:2025jya,Sano:2025ird,Piotrak:2025zhy,Lopez:2025arw,Cielo:2025ibc,Christie:2025knc,Micheli:2025yux,Choudhury:2025ssc} in inflation, related infrared effects \cite{Tsamis:1994ca,Tsamis:1996qm,Tsamis:2005hd,Burgess:2015ajz, shandera2018open}, as well as reheating and other non-equilibrium phases of early-universe evolution \cite{calzetta1989dissipation, morikawa1990dissipation, Arnold:1992rz, Yokoyama:2004pf, Yokoyama:2005dv, Mukaida:2012bz, Drewes:2014pfa, boyanovsky2015effective, Ai:2021gtg,Brax:2025olx}. Related developments include the extension of semiclassical gravity to incorporate quantum stress-energy fluctuations via the influence functional approach to the Einstein-Langevin equation \cite{calzetta1989dissipation, morikawa1990dissipation, Hu:2008rga}, and more recent applications of the Schwinger-Keldysh formalism to graviton decoherence during reheating \cite{Takeda:2025cye}. In \cite{Burrage:2018pyg}, the open system effects of light scalars are derived yielding a non-Markovian master equation with decoherence and diffusion. Ref.~\cite{Kading:2022jjl,Kading:2025cwg} compute reduced density matrices directly using a perturbative path-integral approach, clarifying non-Markovian effects and their relation to master equations.

A comprehensive account of its application to open quantum systems with global and discrete symmetries is given in Ref.~\cite{sieberer2016keldysh}. Ref.~\cite{Hongo:2018ant,Hongo:2019qhi,Landry:2019iel,Minami:2015uzo, Hidaka:2019irz,Akyuz:2023lsm} emphasise symmetry breaking and Goldstone EFTs. Our concern here is, however, the opposite situation: open quantum systems in the presence of gauge symmetries, be they unbroken or spontaneously broken. The Schwinger-Keldysh or real-time formalism has a well-developed application to gauge theories \cite{Landsman:1986uw,Calzetta:2008iqa,DeWitt:2003pm}.  Scalar QED has long served as a convenient playground for investigating these issues and for developing techniques applicable to non-equilibrium gauge theories \cite{Weldon:1982aq,braaten1990deducing, Kraemmer:1994az, Blaizot:1995kg, Flechsig:1995ju, thoma1996damping, le2000thermal, kapusta2007finite, scherer2024photon,Caron-Huot:2007cma, Gorda:2022fci, ferreira2024power, ekstedt2023two, Boyanovsky:1998pg, Boyanovsky:1999jh, Breuer:2002pc, Miyamoto:2013gna, shi2018simulations}. More recent work has examined the 1PI effective action for photons and the associated causal structure and positivity constraints \cite{Creminelli:2024lhd}. See also \cite{Li:2026qms} which shows that Lindblad dynamics are compatible with gauge invariance, despite the nonconservation of a global charge.

Since QED is an Abelian theory, gauge fixing is straightforward, and it is usually unnecessary to introduce ghosts or invoke the machinery of the Becchi-Rouet-Stora-Tyutin (BRST) formalism \cite{Becchi:1975nq,Tyutin:1975qk}. Indeed, the first concrete application of Schwinger's formalism was a gauge theory, QED \cite{Mahanthappa:1962ex,Bakshi:1963bn}. Nevertheless, we will argue that it is worth doing so since the underlying BRST symmetry provides a strong constraint on the resulting open systems, in particular the types of interactions that arise between the two branches. Furthermore, the utility of the BRST formalism makes its implications immediately generalisable to non-Abelian and gravitational theories.

\subsection{In-in formalism}

For a field theory with no gauge symmetries, the CTP/Schwinger-Keldysh/in-in/real-time formalism is straightforward. For every field we introduce two copies of the field $\phi^{\pm}$ living on the two branches of a CTP. For a closed system the `in-in' action is the difference of the action for each branch $S_{\text{in-in}}= S[\phi^+]-S[\phi^-]$ to account for the change in direction of the flow of time between each branch. In the path integral formulation we are then instructed to perform the path integral over both fields with the final time $\tf$ boundary condition that $\phi^{+}(\tf)=\phi^-(\tf)$. The precise choice of the final time is arbitrary, provided that it is in the future of all evaluated correlation functions. The fact that it is arbitrary is itself a statement of causality, interactions in the future of the observations cannot influence the observations themselves, {\it i.e.} in-in correlation functions evaluated at finite times $t_j$ are independent of all interactions for times $t>{\rm Max}\{t_j\}$. 
As in the more familiar in-out formalism, the initial state can be captured either by a boundary condition at the initial time or via an extended $i \epsilon$ prescription which we elucidate in Sec.~\ref{BRSTinSK}.

\subsection*{In-in for global symmetries}

When considering a field theory with a global symmetry, at first sight it might appear that the in-in action carries two copies of the global symmetry, suggesting the presence of two conserved Noether charges. However, the final time boundary conditions demand that the parameters for the two independent global transformations are identical. For example, for a $U(1)$ symmetry of a complex scalar, the naive two global transformations are
\be
\delta \Phi_{\pm} = i \theta_{\pm} \Phi_{\pm} \, .
\ee
Demanding invariance of the boundary conditions leads to
\be
\delta (\Phi_{+}(\tf)-\Phi_-(\tf)) = i \theta_+  \Phi_{+}(\tf)-i \theta_- \Phi_{-}(\tf)= i ( \theta_+-\theta_-) \Phi_{+}(\tf)=0 \, ,
\ee
hence imposing $\theta_+ -\theta_-$=0. In other words, only the `diagonal' (retarded) global symmetry is preserved. This argument extends to any global symmetry, even those which are spacetime-dependent when we account for the arbitrariness of $\tf$. For example, for a Galileon theory with transformations
\be
\delta \pi_{\pm}=v^{\pm}_{\mu}x^{\mu} \, , 
\ee
then the boundary conditions imply
\be
\delta (\pi_{+}(\tf)-\pi_-(\tf))=(v^+_0-v^-_0) \tf+(v^+_i-v^-_i)  x^i=0 \, ,
\ee
which given the arbitrariness of $\tf$ implies $v_{\mu}^+=v_{\mu}^-$.
The significance of this is that when constructing an open EFT for some subset of fields in a closed system which exhibits an unbroken global symmetry, it is only necessary to impose that the open in-in EFT action or Feynman-Vernon influence functional is invariant under the diagonal global symmetry. Thus for example, if we consider a charged scalar $\Phi$ as an open system, we may expect terms such as $\Phi_+^* \Phi_-$ to arise in the influence functional. A simple example of this will be given later.

\subsection*{In-in for gauge symmetries}

What happens when the symmetry is gauged? A naive first guess might be that the influence functional of an open EFT may include terms which are only invariant under the diagonal (retarded) gauge symmetry. However, such an assumption is not justified since the coupling between the branches in the in-in path integral occurs only at the final time $\tf$ and the boundary conditions are satisfied for independent gauge transformations on each branch that satisfy only the requirement that $\lambda_+(\tf)=\lambda_-(\tf)$. However, since $\tf$ is arbitrary, both gauge symmetries are effectively independent, suggesting that the in-in action has twice the gauge symmetry of its equivalent in-out form. This leads to a conundrum: how can a system which has two copies of gauge invariance only preserve the diagonal copy of the global symmetry? 

This rather confusing situation is resolved easily by remembering that it is not possible to quantise a gauge theory in path integral form without effectively fixing the gauge and we cannot construct the in-in formalism before addressing how the gauge symmetry is dealt with. There are two main ways to do this:

\begin{itemize}

\item Begin with the in-out action and fix a gauge, solve the constraints, and reduce the action to physical degrees of freedom only. The in-in formalism then proceeds as in a non-gauge theory.

\item Introduce gauge-fixing terms and replace the gauge symmetry with a global BRST symmetry. The in-in formalism then proceeds as in a non-gauge theory with a global symmetry.

\end{itemize}

The former approach whilst correct is notoriously cumbersome due to the non-local form of the resulting action. Furthermore, it is plagued by the fact that most gauge choices do not completely fix the gauge. It is worth noting that this is, however, the method essentially used (at least for the time diffeomorphisms) in cosmological perturbation theory.

In this paper, we shall consider the second approach, gauge fixing by the Faddeev-Popov-DeWitt method, emphasising the underlying BRST symmetry. The way to do this for gauge theories is well-known, and for example is used in the context of real-time finite temperature QCD \cite{Landsman:1986uw,Kobes:1984vb,Kobes:1985kc,Kobes:1986za,Hata:1980yr,Ojima:1981ma,Czajka:2014eha} where in particular it is important to specify the thermal prescription of the ghosts for which there are competing possibilities.
Our concern here is however its implications for the structure of open effective theories that result from integrating or tracing out inaccessible degrees of freedom, more precisely its implications for the Feynman-Vernon influence functional.\footnote{The influence functional for spinor QED obtained from integrating out the charged matter is obtained in \cite{Shaisultanov:1995ss} for specific backgrounds using Bogoliubov coefficients.}
As noted above, if we choose to quantise a theory in the BRST formalism, where there exists a global BRST symmetry, then according to the above arguments only the diagonal BRST symmetry is consistent with the in-in boundary conditions. The symmetry is unaffected by integrating out or tracing out degrees of freedom, at least in its action on the gauge fields. Furthermore, every physical state must be BRST invariant, and so inclusion of information in the initial state/density operator does not spoil the BRST symmetry\footnote{This is quite different than a physical global symmetry $G$ where the initial state is not required to be invariant.}. We thus conclude that the resulting open in-in EFT or equivalently influence functional is necessarily invariant under the diagonal (retarded) BRST symmetry. 

Having established that the influence functional is invariant only under the diagonal BRST symmetry, we may be forgiven for thinking that it is possible to generate interactions between the two branches of the CTP that do not preserve the advanced (off-diagonal) gauge invariance. The remarkable power of the BRST formalism is that because of the presence of two independent FPDW ghosts, even the diagonal BRST symmetry is sufficient to guarantee two copies of gauge symmetry realised by the influence functional, at least in the Abelian case, regardless of the state chosen. This provides a significant restriction on bottom-up approaches to constructing open EFTs.

Global transformations are a special limit of local ones, and so any in-in formalism describing gauge theories should recover in the appropriate limit the expectations for an open EFT with global symmetries. Once again the diagonal BRST symmetry does the job, it is straightforward to see that the decoupling limit of a global BRST symmetry transformation enforces the required diagonal global symmetry only.

Returning to our conundrum - how can there be operators in the in-in theory which are invariant under two copies of gauge transformations, both retarded and advanced, but only preserve the diagonal (retarded) global symmetry? We will see in concrete examples that the answer differs depending on whether we have spontaneous symmetry breaking (SSB) or not. 

When there is no SSB, gauge invariance for the advanced gauge transformations is maintained by the emergence of Wilson lines that pass via the final time surface whose origin is the propagators of the charged states that communicate the dissipative/noise effects. This apparently acausal situation is dictated by the time reversed propagation of advanced fields. More precisely it occurs because interactions between the two branches which are naively at the same time, are actually at different times on the CTP and the Wilson lines needed to ensure gauge invariance are extended along the CTP, inevitably passing through the final time surface. In the decoupling limit the Wilson line disappears leaving an interaction between the two branches that can break the advanced global symmetry.

When there is SSB, the situation is much simpler. The gauge symmetries on both branches are spontaneously broken, leading to the emergence of both retarded and advanced \stu/Goldstone fields. It is then straightforward to write interactions that are gauge invariant using the \stu fields, but preserve only a residual retarded global symmetry.

Recent approaches to constructing open EFTs for gauge symmetries have allowed additional interactions that break advanced gauge invariance but preserve a modified version of the advanced gauge symmetry \cite{Salcedo:2024nex, Lau:2024mqm, Salcedo:2025ezu, Christodoulidis:2025ymc}. The above arguments show clearly that such a possibility cannot arise from any gauge invariant UV completion/closed system. In Sec.~\ref{sec:fakedissipation} we explain the resolution, a closed system whose action is time-dependent can lead to dissipative-like terms in the equations of motion (such as the Hubble damping term for a scalar in FLRW). If the in-in effective action for such a system is written in incorrect variables, for a gauge theory it can appear to exhibit a modified gauge transformation. However, on correctly rewriting in terms of the true variables, advanced gauge invariance is recovered.

The manuscript is organised as follows. We begin with a description of the in-in formalism and BRST gauge-fixing in open gauge theories. Important ingredients in this construction, such as the $i\epsilon$ prescription, Wilson lines, and the decoupling limit, are discussed in Sec.~\ref{BRSTinSK}. In Sec.~\ref{TopDown} we then move to explicit top-down constructions of open gauge theories. In the following sections we study in more detail thermal scalar QED in Sec.~\ref{sec:thermal}, the Abelian Higgs-Kibble model in Sec.~\ref{sec:SSB}, and its \`a la Caldeira-Leggett version in Sec.~\ref{SSBII}. Finally, in Sec.~\ref{BottomUp} we present a bottom-up framework for open EFTs of electromagnetism directly at the level of the in-in action. Many of the gory details are left to the Appendices.

\section{Gauge Fixing and BRST Doubling }
\label{BRSTinSK}

The Abelian Higgs-Kibble model provides perhaps the simplest setting to study the interplay between gauge symmetry and matter fields in situations where the gauge symmetry is both unbroken and broken. The theory describes a complex scalar field $\Phi$ charged under a $U(1)$ gauge field, with action
\begin{eqnarray} \label{QEDaction}
S & = & - \int \exd^4 x\; \bigg[ \frac{1}{4} F_{\mu\nu} F^{\mu\nu} + ( D_{\mu} \Phi )^{\ast} D^\mu \Phi + V(\Phi^{\ast}\Phi) \bigg] \, ,
\end{eqnarray}
where the field strength and covariant derivative are
\begin{equation} \label{fieldstrength}
F_{\mu\nu} = \partial_\mu A_{\nu} - \partial_\nu A_{\mu} \, , \quad D_{\mu} = \partial_\mu - i q A_{\mu} \, .
\end{equation}
Scalar QED is the special case for which $V=m^2 |\Phi|^2$. Being a gauge theory, the action (\ref{QEDaction}) is invariant under local $U(1)$ gauge transformations
\begin{eqnarray} \label{gaugetrans}
\Phi(x) \to e^{i q \lambda(x)} \Phi(x)\, , \qquad \mathrm{and} \qquad A_{\mu}(x) \to A_{\mu}(x) + \partial_\mu \lambda(x) \ ,
\end{eqnarray}
parametrised by an arbitrary function $\lambda(x)$. For the purposes of this section, we leave the potential $V(\Phi^{\ast}\Phi)$ unspecified, noting only that it depends on $\Phi^{\ast}\Phi$ and is therefore gauge invariant.

\subsection{Gauge Fixing and BRST symmetry}
\label{sec:BRST}

The gauge freedom of (\ref{QEDaction}) is of course a redundancy in the description, and so one must fix a gauge in order to perform calculations. A standard choice is to add a gauge-fixing term
\begin{equation} \label{gaugefixS}
S_{\mathrm{GF}}[A] \equiv  -\frac{1}{2\xi}  \int \exd^4 x\; (\partial_{\mu} A^{\mu})^2 \ ,
\end{equation}
to the total action which enforces the 't Hooft average of Lorenz gauge
\begin{equation} \label{lorenz}
\partial_\mu A^{\mu} = 0 \ , 
\end{equation}
together with the associated Faddeev-Popov-DeWitt (FPDW) ghosts $c$ and $\bar{c}$.
The full gauge fixed version of the Abelian Higgs-Kibble action (\ref{QEDaction}) is 
\begin{eqnarray} \label{QEDactionBRST}
S_{\mathrm{BRST}} & = & - \int \exd^4 x\; \bigg[ \frac{1}{4} F_{\mu\nu} F^{\mu\nu} + ( D_{\mu} \Phi )^{\ast} D^\mu \Phi + V(\Phi^{\ast} \Phi) + \frac{1}{2\xi} (\partial_{\mu} A^{\mu})^2 + \bar{c} \hspace{0.3mm} \Box c \bigg] \, .
\end{eqnarray}
Because the gauge group $U(1)$ is Abelian, these ghosts do not couple to the physical fields. 
The action is invariant under the global BRST transformation
\begin{equation} \label{BRST_trans}
  \begin{split}
  A_{\mu}(x) & \to A_{\mu}(x) + \eta \hspace{0.3mm} \partial_\mu c(x)\, , \\
  \Phi(x) & \to \Phi(x) + i q \eta \hspace{0.3mm}  c(x) \Phi(x)\, ,
  \end{split}
\hspace{20mm}
  \begin{split}
    c(x) & \to c(x)\, , \\
    \overline{c}(x) & \to \overline{c}(x) - \tfrac{\eta}{\xi} \hspace{0.3mm}  \partial_\mu A^{\mu}(x)\, ,
  \end{split}
\end{equation}
where $\eta$ is a Grassmann number satisfying $\eta^2 = 0$. This transformation mirrors the original local gauge symmetry (\ref{gaugetrans}), but with the gauge parameter replaced by $\lambda(x) \to \eta c(x)$. The essential point is that the transformation is controlled by a single {\it global} Grassmann parameter $\eta$, while the local $x$-dependence is carried by the ghost field $c(x)$.

The BRST charge that generates the transformations in Eq.~(\ref{BRST_trans}) is given by
\begin{equation} \label{BRST_charge}
Q_{\mathrm{BRST}} = \int \exd^3\mathbf{x} \; \bigg[ \Pi_{A}^{i} \partial_i c + i q \big( \Pi_{\Phi}  \Phi - \Pi_{\Phi^{\ast}}  \Phi^{\ast} \big) c + \frac{1}{\xi} \Pi_{\bar{c}} \big( \partial_{\mu} A^{\mu} \big) \bigg] \, ,
\end{equation}
where $(\Pi_{A}, \Pi_{\Phi}, \Pi_{\Phi^{\ast}},  \Pi_{c}, \Pi_{\bar{c}}) = ( \dot{A}, \dot{\Phi}, \dot{\Phi}^{\ast} ,  - \dot{c}, +\dot{c} )$ are the canonical momenta conjugate to $\boldsymbol{f} = ( A, c , \bar{c},  \Phi, \Phi^{\ast})$.
The presence of $Q_{\mathrm{BRST}}$ naturally partitions the Hilbert space into three sectors, exact, closed, and cohomology,
where the physical states are those $ | \psi \rangle$ that are annihilated by the charge (such that $Q_{\mathrm{BRST}} | \psi \rangle =0$) which ensures that unphysical photon polarisations are automatically excluded. More generally, a density matrix $\rho$ is physical when $ Q_{\mathrm{BRST}} \rho =  \rho  Q_{\mathrm{BRST}}=0$ (see \cite{Kaplanek:2026kpp} for a more detailed discussion on this).

With the BRST action defined above, one can now properly quantise the theory. To develop the path integral formalism we work in Schr\"odinger-picture. Each classical field has a corresponding Schr\"odinger-picture operator $\hat{A}_{\ssS}^{\mu}(\mathbf{x})$, $\hat{\Phi}_{\ssS}(\mathbf{x})$, $\hat{c}_{\ssS}(\mathbf{x})$ and $\hat{\bar{c}}_{\ssS}(\mathbf{x})$ which all have their own respective field eigenstates
\begin{eqnarray} \label{eigenstatesAPhicc}
&& \hat{A}^\mu_{\ssS}(\mathbf{x}) | \mathrm{A} \rangle
  = \mathrm{A}^{\mu}(\mathbf{x}) | \mathrm{A} \rangle \, , \\
&&\hat{\Phi}_{\ssS}(\mathbf{x}) | \Upphi \rangle
  = \Upphi(\mathbf{x}) | \Upphi \rangle \, , \\
&&\hat{c}_{\ssS}(\mathbf{x}) | \mathrm{c}, \bar{\mathrm{c}} \rangle
  = \mathrm{c}(\mathbf{x}) | \mathrm{c}, \bar{\mathrm{c}} \rangle \, , \\
&&\hat{\bar{c}}_{\ssS}(\mathbf{x}) |  \mathrm{c} , \bar{\mathrm{c}}\rangle
  = \bar{\mathrm{c}}(\mathbf{x}) |  \mathrm{c} , \bar{\mathrm{c}}\rangle \, .
\end{eqnarray}
and can be related to their Heisenberg picture operators, now using the Hamiltonian $H_{\mathrm{BRST}}$ corresponding to the BRST-invariant action (\ref{QEDactionBRST}) as usual {\it eg.}~$\hat{A}^\mu(x) = \hat{U}(t,\ti) \hat{A}^\mu_{\ssS}(\mathbf{x})  \hat{U}^{\dagger}(t,\ti)$, where we assume that the two pictures agree at the initial time $t_{\mathrm{i}}$. Note that, despite being fermionic, ghosts still admit a Schr\"odinger representation because their equations of motion are second order (see \cite{Kaplanek:2026kpp} for a detailed discussion). Moreover, in contrast to physical fermions, ghosts obey periodic boundary conditions \cite{Hata:1980yr}. This choice is required to maintain BRST symmetry, as ghosts mix with the gauge fields.
One then notes the standard path integral identity, but for the BRST theory
\begin{eqnarray} \label{pathintegral}
&& \langle \mathrm{A}_{\mathrm{f}} \hspace{0.4mm} \mathrm{c}_{\mathrm{f}}  \hspace{0.4mm}  \bar{\mathrm{c}}_{\mathrm{f}}  \hspace{0.4mm}  \Upphi_{\mathrm{f}}  \hspace{0.4mm}  \Upphi^{\ast}_{\mathrm{f}} | \hat{U}(t_{\mathrm{f}},t_{\mathrm{i}}) | \mathrm{A}_{\mathrm{i}}  \hspace{0.4mm}  \mathrm{c}_{\mathrm{i}}  \hspace{0.4mm}  \bar{\mathrm{c}}_{\mathrm{i}}  \hspace{0.4mm}  \Upphi_{\mathrm{i}} \hspace{0.4mm}  \Upphi^{\ast}_{\mathrm{i}} \rangle \\
&& \hspace{20mm} = \int_{\mathrm{A}_{\mathrm{i}} }^{\mathrm{A}_{\mathrm{f}} } \mathcal{D}[A] \int_{\mathrm{c}_{\mathrm{i}} }^{\mathrm{c}_{\mathrm{f}} } \mathcal{D}[c] \int_{\bar{\mathrm{c}}_{\mathrm{i}} }^{\bar{\mathrm{c}}_{\mathrm{f}} } \mathcal{D}[\bar{c}] \int_{\Upphi_{\mathrm{i}} }^{\Upphi_{\mathrm{f}} } \mathcal{D}[\Phi] \int_{\Upphi^{\ast}_{\mathrm{i}} }^{\Upphi^{\ast}_{\mathrm{f}} } \mathcal{D}[\Phi^{\ast}]  \; \mu \; e^{i S_{\mathrm{BRST}}[A, c, \bar{c}, \Phi, \Phi^{\ast}; t_{\mathrm{i}}, t_{\mathrm{f}}] }  \notag \, . 
\end{eqnarray}
The limits of the integrals serve to indicate that the path integral is taken over all field functions which begin and end at the initial and final time surface on the spatial field eigenstates.
The path integration gives rise to the transition amplitude  ending up in the field eigenstate $| \mathrm{A}_{\mathrm{f}} \hspace{0.4mm} \mathrm{c}_{\mathrm{f}}  \hspace{0.4mm}  \bar{\mathrm{c}}_{\mathrm{f}}  \hspace{0.4mm}  \Upphi_{\mathrm{f}}  \hspace{0.4mm}  \Upphi^{\ast}_{\mathrm{f}} \rangle$ at time $t_{\mathrm{f}}$ given that one started in $| \mathrm{A}_{\mathrm{i}}  \hspace{0.4mm}  \mathrm{c}_{\mathrm{i}}  \hspace{0.4mm}  \bar{\mathrm{c}}_{\mathrm{i}}  \hspace{0.4mm}  \Upphi_{\mathrm{i}} \hspace{0.4mm}  \Upphi^{\ast}_{\mathrm{i}} \rangle$ at time $t_\mathrm{i}$. Note that we have included a measure factor $\mu = \mu[A, c, \bar{c}, \Phi, \Phi^{\ast}]$ which is technically necessary when working with covariant path integrals. This is discussed in detail in Appendix~\ref{app:measure}. The role of the measure in the in-in formalism will become apparent later.

In standard quantum field theory, path integral identities such as the one above can be used to compute so-called in-out correlation functions. This is most straightforwardly implemented by projecting onto the BRST vacuum state $|\mathrm{vac}\rangle$ at past and future infinity, in practise by means of the $i \epsilon$  prescription, and defining the in-out generating functional (for arbitrary external sources $J_{\mathfrak{a}}$ with $\mathfrak{a}$ labelling each field)
\begin{small}
\begin{eqnarray} \label{Z_inout}
Z[ J_A, J_{\Phi}, J_{\Phi^{\ast}}, J_{c}, J_{\overline{c}} \; ] = \int_{\mathrm{vac}}^{\mathrm{vac}} \mathcal{D}[A,c,\bar{c},\Phi,\Phi^{\ast}] \; \mu \, e^{i S_{\mathrm{BRST}}[A, c, \bar{c}, \Phi, \Phi^{\ast}] + i \int \mathrm{d}^4 x \; \big( J_{A} \cdot A + J_{\Phi} \Phi + J_{\Phi^{\ast} }  \Phi^{\ast} + J_{c} c + J_{\bar{c}} \overline{c} \big) }  \, ,
\end{eqnarray}
\end{small}\ignorespaces
where the notation $\int_{\mathrm{vac}}^{\mathrm{vac}}$ indicates projection onto the in- and out-vacua at the path integral boundaries.
This functional usefully generates time-ordered vacuum correlation functions, for example, the Feynman propagator
\begin{eqnarray} \label{Z_Feynman}
(-i)^2 \frac{\delta^2 Z}{\delta J^\mu_A(x) \delta J^\nu_A(y) } \bigg|_{J_{\mathfrak{a}}=0} = \langle 0 | \mathcal{T}\{ \hat{A}_{\mu}(x) \hat{A}_{\nu}(y) \} | 0 \rangle \, .
\end{eqnarray}
Using this and higher-point time-ordered correlators, one can extract scattering amplitudes via the LSZ reduction formula.

\subsection{BRST in the In-In Formalism}
\label{sec:InIn}
 
There is a simple way to generalize the above construction to account for the time evolution of more complicated states, which in the most general case can be described by a density matrix $\rho(t)$ that begins in some initial state 
\begin{equation}
\rho(\ti) \equiv \rho_{\mathrm{i}} \ .
\end{equation}
This initial state may correspond to a pure state (such as the vacuum) or a more general mixed state, as long as it satisfies the basic properties of a density matrix: (i) $\rho^{\dagger} = \rho$, (ii) $\mathrm{Tr}[\rho] = 1$, and (iii) $\rho \geq 0$.

\begin{figure}[h]
\begin{center}
\includegraphics[width=155mm]{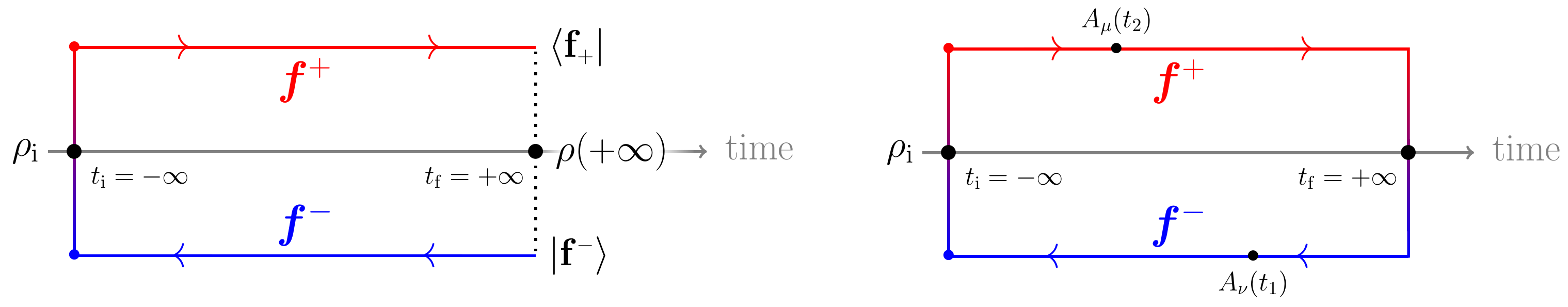}
\caption{\small The closed-time-path interpretation of the late-time density matrix components in Eq.~(\ref{rho_inin}) (left) and the in-in generating functional $Z_{\text{in-in}}$ in Eq.~(\ref{Zinin}) (right). In both cases there is a sum over the components of the initial state $\rho_{i}$ together with unitary time-evolution operators to and from the initial conditions. In the case of $Z_{\text{in-in}}$ there is also a sum over final states, producing a closed time path.} \label{Fig:TPvsCTP}
\end{center}
\end{figure}

The time evolution of the density matrix is determined by the unitary time-evolution operator,
\begin{equation}
\rho(t_{\mathrm{f}}) = \hat{U}(t_{\mathrm{f}}, \ti) \; \rho_{\mathrm{i}} \; \hat{U}^{\dagger}(t_{\mathrm{f}}, \ti) \ ,
\end{equation}
as dictated by the von Neumann equation for $\rho$. The evolution of the density matrix can be similarly expressed in terms of path integrals. The first step is to consider the matrix elements of the late-time density matrix and insert complete sets of states between the initial state and the time-evolution operators:
\begin{small}
\begin{eqnarray} \label{rhotf}
\langle \mathrm{A}_{+} \mathrm{c}_{+} \bar{\mathrm{c}}_{+} \Upphi_{+} \Upphi^{\ast}_{+} | \rho(t_{\mathrm{f}}) | \mathrm{A}_{-} \mathrm{c}_{-} \bar{\mathrm{c}}_{-} \Upphi_{-}  \Upphi^{\ast}_{-} \rangle & =: & \langle \mathbf{f}_{+} | \rho(t_{\mathrm{f}}) | \mathbf{f}_{-} \rangle \\ 
& = & \langle \mathbf{f}_{+} | \hat{U}(t_{\mathrm{f}}, \ti) \rho_{\mathrm{i}} \hat{U}^{\dagger}(t_{\mathrm{f}}, \ti) | \mathbf{f}_{-} \rangle  \\
 & = & \int \exd[\mathbf{f}_{+\mathrm{i}},\mathbf{f}_{-\mathrm{i}}] \; \langle \mathbf{f}_{+} | \hat{U}(t_{\mathrm{f}}, \ti)  | \mathbf{f}_{+\mathrm{i}} \rangle \langle   \mathbf{f}_{+\mathrm{i}} |  \rho_{\mathrm{i}}  | \mathbf{f}_{-\mathrm{i}} \rangle  \langle \mathbf{f}_{-\mathrm{i}} | \hat{U}^{\dagger}(t_{\mathrm{f}}, \ti) | \mathbf{f}_{-} \rangle \, , \qquad \quad
\end{eqnarray}
\end{small}\ignorespaces
where we have introduced the compact notation $\boldsymbol{f} = ( A, c, \overline{c}, \Phi, \Phi^{\ast})$ for the fields and $\mathbf{f}_j = (\mathrm{A}_j,\mathrm{c}_j,\overline{\mathrm{c}}_j, \Upphi_j, \Upphi_j^{\ast})$ for their corresponding eigenstates, to avoid clutter. Using Eq.~(\ref{pathintegral}), the above can be rewritten directly in terms of path integrals as
\begin{equation}  \label{rho_inin}
\langle \mathbf{f}_{+} | \rho(\tf) | \mathbf{f}_{-} \rangle = \int \exd[\mathbf{f}_{+\mathrm{i}},\mathbf{f}_{-\mathrm{i}}]\; \langle \mathbf{f}_{+\mathrm{i}} |  \rho_{\mathrm{i}}  | \mathbf{f}_{-\mathrm{i}} \rangle \int_{\mathbf{f}_{+\mathrm{i}}}^{\mathbf{f}_{+}} \mathcal{D}[\boldsymbol{f}_{+}]  \int_{\mathbf{f}_{-\mathrm{i}}}^{\mathbf{f}_{-}} \mathcal{D}[\boldsymbol{f}_{-}]\; \mu \, e^{i S_{\mathrm{BRST}}[\boldsymbol{f}_{+}] - i S_{\mathrm{BRST}}[\boldsymbol{f}_{-}]} \, .
\end{equation}
This expression shows that two copies of the action are required to fully describe the late-time density matrix, with two independent sets of field variables $\boldsymbol{f}_{\pm} = (A_{\pm},c_\pm,\overline{c}_\pm, \Phi_\pm, \Phi_\pm^{\ast})$. In addition, one must sum over the matrix elements of the initial density matrix in general. This picture can be interpreted in terms of a time path that is disconnected at the final time, as illustrated in the left panel of Fig.~\ref{Fig:TPvsCTP}.

Note that to describe expectation values of operators in a general initial state (and therefore be useful), it is natural to define the in-in generating functional,
\begin{eqnarray} 
Z_{\text{in-in}}[\boldsymbol{J}^{+}, \boldsymbol{J}^{-}] & = & \int \exd[\mathbf{f}, \mathbf{f}_{+\mathrm{i}},\mathbf{f}_{-\mathrm{i}}]\; \langle \mathbf{f}_{+\mathrm{i}} |  \rho_{\mathrm{i}}  | \mathbf{f}_{-\mathrm{i}} \rangle \label{Zinin} \\
&& \times \int_{\mathbf{f}_{+\mathrm{i}}}^{\mathbf{f}} \mathcal{D}[\boldsymbol{f}_{+}]  \int_{\mathbf{f}_{-\mathrm{i}}}^{\mathbf{f}} \mathcal{D}[\boldsymbol{f}_{-}]\; \mu \, e^{i S_{\mathrm{BRST}}[\boldsymbol{f}_{+}] - i S_{\mathrm{BRST}}[\boldsymbol{f}_{-}] +i \int \mathrm{d}^4 x \; ( \boldsymbol{J}^{+} \cdot \boldsymbol{f}_{+} - \boldsymbol{J}^{-} \cdot \boldsymbol{f}_{-} )}\, , \notag
\end{eqnarray}
where sources $\boldsymbol{J}^{\pm} = \{ J^{\pm}_{\mathfrak{a}} \} = (J^{\pm}_{A}, J_{c} , J^{\pm}_{\overline{c}}, J^{\pm}_{\Phi},  J^{\pm}_{\Phi^{\ast} } ) $ are introduced on both branches. Crucially, one sums not only over the initial eigenstates $\mathbf{f}_{\pm\mathrm{i}}$ weighted by the density matrix, but {\it also} over diagonal final field eigenstates $\mathbf{f} = \mathbf{f}_{+} = \mathbf{f}_{-}$. This enforces closure of the closed time-path (CTP) at late times, as illustrated in the right panel of Fig.~\ref{Fig:TPvsCTP}, and ensures that physical observables correspond to traces\footnote{Equivalently, the in-in generating functional may be expressed as a trace over quantum operators such that $Z_{\text{in-in}}[\boldsymbol{J}^{+}, \boldsymbol{J}^{-}]
= \mathrm{Tr}\left[ \mathscr{U}(\boldsymbol{J}^{+}) \rho_{\mathrm{i}} \mathscr{U}^{\dagger}(\boldsymbol{J}^{-}) \right]$ where $\mathscr{U}[\boldsymbol{J}]$ is the sourced time-evolution operator ({\it i.e.}~the usual unitary evolution modified by the source term $\boldsymbol{J}\cdot\boldsymbol{f}$). In the absence of sources this reduces to the ordinary operator $\mathscr{U}(\mathbf{0}) = U(\tf,\ti)$. This makes it clearest why $Z_{\text{in-in}}[\boldsymbol{J}, \boldsymbol{J}]  = 1$.}. Note that consistency with the normalisation of the density matrix requires $Z_{\text{in-in}}[\boldsymbol{J}, \boldsymbol{J}]  = 1$ which follows directly from $\mathrm{Tr}[\rho] = 1$.

This formalism is extremely useful because it allows one to compute correlation functions of Heisenberg-picture field operators for arbitrary initial states. For example, differentiating with respect to sources on different branches yields the Wightman function of the photon,
\begin{eqnarray} 
\frac{\delta^2 Z_{\text{in-in}}}{\delta J^{-\mu}_A(x) \delta J^{+\nu}_A(y) } \bigg|_{J^{\pm}_{\mathfrak{a}}=0} = \mathrm{Tr}\big( \hat{A}_{\mu}(x) \hat{A}_{\nu}(y) \rho_{\mathrm{i}} \big) \ ,
\end{eqnarray}
which involves no time ordering {\it cf.}~Eq.~(\ref{Z_Feynman}). By contrast, differentiating with respect to sources on the same branch produces time-ordered operators in the usual way. The key point is that the in-in contour generates correlation functions for arbitrary initial states, in contrast to the in-out formalism which assumes a specified final vacuum. This makes the in-in formalism essential in cosmology, where the final state is not known and the dynamics are intrinsically non-equilibrium.

Of course, the BRST symmetry of the theory must be manifest in all correlators. This takes some care to understand in the in-in formalism, since the observables are defined along a connected contour. If we ignore the final time boundary conditions, then the action is invariant under two separate BRST transformations, one for each branch
\begin{equation} \label{inin_BRSTtrans}
  \begin{split}
  A_{\pm\mu}(x) & \to A_{\pm \mu}(x) + \eta_{\pm} \hspace{0.3mm} \partial_\mu c_\pm(x) \, , \\
  \Phi_\pm(x) & \to \Phi_\pm(x) + i q \eta_{\pm} \hspace{0.3mm}  c_\pm(x) \Phi_\pm(x) \, ,
  \end{split}
\hspace{20mm}
  \begin{split}
    c_\pm(x) & \to c_\pm(x) \, , \\
    \overline{c}_\pm(x) & \to \overline{c}_\pm(x) - \tfrac{\eta_{\pm}}{\xi} \hspace{0.3mm}  \partial_\mu A_\pm^{\mu}(x) \, ,
  \end{split}
\end{equation}
with $\eta_{\pm}$ the global Grassmann parameter for each branch.
However, following the arguments given in the introduction, the in-in formalism requires that we identify the fields at the final time such that $\boldsymbol{f}_{\pm}(\tf, \mathbf{x}) = \mathbf{f}(\mathbf{x})$. 
This boundary condition is only preserved by the diagonal subgroup of BRST transformations for which $\eta_{\pm}=\eta$ is the same for both branches. 

We can summarize the diagonal (retarded) BRST transformations as $\mathbf{f} \rightarrow \mathbf{f}+\eta \hat s \mathbf{f}$ where:
\ba \label{inin_BRSTtrans2}
&&  \hat s A_{\pm\mu}(x) = \partial_\mu c_\pm(x) \, , \\
&&  \hat s \Phi_\pm(x)= i q  c_\pm(x) \Phi_\pm(x) \, , \\
&& \hat s    c_\pm(x) =0 \, ,\\
&& \hat s  \overline{c}_\pm(x) =- \tfrac{1}{\xi}   \partial_\mu A_\pm^{\mu}(x) \, .
\ea
To understand the implications of this diagonal BRST symmetry, it is useful to separate the action into two categories of terms. The first category includes the gauge-fixing and ghost contributions (BRST-exact):
\begin{eqnarray} 
S_{\mathrm{BRST}} & \supset & - \int \exd^4 x\; \bigg[ \frac{1}{2\xi} (\partial_{\mu} A^{\mu})^2 + \bar{c} \hspace{0.3mm} \Box c \bigg] \, .
\end{eqnarray}
These terms are constructed so that their BRST variations in Eq.~(\ref{inin_BRSTtrans}) automatically cancel. Their invariance is built in and does not depend on any other properties of the theory. The second category consists of the physical gauge-invariant terms, including the kinetic and interaction terms for the photon and matter fields (non-BRST exact):
\begin{eqnarray} 
S_{\mathrm{BRST}} & \supset & - \int \exd^4 x\; \bigg[ \frac{1}{4} F_{\mu\nu} F^{\mu\nu} + ( D_{\mu} \Phi )^{\ast} D^\mu \Phi + V( \Phi^{\ast} \Phi) \bigg] \ .
\end{eqnarray}
Unlike the gauge-fixing and ghost terms, whose invariance is guaranteed by design, the physical contributions are invariant under BRST only because they are built from gauge-invariant combinations of the fields.

\subsubsection{Keldysh (r/a) basis}
\label{sec:rabasis}

Following Keldysh, it turns out to be particularly useful to utilize the retarded and advanced fields defined via
\begin{eqnarray}
\boldsymbol{f}_{\mathrm{r}} = \frac{ \boldsymbol{f}_{+} + \boldsymbol{f}_{-} }{2} \qquad \mathrm{and}\, , \qquad \boldsymbol{f}_{\mathrm{a}} = \boldsymbol{f}_{+} - \boldsymbol{f}_{-} \, ,\label{rabasis_def}
\end{eqnarray}
and one defines analogously the sources $\boldsymbol{J}^{\mathrm{r}} =  \frac{ \boldsymbol{J}^{+} + \boldsymbol{J}^{-} }{2}$  and $\boldsymbol{J}^{\mathrm{a}} = \boldsymbol{J}^{+} - \boldsymbol{J}^{-}$. These variables rewrite the in-in generating functional (\ref{Zinin}) as
\begin{eqnarray}  
Z_{\text{in-in}}[\boldsymbol{J}^{\mathrm{r}}, \boldsymbol{J}^{\mathrm{a}}] & = & \int \exd[\mathbf{f}, \mathbf{f}_{+\mathrm{i}},\mathbf{f}_{-\mathrm{i}}]\; \langle \mathbf{f}_{+\mathrm{i}} |  \rho_{\mathrm{i}}  | \mathbf{f}_{-\mathrm{i}} \rangle \label{Zinin_ra} \\
&& \times \int_{ ( \mathbf{f}_{+\mathrm{i}} + \mathbf{f}_{-\mathrm{i}} ) / 2 }^{\mathbf{f}} \mathcal{D}[\boldsymbol{f}_{\mathrm{r}}]  \int_{\mathbf{f}_{+\mathrm{i}} - \mathbf{f}_{-\mathrm{i}}}^{\mathbf{0}} \mathcal{D}[\boldsymbol{f}_{\mathrm{a}}]\; \mu \, e^{i S_{\mathrm{BRST}}[\boldsymbol{f}_{\mathrm{r}} + \frac{\boldsymbol{f}_{\mathrm{a}}}{2} ] - i S_{\mathrm{BRST}}[\boldsymbol{f}_{\mathrm{r}} - \frac{\boldsymbol{f}_{\mathrm{a}}}{2}] 
+i \int \mathrm{d}^4 x \; ( \boldsymbol{J}^{\mathrm{r}} \cdot \boldsymbol{f}_{\mathrm{a}} + \boldsymbol{J}^{\mathrm{a}} \cdot \boldsymbol{f}_{\mathrm{r}} )} \ . \notag
\end{eqnarray}
This form is generally much easier to work with in practice. The reason is clearest when considering perturbations around a quadratic theory. Expanding to leading order in advanced fields, one finds
\begin{eqnarray}
S_{\mathrm{BRST}}[\boldsymbol{f}_{\mathrm{r}} + \tfrac{\boldsymbol{f}_{\mathrm{a}}}{2} ]  - S_{\mathrm{BRST}}[\boldsymbol{f}_{\mathrm{r}} - \tfrac{\boldsymbol{f}_{\mathrm{a}}}{2} ] \simeq \int \exd^4 x  \; \bigg[ \boldsymbol{f}_{\mathrm{a}} \cdot \frac{\delta S[\boldsymbol{f}_\mathrm{r}]}{\delta \boldsymbol{f}_\mathrm{r}} + (\text{cubic in }\boldsymbol{f}_{\mathrm{a}}) \bigg] \, ,
\end{eqnarray}
where higher-order terms always appear in odd powers of $\boldsymbol{f}_{\mathrm{a}}$, starting at cubic order. For a quadratic theory, truncating these higher-order terms implies that the advanced field $\boldsymbol{f}_{\mathrm{a}}$ has no kinetic term and therefore, does not propagate. Instead, it appears linearly, dotted with the equation of motion for the retarded field, and effectively acts as a Lagrange multiplier enforcing the retarded field's dynamics. At the end, evaluating the path integral in this formulation naturally produces propagators with a clear causal structure. For instance, one finds
\begin{eqnarray} 
- \frac{\delta^2 Z_{\text{in-in}}}{\delta J^{\mathrm{a}\hspace{0.2mm}\mu}_A(x) \delta J^{\mathrm{r}\hspace{0.2mm}\nu}_A(y) } \bigg|_{J^{\pm}_{\mathfrak{a}}=0} = \theta(x^{0} - y^0)  \mathrm{Tr}\big( [  \hat{A}_{\mu}(x) , \hat{A}_{\nu}(y) ] \rho_{\mathrm{i}} \big) \ ,
\end{eqnarray}
which is the retarded propagator for the photon. This makes the retarded/advanced basis extremely convenient for calculations. However, the boundary conditions for these variables are subtle and require careful handling to avoid confusion. There is extensive discussion of these variables and their use in calculations in non-equilibrium field theory texts ({\it eg.}~as in \cite{kamenev2023field}). 

The same principle discussed earlier for the $\pm$ fields carries over directly to the retarded/advanced basis. For the gauge field, the BRST transformations take the form
\begin{equation}
 A_{\mathrm{r}\mu} \to  A_{\mathrm{r}\mu} +  \eta \; \partial_{\mu} c_{\mathrm{r}}\, , \qquad \mathrm{and} \qquad A_{\mathrm{a}\mu} \to A_{\mathrm{a}\mu} + \eta  \; \partial_{\mu} c_{\mathrm{a}}\, ,
\end{equation}
emphasizing that the retarded and advanced fields must transform in the obvious way. It is crucial that this transformation is not deformed: any modification of the advanced-field BRST transformation would violate invariance of the in-in action and disrupt the identification of physical observables (including fundamental relations like the Ward identities). The advanced field acts as a linear Lagrange multiplier that enforces the retarded dynamics, but its BRST transformation must remain consistent with the diagonal global symmetry, exactly as in the $\pm$ variables.

\subsection{$i \epsilon$ prescription in non-gauge theories}
\label{sec:iep_nongauge}

The boundary conditions in the in-in formalism as stated so far are painful to implement in practise, as it is necessary to first track the evolution from an initial time $\ti$ to some finite time $\tf$ in the future, ensure matching between the two branches, and in addition to impose initial conditions at time $\ti$ associated with the specific initial state.  Even for a free theory, these steps are cumbersome.

Since the final time is arbitrary, we may always send $\tf \rightarrow \infty$. In many physical applications, we may in effect assume $\ti =-\infty$ and doing so simplifies calculations as in the $S$-matrix in-out formalism.
Fortunately, as in the in-out formalism, there is a simple way to bypass this procedure of imposing initial and final boundary conditions by using an $i \epsilon$ prescription which is designed to pick out the correct limit. 

To see how this works, let us first consider the example of a real massive scalar field $\phi(x)$ in Minkowski spacetime. Given the two branches of the CTP, the free Feynman propagator in a general mixed state can be packaged into a $2\times 2$ matrix of propagators
\be
{\bf G_0} = 
\begin{pmatrix}
G_{++} & G_{+-} \\
G_{-+} & G_{--}
\end{pmatrix} \, ,
\ee
which in explicit operator language are
\ba
&& G_{++ }(x,y) = \Tr[\rho {\cal T}\hat \phi(x) \hat \phi(y)] \, ,\\
&& G_{+- }(x,y) = \Tr[\rho \hat \phi(y) \hat \phi(x)] \, ,\\
&& G_{-+ }(x,y) = \Tr[\rho \hat \phi(x) \hat \phi(y)] \, ,\\
&& G_{-- }(x,y) = \Tr[\rho \bar {\cal T}\hat \phi(x) \hat \phi(y)]\, ,
\ea
with $\hat \phi(x)$ the free field operator
\be
\hat \phi(x) =\int \frac{\d^4 k}{(2 \pi)^4} \; \theta(k^0) 2 \pi \delta(k^2+m^2) \( e^{ik.x} \hat a_{{\bf k}}+e^{-ik.x} \hat a^{\dagger}_{{\bf k}} \) \, ,
\ee
and $\theta(k^0)$ is the Heaviside step function.
If the state $\rho$ is chosen to preserve translation invariance then we may write
\be
{\bf G_0}(x,y)=\int \frac{\d^4 k}{(2 \pi)^4}\,  e^{ik.(x-y)}{\bf G_0}(k) \, ,
\ee
and a straightforward calculation shows that
\be
{\bf G_0}(k) = 
\begin{pmatrix}
-\frac{i}{k^2+m^2-i \epsilon}+2 \pi \delta(k^2+m^2) \mathfrak{n}(k) & \theta(-k^0) 2 \pi \delta(k^2+m^2) +2 \pi \delta(k^2+m^2) \mathfrak{n}(k)\\
\theta(k^0) 2 \pi \delta(k^2+m^2) +2 \pi \delta(k^2+m^2) \mathfrak{n}(k)& \frac{i}{k^2+m^2+i \epsilon}+2 \pi \delta(k^2+m^2) \mathfrak{n}(k) 
\end{pmatrix}\, , \label{matrixexcitedstate}
\ee
with $\mathfrak{n}(k)=\theta(k^0)n({\bf k})+\theta(-k^0)n({-\bf k})$, and 
\be
n({\bf k}) 2 \omega_k (2 \pi)^3 \delta^3({\bf k} - {\bf k}') = \Tr(\rho \hat a^{\dagger}_{{\bf k}} \hat a_{{\bf k}'} ) \, ,
\ee
encoding the spectrum of excited particles in the state $\rho$ (where $\omega_{k} = \sqrt{\mathbf{k}^2 + m^2}$). 
Here the form of the $i \epsilon$ can be inferred from the time ordering and anti-time ordering operations.

Anticipating the discussion on the path integral measure in Appendix~\ref{app:measure} we will also here define the CTP analogue of the advanced propagator via
\be \label{advancedpropagator}
{\bf G}_{0 A}={\bf G}_0-\begin{pmatrix}
G_{-+} & G_{-+} \\
G_{-+} & G_{-+}
\end{pmatrix}=\begin{pmatrix}
G_{A} & -\Delta \\
0 & G_{R} 
\end{pmatrix}\, ,
\ee
where $G_R$ is the usual free theory retarded Green's function, $G_A$ the advanced and $\Delta$ is the free theory commutator $\Delta = G_{-+}-G_{+-}=G_R-G_A$. 
In our conventions $G_R(x,y)=\theta(x^0-y^0) \Delta(x,y)$, $G_A(x,y)=-\theta(y^0-x^0) \Delta(x,y)$.
In momentum space this is
\be
{\bf G}_{0 A}(k)=\begin{pmatrix}
-\frac{i}{k^2+m^2+i \epsilon \sigma(k^0)} & -\sigma(k^0) 2 \pi \delta(k^2+m^2) \\
 0 & \frac{i}{k^2+m^2-i \epsilon \sigma(k^0)} 
\end{pmatrix} \, ,
\ee
with $\sigma(k^0)=\theta(k^0)-\theta(-k^0)$
which illustrates that the advanced propagator is independent of the state.
Now given the definition of the delta function implicit in the Sokhotski-Plemelj relations
\be \label{SP_relations}
\frac{1}{x\mp i \epsilon}= P\(\frac{1}{x} \)\pm i \pi \delta(x) \, ,
\ee
we infer the identity $(x-i \epsilon) (x+i \epsilon) \pi \delta(x) = \epsilon$ and so
\be
(k^2+m^2-i \epsilon) (k^2+m^2+i \epsilon) \pi \delta(k^2+m^2)= \epsilon \, .
\ee
The determinant of the momentum space propagator matrix is independent of the choice of state 
\be
\det[{\bf G_0}(k)]=\frac{1}{(k^2+m^2-i \epsilon) (k^2+m^2+i \epsilon) } \, ,
\ee
and so we infer that the inverse momentum space propagator is 
\be
{\bf G}_{\bf{0}}^{-1}(k) = 
\begin{pmatrix}
i(k^2+m^2-i \epsilon)+2 \epsilon \mathfrak{n}(k) & -\theta(-k^0) 2 \epsilon  -2 \epsilon \mathfrak{n}(k)\\
-\theta(k^0) 2 \epsilon  -2 \epsilon \mathfrak{n}(k)& -i(k^2+m^2+i \epsilon)+2 \epsilon \mathfrak{n}(k)
\end{pmatrix} \, .
\ee
We also note in passing that
\be \label{detrelation}
\det[{\bf G_0}] =\det[{\bf G_{0 A}}] \, ,
\ee
a result that is relevant in the discussion of the path integral measure in Appendix~\ref{app:measure}.
The in-in action for the free field is then
\ba
S_{\text{in-in}}&=& \frac{i}{2}\int \frac{\d^4 k}{(2 \pi)^4} \, \begin{pmatrix} \phi_+(-k) & \phi_-(-k)\end{pmatrix} {\bf G}_{\bf{0}}^{-1}(k)\begin{pmatrix} \phi_+(k) \\ \phi_-(k)\end{pmatrix} \\
&=& \int \d^4 x \, \( -\frac{1}{2}(\partial \phi_+)^2-\frac{1}{2}m^2 \phi_+^2+\frac{1}{2}(\partial \phi_-)^2+\frac{1}{2}m^2 \phi_-^2\)  +S_{\rm i \epsilon} \, ,
\ea
where we have separated out the $i \epsilon$ terms which are best written in Keldysh basis
\be
S_{\rm i \epsilon}=\int \frac{\d^4 k}{(2 \pi)^4} \; i \epsilon \[ \phi_{\mathrm{a}}(-k) \frac{k^0}{|k^0|} \phi_{\mathrm{r}}(k)+\(\frac{1}{2}+ \mathfrak{n}({\bf k})\) |\phi_{\mathrm{a}}(k)|^2 \] \, ,
\ee
having used
\be
\theta(k^0)= \frac{1}{2}+\frac{1}{2} \frac{k^0}{|k^0|} \, .
\ee
To put this in its best form in position space, it is helpful to rescale $\epsilon \rightarrow 2|k^0| \epsilon$, and then the full in-in action for a free field is
\be
S_{\text{in-in}} =\int \d^4 x  \;  \phi_{\mathrm{a}}(x)[\Box-m^2-2\epsilon \partial_t] \phi_{\mathrm{r}}(x) +i \epsilon \int \d^4 x\int \d^4 y\; \phi_{\mathrm{a}}(x) K_{\phi}(x,y) \phi_{\mathrm{a}}(y) \, ,
\ee
or equivalently 
\ba
S_{\rm i \epsilon} & = & -\epsilon \int \d^4 x    \; \big(\phi_+(x)-\phi_-(x)\big) \partial_t \big( \phi_+(x)+\phi_-(x) \big) \\
&&  +i \epsilon \int \d^4 x\int \d^4 y\; \big( \phi_+(x)-\phi_-(x)\big) K_{\phi}(x,y) \big(\phi_+(y)-\phi_-(y)\big) \notag
\ea
with
\be
K_{\phi}(x,y) = \int \frac{\d^4 k}{(2 \pi)^4} e^{ik.(x-y)}\(1+ 2\mathfrak{n}(k)\)\omega_k \, .
\ee
Here we have replaced $|k^0|$ with $\omega_k = \sqrt{{\bf k}^2+m^2}$ which is allowed since the $i \epsilon$ terms only affect the on-shell parts of the propagator.
The first term in $S_{\rm i \epsilon}$ enforces the necessary $i \epsilon$ terms to determine the retarded and advanced propagators. The second term encodes information about the (generally mixed) initial state, contained in the Keldysh propagator. For a thermal state we have for example,
\be
K_{\phi}^{\beta}(x,y) = \int \frac{\d^4 k}{(2 \pi)^4} e^{ik.(x-y)} \coth(\beta \omega_k/2)\omega_k \, .
\ee

The virtue of the $i \epsilon$ formalism is that it allows us to encode the state directly into the action without having to worry about the boundary conditions at $\ti$ and $\tf$. Specifically we may now declare that
\ba
&& \lim_{\substack{\ti \rightarrow -\infty \\ \tf \rightarrow +\infty}} \int \d [\upphi,\upphi_{+\mathrm{i}},\upphi_{-\mathrm{i}}] \int_{ \upphi_{+\mathrm{i}}}^{ \upphi} \mathcal{D}[\phi_+] \int_{\upphi_{-\mathrm{i}}}^{\upphi} \mathcal{D}[\phi_-] \; \langle \upphi_{+\mathrm{i}} | {\cal \rho} | \upphi_{-\mathrm{i}} \rangle \mu e^{i S_{\text{in-in}}[\phi_+,\phi_-;\ti,\tf]} \notag\\
&& \qquad \equiv \; \int \mathcal{D}[\phi_+] \int \mathcal{D}[\phi_-] \, \mu \, e^{i S_{\text{in-in}}[\phi_+,\phi_-]+iS_{i \epsilon}} \, ,
\ea
where $|\upphi\rangle$ are field eigenstates of the field operator in the sense of Eq.~(\ref{eigenstatesAPhicc}) and where the RHS path integral is computed assuming no boundary terms so that we may freely integrate by parts.
This works because the $i \epsilon$ terms by construction will reproduce the correct propagators for the free field. As long as the initial state is Gaussian, the only content of the choice of the state in the interacting theory is a modification of the free propagators, all vertices are left unchanged.

\subsection{$i \epsilon$ prescription in charged matter in gauge theories}

Let us now generalize the previous discussion to gauge theories. Suppose we are now interested in applying this procedure to a massive complex scalar $\Phi$ charged under a $U(1)$ gauge field. 
The free field operator is now
\be
\hat \Phi(x) =\int \frac{\d^4 k}{(2 \pi)^4} \; \theta(k^0) 2 \pi \delta(k^2+m^2) \( e^{ik.x} \hat a_{{\bf k}}+e^{-ik.x} \hat b^{\dagger}_{{\bf k}} \) \, ,
\ee
and so assuming no symmetry breaking, for a translation invariant state we have two different distributions
\ba
&& n_+({\bf k}) 2 \omega_k (2 \pi)^3 \delta^3({\bf k} - {\bf k}') = \Tr(\rho \hat a^{\dagger}_{{\bf k}} \hat a_{{\bf k}'} ) \, ,  \\
&& n_-({\bf k}) 2 \omega_k (2 \pi)^3 \delta^3({\bf k} - {\bf k}') = \Tr(\rho \hat b^{\dagger}_{{\bf k}} \hat b_{{\bf k}'} ) \,,
\ea
with $\Tr(\rho \hat a_{{\bf k}} \hat a_{{\bf k}'} )=\Tr(\rho \hat b_{{\bf k}} \hat b_{{\bf k}'} )=\Tr(\rho \hat a_{{\bf k}} \hat b^{\dagger}_{{\bf k}'} )= 0$. 
We define the analogous propagators by
\ba
&& G_{++ }(x,y) = \Tr[\rho {\cal T}\hat \Phi(x) \hat \Phi^{\dagger}(y)] \, ,\\
&& G_{+- }(x,y) = \Tr[\rho \hat \Phi^{\dagger}(y) \hat \Phi(x)] \, ,\\
&& G_{-+ }(x,y) = \Tr[\rho \hat \Phi(x) \hat \Phi^{\dagger}(y)] \, ,\\
&& G_{-- }(x,y) = \Tr[\rho \bar {\cal T}\hat \Phi(x) \hat \Phi^{\dagger}(y)]\, .
\ea
with combinatons of the form $\Tr[\rho \hat \Phi(y) \hat \Phi(x)]$ vanishing since we here assume no SSB. In momentum space the matrix of propagators ${\bf G_0}(k)$ is then precisely of the form of Eq.~(\ref{matrixexcitedstate}) but with
\be
\mathfrak{n}(k) = \theta(k^0) n_+({\bf k}) +\theta(-k^0) n_-(-{\bf k}) \, .
\ee
A naive application of the above procedure would lead to a free theory action
\ba
S_{\text{in-in}} &=& \int \d^4 x \; \bigg( \Phi^*_{\mathrm{a}}(x)[\Box-m^2-2\epsilon \partial_t]  \Phi_{\mathrm{r}}(x) +\Phi^*_{\mathrm{r}}(x)[\Box-m^2+2\epsilon \partial_t] \Phi_{\mathrm{a}}(x) \bigg) \\
&\ &+2 i \epsilon \int \d^4 x\int \d^4 y \; \Phi^*_{\mathrm{a}}(x) K_{\Phi}(x,y) \Phi_{\mathrm{a}}(y) \notag \, ,
\ea
with
\be
\label{eq:KPhi}
K_{\Phi}(x,y) = \int \frac{\d^4 k}{(2 \pi)^4} \, e^{ik.(x-y)}\big(1+ 2\mathfrak{n}(k)\big)\omega_k \, .
\ee
This is correct in the global limit for which the gauge fields are set to zero. One might think that the gauge case could be obtained simply by replacing ordinary derivatives by covariant derivatives.
The problem is that the $i \epsilon$ terms are not gauge invariant because this action couples fields on two different branches. The problem can be seen with a simple coupling of the form
\be \label{eq:examplecoupling}
\Phi_-^*(x) \Phi_+(x) \, . 
\ee
Despite the common label $x$ these two fields are not at the same spacetime point on the CTP. The consequence of this is that under a gauge transformation this combination transforms as
\be
\Phi_-^*(x) \Phi_+(x) \rightarrow e^{i q \lambda_+(x)-i q \lambda_-(x)} \Phi_-^*(x) \Phi_+(x)\, . 
\ee
This in turn means that under a diagonal BRST transformation 
\be
\hat s\(\Phi_-^*(x) \Phi_+(x)\) =i q  (c_{+}(x)-c_-(x))  \Phi_-^*(x) \Phi_+(x) 
\ee
which means that these interactions are not BRST invariant. Similarly one can show that even if we replace ordinary derivatives by covariant derivatives in the above expression
\be
\hat s S_{\rm i \epsilon} \neq 0 \, ,
\ee
for the same reason.
This is of course a familiar problem in gauge theories. For example the two point function of a complex scalar is itself not gauge invariant, and this complicates the discussion of composite operators. A standard solution, used for example in computing expectation values of currents via point splitting, is to insert a Wilson line to render objects gauge invariant. 

In the present context, we have two gauge symmetries which may be decomposed into an advanced and retarded one $\lambda_{\pm}=\lambda_r\pm \frac{1}{2} \lambda_\mathrm{a}$ with $A^{\pm}_{\mu}=A^{\mathrm{r}}_{\mu}\pm\frac{1}{2} A^{\mathrm{a}}_{\mu}$ for which
\be
A^{\mathrm{r},\mathrm{a}}_{\mu} \rightarrow A^{\mathrm{r},\mathrm{a}}_{\mu} + \partial_{\mu} \lambda_{\mathrm{r},\mathrm{a}}(x) \, .
\ee
In the example \eqref{eq:examplecoupling} it is only the advanced gauge symmetry that is being broken by coupling the two branches. The retarded gauge field is the one that evolves causally, and to deal with its gauge invariance for general interactions it is sufficient to replace ordinary derivatives by covariant derivatives with respect to the retarded gauge field. By contrast, the advanced gauge field evolves in a time reversed manner from the final time with the boundary condition $A_{\mu}^\mathrm{a}(\tf)=0$. Because of this it is natural to insert a Wilson line which extends from the final time to the point of field insertions (see Appendix \ref{app:Wilson} to see why this is inevitable). Specifically we may perform a field redefinition
\be
\Phi_{\pm}(x) = e^{\mp\frac{1}{2}i q\int_{x}^{x_\mathrm{f}} \d z^{\mu} A^{\mathrm{a}}_{\mu}(z) }\widetilde \Phi_{\pm}(x) =e^{\mp\frac{1}{2}i q (x_\mathrm{f}^{\mu}-x^{\mu}) \int_0^1 \d s\, A^{\mathrm{a}}_{\mu}(x+s(x_\mathrm{f}-x))}\widetilde \Phi_{\pm}(x) \, ,
\ee
where the Wilson line ends at the final time surface $x_\mathrm{f}^0=\tf$ so that the in-in boundary condition is preserved
\be
\Phi_+(\tf) =  \Phi_-(\tf) \quad \implies \quad \widetilde \Phi_+(\tf) = \widetilde \Phi_-(\tf) \, .
\ee
Under a retarded (or diagonal) gauge transformation it turns out $\widetilde \Phi_{\pm}$ transforms covariantly such that
\be
\widetilde \Phi_\pm(x) \rightarrow e^{i q \lambda_\mathrm{r}(x)}\widetilde \Phi_\pm(x) \, ,
\ee
since the original fields transform as $\Phi_\pm(x) \rightarrow e^{i q \lambda_\mathrm{r}(x)} \Phi_\pm(x)$ in this case. Under an advanced gauge transformation one finds that $\widetilde \Phi_{\pm}$ is invariant since
\be
\widetilde \Phi_\pm(x) \rightarrow e^{ \pm i\frac{1}{2} q \lambda_\mathrm{a}(\tf)}\widetilde \Phi_\pm(x)=\widetilde \Phi_\pm(x) \, ,
\ee
given than the gauge transformations must respect the in-in boundary condition $\lambda_\mathrm{a}(\tf)=\lambda_+(\tf)-\lambda_-(\tf)=0$, and since the original fields transform as $\Phi_\pm(x) \rightarrow e^{\pm i \frac{1}{2} q \lambda_\mathrm{a}(x)} \Phi_\pm(x)$. Stated differently, under a diagonal BRST transformation 
\ba
&& \hat s A_{\mu}^\mathrm{r} = \partial_{\mu} c_\mathrm{r}(x) \ , \qquad \hat s \widetilde \Phi_{\pm}(x) = i q c_{\mathrm{r}}(x) \widetilde \Phi_{\pm}(x) \ , \qquad \hat s \widetilde \Phi^*_{\pm}(x) = -i q c_{\mathrm{r}}(x) \widetilde \Phi^*_{\pm}(x) \, ,
\ea
because of the final time boundary condition $c_\mathrm{a}(\tf)=0$. These are just the standard BRST transformations for the retarded gauge transformations and so it follows that any expression built out of $\widetilde \Phi_{\pm}$ which is invariant under the retarded gauge symmetry alone is BRST invariant. For example $\widetilde \Phi^*_-(x) \widetilde \Phi_+(x) $ is BRST invariant even though $\Phi^*_-(x)  \Phi_+(x) $ is not.

Hence, to guarantee that the state specified via the $i \epsilon$ prescription is BRST invariant, which is a necessary requirement for any physical state, we can simply use the naive expression and replace $\Phi_{\pm}$ with $\widetilde \Phi_{\pm}$ and ordinary derivatives by covariant ones with respect to the retarded gauge field, {\it eg.}~for a complex scalar in a Gaussian mixed state we could choose 
\be
S_{i \epsilon}=- 2\epsilon \int \d^4 x \; \(  \widetilde \Phi^*_\mathrm{a}(x) D_t[A_\mathrm{r}]  \widetilde \Phi_\mathrm{r}(x) -\tilde \Phi^*_\mathrm{r}(x) D_t[A_r]  \widetilde \Phi_\mathrm{a}(x)\) +2i \epsilon \int \d^4 x\int \d^4 y \; \widetilde \Phi^*_\mathrm{a}(x) K_{\widetilde \Phi}(x,y) \widetilde \Phi_\mathrm{a}(y) \, ,
\ee
with $K_{\tilde \Phi}(x,y)$ transforming covariantly under retarded gauge transformations
\be
K_{\widetilde \Phi}(x,y)  \rightarrow e^{-i q \lambda_\mathrm{r}(x)+i q \lambda_\mathrm{r}(y)} K_{\widetilde \Phi}(x,y) \, .
\ee
An alternative and in practice more useful procedure is to also insert a Wilson line for the retarded gauge symmetry but to connect it to the initial time $\ti$ surface. Performing a further field redefinition
\be
\widetilde \Phi_{\pm}(x) = e^{i q\int_{x_\mathrm{i}}^{x} \d z^{\mu}\, A^{\mathrm{r}}_{\mu}(z) }\check \Phi_{\pm}(x) =e^{i q (x^{\mu}-x_\mathrm{i}^{\mu}) \int_0^1 \d s \, A^{\mathrm{r}}_{\mu}(x_\mathrm{i}+s(x-x_\mathrm{i}))}\check \Phi_{\pm}(x) \, .
\ee
Now $\check \Phi_{\pm}$ only transforms under retarded gauge transformations evaluated at $x_\mathrm{i}$
\be
\check \Phi_\pm(x) \rightarrow e^{i q \lambda_\mathrm{r}(x_\mathrm{i})}\check \Phi_\pm(x) \, ,
\ee
so that we may choose
\be
S_{i \epsilon}=- 2\epsilon \int \d^4 x  \(  \check \Phi^*_\mathrm{a}(x) \partial_t \check \Phi_\mathrm{r}(x) -\check \Phi^*_\mathrm{r}(x) \partial_t  \check \Phi_\mathrm{a}(x)\) +2 i \epsilon \int \d^4 x\int \d^4 y \; \check \Phi^*_\mathrm{a}(x) K_{\Phi}(x,y) \check \Phi_\mathrm{a}(y) \, ,
\ee
with $K_{\Phi}$ given in \eqref{eq:KPhi}. Since the field redefinition takes the form of a gauge transformation, we can choose to reabsorb it in a redefinition of $A_{\mu}^{\mathrm{r},\mathrm{a}}(x) \rightarrow \widetilde A_{\mu}^{\mathrm{r},\mathrm{a}}$.
Collectively what this amounts to is working in Fock-Schwinger gauge for the advanced fields defined at the point $x_\mathrm{f}$ and Fock-Schwinger gauge for the retarded fields defined at the point $x_{\mathrm{i}}$.
With this prescription we clearly have BRST invariance
\be
\hat s S_{i \epsilon} =0 \, ,
\ee
and causality is maintained in that the Fock-Schwinger gauge retarded (advanced) gauge fields only depend on the past (future)
\ba
&& \widetilde A_{\mu}^\mathrm{r}(x) = A_{\mu}^\mathrm{r}(x) - \partial_{\mu} \( (x^{\nu}-x_\mathrm{i}^{\nu}) \int_0^1 \d s\, A^{\mathrm{r}}_{\nu}(x_\mathrm{i}+s(x-x_\mathrm{i}))\) \, , \\
&& \widetilde A_{\mu}^\mathrm{a}(x) = A^\mathrm{a}_{\mu}(x) + \partial_{\mu} \( (x_\mathrm{f}^{\nu}-x^{\nu}) \int_0^1 \d s\, A^{\mathrm{a}}_{\nu}(x+s(x_\mathrm{f}-x))\) \, .
\ea
Furthermore, in the decoupling limit $q \rightarrow 0$ this reduces to the obvious $i \epsilon$ prescription for a complex scalar with a global $U(1)$ symmetry. A similar procedure may be applied to any charged matter, {\it eg.}~a Dirac spinor.

\subsection{$i \epsilon$ prescription for the photon}

Finally we must deal with the $i \epsilon$ prescription for the photon itself. In the Lorentz gauge with $\xi =1$ (Feynman-'t Hooft gauge), each component of the gauge field $A_{\mu}$ behaves as a scalar field with the only caveat that $A_0$ has a wrong sign kinetic term. It is clear that we need to introduce $i \epsilon$ terms similar to the scalar case to describe a generic photon state, but we also need to introduce similar terms for the FPDW ghosts in order to preserve BRST invariance. This is a well known issue in finite temperature QCD where it matters since the ghosts interact and it is important to provide a prescription for their thermal propagator \cite{Landsman:1986uw,Kobes:1984vb,Kobes:1985kc,Kobes:1986za,Hata:1980yr,Ojima:1981ma}.

By analogy with the scalar case, we can fix the form of the $i\epsilon$ terms largely by symmetry. Specifically if we make the ansatz
\ba
\label{eq:photoniepsilon}
S_{i \epsilon} &=& - 2\epsilon \int \d^4 x \, \(  A_{\mathrm{a}\mu}(x) \partial_t   A^{\mu}_{\mathrm{r}}(x) -\bar c_{\mathrm{a}}(x) \partial_t c_{\mathrm{r}}(x)+\bar c_{\mathrm{r}}(x) \partial_t  c_{\mathrm{a}}(x)\)  \nn \\
&& +i \epsilon \int \d^4 x\int \d^4 y\, \big( A_{\mathrm{a} \mu}(x) K_A^{\mu\nu}(x,y) A_{\mathrm{a}\nu}(y) + c_{\mathrm{a}}(y) K_c(x,y) \bar c_{\mathrm{a}}(x) \big) \, .
\ea
We then have
\ba
\hat s S_{i \epsilon}&=&- 2\epsilon \int \d^4 x \, \big(  \partial_{\mu}c_{\mathrm{a}}(x) \partial_t   A^{\mu}_{\mathrm{r}}(x)+ A_{\mathrm{a}\mu}(x) \partial_t \partial^{\mu} c_\mathrm{r}(x)+ \partial_{\mu} A^{\mu}_\mathrm{a}(x) \partial_t c_\mathrm{r}(x)-\partial_{\mu} A^{\mu}_\mathrm{r}(x) \partial_t  c_\mathrm{a}(x) \big)  \nn \\
&& + i \epsilon \int \d^4 x\int \d^4 y\, \big( 2 \partial_{\mu} c_\mathrm{a}(x) K_A^{\mu\nu}(x,y) A_{\nu \mathrm{a}}(y) +  c_\mathrm{a}(y) K_c(x,y)  \partial_{\mu} A^{\mu}_\mathrm{a}(x) \big)\; . \qquad 
\ea
Now freely integrating by parts we have
\be
\hat s S_{i \epsilon}=- 2i \epsilon \int \d^4 x\int \d^4 y \, c_\mathrm{a}(x) \( 2 \partial_{x\mu} K_A^{\mu\nu}(x,y) + \partial_x^{\nu} K_c(x,y) \) A_{\nu \mathrm{a}}(y)\, 
\ee
and hence this is BRST invariant $\hat s S_{i \epsilon}=0$ provided only that
\be
 2 \partial_{x\mu} K_A^{\mu\nu}(x,y) + \partial_x^{\nu} K_c(x,y)  =0 \, ,
\ee
or in Fourier space
\be \label{BRSTcond1}
 2 k_{\mu} K_A^{\mu\nu}(k)+k^{\nu} K_c(k)=0\, .
\ee
Because physical states are defined only up to exact BRST forms there is an inevitable ambiguity in the precise specification of the state. 
A natural physical choice is the Landshoff-Rebhan prescription \cite{Landshoff:1992ne} which in effect states that the ghosts being unphysical should be kept in vacuum, and transverse polarisations should be excited. This corresponds to the choice (in Fourier space)
\be
K_A^{\mu \nu} (k)=\eta^{\mu\nu} + 2 T^{\mu\alpha}(k) \mathfrak{n}_{\alpha \beta}(k) T^{\beta\nu}(k) \,  , \quad K_c(k) = -2 \, ,
\ee
where we have defined the projection operator onto physical states with $k^2=0$
\be
T^{\mu\nu}(k)=\eta^{\mu\nu}-\frac{k^{\mu} u^{\nu} +u^{\mu} k^{\nu}}{u.k}+u^2 \frac{k^{\mu}k^{\nu}}{(u.k)^2} \, ,
\ee
satisfying $u_{\mu} T^{\mu\nu}(k)=0$ as well as for $k^2=0$
\be
k_{\mu} T^{\mu\nu}=0  \quad \mathrm{and} \quad T^{\mu\alpha}T_{\ \alpha}^{\nu}=T^{\mu\nu} \, ,
\ee
with $u$ an arbitrary reference timelike vector. Note that with this choice \eqref{BRSTcond1} only holds on-shell $k^2=0$ but this is sufficient since the $i \epsilon$ terms only affect the on-shell structure.

An alternative possibility which is common in the finite temperature QCD literature \cite{Kobes:1984vb,Kobes:1986za,Kobes:1985kc,Hata:1980yr,Ojima:1981ma} is to give the ghosts the same spectrum as the gauge fields to maximize symmetry of the propagators. For example, a thermal state can be described democratically with the choice
\be
K_A^{\mu \nu} (k) = \eta^{\mu\nu}(1+2 \mathfrak{n}_{\beta}(k)) \, ,\quad K_c(k)=-2 (1+ 2 \mathfrak{n}_{\beta}(k)) \, ,
\ee
with
\be
\mathfrak{n}_{\beta}(k) = \frac{1}{e^{\beta \omega_k}-1} \, .
\ee 
The price of this elegant choice, is the need to excite unphysical degrees of freedom.

It is worth noting that in the language of the in-in formalism, the $i \epsilon$ terms enter in two ways. The first are local, linear in first time derivatives and act like dissipation, {\it i.e.}~they take the form of a friction term in the equations of motion. The second are in general non-local and couple the two advanced fields, and act like noise in an effective Langevin equation for the retarded fields. When we account for open system effects obtained from integrating out other fields, we inevitably generate finite contributions to both of these terms, which in effect swamp the original $i \epsilon$ terms. However, we shall see that their form is covariant in the manner outlined above.

\subsection{Influence functional}
\label{sec:IF_def}

With the machinery of the in-in formalism in place, we can now define the influence functional and extend the above arguments to it. As in any open-systems calculation, one must first decide what constitutes the system and what constitutes the environment. When there is no symmetry breaking we take the charged matter to be the environment, and split the BRST action (\ref{QEDactionBRST}) as
\begin{equation}
S_{\mathrm{BRST}}[A,c, \overline{c}, \Phi, \Phi^{\ast}] = S_{\mathcal{S}}[A,c,\overline{c}] + S_{\mathcal{E}}[\Phi, \Phi^{\ast}] + S_{\mathrm{int}}[A,\Phi, \Phi^{\ast}] \, .
\end{equation}
The key point is that the system variables $A, c, \overline{c}$ are separated from the environment variables $\Phi, \Phi^{\ast}$, and the two sectors communicate only through $S_{\mathrm{int}}[A,\Phi,\Phi^{\ast}]$, which we perturb in. In an Abelian theory, like QED, the interaction term does not involve the ghosts $c$ and $\overline{c}$. Later sections present explicit examples of this partition, but for now we keep the definition schematic, emphasizing only that the environment consists of the charged matter degrees of freedom.

Once the partition between system and environment is defined, one can define the reduced density matrix $\varrho_{\mathrm{red}}$ as the partial trace of the full density matrix over the environment degrees of freedom:
\begin{equation}
\varrho_{\mathcal{S}} \equiv \TrE \big[ \rho \big] \, .
\end{equation}
Here, the partial trace over the environment corresponds to tracing over all charged matter field eigenstates. In terms of matrix components, this reads at the final time $t_{\mathrm{f}}$:
\begin{equation} \label{ParTr}
\langle \mathrm{A}_{+} \mathrm{c}_{+} \bar{\mathrm{c}}_{+} | \varrho_{\mathcal{S}}(t_{\mathrm{f}})  | \mathrm{A}_{-} \mathrm{c}_{-} \bar{\mathrm{c}}_{-} \rangle = \int \exd[\Upphi,\Upphi^{\ast}] \; \langle \mathrm{A}_{+} \mathrm{c}_{+} \bar{\mathrm{c}}_{+} \Upphi \Upphi^{\ast} | \rho(t_{\mathrm{f}}) | \mathrm{A}_{-} \mathrm{c}_{-} \bar{\mathrm{c}}_{-} \Upphi  \Upphi^{\ast} \rangle \, .
\end{equation}
This expresses the reduced density matrix by tracing over the charged matter states in the matrix components of Eq.~(\ref{rhotf}). The goal is then to use the in-in formalism results from \S\ref{sec:InIn} to compute the components of $\varrho_{\mathrm{red}}$.
One additional simplifying assumption we make is that the initial state $\rho_{\mathrm{i}}$ at time $\ti$ factorizes across system and environment,
\begin{equation} \label{uncorr}
\rho_{\mathrm{i}} = \varrho_{\mathcal{S}\mathrm{i}} \otimes \varrho_{\mathcal{E}\mathrm{i}} \, ,
\end{equation}
or in components
\begin{small}
\begin{equation}
 \langle \mathrm{A}_{+\mathrm{i}} \mathrm{c}_{+\mathrm{i}} \bar{\mathrm{c}}_{+\mathrm{i}} \Upphi_{+\mathrm{i}} \Upphi_{+\mathrm{i}}^{\ast} | \rho_{\mathrm{i}} | \mathrm{A}_{-\mathrm{i}} \mathrm{c}_{-\mathrm{i}} \bar{\mathrm{c}}_{-\mathrm{i}} \Upphi_{-\mathrm{i}} \Upphi_{-\mathrm{i}}^{\ast} \rangle =  \langle \mathrm{A}_{+\mathrm{i}} \mathrm{c}_{+\mathrm{i}} \bar{\mathrm{c}}_{+\mathrm{i}} | \varrho_{\mathcal{S}\mathrm{i}} | \mathrm{A}_{-\mathrm{i}} \mathrm{c}_{-\mathrm{i}} \bar{\mathrm{c}}_{-\mathrm{i}} \rangle  \cdot \langle \Upphi_{+\mathrm{i}} \Upphi_{+\mathrm{i}}^{\ast} | \varrho_{\mathcal{E}\mathrm{i}} | \Upphi_{-\mathrm{i}} \Upphi_{-\mathrm{i}}^{\ast} \rangle \ . \
\end{equation}
\end{small}\ignorespaces
That is, the initial state is uncorrelated between system and environment. In field theories, this is typically a benign assumption: vacuum states and most states of interest naturally factorise in this way, and it holds in all examples considered in this work (see also \cite{Grabert:1988yt,  colla2022initial}). Furthermore, any non-zero entanglement could be regarded as being generated by interactions.

This factorisation allows us to use the in-in formula (\ref{rho_inin}) to express the components of the reduced density matrix as
\begin{small}
\begin{eqnarray} 
&& \langle \mathrm{A}_{+} \mathrm{c}_{+} \bar{\mathrm{c}}_{+} | \varrho_{\mathcal{S}}(+\infty) | \mathrm{A}_{-} \mathrm{c}_{-} \bar{\mathrm{c}}_{-} \rangle \notag \\
&& \qquad = \int \exd[\mathrm{A}_{\pm\mathrm{i}} , \mathrm{c}_{\pm\mathrm{i}} , \bar{\mathrm{c}}_{\pm \mathrm{i}} ]\; \langle \mathrm{A}_{+\mathrm{i}} \mathrm{c}_{+\mathrm{i}} \bar{\mathrm{c}}_{+\mathrm{i}} | \varrho_{\mathcal{S}\mathrm{i}} | \mathrm{A}_{-\mathrm{i}} \mathrm{c}_{-\mathrm{i}} \bar{\mathrm{c}}_{-\mathrm{i}} \rangle \\
&& \qquad \ \ \times \int_{ \mathrm{A}_{+\mathrm{i}} \mathrm{c}_{+\mathrm{i}} \bar{\mathrm{c}}_{+\mathrm{i}} }^{ \mathrm{A}_{+} \mathrm{c}_{+} \bar{\mathrm{c}}_{+} } \mathcal{D}[A_{+},c_{+}, \overline{c}_{+}]  \int_{ \mathrm{A}_{-\mathrm{i}} \mathrm{c}_{-\mathrm{i}} \bar{\mathrm{c}}_{-\mathrm{i}} }^{ \mathrm{A}_{-} \mathrm{c}_{-} \bar{\mathrm{c}}_{-} } \mathcal{D}[A_{-},c_{-}, \overline{c}_{-}] \; \tilde \mu \, e^{i S_{\mathcal{S}}[ A_{+},c_{+},\overline{c}_{+} ] - i S_{\mathcal{S}}[A_{-},c_{-},\overline{c}_{-}] + i S_{\mathrm{IF}}[A_{+} , A_{-} ] } \, , \notag
\end{eqnarray}
\end{small}\ignorespaces
where we define the {\it influence functional} $S_{\mathrm{IF}}$ by
\begin{small}
\begin{eqnarray} \label{SIF_def}
&& \tilde \mu \, e^{ iS_{\mathrm{IF}}[A_{+}, A_{-} ] } \notag \\
&& \qquad = \int \exd[\Upphi, \Upphi^{\ast}, \Upphi_{\pm\mathrm{i}} ,  \Upphi^{\ast}_{\pm \mathrm{i}} ]\; \langle \Upphi_{+\mathrm{i}}\Upphi^{\ast}_{+\mathrm{i}} | \varrho_{\mathcal{E}\mathrm{i}} | \Upphi_{-\mathrm{i}}\Upphi^{\ast}_{-\mathrm{i}}   \rangle \\
&& \qquad \ \ \times \int_{ \Upphi_{+\mathrm{i}} \Upphi^{\ast}_{+\mathrm{i}} }^{ \Upphi \Upphi^{\ast} } \mathcal{D}[\Phi_{+} ,  \Phi_{+}^{\ast} ]  \int_{ \Upphi_{-\mathrm{i}} \Upphi^{\ast}_{-\mathrm{i}} }^{ \Upphi \Upphi^{\ast} } \mathcal{D}[\Phi_{-} ,  \Phi_{-}^{\ast} ]\; \mu \, e^{i S_{\mathcal{E}}[ \Phi_{+},\Phi^{\ast}_{+} ] + i S_{\mathrm{int}}[  A_{+}, \Phi_{+},\Phi^{\ast}_{+} ] - i S_{\mathcal{E}}[ \Phi_{-},\Phi^{\ast}_{-} ] - i S_{\mathrm{int}}[ A_{-}, \Phi_{-},\Phi^{\ast}_{-} ]  } \, . \notag
\end{eqnarray}
\end{small}\ignorespaces
Crucially, the final time boundary conditions in (\ref{SIF_def}) must be identified with each other to implement the partial trace over the environment from Eq.~(\ref{ParTr}). This identification preserves the doubled BRST symmetry described earlier, ensuring that the influence functional respects the same diagonal subgroup structure as the full in-in action. Note that we have included a distinct measure $\tilde \mu$ which is a function of the open system fields. This is the part of the original measure $\mu$ which survives on tracing out the charged matter. See Appendix~\ref{app:measure} for more discussion on this subtle point.

Some comments are in order. First, integrating out the charged matter in Eq.~(\ref{SIF_def}) generates an influence functional in which each branch of the gauge field interacts not only with itself but also, in general, with the opposite branch. As we will see in later examples, these inter-branch interactions are typically highly non-local. The notable exception occurs in special limits, such as the standard decoupling limit in effective field theory, where the integrated-out fields become very heavy and the $+$ and $-$ branches decouple.

Second, the reduced density matrix itself is an interesting object to study. This means that the influence functional can be used to compute quantities such as the R\'enyi or von Neumann entropy, providing a measure of the entanglement between the system and the environment. 

Third, as the theory is Abelian, the ghosts do not interact with the physical fields and, therefore, do not enter the influence functional whatsoever. In a non-Abelian theory this would not be the case.

Finally, the influence functional also encodes all the information needed to compute system correlators. Analogously to the in-in generating functional introduced earlier, one can define the system generating functional $Z_{\mathcal{S}}$ as
\begin{small}
\begin{eqnarray}
&& Z_{\mathcal{S}}[J^{+}_{A},  J^{+}_{c} , J^{+}_{\bar{c}}, J^{-}_{A},  J^{-}_{c} , J^{-}_{\bar{c}} ]  \\
&& \quad = \int \exd[\mathrm{A} , \mathrm{c} , \bar{\mathrm{c}} , \mathrm{A}_{\pm\mathrm{i}} , \mathrm{c}_{\pm\mathrm{i}} , \bar{\mathrm{c}}_{\pm \mathrm{i}} ]\; \langle \mathrm{A}_{+\mathrm{i}} \mathrm{c}_{+\mathrm{i}} \bar{\mathrm{c}}_{+\mathrm{i}} | \varrho_{\mathcal{S}\mathrm{i}} | \mathrm{A}_{-\mathrm{i}} \mathrm{c}_{-\mathrm{i}} \bar{\mathrm{c}}_{-\mathrm{i}} \rangle  \int_{ \mathrm{A}_{+\mathrm{i}} \mathrm{c}_{+\mathrm{i}} \bar{\mathrm{c}}_{+\mathrm{i}} }^{ \mathrm{A} \mathrm{c} \bar{\mathrm{c}} } \mathcal{D}[A_{+},c_{+}, \overline{c}_{+}]  \int_{ \mathrm{A}_{-\mathrm{i}} \mathrm{c}_{-\mathrm{i}} \bar{\mathrm{c}}_{-\mathrm{i}} }^{ \mathrm{A} \mathrm{c} \bar{\mathrm{c}} } \mathcal{D}[A_{-},c_{-}, \overline{c}_{-}] \notag \\
&& \quad \ \times \; \tilde \mu \, e^{i S_{\mathcal{S}}[ A_{+},c_{+},\overline{c}_{+} ] - i S_{\mathcal{S}}[A_{-},c_{-},\overline{c}_{-}] + i S_{\mathrm{IF}}[A_{+},c_{+},\overline{c}_{+} , A_{-}, c_{-}, \overline{c}_{-} ] +i \int \mathrm{d}^4 x \; ( J_{A}^{+} \cdot A_+ + J_{c}^{+} c_{+} + J^{+}_{\bar{c}} \overline{c}_{+} - J_{A}^{-} \cdot A_- - J_{c}^{-} c_{-} - J^{-}_{\bar{c}} \overline{c}_{-} )  } \notag \, .
\end{eqnarray}
\end{small}\ignorespaces
This functional also connects the final eigenstates ({\it c.f.}~Eq.~(\ref{Zinin})) and generates all system correlators for the same reason as $Z_{\text{in-in}}$. In fact, it is straightforward to see that these objects are directly related by
\begin{eqnarray}
Z_{\mathcal{S}}[J^{+}_{A},  J^{+}_{c} , J^{+}_{\bar{c}}, J^{-}_{A},  J^{-}_{c} , J^{-}_{\bar{c}} ] = Z_{\text{in-in}}[J^{+}_{A},  J^{+}_{c} , J^{+}_{\bar{c}}, 0 , 0, J^{-}_{A},  J^{-}_{c} , J^{-}_{\bar{c}}, 0 , 0 ] \, ,
\end{eqnarray}
{\it i.e.}~$Z_{\text{in-in}}$ with the environmental sources set to zero exactly gives $Z_{\mathcal{S}}$. This makes it explicit that the influence functional is in a sense only ``half'' a calculation: it provides the intermediate step needed to arrive at the generating functional for system correlators.

\subsection{BRST invariance of Influence Functional}

We now come to the central point which is relevant for bottom up constructions. Although in the path integral formalism we have fixed a particular gauge on both branches, the in-in action for the closed/UV system is BRST invariant under the diagonal subgroup for which both branches transform with the same global Grassmann parameter. The influence functional arises by tracing or integrating out subsets of degrees of freedom that may be in an arbitrary excited state. Nevertheless, as long as the state is BRST invariant, the influence functional is by itself BRST invariant, {\it i.e.}
\be
\hat s S_{\rm IF}[A_+,A_-, \dots] =0 \, ,
\ee
where $\dots$ indicates other degrees of freedom which may not have been traced over. The argument for this is easiest to state when the in-in boundary conditions are encoded in $i \epsilon$ terms. We have already discussed how BRST invariant states can be encoded in BRST invariant $i \epsilon$ terms. We can split the BRST transformations from Eq.~(\ref{inin_BRSTtrans2}) up as
\be
\hat s = \hat s_{\cal S}+  \hat s_{\cal E} \, ,
\ee
where $\hat s_{\cal S}$ are the BRST transformations of the open system fields, and $\hat s_{\cal E}$ those transformations for the environment fields, for example
\begin{equation}
    \hat s_{\cal E} A_{\pm \mu}(x) = \hat s_{\cal E} c_{\pm}(x)  = \hat s_{\cal E} \overline{c}_{\pm}(x)  = 0 \, , \qquad \mathrm{and} \qquad 
    \hat s_{\cal E} \Phi_{\pm }(x) = i q c_{\pm}(x) \Phi_{\pm }(x) \, ,
\end{equation}
and similarly for $\hat{s}_{\mathcal{S}}$. 
Schematically we have\footnote{We will drop explicit mention of the measure, but it should be included - specifically the $\tilde \mu$ and $\mu$ measure should also be BRST invariant in the sense $\hat s_{\cal S} \tilde \mu=0$ and $\hat s \mu=0$.}
\be
e^{i S_{\mathcal{S}}+i S_{\rm IF}[A_+,A_-, \dots]} = \int {\cal D}[\Phi_+ , \Phi^*_+,\Phi_- , \Phi^*_-] \; e^{i S_{\mathrm{BRST}}+ i S_{i \epsilon} } \, ,
\ee
so that (given $\hat s_{\cal S} S_{\mathcal{S}}=0$)
\be
  \hat s_{\cal S} S_{\rm IF}[A_+,A_-, \dots] = e^{-i S_{\mathcal{S}}-i S_{\rm IF}[A_+,A_-, \dots]} \int {\cal D}[\Phi_+ , \Phi^*_+,\Phi_- , \Phi^*_-] \; \( \hat s_{\cal S} ( S_{\mathrm{BRST}}+  S_{i \epsilon} )\) \, e^{i S_{\mathrm{BRST}}+ i S_{i \epsilon} } \, . 
\ee
By performing a field redefinition on the environment fields of the form $\Phi \rightarrow \Phi+ \eta \hat s_{\cal E} \Phi$ we infer the Schwinger-Dyson equations
\be
\int {\cal D}[\Phi_+ , \Phi^*_+,\Phi_- , \Phi^*_-] \; \( \hat s_{\cal E} (S_{\mathrm{BRST}} +  S_{i \epsilon}) \) \; e^{i S_{\mathrm{BRST}}+ i S_{i \epsilon} } =0 \, , 
\ee
which only uses invariance of the measure under the field redefinitions. Combining the previous two relations, we have 
\be
  \hat s_{\cal S} S_{\rm IF}[A_+,A_-, \dots] = e^{-i S_{\mathcal{S}}-i S_{\rm IF}[A_+,A_-, \dots]} \int {\cal D}[\Phi_+ , \Phi^*_+,\Phi_- , \Phi^*_-] \(\hat s (S_{\mathrm{BRST}}+  S_{i \epsilon}) \) e^{i S_{\mathrm{BRST}}+ i S_{i \epsilon} } \, ,
\ee
and hence
\be
\hat s \(S_{\mathrm{BRST}}+  S_{i \epsilon} \) =0 \qquad \Rightarrow \qquad  \hat s_{\cal S} S_{\rm IF}[A_+,A_-, \dots]\equiv\hat s S_{\rm IF}[A_+,A_-, \dots] =0 \, .
\ee
It is important to stress that the choice of state, and specifically whether or not there is SSB, does not affect this statement since all physical states, be they pure or mixed, are BRST invariant.

In the simple situation in which only the photon remains in the influence functional, we note that it is independent of ghosts due to the Abelian nature of the theory and so:
\be
\hat s S_{\rm IF}[A_+,A_-]= \int \d^4 x \[ \partial_{\mu} c_+ \frac{\delta S_{\rm IF}}{\delta A_{+\mu}} +\partial_{\mu} c_- \frac{\delta S_{\rm IF}}{\delta A_{-\mu}}\] = -\int \d^4 x \[  c_+ \partial_{\mu}\frac{\delta S_{\rm IF}}{\delta A_{+\mu}}+ c_- \partial_{\mu}\frac{\delta S_{\rm IF}}{\delta A_{-\mu}}\] =0 \, .\ee
This condition must be true for any configuration of ghosts and hence BRST invariance can only be achieved when
\be
\partial_{\mu}\frac{\delta S_{\rm IF}}{\delta A_{+\mu}}=0\, ,  \qquad \mathrm{and} \qquad  \partial_{\mu}\frac{\delta S_{\rm IF}}{\delta A_{-\mu}}=0 \, ,
\ee
which is the statement that the influence functional is gauge invariant under two independent copies of gauge transformations, one on each branch. It is remarkable that only the diagonal BRST symmetry alone is sufficient to ensure this, but it is the case because the two distinct FPDW ghosts encode local information on the two separate gauge transformations.

In situations where the gauge symmetry is spontaneously broken, it is natural to keep the \stu field (Goldstones in the global limit)  $\chi_{\pm}$ as part of the open system. The \stu fields will by construction have a simple linear gauge transformation
\be \label{chi_gtrans}
\chi_{\pm} \rightarrow \chi_{\pm} +\alpha \lambda_{\pm}(x) \, ,
\ee
for some constant $\alpha$ determined by its normalisation. Hence their BRST transformations are
\be
\hat s \chi_{\pm}= \alpha c_{\pm}(x) \, .
\ee
Thus repeating the above argument we have 
\ba
\hat s S_{\rm IF}[A_+,A_-,\chi_+,\chi_-]&=& \int \d^4 x \[ \partial_{\mu} c_+ \frac{\delta S_{\rm IF}}{\delta A_{+\mu}} +\partial_{\mu} c_- \frac{\delta S_{\rm IF}}{\delta A_{-\mu}}+ \alpha \chi_+\frac{\delta S_{\rm IF}}{\delta \chi_{+}}+\alpha \chi_-\frac{\delta S_{\rm IF}}{\delta \chi_{-}} \] \\
&=& -\int \d^4 x \[  c_+ \( \partial_{\mu} \frac{\delta S_{\rm IF}}{\delta A_{+\mu}}-\alpha \frac{\delta S_{\rm IF}}{\delta \chi_{+}}\)+ c_- \(\partial_{\mu}\frac{\delta S_{\rm IF}}{\delta A_{-\mu}}-\alpha\frac{\delta S_{\rm IF}}{\delta \chi_{-}} \)\] =0 \, , \qquad \quad 
\ea
and so 
\be
\partial_{\mu} \frac{\delta S_{\rm IF}}{\delta A_{+\mu}}-\alpha \frac{\delta S_{\rm IF}}{\delta \chi_{+}}=0 \qquad \mathrm{and} \qquad \partial_{\mu}\frac{\delta S_{\rm IF}}{\delta A_{-\mu}}-\alpha\frac{\delta S_{\rm IF}}{\delta \chi_{-}} =0 \, ,
\ee
which is again just a statement that the influence functional is invariant under two standard copies of gauge invariance.

Later we work out explicit examples of $S_{\mathrm{IF}}$ in spinor QED/scalar QED/Abelian Higgs-Kibble, both in the broken and unbroken phases. These examples demonstrate concretely that gauge invariance remains intact for the influence functional, upheld by the same diagonal BRST structure that governs the full in-in formalism.

\subsection{Decoupling Limit and Global Symmetries}
\label{GlobalLimit}

Global symmetries are of course a special case of local ones, and so it is useful to understand how the more familiar open EFT formalism for global symmetries arises from their local counterpart. Since the local one is governed entirely by the diagonal BRST symmetry, it must be the case that there is a limit of the BRST transformation that encodes the necessary global symmetry. This is the decoupling limit $q \rightarrow 0$ for which the photon decouples from charged matter. 

Consider the gauge symmetry present before gauge fixing
\be
A'_{\mu}(x)  =A_{\mu}(x)+ \partial_{\mu}  \lambda(x) \, , \quad \Phi'(x) = e^{i q \lambda(x)} \Phi(x)  \, .
\ee
To take the decoupling limit and preserve the global symmetry we decompose the gauge transformation into a constant global part $\theta$ and a local part $\xi(x)$ in the manner
\be
\lambda(x) = \frac{1}{q} \theta + \xi(x) \, .
\ee
In the limit $q \rightarrow 0$, the original local symmetry separates into a decoupled global transformation and a local transformation 
\be
A'_{\mu}(x)  =A_{\mu}(x)+ \partial_{\mu}  \xi(x) \, , \quad \Phi'(x) = e^{i \theta} \Phi(x)  \, .
\ee
In this way we see that any operator invariant under gauge transformations reduces in the decoupling limit to an operator invariant under the global symmetry, and if it depends on the gauge field will be separately gauge invariant without needing to transform the charged fields.

Now let us see how this same idea is realised after gauge fixing on the diagonal BRST transformation from  Eq.~(\ref{inin_BRSTtrans2}). Since the FPDW ghosts follow the form of the gauge transformations, it is natural to split the $c$-ghosts fields up into a constant vev and a fluctuation
\be
c_{\pm}(x) =\frac{1}{q}c_0+C_{\pm}(x) \, .
\ee
The constant ghost vev has no physical meaning, {\it i.e.} it does not correspond to a physical state, but it is a consistent solution and is here being used only to probe the theory.
The in-in boundary conditions require the same vev on each branch such that $c_+(\tf)=c_-(\tf)$ implies $C_+(\tf)=C_-(\tf)$.
With this decomposition in the decoupling limit the BRST transformation becomes
\be
\hat{s} A_{\pm\mu} = \partial_{\mu} C_{\pm}\ , \, \qquad \hat{s} C_{\pm}=0\ , \qquad \hat{s} \overline{c}_{\pm} = - \tfrac{1}{\xi}  \partial_{\mu} A_{\pm}^{\mu} \ , \qquad  \hat{s} \Phi^{\pm} = i c_0 \Phi^{\pm} \, .
\ee
The action of the diagonal BRST transformation on the matter field reduces in the limit to a diagonal global $U(1)$ transformation with Grassmann even parameter $\theta^{+}=\theta^-=\eta c_0$. The fact that this Grassmann parameter has zero body ({\it i.e.}~is not a classical number) does not matter, the requirement of invariance of the path integral under any constant Grassman even parameter leads to the same consequences, and the point of most relevance is that we only recover the diagonal global symmetry.

Once again, the diagonal BRST transformation implies in the decoupling limit the diagonal global transformation alone, despite it ensuring two copies of gauge symmetry away from the decoupling limit. Hence, the construction of an open in-in local gauge theory is consistent with the construction of its global counterpart.

\subsection{Influence Functional as a Generating Functional}
\label{sec:IFisGF}

In situations where there is spontaneous symmetry breaking, there will be a pair of fields $\chi_{\pm}$ that act as \stu fields which parameterize the departure from unitary gauge on each branch and reduce in the global limit to the Goldstone modes. It is natural then not to integrate out these fields. The environment sector will then be built out of unbroken charged matter $\varphi$ and Higgs-like fields $H$, and so we define an influence functional $\widetilde{S}_{\mathrm{IF}}$ for the photon and \stu fields as
\begin{small}
\ba \label{SIF_def2}
&& \widetilde{\mu} e^{ i \widetilde{S}_{\mathrm{IF}}[A_{+}, A_{-} ,\chi_+,\chi_-] } = \int \exd[\mathrm{H},\upvarphi, \upvarphi^{\ast}, \mathrm{H}_{\pm \mathrm{i}},\upvarphi_{\pm\mathrm{i}} ,  \upvarphi^{\ast}_{\pm \mathrm{i}} ]\; \langle \mathrm{H}_{+\mathrm{i}}\upvarphi_{+\mathrm{i}}\upvarphi^{\ast}_{+\mathrm{i}} | \varrho_{\mathcal{E}\mathrm{i}} | \mathrm{H}_{-i}\upvarphi_{-\mathrm{i}}\upvarphi^{\ast}_{-\mathrm{i}} \rangle  \int_{\mathrm{H}_{+\mathrm{i}}\upvarphi_{+\mathrm{i}} \upvarphi^{\ast}_{+\mathrm{i}} }^{ \mathrm{H} \upvarphi \upvarphi^{\ast} } \mathcal{D}[H_+,\varphi_{+} ,  \varphi_{+}^{\ast} ] \nn \\
&& \int_{ \mathrm{H}_{-\mathrm{i}} \upvarphi_{-\mathrm{i}} \upvarphi^{\ast}_{-\mathrm{i}} }^{ \mathrm{H} \upvarphi \upvarphi^{\ast} } \mathcal{D}[H_-,\varphi_{-} ,  \varphi_{-}^{\ast} ]  \mu \, e^{i S_{\mathcal{E}}[ H_+,\varphi_{+},\varphi^{\ast}_{+} ] + i S_{\mathrm{int}}[ H_+, A_{+}, \varphi_{+},\varphi^{\ast}_{+} ] - i S_{\mathcal{E}}[ H_-,\varphi_{-},\varphi^{\ast}_{-} ] - i S_{\mathrm{int}}[ H_-,A_{-}, \varphi_{-},\varphi^{\ast}_{-} ]  } \ .
\ea
\end{small}\ignorespaces
The previously defined influence functional in which all matter is integrated out can then be constructing by performing the path integral over the \stu fields, which will lead to a typically highly non-local result
\ba
\check \mu[A_{+}, A_{-} ] \,  e^{ iS_{\mathrm{IF}}[A_{+}, A_{-} ] } &= & \int \d[ \upchi, \upchi_{\pm\mathrm{i}}] \; \langle \upchi_{+\mathrm{i}} | \varrho_{\chi\mathrm{i}} | \upchi_{-\mathrm{i}} \rangle \nn \\
&& \int_{\upchi_{+\mathrm{i}}}^{\upchi} {\cal D}[\chi_+] \int_{\upchi_{-\mathrm{i}}}^{\upchi} {\cal D}[\chi_-] \;   \widetilde \mu[A_{+}, A_{-} ,\chi_+,\chi_-] \,
e^{ i \widetilde{S}_{\mathrm{IF}}[A_{+}, A_{-} ,\chi_+,\chi_-] } 
\ea
for some choice of initial state $\varrho_{\chi\mathrm{i}}$ for the \stu field. In computing these expressions, we regard the gauge fields as fixed background values. The effective action that enters the exponential does not exhibit any gauge or BRST symmetry with $A_{\pm}$ fixed, but it does still exhibit a global symmetry modulo the contribution from the initial state. Hence we may regard this as the in-in generating functional for a theory with a global symmetry which is spontaneously broken. This is an object that has been considered in recent works on effective field theories of dissipative fluids \cite{Crossley:2015evo,Liu:2018kfw}. For example, in the notation of \cite{Crossley:2015evo} (see Eq.~(1.5) there) we identify
\ba 
iS_{\mathrm{IF}}[A_{+}, A_{-} ] \equiv W[A_1,A_2]\, , \\
i \widetilde{S}_{\mathrm{IF}}[A_{+}, A_{-} ,\chi_+,\chi_-] \equiv i I[B_1,B_2]  \,  ,\label{SIF_to_I}
\ea
where $B_{\pm} = A_{\pm} - \alpha^{-1}\partial \chi_{\pm}$ for some normalisation constant $\alpha$. 
This interpretation gives an alternative understanding of why the influence functional has two copies of gauge invariance, as it is a consequence of Noether's theorem for said global symmetry as discussed in Appendix~\ref{App:Noether}. This fact is used as a key ingredient in bottom-up constructions of EFTs for dissipative fluids \cite{Crossley:2015evo,Liu:2018kfw} (see also \cite{Baggioli:2023tlc}).

\subsection{Apparent dissipation}
\label{sec:fakedissipation}

Before proceeding, we clarify a confusion that has arisen in the literature regarding the rules for constructing bottom-up open EFTs. As we have outlined, any open EFT for a gauge system must be invariant under the diagonal BRST symmetry, and in the Abelian case that guarantees two copies of gauge invariance, both retarded and advanced, for the resulting influence functional. In \cite{Salcedo:2024nex} it is argued that the naive advanced gauge transformation is violated, and a modified dissipation dependent transformation arises at least at quadratic order. This claim is clearly in contradiction with our discussion. Since the underlying gauge symmetries are determined by the original UV/closed theory, their action on the gauge fields cannot be modified by integrating/tracing out degrees of freedom. 

To understand the resolution, let us first consider the example of the equation of motion of a massive scalar field with a friction term
\be
\label{examplediss1}
\ddot \phi+ \gamma \dot \phi - \nabla^2 \phi + m^2 \phi =0 \, .
\ee
 If we define energy in the conventional sense for a scalar $E=\frac{1}{2}\int \d^3 \mathbf{x} \( \dot \phi^2+(\nabla \phi)^2+m^2 \phi^2 \) $ then this clearly dissipates energy
 \be
\frac{\d E}{\d t } = - \gamma \int \d^3 \mathbf{x}\,  \dot \phi^2 <0\, .
 \ee
 For this reason, it is usually stated that this equation does not derive from an action. However, it is easy to write an action that does lead to this equation of motion, provided that we make it time-dependent, specifically
 \be
S = \int \d^4 x \, e^{\gamma t } \(-\frac{1}{2} (\partial \phi)^2 -\frac{1}{2}m^2 \phi^2 \) \, .
 \ee
 This is a closed system and so does not exhibit dissipation in the usual sense of losing energy to another degree of freedom/system. Rather, energy is lost because it was never conserved, the Hamiltonian being time-dependent. If our goal is to describe a dissipative scalar in Minkowski, this is clearly the wrong action and should not be used as a basis for quantisation. However, if we are dealing with a genuinely time-dependent closed system, this may be the correct action. A concrete example of this type is a scalar field in FRW whose action is time-dependent
 \be
S = \int \d^4 x \, a^3(t) \(\frac{1}{2}\dot \phi^2-\frac{1}{2 a^2(t)} (\nabla \phi)^2 -\frac{1}{2}m^2 \phi^2 \) \, ,
 \ee
 whose equation of motion includes the Hubble damping term, which mimics dissipation
 \be
\ddot \phi +\frac{3\dot{a}(t)}{a(t)} \dot \phi -\frac{1}{a^2(t)} \nabla^2 \phi+m^2 \phi=0 \, .
 \ee
We shall refer to this situation as `apparent dissipation', to emphasise the distinction with dissipation in genuine open systems. Apparent dissipation occurs when a closed (time-dependent) system mimics the dissipative behaviour of an open system. 

The above example \eqref{examplediss1} exhibits dissipation through a spatially constant damping term. One generalisation is to allow $\gamma$ to be a spatially non-local function $\gamma(  -\nabla^2)$ so that the equation of motion is 
\be
\ddot \phi+ \gamma(-\nabla^2)\dot \phi - \nabla^2 \phi + m^2 \phi =0 \, .
\ee
This is better written in energy-momentum space $k = (\omega,\mathbf{k})$ where
\be
[-\omega^2 -i \omega \gamma({\bf k}^2) +{\bf k}^2 +m^2] \phi(k) =0 \, , 
\ee
which clearly leads to a modified dispersion relation. As before we can write an action for such a system, albeit a time-dependent spatially non-local one
 \ba
S &=& \frac{1}{2}\int \d^4 x \,  \(- \partial_{\mu} \phi \, e^{\gamma(-\nabla^2) t } \partial^{\mu}\phi -m^2 \phi e^{\gamma(-\nabla^2) t } \phi \) \, , \\
&=&\frac{1}{2}
\int \d t  \int \frac{\d^3 \mathbf{k}}{(2 \pi)^3}\,  e^{\gamma({\bf k}^2) t }\(   |\dot \phi_{\mathbf{k}}(t)|^2 -({\bf k}^2+m^2) |\phi_{\mathbf{k}}(t)|^2\) \, .
 \ea
A straightforward generalisation of such a closed EFT whose equations of motion preserve time translation invariance, but the action is time-dependent is
\be
S = 
\frac{1}{2}\int \d t  \int \frac{\d^3 \mathbf{k}}{(2 \pi)^3}\,   \sum_{n,m=0}^{\infty} c_{nm}({\bf k}^2)  \partial_t^n \phi_{-\mathbf{k}}(t)\, e^{\gamma({\bf k}^2) t} \partial_t^m \phi_{\mathbf{k}}(t)  \, ,
\ee
where the coefficients $c_{nm}({\bf k}^2)$ are Hermitian ($c^*_{nm}=c_{mn}$) analytic function of ${\bf k}^2$ and $\gamma({\bf k}^2)$ is similarly a real analytic function of ${\bf k}^2$. The resulting equation of motion  in momentum space is
\be
\sum_{n,m=0}^{\infty} c_{nm}({\bf k}^2 ) (-\partial_t - \gamma({\bf k}^2 ))^n \partial_t^m \phi_{\mathbf{k}}(t) =0 \, ,
\ee
or equivalently in energy-momentum space 
\be
\sum_{n,m=0}^{\infty} c_{nm}({\bf k}^2) (i \omega - \gamma({\bf k}^2))^n (-i\omega)^m \phi(k) =0 \, .
\ee
As promised, the equations of motion are time translation invariant. By assuming analyticity of $c_{nm}({\bf k}^2)$ and $\gamma({\bf k}^2)$ they can be Taylor expanded in energy/momenta consistent with an EFT interpretation of their origin from some unspecified UV completion.

\subsection{Apparent dissipative Gauge theories}

Now let us return to thinking about gauge theories. A simple example of an apparent dissipative gauge theory is provided by the action
\be
S= \int \d^4 x \, e^{\gamma t} \(-\frac{1}{4} F_{\mu \nu} F^{\mu\nu}+J_{\mu}A^{\mu} \) \, .
\ee
The resulting classical equation of motion is
\be
\partial_{\mu}F^{\mu \nu} + \gamma u_{\mu} F^{\mu \nu} = - J^{\nu} \, ,
\ee
with $u_{\mu}=(1,0,0,0)$. Gauge invariance $A_{\mu} \rightarrow A_{\mu}+\partial_{\mu} \lambda$ demands that the current is conserved in the sense
\be
\partial_{\mu}( e^{\gamma t} J^{\mu}) = 0 \qquad  \Rightarrow \qquad \partial_{\mu}J^{\mu}+\gamma u_{\mu} J^{\mu}=0 \, .
\ee
Now let us consider the in-in version of this action 
\be
S_{\text{in-in}} = \int \d^4 x \, e^{\gamma t} \(-\frac{1}{4} F^+_{\mu \nu} F_+^{\mu\nu}+J^+_{\mu}A_+^{\mu} +\frac{1}{4} F^-_{\mu \nu} F_-^{\mu\nu}-J^-_{\mu}A_-^{\mu} \) \, ,
\ee
where we have suppressed information on the boundary conditions. Rewriting this in terms of Keldysh variables we have
\be
S_{\text{in-in}} = \int \d^4 x \, e^{\gamma t} \(-\frac{1}{2} F^\mathrm{a}_{\mu \nu} F_\mathrm{r}^{\mu\nu}+J^\mathrm{r}_{\mu}A_\mathrm{a}^{\mu} +J^\mathrm{a}_{\mu}A_\mathrm{r}^{\mu} \) \, .
\ee
As it stands, this action makes manifest its time dependence through the $e^{\gamma t} $ factor. However, we can hide this by performing a field redefinition on the advanced field and current
\be
A^\mathrm{a}_{\mu}= e^{-\gamma t} a_{\mu}\, , \qquad \mathrm{and} \qquad J^\mathrm{a}_{\mu} = e^{-\gamma t} j^\mathrm{a}_{\mu} \, ,
\ee
so that the resulting action is
\be
S_{\text{in-in}} = \int \d^4 x \, \(-\frac{1}{2}(B_{\mu \nu} - 2 \gamma u_{\mu} a_{\nu} ) F_\mathrm{r}^{\mu\nu}+J^\mathrm{r}_{\mu}a^{\mu} +j^\mathrm{a}_{\mu}A^{\mu} \) \, ,
\ee
with $B_{\mu \nu}=\partial_{\mu }a_{\nu}-\partial_{\nu} a_{\mu}$.
Because of the field redefinition, the action is no longer gauge invariant under the advanced gauge transformations applied to $a_{\mu}$, but the source-free terms are invariant under the modified advanced gauge transformation
\be
\delta a_{\mu} = \partial_{\mu} \lambda - \gamma u_{\mu} \lambda \, ,
\ee
or in energy-momentum space (with a $-k$ convention to be consistent with below)
\be
\delta a_{\mu}(-k) = i v_{\mu} \lambda(-k) \, ,
\ee
with $v^{\mu}=( i \gamma_2, -{\bf k})$ and $ \gamma_2 = \gamma -i \omega$.
This a special case of the modified gauge transformation identified in \cite{Salcedo:2024nex}.
This modified gauge transformation applied to the source term imposes the modified current conservation
\be
\partial_{\mu} J_\mathrm{r}^{\mu}+\gamma u_{\mu} J_{\mathrm{r}}^{\mu}=0 \, .
\ee
If the current includes both a classical source and noise then similarly the noise satisfies a constraint that the total classical noise current is conserved in this modified sense.

The modified gauge transformation is of course just the usual gauge transformation of the correctly identified gauge field 
\be
\delta A_{\mu}^{\mathrm{a}} = \partial_{\mu} ( e^{-\gamma t}  \lambda) \, ,
\ee
and the `noise constraint' on auxiliary fields represented through $\xi^\mu$ is just the usual requirement of gauge invariance
\be
\partial_{\mu} \( e^{\gamma t}(J_r^{\mu} +\xi^{\mu}) \) =0\, .
\ee
Thus in reality it never was the case that the gauge symmetry was modified.

\subsection{Recovering Gauge Invariance}
\label{sec:RecFake}

In \cite{Salcedo:2024nex} the authors propose the following in-in action (discarding noise terms) to describe the linear response theory for photons propagating through a time-independent homogenous and isotropic medium
\be \label{eq:IFB}
S_{\text{in-in}} = \int \frac{\d^4 k}{(2 \pi)^4} \bigg( a^0(-k) i k_i F_\mathrm{r}^{0i}(k)+ a_i(-k) \(\gamma_2(k) F_\mathrm{r}^{0i}(k) - \gamma_3(k) i k_j F_\mathrm{r}^{ij}(k) + \gamma_4(k) \epsilon^{ijk} F^\mathrm{r}_{jk}(k) \) \bigg) \, ,
\ee
where the functional dependence of the coefficients $\gamma_i(k)$ means $\gamma_i(\omega,{\bf k}^2)$.
Denoting
\be
\gamma_2(k)=\hat \Gamma(k) -i \omega \, ,
\ee
we can rewrite this in a more gauge invariant manner
\begin{small}
\be
S_{\text{in-in}} = \int \frac{\d^4 k}{(2 \pi)^4} \( a_i(-k) \hat \Gamma(k) F_\mathrm{r}^{0i}(k) -B_{0i}(-k) F_\mathrm{r}^{0i}(k)+ \frac{1}{2} B_{ij}(-k) \gamma_3(k) F_\mathrm{r}^{ij}(k) + \gamma_4(k) \epsilon^{ijk} a_i(-k) F^r_{jk}(k) \) \, ,
\ee
\end{small}\ignorespaces
where we recognize the last term as Chern-Simons like.
The only term that is not gauge invariant in the conventional sense is the first $\hat \Gamma(k)$ term. This action is invariant under the modified advanced gauge transformation
\be
\label{modgauge1}
\delta a_{\mu}(k) = i k_{\mu} \lambda(k) - \hat \Gamma(-k) u_{\mu} \lambda(k) \, ,
\ee
where $\delta B_{\mu\nu}(k) = - i \hat{\Gamma}(-k) ( k_\mu u_\nu - u_\mu k_{\nu} ) \lambda(k)$. As we have outlined already, such an open system could not arise from actually integrating/tracing out degrees of freedom from a closed system. The resolution is that the advanced gauge field has been misidentified. Using an expansion
\be
\hat \Gamma(-k) = \sum_{n=0}^{\infty }(-i \omega)^n \Gamma_n({\bf k}^2) \, ,
\ee
 then in mixed representation
\be
\delta a_{0}(t,{\bf k}) =  \partial_t \lambda(t,{\bf k}) - \sum_{n=0}^{\infty } \Gamma_n({\bf k}^2)  \partial_t^n \lambda(t,{\bf k})\, .
\ee
Defining
\be
\lambda(t,{\bf k}) = e^{\alpha({\bf k}^2)t} \Lambda(t,{\bf k})\, ,
\ee
where $\alpha({\bf k}^2)$ is the solution of 
 \be
\alpha({\bf k}^2)- \sum_{n=0}^{\infty } \Gamma_n({\bf k}^2)  \alpha^n({\bf k}^2) =0 \, ,
 \ee
then we have
 \ba
\delta a_{0}(t,{\bf k}) &=& e^{\alpha({\bf k}^2)t}  \left[  \partial_t \Lambda(t,{\bf k}) - \sum_{n=1}^{\infty } \Gamma_n({\bf k}^2) ((\alpha+ \partial_t)^n-\alpha^n) \Lambda(t,{\bf k}) \right]\, \\
&=& e^{\alpha({\bf k}^2)t}  \left[  \partial_t \Lambda(t,{\bf k}) - \sum_{n=1}^{\infty } \sum_{m=1}^{n }\Gamma_n({\bf k}^2) \frac{n!}{m!(n-m)!} \alpha^{n-m}\partial_t^m  \Lambda(t,{\bf k}) \right]\, .
 \ea
It is easy to see that there is a redefinition of the $a_{\mu}$ fields of the form
 \ba
&& a_i(t,{\bf k}) = e^{\alpha({\bf k}^2)t}A^{\mathrm{a}}_i(t,{\bf k}) \, ,\\
&& a_0(t, {\bf k}) = e^{\alpha({\bf k}^2)t} \(  1 - \sum_{n=1}^{\infty } \sum_{m=1}^{n }\Gamma_n({\bf k}^2) \frac{n!}{m!(n-m)!} \alpha^{n-m}\partial_t^{m-1}  \) A^{\mathrm{a}}_0(t,{\bf k})\, ,
 \ea
for which the modified gauge transformation \eqref{modgauge1} becomes
 \be
\delta A^{\mathrm{a}}_{\mu} = \partial_{\mu} \Lambda \, .
 \ee
Thus, as long as we are working with a low energy expansion in derivatives, it is always possible to perform field redefinitions so that the modified advanced gauge transformation is actually just the conventional one, albeit for a system with a time-dependent Hamiltonian. Once this is done the influence functional \eqref{eq:IFB} becomes clearly invariant under two copies of gauge transformations, and the constraint on the noise is just the usual one implied by advanced gauge invariance.
Thus the correct interpretation of \eqref{eq:IFB} is that it describes in general a time-dependent open system with both real and apparent dissipation, written in variables that do not manifest the two copies of gauge invariance.

\section{Evaluating the Influence Functional}
\label{TopDown}

In this section we will formally evaluate the influence functional in a number of interesting cases, leaving more detailed calculations for specific situations to later sections. In all cases we shall assume for simplicity that the field being integrated out is in a Gaussian initial state so that the entire effect of the state may be incorporated in the free propagators via the $i \epsilon$ prescription. Vacuum, thermal and other mixed states are all treated in the same manner.

\subsection{Gauged Caldeira-Leggett model}

\label{sec:CLmodel}

To illustrate the issues with open system gauge theories let us consider a gauged (non-relativistic) field theory version of the Caldeira-Leggett model \cite{Caldeira:1982iu} which is sufficiently simple to be tractable. The system we shall consider is a $U(1)$ gauge field, a massive charged relativistic scalar $\Phi$ coupled to a non-relativistic bath of charged scalars $\varphi_I$ with action (before gauge fixing)\footnote{The global version of this model is considered in \cite{Akyuz:2025bco}.}
\be
S= \int \d^4 x\left[-\frac{1}{4} F_{\mu\nu}^2 -|D[A] \Phi|^2 - m^2 |\Phi|^2 +\sum_{I} \( |D_t[A] \varphi_I|^2 - \Gamma_I^2 |\varphi_I|^2 +g_I(\varphi^*_I \Phi+\Phi^* \varphi_I )\) \right] \, .
\ee
The key simplifying assumption is that $\varphi_I$ only have time derivatives which allows us to determine their propagators exactly. By construction, this is manifestly gauge invariant with the scalars all transforming the same way 
\be
\Phi \rightarrow e^{i q \lambda} \Phi , \quad \varphi_I \rightarrow e^{i q \lambda} \varphi_I \, . 
\ee
Let us consider the influence functional obtained from integrating out the scalars $\varphi_I$ which are specified in some Gaussian (not necessarily thermal) state
\ba
&& e^{i S_{\mathrm{IF}}[A_+,A_-,\Phi_+,\Phi_-,\Phi_+^*,\Phi_-^*]} =\int {\cal D}[\varphi_I^+,\varphi_I^-]\\
&& \frac{\mu}{\tilde \mu} \, e^{i \int \d^4 x \sum_{I} \( |D_t[A_+] \varphi^+_I|^2 - \Gamma_I^2 |\varphi^+_I|^2 +g_I(\varphi^{+*}_I \Phi^++\Phi^{+*} \varphi^+_I) -|D_t[A_-] \varphi^-_I|^2 + \Gamma_I^2 |\varphi^-_I|^2 -g_I(\varphi^{-*}_I \Phi^-+\Phi^{-*} \varphi^-_I)\)+i S_{i \epsilon}} \, .\nn
\ea
What makes the gauge theory different than the global limit is that the scalars on each branch are coupled to a different gauge field. However, in this simplistic model, only via $A_0^{\pm}$. Since it is always possible to choose a gauge for which $A_0^{\pm}=0$, the dependence on $A_{\pm}$ in general is determined entirely by the gauge transformation necessary to get to $A_0^{\pm}=0$. To specify this we need to separate the discussion for the advanced and retarded gauge transformations. To set $A_0^{\mathrm{a}}=0$ we need only solve
\be
\partial_t \lambda^{\mathrm{a}}(x) = A_0^{\mathrm{a}}(x)\, .
\ee
The solution of this is unique because we require that advanced gauge transformations vanish at $t=\tf$, hence
\be
\lambda^{\mathrm{a}}(t,{\bf x}) = -\int^{\tf}_{t} \d t' A_0^{\mathrm{a}}(t',{\bf x}) \, .
\ee
Following our discussion earlier this corresponds to performing a field redefinition via a Wilson line which ends on the final time surface
\be
\varphi_I^{\pm}(x) = U_{\pm}(x)\tilde \varphi_I^{\pm}(x) \, ,
\ee
with
\be
U_{\pm}(x)=e^{\mp \frac{i}{2}q \int^{\tf}_{t} \d t' A_0^{\mathrm{a}}(t',{\bf x}) } \, .
\ee
The apparent acausality of this redefinition is a consequence of the in-in boundary condition at late times that implies that natural time evolution $A^{\mathrm{a}}$ is reversed.

Once this is done, the influence functional becomes 
\ba
&& e^{i S_{\mathrm{IF}}[A_+,A_-,\Phi_+,\Phi_-,\Phi_+^*,\Phi_-^*] }=\int {\cal D}[\tilde \varphi_I^+,\tilde \varphi_I^-]\\
&& \frac{\mu}{\tilde \mu} \, e^{i \int \d^4 x\sum_{I} \( |D_t[A_\mathrm{r}] \tilde \varphi^+_I|^2 - \Gamma_I^2 |\tilde \varphi^+_I|^2 +g_I(\tilde \varphi^{+*}_I \tilde \Phi^++\tilde \Phi^{+*} \tilde \varphi^+_I) -|D_t[A_\mathrm{r}] \tilde \varphi^-_I|^2 + \Gamma_I^2 |\tilde \varphi^-_I|^2 -g_I(\tilde \varphi^{-*}_I \tilde \Phi^-+\tilde \Phi^{-*} \tilde \varphi^-_I)\)+i S_{i \epsilon}} \, ,\nn
\ea
with $ \Phi^{\pm}(x) = U_{\pm}(x) \tilde \Phi^{\pm}(x)$.
In this form, the scalars $\tilde \varphi_I$ are coupled to the same gauge field $A^\mathrm{r}$ on each branch and so we may follow the usual procedure for quantisation in a background gauge field. The Gaussian path integral (for Gaussian initial state encoded in the $i \epsilon$ terms) will give 
\ba
 S_{\rm IF}[A_+,A_-,\Phi_+,\Phi_-,\Phi_+^*,\Phi_-^*] &=&- i \sum_I \Tr \log {{\bf J}_I[A_\mathrm{r}]} + i \sum_I \Tr \log {{\bf J}_{I A}[A_\mathrm{r}]} \\
&&+  i \sum_I g_I^2 \int \d^4 x \int \exd^4 y \begin{pmatrix} \tilde \Phi^*_+(x) & -\tilde \Phi^*_-(x) \end{pmatrix} {\bf J}_I[A_\mathrm{r}](x,y) \begin{pmatrix} \tilde \Phi_+(y) \\ -\tilde \Phi_-(y) \end{pmatrix}  \, , \nn
\ea
with a matrix of CTP Feynman propagators
\be
{\bf J}_I(x,y)=\begin{pmatrix}  J_I^{++}(x,y) & J_I^{+-}(x,y) \\
  J_I^{-+}(x,y) & J_I^{--}(x,y) \, 
  \end{pmatrix} \, ,
\ee
with
\ba
&&  J_I^{++}(x,y) = \Tr[\rho {\cal T}\hat \varphi_{I}(x) \hat{{\varphi}}^{\dagger}_{I}(y)] \, ,\\
&&  J_I^{+-}(x,y) = \Tr[\rho \hat{{\varphi}}^{\dagger}_{I}(y)\hat{\varphi}_{I}(x)] \, ,\\
&&  J_I^{-+}(x,y) = \Tr[\rho \hat \varphi_{I}(x) \hat{{\varphi}}_{I}^{\dagger}(y)] \, ,\\
&&  J_I^{--}(x,y) = \Tr[\rho \bar {\cal T}\hat \varphi_{I}(x) \hat{{\varphi}}_{I}^{\dagger}(y)] \, ,
\ea
and to account for the path integral measure as discussed in Appendix~\ref{app:measure} a matrix of advanced propagators defined via
\be
{\bf J}_{ I A}(x,y)={\bf J}_I(x,y)-\begin{pmatrix}  J_I^{-+}(x,y) & J_I^{-+}(x,y) \\
  J_I^{-+}(x,y) & J_I^{-+}(x,y) \, 
  \end{pmatrix} \, .
\ee
The operator expectation values on the RHS are for a free field in a background gauge field $A^\mathrm{r}_{\mu}$ and hence the operators satisfy the equations of motion
\be
D_t[A_\mathrm{r}]^2 \hat{\tilde \varphi}_I(x)  =- \Gamma_I^2 \hat{\tilde \varphi}_I(x) \, .
\ee
As before this equation only depends on $A^\mathrm{r}_0$ and its general solution will be determined by the gauge transformation that sets $A^\mathrm{r}_0=0$. However, since the retarded field should be evolved causally rather than tying it to the final time, we should tie it to the initial time at which the state $\rho$ is specified.
In other words, we now solve
\be
\partial_t \lambda^\mathrm{r}(x) = A_0^\mathrm{r}(x)\, ,
\ee
with 
\be
\lambda^\mathrm{r}(t,{\bf x}) = \int^{t}_{\ti} \d t' A_0^\mathrm{r}(t',{\bf x})+\lambda^\mathrm{r}_\mathrm{i}({\bf x}) \, . 
\ee
Unlike for the advanced symmetry, there is remaining gauge freedom $\lambda^\mathrm{r}_{\mathrm{i}}({\bf x})$ associated with time independent retarded gauge transformations which may be viewed as acting on the initial time surface.

The solution of the operator equations is
\be
\hat{\tilde \varphi}_I(x) = V(x) \frac{1}{\sqrt{2\Gamma_I}}\(e^{-i \Gamma_I t}\hat a_I({\bf x})+e^{i \Gamma_I t } \hat b_I^{\dagger}({\bf x}) \) \, ,
\ee
with Wilson line
\be
V(x)=e^{ iq\int^{t}_{\ti} \d t' A_0^\mathrm{r}(t',{\bf x}) } \,, 
\ee
and non-zero canonical commutation relations
\be
[\hat a_I({\bf x}) ,\hat a^{\dagger}_J({\bf y})] = \delta_{IJ} \delta^3({\bf x}-{\bf y}) \, , \quad [\hat b_I({\bf x}) ,\hat b^{\dagger}_J({\bf y})] = \delta_{IJ} \delta^3({\bf x}-{\bf y}) \, .
\ee
Now, a Gaussian initial state is completely determined by specifying expectation values such as $\Tr[\rho \hat a^{\dagger}_I({\bf x}) \hat a_J({\bf y})]$. In general, however, this expression is not BRST invariant because there is still the residual retarded gauge symmetry $\lambda_\mathrm{r}({\bf x})$ on the initial time surface. Nevertheless, the physical state $\rho$ must be BRST invariant. As is usual, we define gauge invariant two point functions by multiplying this expression by a Wilson line, in this case defined on the straight line from ${\bf x}$ to ${\bf y}$ at time $\ti$. In other words
\be
\Tr[\rho \hat a^{\dagger}_I({\bf x}) \hat a_J({\bf y})] = \delta_{IJ} W({\bf x}, {\bf y}) H_I(\bf{x},\bf{y}) \, ,
\ee
where $H_I(\vec{x},\vec{y})$ is now gauge invariant and in effect independent of $A_\mathrm{r}$ entirely since it is just part of specifying the initial state, and 
\be
W({\bf x}, {\bf y})=e^{i q \int_0^1 \d s \tfrac{\d z^i}{\d s} A_i(\ti,{\bf z}(s))} \, ,
\ee
is the Wilson line defined on the curve 
\be
z(s) =(\ti, y^i+s(x^i-y^i)) \, .
\ee
Putting this together, the dependence of the propagators ${\bf J}_I(x,y)$ on the retarded gauge field can be factored out via a combined Wilson line
\be
{\bf J}_I(x,y)= {\cal W}^\mathrm{r}_{C_1}(x,y)  {\bf D}_I(x-y) \, ,
\ee
with the Wilson line $C_1$ defined on contour extending from $y$ to $(\ti, {\vec y})$ to $(\ti, {\vec x})$ to $x$ in straight line segments.
\be
{\cal W}^\mathrm{r}_{C_1}(x,y)  = e^{i q      \int_{C_1}A^\mathrm{r}_{\mu}(z) \d z^{\mu}}=V(x) W({\bf x},{\bf y}) V(y)^* \, .
\ee
Now denoting the matrix of Wilson lines associated with the advanced gauge field
\be
{\bf U}(x) = \begin{pmatrix} U_+(x) &0 \\ 0 & U_{-}(x)\end{pmatrix} \, ,
\ee
then in terms of the original field variables, the influence functional is
\ba
&&  S_{\rm IF}[A_+,A_-,\Phi_+,\Phi_-,\Phi_+^*,\Phi_-^*] = -i   \sum_I \Tr \log {\bf D}_I+i   \sum_I \Tr \log {{\bf D}_{I A}} \\
&& \qquad + i \sum_I g_I^2 \int \d^4 x \int \d^4 y  \begin{pmatrix}  \Phi^*_+(x) & -\Phi^*_-(x) \end{pmatrix} {\cal W}^\mathrm{r}_{C_1}(x,y) {\bf U}(x){\bf D}_I(x,y) {\bf U}^{\dagger}(y)\begin{pmatrix} \Phi_+(y) \\ -\Phi_-(y) \end{pmatrix} \, , \nn
\ea
with ${\bf D}_I(x,y)$ the free propagators for a global scalar and ${\bf D}_{I A}$ the equivalent CTP advanced propagators.
In the usual situation for which the initial state is defined as $\ti \rightarrow -\infty$ we may assume $A_i(\ti) \rightarrow 0$. Then a natural choice for the form of the propagator would be that for a translation invariant state would take the form in momentum space
\begin{small}
\be
{\bf D}_I(\omega ,{{\bf k}}) = \begin{pmatrix}
-\frac{i}{-\omega^2+\Gamma_I^2-i \epsilon}+2 \pi \delta(-\omega^2+\Gamma_I^2) n_I(k) & \theta(-\omega) 2 \pi \delta(-\omega^2+\Gamma_I^2) +2 \pi \delta(-\omega^2+\Gamma_I^2) n_I(k)\\
\theta(\omega) 2 \pi \delta(-\omega^2+\Gamma_I^2) +2 \pi \delta(-\omega^2+\Gamma_I^2) n_I(k)& \frac{i}{-\omega^2+\Gamma_I^2+i \epsilon}+2 \pi \delta(-\omega^2+\Gamma_I^2) n_I(k) 
\end{pmatrix} \, ,
\ee
\end{small}
with 
\be
n_I(k)=\theta(\omega) n_{I+}({\bf k})+\theta(-\omega)n_{I-}( -{\bf k}) \, .
\ee
This example nicely illustrates how the open system for the gauge theory connects with global limit. As a gauge fixed gauge theory the total action respects the diagonal BRST symmetry because the influence functional is invariant under two copies of gauge symmetries by virtue of the Wilson lines which emerge naturally from the background field propagators. However, in the global/decoupling limit $q \rightarrow 0$, the photon decouples and the influence functional are only invariant under the retarded global $U(1)$ symmetry with the advanced global symmetry explicitly broken
\ba
&& \lim_{q \rightarrow 0} S_{\rm IF}[A_+,A_-,\Phi_+,\Phi_-,\Phi_+^*,\Phi_-^*] = \\
&& i \sum_I g_I^2 \int \d^4 x \int \d^4 y \begin{pmatrix}  \Phi^*_+(x) & -\Phi^*_-(x) \end{pmatrix} { {\bf D}_I(x,y) \begin{pmatrix} \Phi_+(y) \\ -\Phi_-(y) \end{pmatrix} }-i   \sum_I \Tr \log {{\bf D}_I}+i   \sum_I \Tr \log {{\bf D}_{I A}} \notag
 \, .
\ea
This is consistent with our expectation that for a global theory only the diagonal symmetry is preserved. Note that this global breaking comes from the $D_I^{+-}$ and $D_I^{-+}$ terms which are present even in vacuum, because even without considering a non-vacuum state the final time boundary conditions break the advanced global symmetry.

\subsection{Spinor QED: Integrating out the photon}

Although not our main interest, most discussions of open systems for QED focus on the effects of photons, {\it eg.} a radiation bath, on the dynamics of electrons. For this it is more appropriate to consider the influence functional for electrons obtained from integrating out the photon. This is extremely easy to do because (a) the gauge field path integral is Gaussian, and (b) the quadratic part of the photon action is independent of the electron. Denoting the QED action by
\be
S[ A, \Psi, \overline{\Psi} ]  =  \int \exd^4 x \; \[ - \frac{1}{4} F_{\mu\nu} F^{\mu\nu} + i \overline{\Psi} \slashed{D} \Psi - m \overline{\Psi} \Psi \] \, ,
\ee
then the influence functional is with the choice $\xi=1$
\ba
e^{i S_{\rm IF}[\Psi_{+}, \overline{\Psi}_{+}, \Psi_{-}, \overline{\Psi}_{-}]} & =& \int {\mathcal D} [A_+,A_-,c_+,c_-,\bar c_+,\bar c_-] \\
&&  \frac{\mu}{\tilde \mu} \; e^{i \int \d^4 x \, \(\tfrac{1}{2} A^+_{\mu} \Box  A_+^{\mu}+  \overline{\Psi}_+ q\slashed{A}_+ \Psi_+-\tfrac{1}{2} A^-_{\mu} \Box  A_-^{\mu}-  \overline{\Psi}_- q\slashed{A}_- \Psi_-+ \bar c_+ \Box c_+-\bar c_- \Box c_-\)+iS_{\rm i \epsilon}} \, ,\nn 
\ea
with $S_{\rm i \epsilon}$ given by \eqref{eq:photoniepsilon}. Note that we have bypassed the need to perform the path integral at the final time $\tf$ by using the $i \epsilon$ prescription. Since the ghosts do not couple, even in a mixed state their path integral can be absorbed into the normalisation of the measure. The coupling between the two branches comes entirely from the $i \epsilon$ terms. Performing the Gaussian integral over the gauge fields we obtain the influence functional 
\be
S_{\rm IF}[\Psi_{+}, \overline{\Psi}_{+}, \Psi_{-}, \overline{\Psi}_{-}] = \frac{i}{2} \int \d^4 x \int \d^4 y \begin{pmatrix}  J^{\mu}_+(x) & -J^{\mu}_-(x)\end{pmatrix}
{\bf D}_{\mu\nu}(x,y)
\begin{pmatrix}  J^{\nu}_+(y) \\-J^{\nu}_-(y)\end{pmatrix} \, ,
\ee
with 
\be
J^{\mu}_{\pm}(x) = \overline{\Psi}_\pm(x) q \gamma^{\mu}\Psi_\pm(x) \, ,
\ee
and
\be
{\bf D}_{\mu\nu}(x,y)=\begin{pmatrix}  D_{\mu\nu}^{++}(x,y) & D_{\mu\nu}^{+-}(x,y) \\
  D_{\mu\nu}^{-+}(x,y) & D_{\mu\nu}^{--}(x,y) \, 
  \end{pmatrix} \, .
\ee
The Green's functions $D_{\mu\nu}(x,y)$ are the free photon propagators in a general mixed state. For a translation invariant state we have in momentum space
\be
 {\bf D}_{\mu\nu}(k) = \begin{pmatrix}
-\frac{i \eta_{\mu\nu}}{k^2-i \epsilon}+ \pi \delta(k^2) (-\eta_{\mu\nu}+K^A_{\mu\nu}(k)) & \pi \delta(k^2) (\eta_{\mu\nu}[2\theta(-k^0)-1]+K^A_{\mu\nu}(k))\\
\pi \delta(k^2) (\eta_{\mu\nu}[2\theta(k^0)-1]+K^A_{\mu\nu}(k)) & \frac{i\eta_{\mu\nu}}{k^2+i \epsilon}+ \pi \delta(k^2)(-\eta_{\mu\nu}+K^A_{\mu\nu}(k)) \end{pmatrix} \, .
\ee
This result is well known in the literature although often appearing in different gauges.

\subsection{Spinor QED: Integrating out the electron}

In the opposite situation, we can consider the influence of a distribution of charged matter, {\it i.e.} a plasma, on the propagation of photons. In path integral terms, our goal is to determine the influence functional for the photon obtained from integrating out the electron in some chosen initial Gaussian state. As for the photon the electron path integral is (a) Gaussian. However, unlike the previous situation, its quadratic term is not independent of the photon ({\it i.e.} (b) no longer holds) and this significantly complicates the calculation of the influence functional, sufficiently so that it is impossible to give a closed form for the exact solution. We can, however, at least provide a formal expression which is suitable for developing approximation schemes. 
 
In the context of the $i \epsilon$ prescription, the influence functional is defined by the path integral
\ba
e^{i S_{\rm IF}[A_+,A_-]} &=&\int {\mathcal D} [\Psi_+,\bar \Psi_+,\Psi_-,\bar \Psi_- ] \\
&& \frac{\mu}{\tilde \mu} \, e^{i \int \d^4 x \( i\overline{\Psi}_+ \slashed{\partial} \Psi_+ - m \overline{\Psi}_+ \Psi_+-i \overline{\Psi}_- \slashed{\partial} \Psi_- + m \overline{\Psi}_- \Psi_- +\overline{\Psi}_+ q\slashed{A}_+ \Psi_+-  \overline{\Psi}_- q\slashed{A}_- \Psi_-\)+iS_{\rm i \epsilon}} \, . \nn
\ea
As usual the ghost path integral decouples and can be disregarded. The remaining $i \epsilon$ terms serve to define the Gaussian integral over the electron fields.
We can define a matrix of propagators which is now spinor valued
\be
{\bf S}_{\alpha \beta}(x,y)=\begin{pmatrix}  S_{\alpha \beta}^{++}(x,y) & S_{\alpha \beta}^{+-}(x,y) \\
  S_{\alpha \beta}^{-+}(x,y) & S_{\alpha \beta}^{--}(x,y)
  \end{pmatrix}  \, . 
\ee
In general these propagators depend on both gauge fields $A_{\pm}$ or equivalently $A_\mathrm{r}$ and $A_\mathrm{a}$. If we set the advanced gauge field to zero, that what we mean by the propagators are the operator expressions
\ba
&& S^{++}_{\alpha \beta}(x,y) = \Tr[\rho {\cal T}\hat \Psi_{\alpha}(x) \hat{\overline{\Psi}}_{\beta}(y)] \, ,\\
&& S^{+- }_{\alpha \beta}(x,y) =- \Tr[\rho \hat{\overline{\Psi}}_{\beta}(y) \hat{\Psi}_{\alpha}(x)] \, ,\\
&& S^{-+ }_{\alpha \beta}(x,y) = \Tr[\rho \hat \Psi_{\alpha}(x) \hat{\overline{\Psi}}_{\beta}(y)] \, ,\\
&& S^{-- }_{\alpha \beta}(x,y) = \Tr[\rho \bar {\cal T}\hat \Psi_{\alpha}(x) \hat{\overline{\Psi}}_{\beta}(y)] \, ,
\ea
where the operator spinor satisfies the a covariant equation in the background gauge field $A_\mathrm{r}$
\be \label{eq:covariantpsi}
[i \slashed{D}[A_\mathrm{r}]-m] \hat \Psi = 0\, .
\ee
The inclusion of the advanced field modifies these expressions to the following
\ba
&& S^{++}_{\alpha \beta}(x,y) = Z^{-1}\Tr[\rho {\bar {\cal T }^*}e^{i \hat S_{\rm int}[A_\mathrm{a}]} {\cal T}^* e^{i \hat S_{\rm int}[A_\mathrm{a}]}\hat \Psi_{\alpha}(x) \hat{\overline{\Psi}}_{\beta}(y)] \, ,\\
&& S^{+- }_{\alpha \beta}(x,y) =- Z^{-1}\Tr[\rho {\bar {\cal T }^*} (e^{i \hat S_{\rm int}[A_\mathrm{a}]}  \hat{\overline{\Psi}}_{\beta}(y) ){\cal T}^*(e^{i \hat S_{\rm int}[A_\mathrm{a}]}  \hat{\Psi}_{\alpha}(x))] \, ,\\
&& S^{-+ }_{\alpha \beta}(x,y) =Z^{-1} \Tr[\rho {\bar {\cal T }^*} (e^{i \hat S_{\rm int}[A_\mathrm{a}]}\hat \Psi_{\alpha}(x)) {\cal T}^* (e^{i \hat S_{\rm int}[A_\mathrm{a}]}\hat{\overline{\Psi}}_{\beta}(y))] \, ,\\
&& S^{-- }_{\alpha \beta}(x,y) = Z^{-1}\Tr[\rho \bar {\cal T}^*(e^{i \hat S_{\rm int}[A_\mathrm{a}]}\hat \Psi_{\alpha}(x)) {\cal T}^*( e^{i \hat S_{\rm int}[A_\mathrm{a}]}\hat{\overline{\Psi}}_{\beta}(y))] \, ,
\ea
with interaction
\be
\hat S_{\rm int} = \int \d^4 x \, q \hat{\overline {\Psi}} \slashed{A}_\mathrm{a} \hat \Psi \, ,
\ee
and normalisation
\be
Z = \Tr[\rho {\bar {\cal T }^*}e^{i \hat S_{\rm int}[A_\mathrm{a}]}  {\cal T}^* e^{i \hat S_{\rm int}[A_\mathrm{a}]}] \, .
\ee
The unusual sign for the exponential in the time reversed part is due to the way the advanced field arises with opposite sign on the time reversed part of the contour, which means in particular that $Z \neq 1$!
Note that in writing these expressions covariantly with the covariant time ordering ${\cal T}^*$, we are making heavy use of Matthews' theorem as discussed in Appendix~\ref{app:measure}.

Once these propagators are specified, the influence functional is formally straightforward to evaluate. In index suppressed notation 
\ba
e^{i S_{\rm IF}[A_+,A_-]} &=& \int {\mathcal D} [\Psi_+,\bar \Psi_+,\Psi_-,\bar \Psi_-] \, \frac{\mu}{\tilde \mu}\exp\( -\begin{pmatrix} \overline \Psi_+ & \overline \Psi_- \end{pmatrix} {{{\bf S}}}^{-1}[A_+,A_-] \begin{pmatrix} \Psi_+ \\  \Psi_- \end{pmatrix} 
\) \\&=& \frac{\det[{\bf S}_A[A_+,A_-]]}{\det[{\bf S}[A_+,A_-]]} \, ,
\ea
so that 
\be
S_{\rm IF}[A_+,A_-] = i \Tr \log {\bf S}[A_+,A_-]-i \Tr \log {\bf S}_A[A_+,A_-]\, ,
\ee
remembering that the trace is over $x$, spinor indices $\alpha$ and the $2\times 2$ matrix indices $\pm$ associated with the two branches of the CTP. Once again, we have included advanced CTP propagators to account for the path integral measure $\mu/\tilde \mu$.

In the standard perturbative approach of constructing the propagators we would first specify the free field propagators defined in the absence of gauge fields by ${\bf S}^0_{\alpha \beta}(x,y)$. 
For a translation and rotation invariant state in momentum space, they take the form
\be
{\bf S}^0(k) = (-\slashed{k}+m)\begin{pmatrix} \frac{-i}{k^2+m^2-i \epsilon} + \mathfrak{n}(k) 2 \pi \delta(k^2+m^2)&  [ \theta(-k^0) + \mathfrak{n}(k) ]  2 \pi \delta(k^2+m^2) \\  [ \theta(k^0) + \mathfrak{n}(k) ] 2 \pi \delta(k^2+m^2) &  \frac{i}{k^2+m^2+i \epsilon} +  \mathfrak{n}(k) 2 \pi \delta(k^2+m^2)\end{pmatrix} 
\ee
with number densities
\be
\mathfrak{n}(k)=\theta(k^0) n_+(|{\bf k}|) - \theta(-k^0) n_-(|{\bf k}|) \, .
\ee
Then since the fermion path integral is Gaussian, the gauge field interaction can be viewed as a perturbative correction to the propagator in the sense
\be
{\bf S}^{-1}={\bf S}_0^{-1}-i q \slashed{\bf A} \, ,
\ee
where
\be
\slashed{\bf A}(x,y) = \begin{pmatrix} {\slashed A}_+(x)&0 \\ 0 & -{\slashed A}_-(x)\end{pmatrix} \delta^4(x-y) \, .
\ee
This can then be resummed into the Schwinger-Dyson equation
\be
{\bf S}[A_+,A_-] =\(1- {\bf S}^0 i q \slashed{\bf A} \)^{-1} {\bf S}^0 \, .
\ee
When computed perturbatively, the gauge invariance/transformation properties of the final answer are disguised at every step and one must rely on a miracle of cancellations of integrals at each order to make gauge invariance manifest. This miracle occurs by virtue of the Ward identities, but it is not manifest term by term. This in turn means that the BRST invariance of the influence functional is not manifest. For our present discussion it is helpful to reorganize the expansions to make gauge properties manifest.

Given the somewhat different nature of the advanced and retarded gauge fields, following the previous section, it is helpful to separate the discussion for each. Due to the boundary conditions of the advanced field it is natural to split a general advanced gauge field up into a gauge part, and the value in Fock-Schwinger gauge defined on the final time surface (see Appendix \ref{app:Wilson}). This is achieved by performing the field redefinition
\be \label{eq:fieldredef1}
\Psi_{\pm}(x) =U_{\pm}(x)\tilde \Psi_{\pm}(x) \, ,
\ee
where
\be
 U_{\pm}(x)=e^{\pm \frac{i}{2}q \int_0^1 \d s\,  (x_\mathrm{f}-x)^{\mu} A^{\mathrm{a}}_{\mu}(x+s(x_\mathrm{f}-x))} \, ,
\ee
with $x_\mathrm{f}=(\tf,\vec x)$. The phase factor can be removed with a field redefinition of the advanced gauge field transforming it into the gauge and BRST invariant combination
\be
\tilde A^{\mathrm{a}}_{\mu}(x) = A^{\mathrm{a}}_{\mu}(x) - \partial_{\mu}\int_0^1 \d s \, (x_\mathrm{f}-x)^{\nu} A^{\mathrm{a}}_{\nu}(x+s(x_\mathrm{f}-x)) \, .
\ee
The field $\tilde \psi_{\pm}(x)$ is invariant under advanced gauge transformations, but transforms covariantly under retarded gauge transformations
\be
\tilde \Psi_{\pm}(x) \rightarrow e^{i q \lambda_\mathrm{r}(x)}\tilde \Psi_{\pm}(x) \, .
\ee
The matrix of two-point functions of $\Psi$ can be rewritten as
\be
{\bf S}[A_+,A_-](x,y) ={\bf U}(x) {\bf \tilde S}[A_\mathrm{r}, \tilde A_\mathrm{a}](x,y){\bf U}(y)^{\dagger} \, ,
\ee
where ${\bf \tilde S}[A_\mathrm{r}, \tilde A_\mathrm{a}](x,y)$ transforms covariantly under retarded gauge transformations
\be
{\bf \tilde S}[A_\mathrm{r}, \tilde A_\mathrm{a}](x,y) \rightarrow e^{iq (\lambda_\mathrm{r}(x)-\lambda_\mathrm{r}(y))}{\bf \tilde S}[A_\mathrm{r}, \tilde A_\mathrm{a}](x,y) \, ,
\ee
and as before
\be
{\bf U}(x) = \begin{pmatrix} U_+(x) &0 \\ 0 & U_{-}(x)\end{pmatrix} \, .
\ee
A fully gauge invariant two-point function $\mathbf{\check S}$ can be identified by factoring out a Wilson line that accounts for the retarded gauge transformation properties 
\be
{\bf \tilde S}[A_\mathrm{r}, \tilde A_\mathrm{a}](x,y) = {\cal W}_{C_3}(x,y) {\bf \check S}[A_\mathrm{r}, \tilde A_\mathrm{a}](x,y) \, ,
\ee
with $C_3$ a straight line from $x$ to $y$.
In particular, the initial state can be defined by specifying the initial value of the gauge invariant propagators
\be
{\bf \check S}[A_\mathrm{r}, \tilde A_\mathrm{a}](\ti,{\bf x},\ti , {\bf y}) = {\bf S}^0({\bf x},{\bf y}) \, .
\ee
In principle, we can imagine that the propagators ${\bf \tilde S}[A_\mathrm{r}]$ in the presence of a background retarded gauge field $A_\mathrm{r}$ have been determined exactly by solving \eqref{eq:covariantpsi}. It is then natural to develop a perturbative expansion in the advanced field.
the resummed Schwinger-Dyson equation for the full dressed propagator is
\be
{\bf S}[A_+,A_-](x,y) = {\bf U}(x)\left[  \(1- {\bf \tilde S}[A_\mathrm{r}] i q {\bf \tilde {\slashed{A}}_\mathrm{a}}\)^{-1} {\bf \tilde S}[A_\mathrm{r}] \right](x,y){\bf U}(y)^{\dagger} \, ,
\ee
with 
\be
{\bf \tilde {\slashed{A}}_\mathrm{a}}(x,y) = \frac{1}{2}\begin{pmatrix} {\tilde {\slashed A}}_\mathrm{a}(x)&0 \\ 0 & {\tilde {\slashed A}}_\mathrm{a}(x)\end{pmatrix} \delta^4(x-y) \, .
\ee
Unlike the previous Schwinger-Dyson equation, the gauge transformation properties under advanced gauge transformations are manifest, and it is also manifestly covariant under retarded gauge transformations.
Using these same contours to give a gauge invariant meaning to the fermionic Gaussian path integral in a background gauge field then the influence functional becomes 
\be \label{eq:gaugeinvariantIF}
S_{\rm IF}[A_+,A_-] = i \Tr \log {\bf \tilde S}[A_\mathrm{r}]-i \Tr \log \(1- {\bf \tilde S}[A_\mathrm{r}] i q {\bf \tilde {\slashed{A}}_\mathrm{a}}\) - i \Tr \log {\bf \tilde S}_A[A_\mathrm{r}]+i \Tr \log \(1- {\bf \tilde S}_A[A_\mathrm{r}] i q {\bf \tilde {\slashed{A}}_\mathrm{a}}\)\, ,
\ee
which is now manifestly BRST invariant since $\tilde A$ and ${\bf \tilde S}[A_\mathrm{r}]$ are. 
The advantage of the form \eqref{eq:gaugeinvariantIF} is that it is manifestly invariant under both gauge symmetries, but the disadvantage is that via $\tilde A_\mathrm{a}$ it is superficially dependent on the final time $\tf$ which would seem to contradict causality. In practice however
Noether's theorem applied at the level of the influence functional (see Appendix \ref{App:Noether}) implies that all terms arising from the difference between $\tilde A_\mathrm{a}$ and $A_\mathrm{a}$ vanish so that
\be
\label{eq:invariant2}
S_{\rm IF}[A_+,A_-] = i \Tr \log {\bf \tilde S}[A_\mathrm{r}]-i \Tr \log \(1- {\bf \tilde S}[A_\mathrm{r}] i q {\bf {\slashed{A}}_\mathrm{a}}\) -i \Tr \log {\bf \tilde S}_A[A_\mathrm{r}]+i \Tr \log \(1- {\bf \tilde S}_A[A_\mathrm{r}] i q {\bf {\slashed{A}}_\mathrm{a}}\)  \, ,
\ee
with 
\be
{\bf {\slashed{A}}_\mathrm{a}}(x,y) = \frac{1}{2}\begin{pmatrix} {{\slashed A}}_\mathrm{a}(x)&0 \\ 0 & {{\slashed A}}_\mathrm{a}(x)\end{pmatrix} \delta^4(x-y) \, .
\ee
We will see this explicitly in perturbative calculations, arising from what amounts to covariant Ward identities. Note that these relations should not be confused with the usual Ward identities, since here they apply to the influence functional which is obtained via a partial path integral rather than the full correlation functions, the latter including photon loops. This distinction becomes particularly relevant in the non-Abelian case where the influence functional will itself depend on ghost fields.

\subsection{Keldysh expansion}

Expanding to second order in the advanced fields we have
\ba
S_{\rm IF}[A_\mathrm{r},A_\mathrm{a}] &=& i \Tr \log {\bf \tilde S}[A_\mathrm{r}] +i \Tr \({\bf \tilde S}[A_\mathrm{r}] i q {\bf \tilde {\slashed{A}_\mathrm{a}}} \)+\frac{i}{2} \Tr \({\bf \tilde S}[A_\mathrm{r}] i q {\bf \tilde {\slashed{A}_\mathrm{a}}} \)^2 + \dots \nn \\
&& -i \Tr \log {\bf \tilde S}_A[A_\mathrm{r}] -i \Tr \({\bf \tilde S}_A[A_\mathrm{r}] i q {\bf \tilde {\slashed{A}_\mathrm{a}}} \)-\frac{i}{2} \Tr \({\bf \tilde S}_A[A_\mathrm{r}] i q {\bf \tilde {\slashed{A}_\mathrm{a}}} \)^2 + \dots \, .
\ea
As discussed in Appendix~\ref{app:measure} the path integral measure plays a crucial role in ensuring the influence functional respects the unitarity condition $S_{\rm IF}[A_\mathrm{r},0]=0$. This is achieved since
\be
\Tr \log {\bf \tilde S}[A_\mathrm{r}]=\Tr \log {\bf \tilde S}_A[A_\mathrm{r}] \, ,
\ee
or equivalently
\be
\det {\bf \tilde S}[A_\mathrm{r}]=\det {\bf \tilde S}_A[A_\mathrm{r}] \, .
\ee
This is easily derived by writing the propagators in Keldysh basis and noting that one of the off-diagonal blocks vanishes so that the determinant is determined by the diagonal blocks, which are composed of the advanced and retarded Green's functions (see for example the analogous discussion around \eqref{detrelation}).
Hence
\be
\label{eq:invariant1}
S_{\rm IF}[A_\mathrm{r},A_\mathrm{a}] =  i \Tr \({\bf \tilde S}[A_\mathrm{r}] i q {\bf \tilde {\slashed{A}_\mathrm{a}}} \)+\frac{i}{2} \Tr \({\bf \tilde S}[A_\mathrm{r}] i q {\bf \tilde {\slashed{A}_\mathrm{a}}} \)^2  - i \Tr \({\bf \tilde S}_A[A_\mathrm{r}] i q {\bf \tilde {\slashed{A}_\mathrm{a}}} \)-\frac{i}{2} \Tr \({\bf \tilde S}_A[A_\mathrm{r}] i q {\bf \tilde {\slashed{A}_\mathrm{a}}} \)^2 + \dots \, .
\ee
To proceed we will drop the explicit mention of the contributions from advanced propagators since as already mentioned, in perturbation theory they serve only to justify the Wick rotation of loop integrals in dimensional regularisation. With this understanding we now take 
\ba
S_{\rm IF}[A_\mathrm{r},A_\mathrm{a}] &=&  i \Tr \({\bf \tilde S}[A_\mathrm{r}] i q {\bf \tilde {\slashed{A}_\mathrm{a}}} \)+\frac{i}{2} \Tr \({\bf \tilde S}[A_\mathrm{r}] i q {\bf \tilde {\slashed{A}_\mathrm{a}}} \)^2 + \dots  .
\ea
A common approximation in the context of the Keldysh expansion is to consider the effectively classical equations of motion for the retarded field that arise from the in-in action. These come from varying the full in-in action with respect to the advanced gauge field and then setting $A_\mathrm{a}=0$ to remove any noise terms. These give
\ba \label{eq:Fret}
\partial^{\nu} F^\mathrm{r}_{\nu\mu}(x)&=&   -\frac{\delta}{\delta A^{\mu}_\mathrm{a}(x)} \( i \Tr \({\bf \tilde S}[A_\mathrm{r}] i q {\bf \tilde {\slashed{A}_\mathrm{a}}} \) \) \\
&=& -J^p_{\mu}(x)+ \int_0^1 \d s \, \frac{1}{(1-s)}(x_\mathrm{f}-x)^{\mu} (\partial^{\nu} J^p_{\nu})\(x_{\rm ret}(s)\) \, ,
\ea
with the effective plasma current
\ba
J^p_{\mu}(x)&=& - \frac{1}{2}q \,  \tr \left[ \gamma_{\mu} (\tilde S^{++}[A_\mathrm{r}](x,x)+\tilde S^{--}[A_\mathrm{r}](x,x))\right] \\
&=& -\frac{1}{2}q \,  \tr \left[ \gamma_{\mu} (\tilde S^{+-}[A_\mathrm{r}](x,x)+\tilde S^{-+}[A_\mathrm{r}](x,x)) \right] \, , 
\ea
with the trace now only over the spinor indices
and
\be
x_{\rm ret}(s)=x-\frac{s}{1-s} (x_\mathrm{f}-x) \, .
\ee
The time ordering cancels in the average of two propagators. 
The unusual non-local term on the RHS of \eqref{eq:Fret} is enforcing that the total current on the RHS to be identically conserved which is a consequence of using a manifestly gauge invariant expansion \eqref{eq:invariant1}. Note that it respects primitive causality since $x_{\rm ret}^0\le x^0$. 

The only reason the conservation of $J^p_{\mu}(x)$ is in question is due to potential divergences. However, the potential divergence comes entirely from the vacuum expectation value and can be incorporated into gauge invariant local counterterms, hence
\be
\partial^{\mu} J^p_{\mu}(x)=0 \, .
\ee
This follows from Noether's theorem as argued in Appendix \ref{App:Noether}.
Hence, we obtain the more familiar equation
\be
\partial^{\mu} F^\mathrm{r}_{\mu\nu}(x)=-J^p_{\nu}(x) \, ,
\ee 
which is what we obtain by working with the form \eqref{eq:invariant2}.
Stated differently, the leading order influence functional is
\be
S_{\rm IF}[A_\mathrm{r},A_\mathrm{a}] = \int \d^4 x \, A^{\mu}_\mathrm{a}(x)J^p_{\mu}[A_\mathrm{r}](x) +{\cal O}(A_\mathrm{a}^2) \, .
\ee
It is common to define a field theory analogue of the Wigner quasi-probability distribution
\be
W_{\alpha \beta}[A_\mathrm{r}](x,p) = \frac{1}{2}\int \d^4 y \, e^{-ip.y} \( \check S^{-+}_{\alpha \beta}[A_\mathrm{r}](x-y/2,x+y/2)+ \check S^{+-}_{\alpha \beta}[A_\mathrm{r}](x-y/2,x+y/2)\) \, ,
\ee
where we have used the fully gauge invariant propagator $\check S$
so that
\be
\int \frac{\d^4 p}{(2 \pi)^4} W_{\alpha \beta}[A_\mathrm{r}](x,p) = \frac{1}{2} \check S_{\alpha \beta}^{-+}[A_\mathrm{r}](x,x)+\frac{1}{2} \check S_{\alpha \beta}^{+-}[A_\mathrm{r}](x,x)=\frac{1}{2}\tilde S_{\alpha \beta}^{-+}[A_\mathrm{r}](x,x)+\frac{1}{2}\tilde S_{\alpha \beta}^{+-}[A_\mathrm{r}](x,x)
\, .
\ee
then the retarded gauge field equations of motion are
\be
\partial^{\mu} F^\mathrm{r}_{\mu\nu}(x)= q \int \frac{\d^4 p}{(2 \pi)^4} \tr\(\gamma_{\nu} W[A_\mathrm{r}](x,p)\) \, .
\ee
The Wigner function $W[A_\mathrm{r}](x,p)$ is the quantum analogue of the classical phase space charge distribution and a quantum transport theory may be developed by considering the equation of motion for $W[A_\mathrm{r}](x,p)$ that follows from the operator equations of motion \eqref{eq:covariantpsi} \cite{Vasak:1987um}. This is advantageous in comparing with the classical limit. We will not take this approach here, but rather construct the propagator directly out of mode functions.

\subsection{Evaluating the plasma current}

In order to actually determine the current $J^p_{\mu}[A_\mathrm{r}]$ and the noise corrections, we need to determine the propagators ${\tilde{\bf S}}[A_\mathrm{r}]$. 
At the operator level this amounts to solving the equation of motion \eqref{eq:covariantpsi}. The general Dirac operators may be written as
\be
\hat \Psi(x) = \int \d \tilde k  \sum_s \left[ u_{k,s}(x) b_s({\bf k}) + v_{k,s}(x) d_s^{\dagger}({\bf k})  \] \, ,
\ee
with
\be
\int \d \tilde k=\int \frac{\d^4 k}{(2 \pi)^4} \,  \theta(k^0) 2 \pi \delta(k^2+m^2) \, ,
\ee
and where $u_{k,s}(x)$ and $v_{k,s}(x)$ satisfy the same equation 
\be
[i \slashed{D}[A_\mathrm{r}]-m] u_{k,s}(x) =0 \, ,   \quad [i \slashed{D}[A_\mathrm{r}]-m] v_{k,s}(x)= 0 \, ,
\ee
with initial condition 
\be
u_{k,s}(\ti,{\bf x}) = u_s({\bf k}) e^{i {\bf k}.{\bf x}} \, , \quad v_{k,s}(\ti,{\bf x}) = v_s({\bf k}) e^{-i {\bf k}.{\bf x}} \, ,
\ee
where $u_s({\bf k})$ and $v_{k,s}$ are the standard Minkowski space momentum space spinors.
With this choice, $b_s({\bf k})$ and $d_s^{\dagger}({\bf k})$ satisfy the standard free Dirac field commutation relations
\be
\{ b_s({\bf k}),b_{s'}^{\dagger}({\bf k}')\}= \{ d_s({\bf k}),d_{s'}^{\dagger}({\bf k}')\}=2\omega_k (2\pi)^3\delta^3({\bf k}-{\bf k}') \delta_{ss'} \, .
\ee
A non-zero gauge field will cause the solutions to scatter so that at finite time they are linear superpositions of positive and negative frequency modes.

Making no other assumptions other than the initial state being Gaussian and factorizable, the general form of the plasma current $J_{\mu}^p(x)$ is
\begin{small}
\ba
J_{\mu}^p(x) &=& q \int \d \tilde k \int \d {\tilde k}' \sum_{s,s'} \( \frac{1}{2}\bar u_{k',s'}(x) \gamma_{\mu} u_{k,s}(x)\Tr[ \rho [ b^{\dagger}_{s'}({\bf k'}), b_s({\bf k})]] +\bar u_{k',s'}(x) \gamma_{\mu}  v_{k,s}(x)\Tr[ \rho b^{\dagger}_{s'}({\bf k'})d_s^{\dagger}({\bf k})] \right. \nn \\
&& \left. +\bar v_{k',s'}(x) \gamma_{\mu}u_{k,s}(x)\Tr[ \rho d_{s'}({\bf k'})b_s({\bf k})] +\frac{1}{2}\bar v_{k',s'}(x) \gamma_{\mu} v_{k,s}(x)\Tr[ \rho [d_{s'}({\bf k'}),d^{\dagger}_s({\bf k})]] \) \, .
\ea
\end{small}\ignorespaces
At present we are interested in states for which there is no symmetry breaking so that $\Tr[ \rho d_{s'}({\bf k'})b_s({\bf k})] =\Tr[ \rho b^{\dagger}_{s'}({\bf k'})d_s^{\dagger}({\bf k})] =0$.

 We can separate this into a manifestly finite part
\be
J_{\mu}^{p,\text{finite}}(x) = q \int \d \tilde k \int \d {\tilde k}' \sum_{s,s'} \( \bar u_{k',s'}(x) \gamma_{\mu} u_{k,s}(x)\Tr[ \rho b^{\dagger}_{s'}({\bf k'})b_s({\bf k})]   -\bar v_{k',s'}(x) \gamma_{\mu} v_{k,s}(x)\Tr[ \rho d^{\dagger}_{s'}({\bf k'})d_s({\bf k})] \) \, , 
\ee
which is easily seen to be conserved by virtue of the equations of motion and a pure vacuum contribution
\be
J_{\mu}^{p,\text{vacuum}}(x) = \frac{1}{2} q \int \d \tilde k \sum_s \( \bar v_{k,s}(x) \gamma_{\mu} v_{k,s}(x)-\bar u_{k,s}(x) \gamma_{\mu} u_{k,s}(x) \) \, .
\ee
the divergences in $J_{\mu}^{p,\text{vacuum}}(x)$ are just the usual ones that arise in the computation of the Euler-Heisenberg Lagrangian which arises from the one-loop vacuum fluctuations in the in-out formalism. Thus
\be
J_{\mu}^{p,\text{vacuum}}(x) =\frac{\delta S_{\text{1-loop vac}}[A_\mathrm{r}]}{\delta A_\mathrm{r}^{\mu}(x)} \, .
\ee
Formally 
\ba
S_{\text{1-loop vac}}[A_\mathrm{r}]& = & \frac{i}{2} \Tr \log[\check S^{++}_{\rm vac}[A_\mathrm{r}]]+\frac{i}{2} \Tr \log[\check S^{--}_{\rm vac}[A_\mathrm{r}]]\\
&=& -\frac{i}{2} \Tr \log [i {\slashed D}[A_\mathrm{r}]-m+i \epsilon]-\frac{i}{2} \Tr \log [i {\slashed D}[A_\mathrm{r}]-m-i \epsilon] \, ,
\ea
and its computation is well known \cite{Schwinger:1951nm}.
This is automatically conserved by gauge invariance
\be
\partial^{\mu}J_{\mu}^{p,\text{vacuum}}(x) =0 \, ,
\ee
the genuine open EFT contribution thus arises from the finite current $J_{\mu}^{p,\text{finite}}(x)$.

\subsection{Eikonal approximation}

To completely determine the propagators we need to know the form of $u_{k,s}(x)$ and $v_{k,s}(x)$. For a general background gauge field this is an impractical task. Fortunately there are many approximation methods. One such approximation is the eikonal one, which is useful to consider since we can see the natural emergence of the Wilson lines, connecting with our earlier discussion.  Rewriting the equations of motion for $u$ in second order form
\be
[-i \slashed{D}[A_\mathrm{r}]-m] [i \slashed{D}[A_\mathrm{r}]-m] u_{k,s}(x) = [-D[A_\mathrm{r}]^2+m^2-\frac{1}{2}q \sigma_{\mu\nu} F^{\mu\nu}(x)]u_{k,s}(x)=0 \, ,
\ee
with the convention $\sigma_{\mu\nu}=\frac{i}{2} [\gamma_{\mu},\gamma_{\nu}]$. The interpretation of $u_{k,s}(x)$ is the wavefunction of a spin 1/2 momentum eigenstate scattering in the background gauge field and so it is natural to factor out of the wavefunction a plane wave part by writing $u_{k,s}(x)=e^{ik.x} U_{k,s}(x)$ with $k^2+m^2=0$ so that
\be
[-2 i k.D[A_\mathrm{r}] -D[A_\mathrm{r}]^2- \frac{1}{2}q \sigma_{\mu\nu} F^{\mu\nu}(x)]U_{k,s}(x) =0 \, .
\ee
The initial condition is now $U_{k,s}(\ti,{\bf x})=u_s({\bf k})$.
The second order covariant derivative term is built out of a term parallel with $k$  {\it i.e.} $(k.D[A_\mathrm{r}])^2/k^2$ and a term orthogonal which in the rest frame $k=(m,0,0,0)$ is the covariant spatial Laplacian. The eikonal approximation amounts to neglecting the $(k.D[A_\mathrm{r}])^2/k^2$ term relative to the first order derivatives $-i k.D[A_\mathrm{r}]$, leaving the covariant spatial Laplacian. The most extreme version of the eikonal approximation amounts to regarding the whole $D[A_\mathrm{r}]^2$ term in this equation as small in comparison to $-ik.D[A_\mathrm{r}]$. For simplicity of analysis we shall consider this case. With this assumption, the Dirac equation reduces to
\be
[-i k.D[A_\mathrm{r}] -\frac{1}{4}q \sigma_{\mu\nu} F^{\mu\nu}(x)]U_{k,s}(x) \simeq 0 \, .
\ee
which can be formally in terms of the initial conditions as
\be
U_{k,s}(x)= e^{iq \int_{\tau_\mathrm{i}}^0 \d \tau \, A_{\mu}(z(\tau)) \tfrac{\d z^{\mu} }{\d \tau} }{\cal P} e^{-\frac{i}{4} q\int_{\tau_\mathrm{i}}^0 \d \tau  \,  \sigma_{\alpha \beta}F^{\alpha \beta}(z(\tau))} u_s({\bf k}) \, ,
\ee
with path
\be
z^{\mu}(\tau)=x^{\mu}+ \tau k^{\mu} \, ,
\ee
designed so that $z^{\mu}(0)=x^{\mu}$ and $z^0(\tau_\mathrm{i})=\ti=t+k^0 \tau_\mathrm{i}$.
The trajectory is clearly that of a free relativistic particle with $\tau$ the proper time as expected \cite{Schwinger:1951nm}. In the limit $\ti \rightarrow -\infty$ we have $\tau_\mathrm{i} \rightarrow -\infty$. Causality is manifest in that $z^0(\tau) \le t$.

This expression simplifies greatly in the limit in which $m^2$ is large in comparison to $q |F|$ for which we may drop the spinor factor
\be
U_{k,s}(x) \simeq  e^{iq \int_{\tau_\mathrm{i}}^0 \d \tau \, A_{\mu}(z(\tau)) \tfrac{\d z^{\mu} }{\d \tau} }  u_s({\bf k}) \,.
\ee
For small spatial momenta $|{\bf k}| \ll m$ we obtain the result 
\be
u_{k,s}(x) \simeq e^{-i m t} V(x) u_s({\bf k}) \, ,
\ee
analogous to the non-relativistic scalar case with $\Gamma$ replaced by $m$ and the inclusion of a spinor wavefunction.  

This procedure can be repeated for the $v_{k,s}(x)$ solutions and so putting this together, the solution of the operator equations for the Dirac field in the background retarded gauge field take the approximate (eikonal) form
\ba
\hat \Psi(x) & \simeq & \int \d \tilde k  \sum_s \left[ e^{i k.x} e^{iq \int_{\tau_i}^0 \d \tau \, A_{\mu}(z(\tau)) \tfrac{\d z^{\mu} }{\d \tau} }{\cal P} e^{-\frac{i}{4} q\int_{\tau_\mathrm{i}}^0 \d \tau    \sigma_{\alpha \beta}F^{\alpha \beta}(z(\tau))} u_s({\bf k}) b_s({\bf k}) \right. \nn \\
&& \left. + e^{-ik.x} e^{iq \int_{\tau_i}^0 \d \tau \, A_{\mu}(z(\tau)) \tfrac{\d z^{\mu} }{\d \tau} }{\cal P} e^{\frac{i}{4} q\int_{\tau_i}^0 \d \tau \,  \sigma_{\alpha \beta}F^{\alpha \beta}(z(\tau))} v_s({\bf k}) d_s^{\dagger}({\bf k})  \] \, .
\ea
We see the natural emergence of a Wilson line for the retarded gauge field that extends from the point $x$ to the initial time which ensures the Dirac fermion has the correct gauge transformation properties.\footnote{Schwinger \cite{Schwinger:1951nm} introduced what we now call Wilson lines for Abelian gauge theories 20 years before Wilson \cite{Wilson:1974sk} demonstrating that they naturally emerge in the calculation of Green's functions in background electromagnetic fields.}

\subsection{Determining the propagators}

Once we know the mode functions $u,v$ to determine the propagators $\tilde {\bf S}[A_\mathrm{r}]$ it is sufficient to know the $\Tr[\rho b^{\dagger} b ]$ etc. combinations. Furthermore, since $A_\mathrm{a}=0$, all of the propagators can be determined from the Wightman function $\tilde S^{-+}[A_\mathrm{r}]_{\alpha \beta}(x,y)$ alone by appropriate time-ordering. Explicitly we have
\ba
\tilde S^{-+}[A_\mathrm{r}](x,y) &=& \Tr \{ \rho \Psi(x) \bar \Psi(y) \} \\
&=& \int \d \tilde k\int \d {\tilde k}' \sum_{s,s'}  \left[ u_{k,s}(x) \bar u_{k',s'}(y) \Tr[\rho b_s({\bf k}) b_{s'}^{\dagger}({\bf k'})]+v_{k,s}(x) \bar v_{k',s'}(y) \Tr[\rho d_s^{\dagger}({\bf k}) d_{s'}({\bf k'})]  \right] \nn \, .
\ea
Evaluating this at the initial time we have
\ba
 {\tilde S}^{-+}(\ti,{\bf x},\ti,{\bf y}) & = & \int \d \tilde k\int \d {\tilde k}' \sum_{s,s'} \left[ e^{-i( \omega_k-\omega_{k'})\ti }e^{i {\bf k}.{\bf x}- i{\bf k}'.{\bf y}} u_{s}({\bf k}) \bar u_{s'}({\bf k}') \Tr[\rho b_s({\bf k}) b_{s'}^{\dagger}({\bf k'})] \nn \right. \\
&& +  \left.    e^{i( \omega_k-\omega_{k'})\ti }e^{-i {\bf k}.{\bf x}+ i{\bf k}'.{\bf y}}v_{s}({\bf k})\bar v_{s'}({\bf k}') \Tr[\rho d_s^{\dagger}({\bf k}) d_{s'}({\bf k'})] \] \, .
\ea
To make gauge invariance manifest we need to accommodate a possible non-zero gauge field on the initial surface and so we identify 
\be \label{eq:initialprop}
\tilde S^{-+}(\ti,{\bf x},\ti,{\bf y})={\cal W}_{C_3}(\ti,{\bf x},\ti,{\bf y}) {\check S}^{-+}({\bf x},{\bf y}) \, ,
\ee
with ${\check S}^{-+}({\bf x},{\bf y})$ now gauge invariant. A suitable choice would for example the initial propagator for a translation invariant state, {\it eg.} 
\be
{\check S}^{-+}({\bf x},{\bf y})=\int \d \tilde k \sum_{s} \left[  e^{i {\bf k}.({\bf x}-{\bf y})} u_{s}({\bf k}) \bar u_{s}({\bf k}) (1-n_{+s}({\bf k}))+e^{-i {\bf k}.({\bf x}-{\bf y})}v_{s}({\bf k})\bar v_{s}({\bf k}) n_{-s}({\bf k})] \] \, ,
\ee
although this is by no means the only consistent choice, and this choice would not be invariant under gauge transformations on the initial time surface. The precise form of $\Tr[\rho b_s({\bf k}) b_{s'}^{\dagger}({\bf k'})]$ etc can be inferred from ${\check S}^{-+}({\bf x},{\bf y})$ by double Fourier transforming \eqref{eq:initialprop} and matching spinor coefficients. Note that in the presence of a retarded gauge field on the initial surface we cannot assume that $\Tr[\rho b_s({\bf k}) b_{s'}^{\dagger}({\bf k'})] \propto \delta^3({\bf k}-{\bf k}')$ since translation invariance may be broken. However, in order for the limit $\ti \rightarrow -\infty$ to be well defined we would require 
\ba
&& \Tr[\rho b_{s'}^{\dagger}({\bf k'}) b_s({\bf k}) ] ={\cal M}^b_{s s'}({\bf k'},{\bf k})  \delta(\omega_k-\omega_{k'}) \, ,\\
&& \Tr[\rho d_{s'}^{\dagger}({\bf k}) d_s({\bf k'}) ]={\cal M}^d_{s s'}({\bf k'},{\bf k}) \delta(\omega_k-\omega_{k'}) \, , 
\ea
so that all temporal oscillatory factors vanish, which is clearly a statement about energy conservation for asymptotic states. With this choice the Wightman function at general times is completely specified by the initial conditions encoded in ${\cal M}^b_{s s'}({\bf k'},{\bf k})$ and ${\cal M}^d_{s s'}({\bf k'},{\bf k})$
\ba
 && {\tilde S}^{-+}(x,y) ={\tilde S}_{\rm vac}^{-+}(x,y)+\\
 && \int \d \tilde k\int \d {\tilde k}' \sum_{s,s'} \left[- u_{k,s}(x) \bar u_{k',s'}(y) {\cal M}^b_{s s'}({\bf k'},{\bf k}) \delta(\omega_k-\omega_{k'}) +   v_{k,s}(x) \bar v_{k',s'}(y) {\cal M}^d_{s s'}({\bf k'},{\bf k}) \delta(\omega_k-\omega_{k'}) \] \, . \nn
\ea
The finite part of plasma current becomes 
\begin{small}
\be
J_{\mu}^{p, \text{finite}}(x) = q \int \d \tilde k \int \d {\tilde k}' \sum_{s,s'} \Big[ \bar u_{k',s'}(x) \gamma_{\mu} u_{k,s}(x){\cal M}^b_{s s'}({\bf k'},{\bf k}) -\bar v_{k',s'}(x) \gamma_{\mu} v_{k,s}(x){\cal M}^d_{s s'}({\bf k'},{\bf k}) \Big] \delta(\omega_k-\omega_{k'}) \, .
\ee
\end{small}\ignorespaces
Denoting the eikonal factor (with $\ti=-\infty$)
\be
{\cal U}_{\pm}(k,x)=e^{iq \int_{-\infty}^0 \d \tau k^{\mu} A_{\mu}(x+k \tau) }{\cal P} e^{\mp \frac{i}{4} q\int_{-\infty}^0 \d \tau    \sigma_{\alpha \beta}F^{\alpha \beta}(x+k \tau)} \, ,
\ee
then at leading order in the eikonal approximation the finite part of the plasma current is 
\ba
J_{\mu}^{p, \text{finite}}(x) &=& q \int \d \tilde k \int \d {\tilde k}' \sum_{s,s'} \left[ \bar u_{s'}({\bf k}'){\cal U}_{+}^{\dagger}(k',x) \gamma_{\mu} {\cal U}_{+}(k,x) u_s({\bf k}){\cal M}^b_{s s'}({\bf k'},{\bf k}) \right. \nn \\
&& \left. -\bar v_{s'}({\bf k}'){\cal U}_{-}^{\dagger}(k',x) \gamma_{\mu} {\cal U}_{-}(k,x) v_s({\bf k}){\cal M}^d_{s s'}({\bf k'},{\bf k}) \right] \delta(\omega_k-\omega_{k'})  \, . 
\ea
When the initial state is translation invariant and diagonal in spins
\be
\Tr[\rho b_{s'}^{\dagger}({\bf k'}) b_s({\bf k}) ]= \delta_{ss'} 2 \omega_k (2\pi)^3 \delta^3({\bf k}-{\bf k}') n_{+s}({\bf k}) \, , \quad \Tr[\rho d_{s'}^{\dagger}({\bf k'}) d_s({\bf k}) ]= \delta_{ss'}  2 \omega_k (2\pi)^3 \delta^3({\bf k}-{\bf k}') n_{-s}({\bf k}) \, ,
\ee
this simplifies greatly to
\begin{small}
\be
J_{\mu}^{p, \text{finite}}(x) = q \int \d \tilde k \sum_s   \(  \bar u_{s}({\bf k}){\cal U}_{+}^{\dagger}(k,x) \gamma_{\mu} {\cal U}_{+}(k,x) u_s({\bf k})n_{+s}({\bf k})  -\bar v_{s}({\bf k}){\cal U}_{-}^{\dagger}(k,x) \gamma_{\mu} {\cal U}_{-}(k,x) v_s({\bf k})n_{-s}({\bf k}) \)  \, . 
\ee
\end{small}\ignorespaces
Together with the usual vacuum contribution which can be calculated by conventional in-out means, this completes the specification of the influence functional at leading order in the Keldysh expansion for arbitrary retarded gauge fields, in the eikonal approximation.

\subsection{Evaluating the Noise}

The plasma current encodes the dissipative effects from an electron plasma on the photon propagation at first order in the Keldysh expansion. The noise (`Brownian motion') induced by the environment is contained in the second and higher order terms in the Keldysh expansion. In the present context this is (truncating at second order)
\be
S_{\rm noise}[A_\mathrm{r},A_\mathrm{a}] = \frac{i}{2} \Tr \({\bf \tilde S}[A_\mathrm{r}] i q {\tilde{\bf  {\slashed{A}}_\mathrm{a}} }\)^2 \, .
\ee
We can write this as
\be
S_{\rm noise}[A_\mathrm{r},A_\mathrm{a}] = \frac{i}{2} \int \d^4 x \int \d^4 y  \tilde A^{\mu}_\mathrm{a}(x) N_{\mu \nu}[A_\mathrm{r}](x,y) \tilde A^{\nu}_\mathrm{a}(y)  \, .
\ee
where the symmetric noise Kernel takes the explicit form 
\begin{small}
\ba
N_{\mu \nu}[A_\mathrm{r}](x,y) &=& -\frac{1}{4} q^2 \left(  \tr\[ \tilde S_{++}[A_\mathrm{r}](y,x) \gamma_{\mu} \tilde S_{++}[A_\mathrm{r}](x,y)\gamma_{\nu} \] +\tr\[ \tilde S_{--}[A_\mathrm{r}](y,x) \gamma_{\mu} \tilde S_{--}[A_\mathrm{r}](x,y)\gamma_{\nu} \]\right.  \nn \\
&& + \left.   \tr\[ \tilde S_{+-}[A_\mathrm{r}](y,x) \gamma_{\mu} \tilde S_{-+}[A_\mathrm{r}](x,y)\gamma_{\nu} \] +\tr\[ \tilde S_{-+}[A_\mathrm{r}](y,x) \gamma_{\mu} \tilde S_{+-}[A_\mathrm{r}](x,y)\gamma_{\nu} \] \right) \, , 
\ea
\end{small}\ignorespaces
with the trace now taken only over the spinor indices. This has the desired symmetry
\be
N_{\mu \nu}[A_\mathrm{r}](x,y)=N_{\nu \mu}[A_\mathrm{r}](y,x) \,,
\ee
implicit in its definition.
By careful consideration of the two possible time orderings, $x  \prec y$, $y \prec x$ the noise can be written entirely in terms of Wightman functions
\be
N_{\mu \nu}[A_\mathrm{r}](x,y) = -\frac{q^2}{2} \left(    \tr\[ \tilde S_{+-}[A_\mathrm{r}](y,x) \gamma_{\mu} \tilde S_{-+}[A_\mathrm{r}](x,y)\gamma_{\nu} \] +\tr\[ \tilde S_{-+}[A_\mathrm{r}](y,x) \gamma_{\mu} \tilde S_{+-}[A_\mathrm{r}](x,y)\gamma_{\nu} \] \right)  \, .
\ee
This noise is manifestly invariant under retarded gauge transformations and further satisfies the Ward-like identity (as a consequence of Noether's theorem \ref{App:Noether})
\be
\partial^{\mu}_x N_{\mu \nu}[A_\mathrm{r}](x,y)= D^{\mu}_x[A_\mathrm{r}] N_{\mu \nu}[A_\mathrm{r}](x,y)=0 \, .
\ee
To explicitly demonstrate this we use the covariant equations for the Green's functions
\ba
 [i \slashed{D}_x[A_\mathrm{r}]-m]\tilde S_{+-}(x,y)=[i \slashed{D}_x[A_\mathrm{r}]-m]\tilde S_{-+}(x,y)&=&0 \, , \\
\tilde S_{+-}(x,y)[-i \overleftarrow{\slashed{D}}_y[A_\mathrm{r}]-m]=\tilde S_{-+}(x,y)[-i \overleftarrow{\slashed{D}}_y[A_\mathrm{r}]-m] &= & 0 \, .
\ea
This ensures that we can replace $\tilde A_\mathrm{a}$ with $A_\mathrm{a}$, {\it i.e.}~makes manifest gauge invariance under advanced gauge transformations
\be
S_{\rm noise}[A_\mathrm{r},A_\mathrm{a}] = \frac{i}{2} \int \d^4 x \int \d^4 y \; A^{\mu}_\mathrm{a}(x) N_{\mu \nu}[A_\mathrm{r}](x,y)  A^{\nu}_\mathrm{a}(y)  \, .
\ee

\subsection{Spinor QED \`a la Caldeira-Leggett}

 Spinor QED itself may be regarded as an open system. A suitable model of an environment are massive charged Dirac fermions $\chi_I$ which are themselves in some non-vacuum state. If these fermions are coupled to the electron they will induce explicit open system effects for the electron already at tree level and for the photon at one-loop level. Consider then the action
\begin{eqnarray}
S[A,\Psi,\overline{\Psi}, \chi_I] = S_{\mathcal{S}}[ A, \Psi,\overline{\Psi} ] + S_{\mathcal{E}}[ \chi_I ]  + S_{\mathrm{int}}[A,\Psi,\overline{\Psi}, \chi_I] \, ,
\end{eqnarray}
where we have the splitting
\begin{eqnarray}
S_{\mathcal{S}}[ A, \Psi,\overline{\Psi} ] & = & \int \exd^4 x \; \bigg[ - \frac{1}{4} F_{\mu\nu} F^{\mu\nu} + i \overline{\Psi} \slashed{D} \Psi - m \overline{\Psi} \Psi \bigg] \, , \\
S_{\mathcal{E}}[ \chi_I ] & = & \sum_I \int \exd^4 x \; \bigg[ i \overline{\chi}_I \slashed{D} \chi_I - M_I \bar{\chi}_I \chi_I \bigg] \, ,\\
S_{\mathrm{int}}[A,\Psi,\overline{\Psi}, \chi_I] & = & \sum_I \int \exd^4 x \; g_I \Big( \overline{\Psi} \chi_I + \overline{\chi}_I \Psi \Big) \, .
\end{eqnarray}
As the above notation implies, the idea is that we integrate out the fermion $\chi_I$, treating it as the environment in the influence functional. We shall continue in this section to consider the case where there is no SSB so that only the $\chi \bar \chi$ propagators are non-zero.

The calculation of the influence functional in this theory is straightforward, since the path integral over the fermion bath is Gaussian, we may perform it explicitly as before, resulting in separate tree level and one-loop contributions 
\begin{eqnarray}
S_{\mathrm{IF}}[A_{+},\Psi_{+},\overline{\Psi}_{+},A_{-},\Psi_{-},\overline{\Psi}_{-}]  & = & \int \exd^4 x \int \exd^4 y \; \sum_I  i g_I^2 \begin{pmatrix} \overline{\Psi}_{+}(x) & -\overline{\Psi}_{-}(x) \end{pmatrix} {\bf S}_I[A_+,A_-](x,y) \begin{pmatrix} \Psi_{+}(y)  \nn \\ -\Psi_{-}(y) \end{pmatrix} \\
&& + i \sum_I \Tr[\log {\bf S}_I[A_+,A_-] ]-i \sum_I \Tr[\log {\bf S}_{I A}[A_+,A_-] ] \, ,
\end{eqnarray}
with ${\bf S}_I[A_+,A_-]$ the matrix of CTP time ordered propagators of $\chi_I$ with CTP Feynman boundary conditions and ${\bf S}_{I A}[A_+,A_-]$ the analogous with advanced boundary conditions. Performing the field redefinition \eqref{eq:fieldredef1} that removes the gauge dependence in the advanced field we can rewrite this as
\begin{eqnarray}
S_{\mathrm{IF}} & = & \int \exd^4 x \int \exd^4 y \; \sum_I  i g_I^2 \begin{pmatrix} \overline{\tilde \Psi}_{+}(x) & -\overline{\tilde \Psi}_{-}(x) \end{pmatrix} \(1-{\tilde{\bf S}}_I[A_\mathrm{r}] i q \tilde{\slashed{A}}_\mathrm{a} \)^{-1}{\tilde{\bf S}}_I[A_\mathrm{r}](x,y) \begin{pmatrix} \tilde \Psi_{+}(y)  \nn \\ -\tilde \Psi_{-}(y) \end{pmatrix} \\
&& + i \sum_I \Tr \[\log {\tilde{\bf S}}_I[A_\mathrm{r}] \] -i\sum_I \Tr \[\log \(1-{\tilde{\bf S}}_I[A_\mathrm{r}] i q \tilde{\slashed{A}}_\mathrm{a} \) \] \, \nn \\
&&- i \sum_I \Tr \[\log {\tilde{\bf S}}_{I A}[A_\mathrm{r}] \] +i\sum_I \Tr \[\log \(1-{\tilde{\bf S}}_{I A}[A_\mathrm{r}] i q \tilde{\slashed{A}}_\mathrm{a} \) \] \, .
\end{eqnarray}
Performing the Keldysh expansion to second order in the advanced variables we have
\begin{eqnarray}
S_{\mathrm{IF}} & = & \frac{1}{2}\int \exd^4 x \int \exd^4 y \; \sum_I  i g_I^2 \begin{pmatrix} \overline{\tilde \Psi}_\mathrm{r}(x) & -\overline{\tilde \Psi}_\mathrm{r}(x) \end{pmatrix} {\tilde{\bf S}}_I[A_\mathrm{r}](x,y)  \begin{pmatrix} \tilde \Psi_\mathrm{a}(y) \\  \tilde \Psi_\mathrm{a}(y) \end{pmatrix}  \nn \\
& & + \frac{1}{2} \int \exd^4 x \int \exd^4 y \; \sum_I  i g_I^2 \begin{pmatrix} \overline{\tilde \Psi}_\mathrm{a}(x) & \overline{\tilde \Psi}_\mathrm{a}(x) \end{pmatrix} {\tilde{\bf S}}_I[A_\mathrm{r}](x,y)   \begin{pmatrix} \tilde \Psi_\mathrm{r}(y) \\  -\tilde \Psi_\mathrm{r}(y) \end{pmatrix} \nn \\
& & + \int \exd^4 x \int \exd^4 y \; \sum_I  i g_I^2 \begin{pmatrix} \overline{\tilde \Psi}_\mathrm{r}(x) & -\overline{\tilde \Psi}_\mathrm{r}(x) \end{pmatrix} {\tilde{\bf S}}_I[A_\mathrm{r}] i q \tilde{\slashed{A}}_\mathrm{a} {\tilde{\bf S}}_I[A_\mathrm{r}](x,y) \begin{pmatrix} \tilde \Psi_\mathrm{r}(y) \\  -\tilde \Psi_\mathrm{r}(y) \end{pmatrix}  \nn \\
& & + \frac{1}{2}\int \exd^4 x \int \exd^4 y \; \sum_I  i g_I^2 \begin{pmatrix} \overline{\tilde \Psi}_\mathrm{a}(x) & \overline{\tilde \Psi}_\mathrm{a}(x) \end{pmatrix} {\tilde{\bf S}}_I[A_\mathrm{r}] i q \tilde{\slashed{A}}_\mathrm{a} {\tilde{\bf S}}_I[A_\mathrm{r}](x,y) \begin{pmatrix} \tilde \Psi_\mathrm{r}(y)  \nn \\ -\tilde \Psi_\mathrm{r}(y) \end{pmatrix} \\
&& + \frac{1}{2} \int \exd^4 x \int \exd^4 y \; \sum_I  i g_I^2 \begin{pmatrix} \overline{\tilde \Psi}_\mathrm{r}(x) & -\overline{\tilde \Psi}_\mathrm{r}(x) \end{pmatrix} {\tilde{\bf S}}_I[A_\mathrm{r}] i q \tilde{\slashed{A}}_\mathrm{a} {\tilde{\bf S}}_I[A_\mathrm{r}](x,y) \begin{pmatrix} \tilde \Psi_\mathrm{a}(y)  \nn \\ \tilde \Psi_\mathrm{a}(y) \end{pmatrix} \\
&& + \int \exd^4 x \int \exd^4 y \; \sum_I  i g_I^2 \begin{pmatrix} \overline{\tilde \Psi}_\mathrm{r}(x) & -\overline{\tilde \Psi}_\mathrm{r}(x) \end{pmatrix} \(  {\tilde{\bf S}}_I[A_\mathrm{r}] i q \tilde{\slashed{A}}_\mathrm{a} \)^2 {\tilde{\bf S}}_I[A_\mathrm{r}](x,y) \begin{pmatrix} \tilde \Psi_\mathrm{r}(y)  \nn \\ -\tilde \Psi_\mathrm{r}(y) \end{pmatrix} \\
&& + i\sum_I \Tr[{\tilde{\bf S}}_I[A_\mathrm{r}] i q \tilde{\slashed{A}}_\mathrm{a} ]+\frac{i}{2}\sum_I \Tr \[\({ \tilde {\bf S}}_I[A_\mathrm{r}] i q \tilde{\slashed{A}}_\mathrm{a} \)^2\] + \dots \, \nn \\
&& - i\sum_I \Tr[{\tilde{\bf S}}_{I A}[A_\mathrm{r}] i q \tilde{\slashed{A}}_\mathrm{a} ]-\frac{i}{2}\sum_I \Tr \[\({ \tilde {\bf S}}_{I A}[A_\mathrm{r}] i q \tilde{\slashed{A}}_\mathrm{a} \)^2\] + \dots \, .
\end{eqnarray}
Although cumbersome, each term is straightforward to calculate for an arbitrary initial state for the fermion bath. Furthermore each term in the expansion is manifestly covariant under retarded gauge transformations and invariant under advanced gauge transformations. Nevertheless if we take the decoupling limit $q \rightarrow 0$ of this expression we obtain
\be
\lim_{q \rightarrow 0} S_{\mathrm{IF}}[A_{+},\Psi_{+},\overline{\Psi}_{+},A_{-},\Psi_{-},\overline{\Psi}_{-}]  = \int \exd^4 x \int \exd^4 y \; \sum_I  i g_I^2 \begin{pmatrix} \overline{\Psi}_{+}(x) & -\overline{\Psi}_{-}(x) \end{pmatrix} {\tilde{\bf S}}_I[0](x,y) \begin{pmatrix}  \Psi_{+}(y)  \nn \\ -\Psi_{-}(y) \end{pmatrix}  \, ,
\ee
which is manifestly invariant under diagonal (retarded) global transformations but breaks the advanced global transformations.

\section{Thermal Scalar QED}
\label{sec:thermal}

To illustrate the in-in calculations for a system with no SSB, in this section we will consider scalar QED with $V(\Phi^{\ast} \Phi) = m^2 \Phi^{\ast} \Phi$ at finite temperature. The natural splitting of the BRST scalar QED action is then clearly 
\begin{eqnarray}
S_{\mathcal{S}}[A,c,\overline{c}] & = & - \int \exd^4 x \; \bigg[ \frac{1}{4} F_{\mu\nu} F^{\mu\nu} + \frac{(\partial_{\mu} A^{\mu})^2}{2\xi}  + \bar{c} \hspace{0.3mm} \Box c \bigg] \, , \label{SS_sec4} \\
S_{\mathcal{E}}[\Phi,\Phi^{\ast}] & = & - \int \exd^4 x \; \bigg[ \partial_{\mu} \Phi^{\ast}  \partial^\mu \Phi +  m^2 \Phi^{\ast} \Phi \bigg] \, ,\label{SE_sec4}
\end{eqnarray} 
and the interaction being
\begin{eqnarray} \label{QEDactionINT}
S_{\mathrm{int}}[A,\Phi] = -  \int \exd^4 x\; \bigg[ i q A^{\mu} \big( \Phi^{\ast} \partial_{\mu} \Phi - \Phi \partial_{\mu} \Phi^{\ast} \big) + q^2 A_{\mu} A^{\mu} \Phi^{\ast} \Phi \bigg] \; . \quad
\end{eqnarray}
We also assume the uncorrelated form of Eq.~(\ref{uncorr}), not specifying the photon initial state $\varrho_{\mathcal{S}\mathrm{i}}$, but taking the charged matter to begin in a thermal state at the initial time. We stress that thermality is not important, and any mixed state could be chosen by utilizing the appropriate occupation number densities $n_{\pm}(k)$. However, the thermal case has been most studied in the literature, usually as a toy model of QCD.

Note that if we were to take the time integrals from a finite initial time, it would be necessary to include interactions arising from the initial state, which can be encoded in an imaginary time segment extension of the CTP \cite{mills1969propagators}. Taking $t_i \rightarrow -\infty$, these contributions factorise from the path integral, and it is sufficient to account for thermality at the level of the free theory Green's functions \cite{Landsman:1986uw}.

We perturb the influence functional in the small-coupling limit $q \ll 1$ using the definition (\ref{SIF_def}). In the present example, integrating over the environment with Eq.~(\ref{SE_sec4}) generates loops built from free ($A_+=A_-=0$) Feynman (time-ordered) and Wightman correlators of the scalar field in the thermal state $\varrho_{\beta}$
together with their complex conjugates. Explicitly,
\begin{eqnarray}
\mathcal{F}^{\beta}(x,y) & \equiv & G_{++}(x,y) \Big|_{A_+=A_-=0, \rho=\varrho_{\beta}}= \int \frac{\exd^4 k}{(2\pi)^4} \;  \mathcal{F}^{\beta}(k) e^{i k \cdot (x - y)} \ , \label{Feynman_def} \\
\mathcal{W}^\beta(x,y) & \equiv & G_{-+}(x,y)\Big|_{A_+=A_-=0, \rho=\varrho_{\beta}} =  \ \int \frac{\exd^4 k}{(2\pi)^4} \; \mathcal{W}^{\beta}(k) e^{i k \cdot (x - y)}  \ , \label{Wightman_def}
\end{eqnarray}
where
\begin{equation} \label{freeprop_thermal}
\mathcal{F}^{\beta}(k) = \frac{- i }{k^2  + m^2 - i \epsilon} + \frac{ 2 \pi \delta(k^2 + m^2) }{ e^{\beta |k_0|}  - 1  } \quad \mathrm{and} \quad \mathcal{W}^{\beta}(k) = 2 \pi \delta(k^2 + m^2) \bigg[ \theta(k^0) + \frac{1}{ e^{\beta |k_0|}  - 1  }   \bigg] \ . 
\end{equation}

\begin{figure}[h]
\begin{center}
\includegraphics[width=120mm]{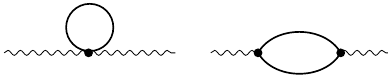}
\caption{\small The topology of loop diagrams in the influence functional (\ref{thermalIF_loops}). Both the left and right diagrams correspond to the loops in Eq.~(\ref{Pibeta}) involving Feynman propagators. Meanwhile the rightmost topology, involving Wightman functions, contributes to the loop in Eq.~(\ref{Nbeta}).} \label{Fig:thermalloops}
\end{center}
\end{figure}

Truncating the influence functional at second order in $q$ yields a functional that is quadratic in the gauge field. For this reason, it is most transparent to express $S_{\mathrm{IF}}$ in momentum space 
\begin{eqnarray}
S_{\mathrm{IF}}[ A_{+}, A_{-} ] & \simeq & \frac{q^2}{2} \int \frac{ \mathrm{d}^4k}{(2\pi)^4} \; \bigg[ - A_{+}^\mu(k) \Pi^{\beta}_{\mu\nu}(k) A_{+}^\nu(-k) - i A_{+}^\mu(k) \mathcal{N}^{\beta}_{\mu\nu}(-k) A_{-}^\nu(-k) \label{thermalIF_loops} \\
&& \hspace{35mm} - i A_{-}^\mu(k) \mathcal{N}^{\beta}_{\mu\nu}(k) A_{+}^\nu(-k) + A_{-}^\mu(k) \Pi^{\beta \ast}_{\mu\nu}(k) A_{-}^\nu(-k) \bigg] + \mathcal{O}(q^4) \, , \notag
\end{eqnarray}
with the thermal loops (see Figure \ref{Fig:thermalloops}) defined as
\begin{eqnarray}
\Pi^{\beta}_{\mu\nu}(k) &\equiv& 2 \eta_{\mu\nu} \int \frac{\exd^4 \ell}{(2\pi)^4}  \big[ \mathcal{F}^{\beta}(\ell) \big] - 2 i \int \frac{\exd^4 \ell}{(2\pi)^4} \big[ \mathcal{F}^{\beta}(\ell) \big] \big[ \mathcal{F}^{\beta}(-\ell-k) \big] ( 2 \ell_{\mu} + k_\mu) \ell_{\nu} \, ,\qquad \label{Pibeta} \\
\mathcal{N}^{\beta}_{\mu\nu}(k) & \equiv  & 2 \int \frac{\exd^4 \ell}{(2\pi)^4}  \mathcal{W}^{\beta}(\ell) \mathcal{W}^{\beta}(-\ell-k) ( 2 \ell_{\mu} + k_\mu) \ell_{\nu} \, ,\label{Nbeta}
\end{eqnarray}
with the derivation detailed in Appendix \ref{App:thermalQED}. 
Note that $\Pi^{\beta}_{\mu\nu}(k)$ is the familiar photon self-energy, and therefore unsurprisingly contains explicit UV divergences identical to those in the vacuum calculation (see Appendix~\ref{App:thermalVac}). By contrast, the loop $\mathcal{N}^{\beta}_{\mu\nu}(k)$ is free of divergences. Consequently, the renormalisation of the one-loop influence functional proceeds in exactly the same way as in the standard in-out formalism: all divergences are absorbed into the field-strength renormalisation proportional to $q^2 F^2$, ensuring that the photon remains massless to all orders in perturbation theory (see (\ref{selfenergy_c}) for the precise condition).

Since the influence functional involves two copies of the gauge symmetry, the Ward identities require that $k^\mu \Pi^{\beta}_{\mu\nu}(k) = k^\mu \mathcal{N}^{\beta}_{\mu\nu}(k) = 0$. For a Lorentz-invariant state, this condition would imply that the loop corrections are proportional to $k^2 \eta_{\mu\nu} - k_\mu k_\nu$, as indeed occurs in the vacuum case (see Appendix~\ref{App:thermalVac}). However, in the present setup the thermal state introduces a preferred rest frame, breaking boost invariance. As a result, the tensor structure $k^2 \eta_{\mu\nu} - k_\mu k_\nu$ decomposes into independent transverse and longitudinal components, satisfying \cite{Weldon:1982aq,Creminelli:2024lhd}
\begin{equation} \label{PLPT_sum}
\mathcal{P}_{\mu\nu}^{\mathrm{L}}(k) + \tfrac{k^2}{|\mathbf{k}|^2} \mathcal{P}_{\mu\nu}^{\mathrm{T}}(k) = k^2 \eta_{\mu\nu} - k_{\mu} k_{\nu} \ .
\end{equation}
To define these projections let $u^{\mu} = \delta^{\mu}_0$ denote the frame vector of the bath (satisfying $u^2 = -1$ and $u \cdot k = -k_0$). One then projects $k^2 \eta_{\mu\nu} - k_{\mu} k_{\nu} $ onto the frame vector $u_{\mu}$ and constructs the longitudinal component
\begin{equation}
\label{longitcomp}
\mathcal{P}^{\mathrm{L}}_{\mu\nu}(k) = k^2 \;  \frac{ ( u_{\mu} - \tfrac{u \cdot k}{k^2} k_\mu ) ( u_{\nu} - \tfrac{u \cdot k}{k^2} k_\nu ) }{ \big( u - \tfrac{u \cdot k}{k^2} k \big)^2 } \ ,
\end{equation}
and the transverse component then follows from (\ref{PLPT_sum}). Both operators are consistent with gauge invariance in the sense that $k^{\mu} \mathcal{P}_{\mu\nu}^{\mathrm{L,T}} = 0$. As matrices these take the simple forms
\begin{equation}
\left[ \begin{matrix} \mathcal{P}^{\mathrm{L}}_{00} & \mathcal{P}^{\mathrm{L}}_{i0} \\
\mathcal{P}^{\mathrm{L}}_{0j} &  \mathcal{P}^{\mathrm{L}}_{ij} \end{matrix} \right] = \left[  \begin{matrix} - |\mathbf{k}|^2 & - k_i k_0  \\
- k_0 k_j  & - k_i k_j  \cdot \frac{k_0^2}{|\mathbf{k}|^2}  \end{matrix}  \right] \qquad \mathrm{and}\, , \qquad \left[ \begin{matrix} \mathcal{P}^{\mathrm{T}}_{00} & \mathcal{P}^{\mathrm{T}}_{i0} \\
\mathcal{P}^{\mathrm{T}}_{0j} &  \mathcal{P}^{\mathrm{T}}_{ij} \end{matrix} \right] = \left[  \begin{matrix} 0 & 0 \\
0 & |\mathbf{k}|^2 \delta_{ij} - k_i k_j  \end{matrix}  \right] \ . \label{PTPL_matrices}
\end{equation}
Of course there are different choices that can be used, but this basis is particularly useful since $\mathcal{P}_{\mu\nu}^{\mathrm{L,T}}$ are orthonormal projections in the sense that
\begin{equation}
\label{orthon}
\eta^{\nu\rho} \mathcal{P}_{\mu \nu}^{\mathfrak{p}}(k) \mathcal{P}_{\rho \sigma}^{\mathfrak{q}}(k) = \delta^{\mathfrak{pq}} \mathcal{P}_{\mu \sigma}^{\mathfrak{p}}(k)\, , \qquad \mathrm{for} \ \mathfrak{p}, \mathfrak{q} = \mathrm{L}, \mathrm{T} \ .
\end{equation}
After renormalisation, the influence functional (\ref{thermalIF_loops}) takes the form
\begin{eqnarray}
\label{IFscalarQEDra}
S_{\mathrm{IF}}[ A_{+}, A_{-} ] & \simeq & \frac{q^2}{2} \sum_{\mathfrak{p} = \mathrm{L}, \mathrm{T}} \int \frac{ \mathrm{d}^4k}{(2\pi)^4} \; \mathcal{P}^{\mathfrak{p}}_{\mu\nu}(k)  \bigg[ - A_{+}^\mu(k) \Pi_{\mathfrak{p}}^{\beta}(k) A_{+}^\nu(-k) - i A_{+}^\mu(k) \mathcal{N}_{\mathfrak{p}}^{\beta}(-k) A_{-}^\nu(-k) \label{thermalIF_ans} \\
&& \hspace{40mm} - i A_{-}^\mu(k) \mathcal{N}_{\mathfrak{p}}^{\beta}(k) A_{+}^\nu(-k) + A_{-}^\mu(k) \Pi^{\beta \ast}_{\mathfrak{p}}(k) A_{-}^\nu(-k) \bigg] + \mathcal{O}(q^4)  \, .\notag
\end{eqnarray}
This expression makes explicit how the renormalised loop corrections decompose into longitudinal and transverse parts, and shows how all interactions between branches respect gauge invariance. The full calculation of the functions $\Pi_{\mathrm{L},\mathrm{T}}^{\beta}(k)$ and $\mathcal{N}_{\mathrm{L},\mathrm{T}}^{\beta}(k)$ are performed in Appendix \ref{App:thermal_corr}, with full expressions given there. Note that one finds explicitly
\begin{eqnarray}
2\mathrm{Im}[\Pi^{\beta}_{\mathrm{L,T}}(k)] + \mathcal{N}^{\beta}_{\mathrm{L,T}}(k) + \mathcal{N}^{\beta}_{\mathrm{L,T}}(-k) & = & 0 \ .
\end{eqnarray}
This relation is the perturbative analogue of the condition discussed in \S\ref{sec:rabasis}, which enforces that the advanced field does not propagate in the retarded/advanced basis. This is equivalent to the requirement $S_{\mathrm{IF}}[\varphi,\varphi] =0$ which ensures preservation of the trace of the reduced density matrix.

One finds that $\mathcal{N}_{\mathrm{L}, \mathrm{T}}^{\beta}$ are purely real. For example,
\begin{small}
\begin{eqnarray}
\mathcal{N}^{\beta}_{\mathrm{L}}(k) & = & - \frac{\theta(-k^2 - 4m^2)}{8\pi |\mathbf{k}|^3} \int_{\Omega_{-}}^{\Omega_{+}}\exd \Omega \; \bigg( \theta(-k^0)  \coth\left( \sfrac{\beta\Omega}{2} \right) (2\Omega - |k_0|)^2 +\frac{ 4 \Omega^2 - 2 |k_0| \Omega }{ [ e^{\beta \Omega} - 1 ] [ e^{\beta ( |k_0|-\Omega) }  - 1 ]  } \bigg) \notag \\
&&- \frac{\theta(k^2) }{4\pi |\mathbf{k}|^3} \int_{\Omega_{+}}^{\infty} \exd \Omega \; \bigg( \frac{\theta(k^0)}{ e^{\beta \Omega} - 1 } + \frac{ \theta(-k^0)}{e^{\beta ( \Omega - |k_0| ) } - 1 } +  \frac{1}{[ e^{\beta \Omega}  - 1 ][ e^{\beta ( \Omega - |k_0|)}  - 1 ] } \bigg) (2\Omega - |k_0|)^2 \, , 
\end{eqnarray}
\end{small}\ignorespaces
where\footnote{These variables often appear in perturbative real-time calculations of thermal effects --- see {\it eg.} \cite{Stanford:2015owe,Takeda:2025cye}.}
\begin{eqnarray} \label{Omegapm}
\Omega_{\pm} \equiv \frac{|k_0|}{2} \pm \frac{|\mathbf{k}|}{2} \sqrt{ 1 + \frac{4m^2}{k^2} }  \ .
\end{eqnarray}
The renormalised self-energies $\Pi_{\mathrm{L}, \mathrm{T}}^{\beta}(k)$ have nonzero real parts for all $k$, and exhibit discontinuities proportional to $\theta(-k^2 - 4m^2)$ and $\theta(k^2)$. The $\theta(-k^2 - 4m^2)$ discontinuity is the usual vacuum pair-production threshold (timelike region), here modified by thermal factors; staying below this threshold ({\it eg.} in a Wilsonian low-energy limit) keeps one in a region of analyticity. By contrast, the $\theta(k^2)$ terms corresponds to processes in the spacelike region $k^2>0$ and is associated with medium-specific effects like Landau damping {\it i.e.} absorption or scattering of modes by the thermal bath. These spacelike discontinuities are intrinsic to finite-temperature effects and are more difficult to avoid in an EFT, because they produce branch cuts at low energies. Finally, note that the influence functional is highly nonlocal, as is generic for open systems; locality is recovered only in special limits (for example, the Wilsonian heavy-mass limit).

Expressing Eq.~(\ref{thermalIF_ans}) in the Keldysh r/a basis defined in Eq.~(\ref{rabasis_def}), one finds
\begin{eqnarray}
\label{ScalarQEDra}
S_{\mathrm{IF}}[ A_{\mathrm{r}}, A_{\mathrm{a}} ] & \simeq & q^2 \sum_{\mathfrak{p} = \mathrm{L}, \mathrm{T}} \int \frac{ \mathrm{d}^4k}{(2\pi)^4} \; \mathcal{P}^{\mathfrak{p}}_{\mu\nu}(k)  \bigg[ - A_{\mathrm{a}}^\mu(k) \mathcal{D}_{\mathfrak{p}}^{\beta}(k) A_{\mathrm{r}}^\nu(-k) + \frac{i}{2} A_{\mathrm{a}}^\mu(k) \mathcal{S}_{\mathfrak{p}}^{\beta}(k) A_{\mathrm{a}}^\nu(-k)  \bigg]   \, ,\label{thermalIF_ans_RA}
\end{eqnarray}
where we have defined the kernels for each  $\mathfrak{p} = \mathrm{L}, \mathrm{T}$
\begin{eqnarray}
\mathcal{D}^{\beta}_{\mathfrak{p}}(k) & \equiv  & \mathrm{Re}[ \Pi_{\mathfrak{p}}^{\beta}(k) ] - \frac{i [ \mathcal{N}^{\beta}_{\mathfrak{p}}(k) - \mathcal{N}^{\beta}_{\mathfrak{p}}(-k) ]}{2} \, , \\
\mathcal{S}^{\beta}_{\mathfrak{p}}(k) & \equiv & \frac{1}{2} \mathrm{Im}[\Pi^{\beta}_{\mathfrak{p}}(k)] - \frac{\mathcal{N}^{\beta}_{\mathfrak{p}}(k) + \mathcal{N}^{\beta}_{\mathfrak{p}}(-k)}{4} \, ,
\end{eqnarray}
where $\mathcal{D}^{\beta}_{\mathfrak{p}}$ and $\mathcal{S}^{\beta}_{\mathfrak{p}}$ are responsible  for the dissipation and  noise respectively.

\subsection{High temperature limit}
\label{nonlocality}

From the previous section it is quite apparent that these kernels are both non-local in space and time. It is natural to ask whether a local description can be recovered in any limit.  A natural regime to examine is the high-temperature limit, where time-local behaviour arises for some thermal systems. Taking $\beta \to 0$ in the expressions above yields (see \cite{Kraemmer:1994az})
\begin{eqnarray}
\mathrm{Re}\big[ \Pi^{\beta}_{\mathrm{L}}(k) \big]  & \simeq & - \frac{1}{3 |\mathbf{k}|^2 \beta^2} \bigg[ \frac{k_0}{2|\mathbf{k}|} \log \Big| \frac{ k_0/|\mathbf{k}| + 1 }{ k_0/|\mathbf{k}| - 1  } \Big| - 1 \bigg] + \mathcal{O}\left(\sfrac{1}{\beta}\right) \, ,\\
\mathrm{Im}\big[ \Pi^{\beta}_{\mathrm{L}}(k) \big] & \simeq & \frac{ \theta(k^2) \pi  }{3 |\mathbf{k}|^3 \beta^3} - \frac{ \theta(k^2) + \theta(-k^2 - 4m^2) }{2 \pi |\mathbf{k}|^2 \beta^2 } \bigg[ \sqrt{ 1 + \sfrac{4m^2}{k^2} } - \frac{|k_0|}{2|\mathbf{k}|} \log \Big| \sfrac{\Omega_{+}}{\Omega_{-}} \Big|  \bigg]  + \mathcal{O}\left(\sfrac{1}{\beta}\right) \, , \\
\mathcal{N}^{\beta}_{\mathrm{L}}(k) & \simeq & - \frac{ \theta(k^2) \pi  }{3|\mathbf{k}|^3 \beta^3}  - \frac{ \theta(k^2) \pi k_0}{12 |\mathbf{k}|^3 \beta^2} + \frac{ \theta(k^2) + \theta(-k^2 - 4m^2) }{2 \pi |\mathbf{k}|^2 \beta^2 } \bigg[ \sqrt{ 1 + \sfrac{4m^2}{k^2} } - \frac{|k_0|}{2|\mathbf{k}|} \log \Big| \sfrac{\Omega_{+}}{\Omega_{-}} \Big|  \bigg]  \, ,\qquad
\end{eqnarray}
\begin{eqnarray}
\mathrm{Re}\big[ \Pi^{\beta}_{\mathrm{T}}(k) \big] & \simeq & \frac{1}{6 |\mathbf{k}|^2 \beta^2} \bigg[ \frac{k_0^2}{|\mathbf{k}|^2} + \left( 1 - \frac{k_0^2}{|\mathbf{k}|^2} \right) \frac{k_0}{2|\mathbf{k}|} \log \Big| \frac{ k_0/|\mathbf{k}| + 1 }{ k_0/|\mathbf{k}| - 1  } \Big| - 1 \bigg] + \mathcal{O}\left(\sfrac{1}{\beta}\right) \, ,\\
\mathrm{Im}\big[ \Pi^{\beta}_{\mathrm{T}}(k) \big] & \simeq & \frac{2k^2}{|\mathbf{k}|^2} \bigg\{ -  \frac{ \theta(k^2)  \pi }{12 |\mathbf{k}|^3 \beta^3} \, ,\\
&& + \frac{ \theta(k^2) + \theta(-k^2 - 4m^2)  }{8 \pi |\mathbf{k}|^2 \beta^2 } \bigg[ \sqrt{ 1 + \frac{4m^2}{k^2} } + \left( 1 + \frac{4m^2}{k^2} - \frac{k_0^2}{|\mathbf{k}|^2} \right) \frac{|\mathbf{k}|}{2|k_0|} \log \Big| \frac{\Omega_{+}}{\Omega_{-}} \Big|  \bigg] \bigg\} + \mathcal{O}\left(\sfrac{1}{\beta}\right) \, , \notag \\
\mathcal{N}^{\beta}_{\mathrm{T}}(k) & \simeq & \frac{2k^2}{|\mathbf{k}|^2} \bigg\{ +  \frac{ \theta(k^2) \pi  }{12 |\mathbf{k}|^3 \beta^3}  - \frac{ \theta(k^2) \pi k_0}{24 |\mathbf{k}|^3 \beta^2} \, ,\\
&& - \frac{ \theta(k^2) + \theta(-k^2 - 4m^2)  }{8 \pi |\mathbf{k}|^2 \beta^2 } \bigg[ \sqrt{ 1 + \frac{4m^2}{k^2} } + \left( 1 + \frac{4m^2}{k^2} - \frac{k_0^2}{|\mathbf{k}|^2} \right) \frac{|\mathbf{k}|}{2|k_0|} \log \Big| \frac{\Omega_{+}}{\Omega_{-}} \Big|  \bigg] \bigg\} + \mathcal{O}\left(\sfrac{1}{\beta}\right) \, ,\notag
\end{eqnarray}
where $\mathrm{Re}[\Pi^{\beta}_{\mathrm{L},T}] \sim \mathcal{O}(\beta^{-2})$ and all other quantities scale as $\mathcal{O}(\beta^{-3})$.
Looking at the results above, it is clear that simply integrating out a weakly interacting scalar charged particle, at one-loop, is not ``complex enough'' to behave as a local thermal bath of the sort customarily encountered in many-body theories. Our propagators have the usual finite-temperature occupation factors and, as expected, the classical Boltzmann contributions survive in the high-temperature limit. However, what is absent in our setup are fast relaxation times generated by frequent collisions: the environment is effectively weakly interacting, so its correlation functions retain long memory and need not decay on a finite microscopic timescale. As a consequence, the photon ``remembers'' what happened over a long history of electron trajectories. In this case, the high-temperature limit does not guarantee a time-local theory, but simply produces classical expressions whose overall scale is controlled by inverse powers of the Boltzmann factor. Gradient expansions are therefore not possible in our case, as the retarded correlator near $\omega \simeq 0$ is expected to exhibit branch cuts, for example due to Landau damping.

On the other hand, one would expect that sufficiently strong interactions with the environment lead to equilibration of the dynamics and, consequently, to effectively local behaviour. This is not typically the case on cosmological scales, where gravity and other fields interact only weakly at low temperatures, thereby maintaining non-equilibrium dynamics. It is therefore more likely that in cosmology we will encounter non-local kernels parametrising open-system effects. In such cases, the retarded propagator is expected to exhibit branch cuts near $\omega \simeq 0$, for example due to free streaming of particles or other weakly interacting regimes. As an example, in hydrodynamics one assumes that the microscopic physics is strongly interacting and in thermal equilibrium. Consequently the low-frequency, long-distance behaviour of its correlators can be organised in a local gradient expansion \cite{Endlich:2012vt}. In general though a genuine derivative expansion need not exist. In some cases, if the non-local kernels are sufficiently localised ({\it i.e.} sharply peaked in time over the relevant frequency range), they may still be approximated as local by retaining only the leading term \cite{Berera:2007qm}.

\subsection{Wilsonian limit}
\label{MassLimit}

In the large mass limit $\beta m \gg 1$, the thermal distributions are exponentially small. This is easy to see by employing the geometric series 
\begin{align}
\frac{1}{e^{\beta \Omega} - 1} 
&= \frac{e^{-\beta \Omega}}{1 - e^{-\beta \Omega}} 
= \sum_{n=1}^{\infty} e^{-n\beta \Omega} \, ,
\end{align}
where the Boltzmann terms rapidly decay. From this, it is clear that in this limit all thermal and open system effects are suppressed.

For  $m \gg \{ |k|, 1/\beta \}$, one drops all terms proportional to $\theta(-k^2 - 4m^2)$ and one recovers standard Wilsonian (vacuum) EFT contributions in the real parts of the kernels, with exponentially suppressed thermal contributions:
\begin{eqnarray}
\mathrm{Re}[\Pi^{\beta}_{\mathrm{L}}(k) \big] & \simeq & - \frac{1}{32} \left[-\frac{k^2}{15 m^2} + \mathcal{O}\left(m^{-4}\right)\right] 
+  e ^{- \beta  m}  \frac{  \sqrt{m}\big(\frac{k^2}{k_0^2} -1 \big)}{\sqrt{2}(\pi\beta)^{3/2}| \mathbf{k}|^2} + \mathcal{O}\left(m^{-\frac{1}{2}}\right)  
\\ \mathrm{Im}[ \Pi^{\beta}_{\mathrm{L}}(k) \big] & \simeq &     \frac{ \theta(k^2) }{8\pi |\mathbf{k}|^3} \, \big(1 + e^{-\beta |k_0|} \big) \;  e^{-\beta m\,\tfrac{|\mathbf{k}|}{|k|}}
 \left[
\frac{4 |\mathbf k|^2 m^2 }{\beta k^2}
+\frac{8 m \mathbf k }{\beta^{2} |k|} + \mathcal{O}\left(m^{0}\right)
\right]
\label{ImPiL} 
\\ \mathrm{Re}[ \Pi^{\beta}_{\mathrm{T}}(k) ] & \simeq &  \frac{k^2}{32 \pi^2 |\mathbf{k}|^2} \left[-\frac{k^2}{15 m^2}  + \mathcal{O}\left(m^{-4}\right)\right] +   e ^{- \beta  m} \bigg[ \frac{\sqrt{m}  \big( k^2 + 5 |\mathbf{k}|^2 \big)}{4 \sqrt{2} \pi^{3/2} \beta^{3/2} |\mathbf{k}|^4}    + \mathcal{O}\big(m^{-\tfrac{1}{2}}\big) \bigg]  
\\ 
\mathrm{Im}[ \Pi^{\beta}_{\mathrm{T}}(k) \big] & \simeq & -  \frac{ \theta(k^2) }{16\pi |\mathbf{k}|^5} \,  \big( 1+ e^{- \beta |k_0|} \big) \;   e^{-\beta m\,\tfrac{|\mathbf{k}|}{|k|}} 
 \left[ \frac{8 m |\mathbf k| |k| }{\beta^{2} }  + \mathcal{O}\left(m^{0}\right)
\right] 
\label{ImPiTW} 
 \\
 \mathcal{N}^{\beta}_{\mathrm{L}}(k) & \simeq &  -
 \frac{\theta(k^2) }{4\pi |\mathbf{k}|^3} e^{- \beta m \frac{|\mathbf{k}|}{|k| }} \left[ \theta(k^0) + e^{ \beta  |k_0| }  \theta(-k^0)  +1  \right] 
  \left[\frac{4 |\mathbf{k}|^2 m^2}{\beta  k^2 } + \frac{8 |\mathbf{k}| m}{\beta ^2  | k| } + \mathcal{O}\left(m^{0}\right)\right]  
\label{NoiseLmW} 
\\ \mathcal{N}^{\beta}_{\mathrm{T}}(k) & \simeq & 
 \frac{ \theta(k^2)  }{8 \pi |\mathbf{k}|^5}
  e^{- \beta m \frac{|\mathbf{k}|}{|k| }} \left[ \theta(k^0) + e^{ \beta  |k_0| }  \theta(-k^0)  +1  \right] 
  \left[\frac{8 m |\mathbf{k}| | k| }{\beta ^2 }+ \mathcal{O}\left(m^{0}\right)\right]  
\label{NoiseTmW} 
\end{eqnarray}
We see that the first terms in $\mathrm{Re} [\Pi^\beta_{\mathrm{L},\mathrm{T}}]$ are pure vacuum contributions, representing a gradient expansion in $k^2/m^2$. All thermal contributions are multiplied by Boltzmann factors of the form $\exp{(-\beta m)}$ and $\exp{(-\beta m |\mathbf k|/|k|)}$. In the Wilsonian regime $\beta m \gg 1$, and so thermal effects are exponentially suppressed for small external momenta $k \ll m$, so that only the vacuum terms, analytic in $k^2$, survive in $\Pi^\beta_{\mathrm{L},\mathrm{T}}$. In this regime the external momenta lie far below the branch cut at $k^2 = 4m^2$, justifying a polynomial expansion in $k^2/m^2$. Any non-local structure therefore resides at length scales $\sim 1/m$, beyond the cut-off of the EFT. In this limit the open EFT reduces to an ordinary closed, local Wilsonian EFT for the photon.

\section{Spontaneous Symmetry Breaking I: Abelian Higgs-Kibble Model}
\label{sec:SSB}

Up to now we have dealt with situations in which the gauge symmetry remains unbroken. When the gauge symmetry is spontaneously broken there is a preferred gauge, unitary gauge, determined by the order parameter of the breaking. Ironically this significantly simplifies the analysis of gauge theories because we can define all observables in that gauge, removing the need for the introduction of Wilson lines. A general gauge can be reached from unitary gauge by means of the introduction of \stu fields.

Although not required, it is natural to include the \stu fields as part of the open EFT, not least because \stu fields together with photon naturally combine into a massive spin 1 particle, but also because doing so removes the most severe non-local interactions. Thus in considering the influence functional rather than integrating out all charged matter, we will integrate out the Higgs-like particle responsible for symmetry breaking and charged matter which does not break the symmetry.
A natural question is whether the role of these St\"uckelberg fields changes when open system effects are significant.
While integrating out the massive modes in the broken phase can lead to more interesting influence functionals, BRST invariance continues to hold manifestly at all orders.

It is significantly easier to understand SSB for bosonic theories and so we now turn to the Abelian Higgs-Kibble model with a symmetry breaking potential
\begin{equation} \label{V_SSB}
V(\Phi^{\ast} \Phi) = \lambda (v^2 - \Phi^{\ast} \Phi)^2 \, ,
\end{equation}
where the constant $v>0$ has dimensions of mass, and the coupling $\lambda > 0$ is dimensionless. 
In this case we instead consider fluctuations near the global minimum of the potential where $|\Phi| \simeq v$
\begin{eqnarray} \label{nearv}
\Phi = \left( v + \frac{1}{\sqrt{2}} \zeta\right) e^{\tfrac{i \chi}{\sqrt{2}v}} \, .
\end{eqnarray}
The covariant derivative now takes the form
\begin{equation} \label{coder_zeta2}
D_{\mu} \Phi = \bigg[ \frac{ \partial_\mu \zeta  }{\sqrt{2}} - i \left( q A_{\mu} - \frac{ \partial_\mu \chi }{\sqrt{2} v} \right) \left( v + \frac{ \zeta }{\sqrt{2}}\right) \bigg] e^{\tfrac{i \chi}{\sqrt{2}v}} \ .
\end{equation}
Clearly the earlier gauge transformation from Eq.~(\ref{gaugetrans}) now is implemented on the charged matter field by varying $\chi(x) \to \chi(x) + \sqrt{2} q v \lambda(x)$ ({\it cf.} Eq.~(\ref{chi_gtrans})) while doing nothing to the Higgs field $\zeta$. We can use this gauge freedom to set $\chi=0$, unitary gauge, where it becomes explicit that the photon is massive. Reintroducing gauge invariance by the \stu mechanism corresponds to reintroducing $\chi$ and so we recognize $\chi$ as the \stu field for the broken gauge symmetry.

Expanding the action using (\ref{nearv}) then generates a mass term for the photon $q^2 v^2 A_\mu A^\mu$ as well as a mixing term $-\sqrt{2}q v A_{\mu} \partial^\mu \chi$. This latter mixing term can be removed by taking an adjusted gauge-fixing term ({\it cf.} Eq.~(\ref{gaugefixS}))
\begin{equation} \label{GF_SSB}
S_{\mathrm{GF}}[A] \to  -\frac{1}{2\xi}  \int \exd^4 x\; (\partial_{\mu} A^{\mu} - \sqrt{2} q v \xi  \chi)^2 \, .
\end{equation}
This choice naturally gives $\chi$ a gauge-dependent mass of $2\xi q^2 v^2$, 
which is expected since $\chi$ is a St\"uckelberg field\footnote{In the decoupling limit, $q\rightarrow 0$, $\chi$ becomes massless and may be identified as the Goldstone mode associated with with breaking of the global symmetry. Away from the decoupling limit the Goldstone-Salam-Weinberg theorem does not apply and there is no problem with $\chi$ being massive.} describing the helicity-0 component of the now-massive photon. The BRST invarant action includes a related mass term for the ghosts, 
%
\begin{eqnarray} 
S_{\mathrm{BRST}} & = & - \int \exd^4 x\; \bigg[ \frac{1}{4} F_{\mu\nu} F^{\mu\nu} + q^2 v^2  A_\mu A^\mu + \frac{( \partial_\mu A^\mu )^2}{2\xi}  + \frac{1}{2} ( \partial \chi )^2 + \xi q^2 v^2  \chi^2 + \bar{c} \big( \Box - 2 \xi q^2 v^2  \big) c \label{SSBaction_first} \\
&& + \frac{1}{2} ( \partial \zeta )^2 + 2 \lambda v^2 \zeta^2 + \sqrt{2} v \lambda \zeta^3 + \frac{\lambda}{4} \zeta^4 + \Big( \partial_\mu \chi - \sqrt{2} v  q  A_{\mu} \Big)  \Big( \partial^\mu \chi - \sqrt{2} v  q  A^{\mu} \Big) \Big(  \frac{ \zeta }{\sqrt{2}v} + \frac{\zeta^2}{ 4 v^2 } \Big) \bigg] \, ,\notag
\end{eqnarray}
%
and is invariant under the BRST symmetry ({\it cf.} Eq.~(\ref{BRST_trans})):
\begin{equation}
  \label{BRST_Higgs}
   \hat s A_{\mu}  = \partial_\mu c \, , \quad \hat s \chi  = \sqrt{2} v q c \, , \quad \hat s \zeta  = 0 \, , \quad \hat{s} c  = 0 \, , \quad  \hat s \bar{c}  = - \tfrac{1}{\xi} (\partial_\mu A^\mu + \sqrt{2} q v \xi \chi ) \, .
\end{equation}
As usual this action takes the simplest form with the choice $\xi=1$ since at quadratic order all the degrees of freedom decouple
\begin{eqnarray} 
S_{\mathrm{BRST}} & = & - \int \exd^4 x\; \bigg[ -\frac{1}{2} A_{\mu}\Box A^{\mu} + q^2 v^2  A_\mu A^\mu  + \frac{1}{2} (\partial \chi)^2 +  q^2 v^2  \chi^2 + \bar{c} \big( \Box - 2  q^2 v^2  \big) c \label{SSBaction_first2} \\
&& + \frac{1}{2} ( \partial \zeta )^2 + 2 \lambda v^2 \zeta^2 + \sqrt{2} v \lambda \zeta^3 + \frac{\lambda}{4} \zeta^4 +  \mathscr{D}_\mu\chi \mathscr{D}^\mu\chi\Big(  \frac{ \zeta }{\sqrt{2}v} + \frac{\zeta^2}{ 4 v^2 } \Big) \bigg] \, , \notag
\end{eqnarray}
where we have defined the covariant derivative of the \stu field
\begin{equation}
\mathscr{D}_\mu\chi \equiv  \partial_\mu \chi - \sqrt{2} v  q A_{\mu} \, ,
\end{equation}
which is invariant under gauge transformations.

\subsection{Integrating out the Higgs}

The goal here is to integrate out the Higgs field, so as to generate an influence functional for both the photon and the St\"uckelberg field. One could directly integrate out $\zeta$ in Eq.~(\ref{SSBaction_first}) but it turns out that this form is slightly complicated to track perturbation theory with. Instead we perform the local field redefinition
\begin{equation} \label{higgsredef}
\zeta \ \to \ H = \zeta+\frac{ \zeta^2 }{2\sqrt{2}v} \, , 
\end{equation}
which casts the BRST action (\ref{SSBaction_first}) into more a useful form --- one in which it is natural to organize the influence functional as an expansion in $1/v$. Although $H$ is technically a composite operator, integrating it out is straightforward, and in dimensional regularisation it is not necessary to track the change in measure.

Since the calculations that follow involve loop corrections and therefore UV divergences, it is necessary to supplement the action (\ref{SSBaction_first}) with the appropriate counterterms. In a renormalizable theory like the Abelian Higgs model, the allowed counterterms are those obtained from the original (unbroken) action through mass and coupling redefinitions or wavefunction renormalisations (see {\it eg.}~\cite{Irges:2017ztc}). This means for example that one is allowed polynomial counterterms $\zeta, \zeta^{2}, \zeta^{3}, \zeta^{4}$ (terminating at quartic order) as well as $(\partial \zeta)^{2}$. After the field redefinition (\ref{higgsredef}), these counterterms are expressed in terms of $H$, and to $\mathcal{O}(v^{-4})$ the operators $H$, $H^{2}$, $H^{3}$, and $H^{4}$ remain linearly independent. In contrast, the kinetic $(\partial \zeta)^{2}$ counterterm is more constrained since:
\begin{equation}
(\partial \zeta)^{2}
= (\partial H)^{2}\Big(1 - \frac{\sqrt{2}\,H}{v} + \frac{2H^{2}}{v^{2}} + \ldots\Big) \ .
\end{equation}
Collecting the relevant contributions, the counterterms necessary for our computation of the influence functional at $\mathcal{O}(v^{-4})$ are specifically 
\begin{equation} \label{Higgscounterterms}
S_{\text{counterterms}} \simeq - \int \exd^4 x \; \bigg[ \; \frac{a_1}{\sqrt{2}v} H  + \frac{a_2}{2v^2} H^2 + \frac{b_2}{2 v^2} ( \partial H )^2 \Big( 1 - \frac{ \sqrt{2} H}{v} + \frac{2 H^2}{v^2} \Big) + \frac{b_4}{v^4} (\mathscr{D}\chi)^4 + \ldots \; \bigg] \, ,
\end{equation}
where we have anticipated the $v$-scaling of each of the above operators.

The system counterterm proportional to $(\mathscr{D}\chi)^{4}$ is an exception to the general discussion above, as it is dimension-8 and therefore appears to be the consequence of a non-renormalizable theory which could not be the case. In fact its appearance is a direct consequence of the fact that $H$ defined in (\ref{higgsredef}) is a composite operator. Two-point functions of $H$ enter the influence functional, and as is standard for correlators of composite operators, additional short-distance divergences arise. These are captured schematically by the operator product expansion (OPE)
\begin{equation} \label{OPE}
\lim_{x \to y} H(x) H(y) = \lim_{x \to y} \left( \zeta(x) +\frac{ \zeta^2(x) }{2\sqrt{2}v}  \right)  \left( \zeta(y) +\frac{ \zeta^2(y) }{2\sqrt{2}v}  \right) \propto \frac{1}{v^2(x-y)^{4}} \mathbb{I} + \mathrm{contact\ terms} + \ldots \, ,
\end{equation}
where we used $\zeta(x)\,\zeta(y) \propto (x-y)^{-2}\mathbb{I} + \ldots$ as $x\to y$. Accordingly, the one-loop corrections to the $HH$ two-point function contain additional divergences scaling as $(k^2)^2$, which correspond to the contact terms in the above OPE. The counterterm $(\mathscr{D}\chi)^{4}$ renormalizes this divergence because, in the influence functional, only the Higgs degrees of freedom are integrated out. In a full calculation where $A_\mu$ and $\chi$ are also integrated out, this extra divergence would be cancelled by a corresponding divergence in the four-point function of $\mathscr{D}_\mu\chi$, and no new fundamental counterterm is required; it appears here only because we perform a partial path integral over a composite operator.

Equipped with the counterterms, we now perform a system-environment splitting treating $H$ as the environment so that
\begin{small}
\begin{eqnarray}
S_{\mathcal{S}}[A,\chi,c,\bar{c}]  & = &  - \int \exd^4 x\; \bigg[ - \frac{1}{2} A_{\mu}\Box A^{\mu} + q^2 v^2  A_\mu A^\mu  + \frac{1}{2} (\partial \chi)^2 +  q^2 v^2  \chi^2 + \bar{c} \big( \Box - 2  q^2 v^2  \big) c + \frac{b_4}{4v^4} ( \mathscr{D}_\mu \chi)^4 \bigg] \, ,\qquad \ \ \label{SSB_SS} \\
S_{\mathcal{E}}[H] & = & - \int \exd^4 x\; \bigg[  \frac{1}{2} (\partial H)^2 + \frac{1}{2} M^2 H^2 - \frac{(\partial H)^2 H}{\sqrt{2}v}  + \frac{(\partial H)^2H^2}{v^2} -\frac{\sqrt{2} (\partial H)^2H^3 }{v^3} + \ldots \bigg] \, , \qquad \qquad\label{SSB_SE}  \\
&&  - \int \exd^4 x\; \bigg[ \frac{a_1}{\sqrt{2}v} H + \frac{b_2}{2v^2} (\partial H)^2 + \frac{a_2}{2v^2} H^2 - \frac{b_2 (\partial H)^2 H}{\sqrt{2}v^3} + \ldots \bigg] \, ,\notag \\
S_{\mathrm{int}}[A,\chi,H] & : = &  - \; \frac{1}{\sqrt{2}v} \int \exd^4 x\; ( \mathscr{D}_\mu \chi) ( \mathscr{D}^\mu \chi ) H \label{SSB_Sint} \, ,
\end{eqnarray}
\end{small}\ignorespaces
with the mass of the Higgs given by
\be
M \equiv 2 \sqrt{\lambda} v  \ .
\ee
When deriving the influence functional now, we modify here the earlier definition (\ref{SIF_def}) in the obvious way so that one integrates out only $H$ (compared to \S\ref{sec:thermal}, this means that one integrates out only half of the degrees of freedom in the complex scalar). 
To be precise
\be
e^{i S_{\mathrm{IF}}[A_+,\chi_+,A_-,\chi_-]} = \int {\cal D}[H_+]\int {\cal D}[H_-] \, \frac{\mu}{\tilde \mu} \, e^{i S_{\mathcal{E}}[H_+]-i S_{\mathcal{E}}[H_-] +i S_{\mathrm{int}}[A_+,\chi_+,H_+]-i S_{\mathrm{int}}[A_-,\chi_-,H_-] +i S_{i \epsilon}} \, . \label{SIF_SSB_DEF}
\ee
The $i \epsilon$ terms needed here are those for the Higgs field $H$ which being a $U(1)$ invariant are those of an ordinary scalar field as described in \S\ref{sec:iep_nongauge}, which assumes that the initial environment state $\varrho_{\mathcal{E}\mathrm{i}}$
is Gaussian. We let this be any mixed state that preserves translation invariance, so that the free 2-point functions for $H$ are
\begin{eqnarray}
\mathcal{F}(x,y) & \equiv & G_{++}(x,y) \Big|_{A_+=A_-=0, \; \rho=\varrho_{\mathcal{E}\mathrm{i}}
}\ = \int \frac{\exd^4 k}{(2\pi)^4} \; \mathcal{F}(k) e^{i k \cdot (x - y)} \ , \label{Fegen_def} \\
\mathcal{W}(x,y) & \equiv & G_{-+}(x,y)\Big|_{A_+=A_-=0, \; \rho=\varrho_{\mathcal{E}\mathrm{i}}
} \ =  \ \int \frac{\exd^4 k}{(2\pi)^4} \; \mathcal{W}(k) e^{i k \cdot (x - y)}  \ , \label{Wegen_def}
\end{eqnarray}
where we have the usual Feynman and Wightman functions on momentum space
\begin{equation}
\mathcal{F}(k) = \frac{- i}{k^2 + M^2- i \epsilon} + 2 \pi \mathfrak{n}(k) \delta(k^2 + M^2)\, , \quad \mathrm{and} \quad \mathcal{W}(k) = 2 \pi \big[ \theta(k^0) + \mathfrak{n}(k) \big] \delta(k^2 + M^2) \, ,\label{FWwithn_MAIN}
\end{equation}
for an arbitrary mode occupation distribution function $\mathfrak{n}(k)$ as in Eq.~(\ref{matrixexcitedstate}).

Since the path integrals in (\ref{SIF_SSB_DEF}) can no longer be done exactly, we need to perturb in some small parameter. A natural choice is to treat the Higgs mass $M$ as a constant (having eliminated the self-coupling $\lambda$ everywhere), and then take the limit of large $v$ so that we are effectively doing a perturbative expansion in $1/v$.  Note that conventional perturbative unitarity of the theory implies that $\lambda \lesssim {\cal O}(1)$, and so we naturally obtain 
\begin{equation}
M \lesssim  v \qquad \qquad (\text{perturbative unitarity}) \, .
\end{equation}
For simplicity we shall assume a strong hierarchy $M\ll v$ ensuring the expansion in $1/v$ is valid. We will further assume the mass of the photon $m_{A}^2 = 2 q^2 v^2 $, is well below that of the Higgs ($M$) which is the traditional regime for considering a low energy EFT for the Goldstone/\stu modes. 
\begin{equation}
q v \ll M \qquad \qquad (\text{photon mass below Higgs mass}) \ .
\end{equation}
Doing so inevitably kills off vacuum interactions which are kinematically forbidden, but in a generic mixed state there will still be non-zero dissipative/noise interactions. We then proceed assuming the hierarchy of scales $q \ll M / v \ll 1$ (equivalently $q^2 \ll \lambda \ll 1$). 

Since the interaction (\ref{SSB_Sint}) is linear in $H$, this means $S_{\mathrm{IF}}[A_+,\chi_+,A_-,\chi_-]$ is simply the in-in connected generating function $W[J^+_H,J^-_H]$ for the composite operator $H$, with the source $J_H=-(\mathscr{D}_\mu \chi) ( \mathscr{D}^\mu \chi )/(\sqrt{2}v)$. At a practical level this means only connected Feynman diagrams for $H$ need to be included in computing it. Perturbing the influence functional in the interaction (\ref{SSB_Sint}) then gives rise to
%
\begin{eqnarray}
&& S_{\mathrm{IF}}[A_{+},\chi_{+},A_{-}, \chi_{-}] \notag \\
& & \quad \simeq - \frac{1}{\sqrt{2}v} \int \exd^4 x\; \bigg[ \big( \mathscr{D}_{+}\chi_{+}(x)\big)^2 \langle H_{+}(x) \rangle_{\mathcal{E}}^{\conn} - \big( \mathscr{D}_{-}\chi_{-}(x)\big)^2 \langle H_{-}(x) \rangle_{\mathcal{E}}^{\conn} \bigg] \label{SSB_SIF_3pt_average}  \\
& & \hspace{7mm} + \frac{i}{4v^2} \int \exd^4 x \int \exd^4 y\; \bigg[ \; \big( \mathscr{D}_{+}\chi_{+}(x) \big)^2 \big( \mathscr{D}_{+}\chi_{+}(y) \big)^2  \langle H_{+}(x) H_{+}(y) \rangle_{\mathcal{E}}^{\conn} \notag \\
&& \hspace{37mm} - \big( \mathscr{D}_{+}\chi_{+}(x) \big)^2 \big( \mathscr{D}_{-}\chi_{-}(y) \big)^2  \langle H_{+}(x) H_{-}(y) \rangle_{\mathcal{E}}^{\conn}  \notag \\
&& \hspace{37mm} - \big( \mathscr{D}_{-}\chi_{-}(x) \big)^2 \big( \mathscr{D}_{+}\chi_{+}(y) \big)^2  \langle H_{-}(x) H_{+}(y) \rangle_{\mathcal{E}}^{\conn}   \notag \\
&& \hspace{37mm}  + \big( \mathscr{D}_{-}\chi_{-}(x) \big)^2 \big( \mathscr{D}_{-}\chi_{-}(y) \big)^2 \langle H_{-}(x) H_{-}(y) \rangle_{\mathcal{E}}^{\conn}  \bigg] \notag \\
&& \hspace{4mm} - \frac{1}{ 12 \sqrt{2} v^3 } \int \exd^4 x \int \exd^4 y \int \exd^4 z\; \bigg[ \; \big( \mathscr{D}_{+}\chi_{+}(x) \big)^2 \big( \mathscr{D}_{+}\chi_{+}(y) \big)^2 \big( \mathscr{D}_{+}\chi_{+}(z) \big)^2 \langle H_{+}(x) H_{+}(y) H_{+}(z) \rangle_{\mathcal{E}}^\conn \notag \\
&& \hspace{49mm} - 3 \big( \mathscr{D}_{+}\chi_{+}(x) \big)^2 \big( \mathscr{D}_{+}\chi_{+}(y) \big)^2 \big( \mathscr{D}_{-}\chi_{-}(z) \big)^2 \langle H_{+}(x) H_{+}(y) H_{-}(z) \rangle_{\mathcal{E}}^\conn \notag \\
&& \hspace{49mm}  + 3 \big( \mathscr{D}_{+}\chi_{+}(x) \big)^2 \big( \mathscr{D}_{-}\chi_{-}(y) \big)^2 \big( \mathscr{D}_{-}\chi_{-}(z) \big)^2\langle H_{+}(x) H_{-}(y) H_{-}(z) \rangle_{\mathcal{E}}^\conn \notag \\
&& \hspace{50mm}   - \big( \mathscr{D}_{-}\chi_{-}(x) \big)^2 \big( \mathscr{D}_{-}\chi_{-}(y) \big)^2 \big( \mathscr{D}_{-}\chi_{-}(z) \big)^2 \langle H_{-}(x) H_{-}(y) H_{-}(z) \rangle_{\mathcal{E}}^\conn  \bigg] \notag \\
&& \hspace{7mm} +\ \text{higher points} \, , \notag
\end{eqnarray}
%
where $\mathscr{D}_{\pm}\chi_{\pm} \equiv  \partial \chi_{\pm} - \sqrt{2} v  e  A_{\pm} $ in obvious notation, and the environmental average is
\begin{equation} \label{SSB_average}
\langle f[ H_{+}, H_{-}] \rangle_{\mathcal{E}} = \int \mathcal{D}[H_{+}] \int \mathcal{D}[H_{-}] \;  f[H_{+}, H_{-}] \; e^{i S_{\cE}[H_{+} ] - i S_{\cE}[H_{-} ]  + S_{i \epsilon}  } \, ,
\end{equation}
with $S_{\cE}$ the self-interacting Higgs action of Eq.~(\ref{SSB_SE}), including counterterms. The superscript ``conn'' denotes the connected contributions to the environmental average, since disconnected contributions precisely cancel. Finally, notice that the powers of $1/v$ appearing in front of each term do not align, so it is essential to truncate the averages consistently to the appropriate order in $1/v$ in the following explicit calculations.

\subsection{Influence Functional at $\mathcal{O}(v^{-2})$}

The $\mathcal{O}(v^{-2})$ terms are straightforward to evaluate. The first step is to eliminate the tadpoles $\langle H_{\pm}(x) \rangle_{\mathcal{E}}^{\conn}$ appearing in Eq.~(\ref{SSB_SIF_3pt_average}). This is done using the linear $a_{1}$ counterterm in the environmental action (\ref{SSB_SE}). At leading order, only a single $x$-independent loop contributes (the leftmost diagram in Fig.~\ref{Fig:SSBTadpoles}, cancelled by Eq.~(\ref{tadpole_renorm_cond})). Even though the environment is in an arbitrary state, any state-dependence in the one-point function must be subtracted so that the renormalised vacuum expectation value $v$ remains fixed.

The only other contributions to consider from Eq.~(\ref{SSB_SIF_3pt_average}) at $\mathcal{O}(v^{-2})$ are the Higgs two-point averages. To leading order these are just the free two-point correlators
\begin{eqnarray}
\langle H_{+}(x) H_{+}(y) \rangle^\conn_{\mathcal{E}} \simeq \mathcal{F}(x,y) + \mathcal{O}(v^{-2})\, , \quad \mathrm{and} \quad \langle H_{-}(x) H_{+}(y) \rangle^\conn_{\mathcal{E}} \simeq \mathcal{W}(x,y) + \mathcal{O}(v^{-2}) \, ,
\end{eqnarray}
from Eqs.~(\ref{Fegen_def}) and (\ref{Wegen_def}). Expressed in position space, the influence functional at this order is then very simply
\begin{eqnarray}
S_{\mathrm{IF}}[A_{+},\chi_{+},A_{-}, \chi_{-}] & \simeq & \frac{i}{4v^2} \int \exd^4 x \int \exd^4 y\; \bigg[ \; \big( \mathscr{D}_{+}\chi_{+}(x) \big)^2 \mathcal{F}(x,y) \big( \mathscr{D}_{+}\chi_{+}(y) \big)^2 \label{SIF_SSB_v2answer} \\
&& \hspace{27mm} - \big( \mathscr{D}_{+}\chi_{+}(x) \big)^2 \mathcal{W}^{\ast}(x,y) \big( \mathscr{D}_{-}\chi_{-}(y) \big)^2 \notag \\
&& \hspace{27mm} - \big( \mathscr{D}_{-}\chi_{-}(x) \big)^2 \mathcal{W}(x,y) \big( \mathscr{D}_{+}\chi_{+}(y) \big)^2 \notag \\
&& \hspace{27mm} + \big( \mathscr{D}_{-}\chi_{-}(x) \big)^2 \mathcal{F}^{\ast}(x,y) \big( \mathscr{D}_{-}\chi_{-}(y) \big)^2  \bigg] + \mathcal{O}(v^{-4}) \, . \notag 
\end{eqnarray}
Even this simple result is quite revealing. It manifestly preserves gauge invariance and exhibits the expected decoupling for a heavy Higgs: in the limit $\Box \ll M^{2}$ one has
\begin{equation} \label{heavy_FW}
\mathcal{F}(x,y) \simeq - \frac{i}{M^2} \delta(x - y) \qquad \mathrm{and} \qquad \mathcal{W}(x,y) \simeq 0 \, ,
\end{equation}
since the factors $\delta(k^{2}+M^{2})$ in Eq.~(\ref{FWwithn_MAIN}) vanish to all orders in $1/M$ at low energies, suppressing any contribution depending on $\mathfrak{n}$ from the excited environmental state. This gives rise to a local Wilsonian EFT of the expected form,
\begin{equation}
\lim_{M^2 \gg \Box} S_{\mathrm{IF}}[A_{+},\chi_{+},A_{-}, \chi_{-}] \simeq \frac{1}{4v^2M^2} \int \exd^4 x \bigg[ \;  \big( \mathscr{D}_{+}\chi_{+}(x) \big)^4 - \big( \mathscr{D}_{-}\chi_{-}(x) \big)^4 \bigg] + \mathcal{O}\left( \frac{\Box^2}{M^4} \right) \ ,
\end{equation}
in the sense that one finds two local copies of the expected EFT with no cross-branch interactions.

\subsection{Influence Functional at $\mathcal{O}(v^{-4})$}
\label{sec:SSB_loops}

The $\mathcal{O}(v^{-4})$ corrections to the influence functional are more involved. At this order the two-point contributions discussed above receive loop corrections.\footnote{In addition, the $x$-independent tadpoles in Eq.~(\ref{SSB_SIF_3pt_average}) acquire two-loop contributions, which must also be cancelled by the $a_{1}$ counterterm; see the two-loop diagrams in Fig.~\ref{Fig:SSBTadpoles}.}, and one also encounters tree-level three-point interactions. One finds that (\ref{SSB_SIF_3pt_average}) becomes
\begin{small}
\begin{eqnarray}
&& S_{\mathrm{IF}}[A_{+},\chi_{+},A_{-}, \chi_{-}] \ \simeq \ \frac{i}{4v^2} \int \exd^4 x \int \exd^4 y\; \bigg[ \ \big( \mathscr{D}_{+}\chi_{+}(x) \big)^2 \big( \mathscr{D}_{+}\chi_{+}(y) \big)^2 \bigg( \mathcal{F}(x,y) +  \frac{\mathcal{Q}_{\mathcal{F}}(x,y)}{v^2} \bigg) \label{SSB_SIF_v4} \\
&& \hspace{61mm} - \big( \mathscr{D}_{+}\chi_{+}(x) \big)^2 \big( \mathscr{D}_{-}\chi_{-}(y) \big)^2 \bigg( \mathcal{W}^{\ast}(x,y) + \frac{\mathcal{Q}_{\mathcal{W}}^{\ast}(x,y)}{v^2} \bigg) \notag \\
&& \hspace{61mm} - \big( \mathscr{D}_{-}\chi_{-}(x) \big)^2 \big( \mathscr{D}_{+}\chi_{+}(y) \big)^2 \bigg( \mathcal{W}(x,y) + \frac{\mathcal{Q}_{\mathcal{W}}(x,y)}{v^2} \bigg)   \notag \\
&& \hspace{61mm} + \big( \mathscr{D}_{-}\chi_{-}(x) \big)^2 \big( \mathscr{D}_{-}\chi_{-}(y) \big)^2 \bigg( \mathcal{F}^{\ast}(x,y) +  \frac{\mathcal{Q}_{\mathcal{F}}^{\ast}(x,y)}{v^2} \bigg) \; \bigg] \notag \\
&& \hspace{20mm} - \frac{i}{ 24 v^4 } \int \exd^4 x \int \exd^4 y \int \exd^4 z\; \bigg[ \  \big( \mathscr{D}_{+}\chi_{+}(x) \big)^2 \big( \mathscr{D}_{+}\chi_{+}(y) \big)^2 \big( \mathscr{D}_{+}\chi_{+}(z) \big)^2 \; \Gamma_{\mathrm{u}}(x,y,z) \notag \\
&& \hspace{58mm} - 3 \big( \mathscr{D}_{+}\chi_{+}(x) \big)^2 \big( \mathscr{D}_{+}\chi_{+}(y) \big)^2 \big( \mathscr{D}_{-}\chi_{-}(z) \big)^2  \; \Gamma_{\mathrm{n}}^{\ast}(x,y,z) \notag \\
&& \hspace{58mm} - 3 \big( \mathscr{D}_{+}\chi_{+}(x) \big)^2 \big( \mathscr{D}_{-}\chi_{-}(y) \big)^2 \big( \mathscr{D}_{-}\chi_{-}(z) \big)^2 \; \Gamma_{\mathrm{n}}(x,y,z) \notag \\
&& \hspace{59.5mm} + \big( \mathscr{D}_{-}\chi_{-}(x) \big)^2 \big( \mathscr{D}_{-}\chi_{-}(y) \big)^2 \big( \mathscr{D}_{-}\chi_{-}(z) \big)^2 \; \Gamma^{\ast}_{\mathrm{u}}(x,y,z) \; \bigg] + \mathcal{O}(v^{-6}) \, ,\notag \quad
\end{eqnarray}
\end{small}\ignorespaces
with the definitions
\begin{equation}
\Gamma_{\mathrm{u,n}}(x,y,z) = \int \frac{\exd^4 k}{(2\pi)^4} \int \frac{\exd^4 p}{(2\pi)^4} \int \frac{\exd^4 q}{(2\pi)^4} \; (2\pi)^4 \delta^{(4)}(k+p+q) \; \gamma_{\mathrm{u,n}}(k,p,q ) \; e^{i k \cdot x + i p \cdot y + i q \cdot z} \, ,
\end{equation}
where:
\begin{eqnarray}
\gamma_{\mathrm{u}}(k,p,q) & = & (k^2 + p^2 + q^2) \Big( \mathcal{F}(k) \mathcal{F}(p) \mathcal{F}(q) - \mathcal{W}(-k) \mathcal{W}(-p) \mathcal{W}(-q) \Big) \, , \qquad \label{GammaU_def}   \\
\gamma_{\mathrm{n}}(k,p,q)  & = & (k^2 + p^2 + q^2) \Big( \mathcal{F}^{\ast}(k) \mathcal{F}^{\ast}(p) \mathcal{W}(-q)  - \mathcal{W}(k) \mathcal{W}(p) \mathcal{F}(q) \Big) \label{GammaN_def} \, .
\end{eqnarray}
As well as
\begin{eqnarray}
\mathcal{Q}_{\mathcal{F},\mathcal{W}}(x,y) = \int \frac{\exd^4 k}{(2\pi)^4} \; \mathscr{Q}_{\mathcal{F},\mathcal{W}}(k) e^{i k \cdot (x -y)} \, , 
\end{eqnarray}
where the functions are loops that are explicitly
\begin{equation}
\left[ \begin{matrix} \QFe(k) & \QWe(-k) \\ \QWe(k)  & \QFeAST(k)  \end{matrix} \right] = \left[ \begin{matrix} \mathcal{F}(k) & \mathcal{W}(-k) \\ \mathcal{W}(k) &  \mathcal{F}^*(k)  \end{matrix} \right] \left[ \begin{matrix} \Sigma_{\mathcal{F}}(k) & \Sigma_{\mathcal{W}}(-k) \\ \Sigma_{\mathcal{W}}(k) &  \Sigma_{\mathcal{F}}^{\ast}(k) \end{matrix} \right] \left[ \begin{matrix} \mathcal{F}(k) & \mathcal{W}(-k) \\ \mathcal{W}(k) &  \mathcal{F}^*(k)  \end{matrix} \right] \, , 
\end{equation}
where the independent {\it partial} self-energies are
\begin{eqnarray}
 \Sigma_{\mathcal{F}}(k) & = &  -  \frac{1}{4} \int \frac{\exd^4 \ell}{(2\pi)^4}  \; \Big( k^2 + (k -\ell)^2 + \ell^2 \Big)^2  \mathcal{F}(\ell)   \mathcal{F}(k - \ell )  \label{SigmaF_main} \\
 &&  - 2 i \int \frac{\exd^4 \ell}{(2 \pi)^4} \big( k^2 + \ell^2 \big)  \mathcal{F}(\ell)   - i \big[ a_2 + b_2 k^2 + b_4 (k^2+M^2)^2 \big] \, , \notag \\
\Sigma_{\mathcal{W}}(k) & = & \frac{1}{4} \int \frac{\exd^4 \ell}{(2\pi)^4}  \; \Big( k^2 + (k -\ell)^2 + \ell^2 \Big)^2 \; \mathcal{W}(\ell) \mathcal{W}(k - \ell )  \, , \label{SigmaW_main}
\end{eqnarray}
with the derivation outlined in Appendix \ref{app:selfen}. Note that the $(\mathscr{D}\chi)^4$ counterterm proportional to $b_4$ in Eq.~(\ref{SSB_SS}) enters the above formula to renormalize the extra divergence from the composite operator (see the text surrounding Eq.~(\ref{OPE})) --- at the level of the above loops, the derivative self-interactions of the Higgs create a divergence proportional to $(k^2)^2$ in the first line of (\ref{SigmaF_main}).
\begin{figure}[h]
\begin{center}
\includegraphics[width=150mm]{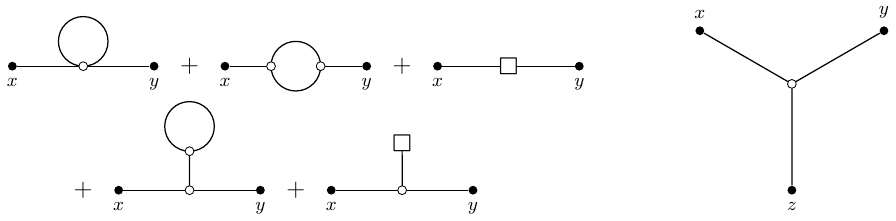}
\caption{\small The topology of the Feynman diagrams contributing to the $\mathcal{O}(v^{-4})$ influence functional in Eq.~(\ref{SSB_SIF_v4}). The diagrams on the left contribute to the Higgs two-point loop corrections, with boxes representing the counterterms in Eq.~(\ref{SSB_SE}). The rightmost diagram represents the Higgs three point vertex.} \label{Fig:v4}
\end{center}
\end{figure}

In the retarded/advanced basis, the influence functional reads instead
\begin{small}
\begin{eqnarray}
&& S_{\mathrm{IF}}[A_{\mathrm{r}},\chi_{\mathrm{r}},A_{\mathrm{a}}, \chi_{\mathrm{a}}] \simeq \frac{i}{2v^2} \int \exd^4 x \int \exd^4 y\; \bigg[ \; \big( \mathscr{D}_{\mathrm{a}}\chi_{\mathrm{a}}(x) \cdot \mathscr{D}_{\mathrm{r}}\chi_{\mathrm{r}}(x) \big) \big( \mathscr{D}_{\mathrm{r}}\chi_{\mathrm{r}}(y) \big)^2 \Big( \mathcal{R}(x,y) +  \frac{\mathcal{Q}_{\mathcal{R}}(x,y)}{v^2} \Big) \\
&& \hspace{55mm} + \big( \mathscr{D}_{\mathrm{r}}\chi_{\mathrm{r}}(x) \big)^2 \big( \mathscr{D}_{\mathrm{a}}\chi_{\mathrm{a}}(y) \cdot \mathscr{D}_{\mathrm{r}}\chi_{\mathrm{r}}(y) \big) \Big( \mathcal{A}(x,y) +  \frac{\mathcal{Q}_{\mathcal{A}}(x,y)}{v^2} \Big) \notag \\
&& \hspace{55mm} + 8 \big( \mathscr{D}_{\mathrm{a}}\chi_{\mathrm{a}}(x) \cdot \mathscr{D}_{\mathrm{r}}\chi_{\mathrm{r}}(x) \big) \big( \mathscr{D}_{\mathrm{a}}\chi_{\mathrm{a}}(y) \cdot \mathscr{D}_{\mathrm{r}}\chi_{\mathrm{r}}(y) \big) \Big( \mathcal{K}(x,y) +  \frac{\mathcal{Q}_{\mathcal{K}}(x,y)}{v^2} \Big) \bigg] \notag \\
&& \qquad - \frac{i}{ 24 v^4 } \int \exd^4 x \int \exd^4 y \int \exd^4 z\; \bigg[ \big( \mathscr{D}_{\mathrm{a}}\chi_{\mathrm{a}}(x) \cdot \mathscr{D}_{\mathrm{r}}\chi_{\mathrm{r}}(x)  \big) \big( \mathscr{D}_{\mathrm{r}}\chi_{\mathrm{r}}(y) \big)^2 \big( \mathscr{D}_{\mathrm{r}}\chi_{\mathrm{r}}(z) \big)^2  \; \Big(  \Gamma_{\mathrm{u}} -  \Gamma^{\ast}_{\mathrm{u}} - 3\Gamma_{\mathrm{n}} - 3\Gamma^{\ast}_{\mathrm{n}} \Big)(x,y,z) \notag \\
&& \hspace{44.5mm} +  \big( \mathscr{D}_{\mathrm{r}}\chi_{\mathrm{r}}(x) \big)^2 \big( \mathscr{D}_{\mathrm{a}}\chi_{\mathrm{a}}(y) \cdot \mathscr{D}_{\mathrm{r}}\chi_{\mathrm{r}}(y)  \big) \big( \mathscr{D}_{\mathrm{r}}\chi_{\mathrm{r}}(z) \big)^2  \; \Big(  \Gamma_{\mathrm{u}} -  \Gamma^{\ast}_{\mathrm{u}} + 3\Gamma_{\mathrm{n}} - 3\Gamma^{\ast}_{\mathrm{n}} \Big)(x,y,z) \notag \\
&& \hspace{44.5mm} + \big( \mathscr{D}_{\mathrm{r}}\chi_{\mathrm{r}}(x) \big)^2 \big( \mathscr{D}_{\mathrm{r}}\chi_{\mathrm{r}}(y) \big)^2 \big( \mathscr{D}_{\mathrm{a}}\chi_{\mathrm{a}}(z) \cdot \mathscr{D}_{\mathrm{r}}\chi_{\mathrm{r}}(z)  \big) \; \Big(  \Gamma_{\mathrm{u}} -  \Gamma^{\ast}_{\mathrm{u}} + 3\Gamma_{\mathrm{n}} + 3\Gamma^{\ast}_{\mathrm{n}} \Big)(x,y,z) \notag \\
&& \hspace{32mm} + \big( \mathscr{D}_{\mathrm{a}}\chi_{\mathrm{a}}(x) \cdot \mathscr{D}_{\mathrm{r}}\chi_{\mathrm{r}}(x)  \big) \big( \mathscr{D}_{\mathrm{a}}\chi_{\mathrm{a}}(y) \cdot \mathscr{D}_{\mathrm{r}}\chi_{\mathrm{r}}(y)  \big) \big( \mathscr{D}_{\mathrm{r}}\chi_{\mathrm{r}}(z) \big)^2  \; \Big(  \Gamma_{\mathrm{u}} +  \Gamma^{\ast}_{\mathrm{u}} + 3\Gamma_{\mathrm{n}} - 3\Gamma^{\ast}_{\mathrm{n}} \Big)(x,y,z) \notag \\
&& \hspace{32mm} + \big( \mathscr{D}_{\mathrm{a}}\chi_{\mathrm{a}}(x) \cdot \mathscr{D}_{\mathrm{r}}\chi_{\mathrm{r}}(x)  \big) \big( \mathscr{D}_{\mathrm{r}}\chi_{\mathrm{r}}(y) \big)^2 \big( \mathscr{D}_{\mathrm{a}}\chi_{\mathrm{a}}(z) \cdot \mathscr{D}_{\mathrm{r}}\chi_{\mathrm{r}}(z)  \big)  \; \Big(  \Gamma_{\mathrm{u}} +  \Gamma^{\ast}_{\mathrm{u}} + 3\Gamma_{\mathrm{n}} + 3\Gamma^{\ast}_{\mathrm{n}} \Big)(x,y,z) \notag \\
&& \hspace{32mm} + \big( \mathscr{D}_{\mathrm{r}}\chi_{\mathrm{r}}(x) \big)^2 \big( \mathscr{D}_{\mathrm{a}}\chi_{\mathrm{a}}(y) \cdot \mathscr{D}_{\mathrm{r}}\chi_{\mathrm{r}}(y)  \big)  \big( \mathscr{D}_{\mathrm{a}}\chi_{\mathrm{a}}(z) \cdot \mathscr{D}_{\mathrm{r}}\chi_{\mathrm{r}}(z)  \big)  \; \Big(  \Gamma_{\mathrm{u}} +  \Gamma^{\ast}_{\mathrm{u}} - 3\Gamma_{\mathrm{n}} + 3\Gamma^{\ast}_{\mathrm{n}} \Big)(x,y,z) \notag \\
&& \qquad + \text{third-order and higher in }\mathscr{D}_{\mathrm{a}}\chi_{\mathrm{a}} \, ,\notag
\end{eqnarray}
\end{small}\ignorespaces
with now $\mathscr{D}_{a ,r}\chi_{a ,r} \equiv  \partial \chi_{a ,r} - \sqrt{2} v  e  A_{a ,r} $ and
where we have kept only terms linear and quadratic in the advanced fields (so as to identify dissipation and noise), and let $\mathcal{R} = \frac{i}{2} \mathrm{Im}[ \mathcal{F} + \mathcal{W} ]$, $\mathcal{A} = \frac{i}{2} \mathrm{Im}[ \mathcal{F} - \mathcal{W} ]$ and $\mathcal{K} = \frac{1}{2} \mathrm{Re}[ \mathcal{F} + \mathcal{W} ]$ be the free retarded, advanced and Keldysh propagators respectively. We have also defined analogously
\begin{eqnarray}
\mathcal{Q}_{\mathcal{R}} \equiv \tfrac{i}{2} \mathrm{Im}[\mathcal{Q}_{\mathcal{F}} + \mathcal{Q}_{\mathcal{W}}] \, , \qquad \mathcal{Q}_{\mathcal{A}} \equiv \tfrac{i}{2} \mathrm{Im}[\mathcal{Q}_{\mathcal{F}} - \mathcal{Q}_{\mathcal{W}}] \qquad \mathrm{and} \qquad \mathcal{Q}_{\mathcal{K}} \equiv \tfrac{1}{2} \mathrm{Re}[\mathcal{Q}_{\mathcal{F}} + \mathcal{Q}_{\mathcal{W}}] \, .
\end{eqnarray}
Note that we have importantly used the relations
\begin{eqnarray}
{[} \mathcal{F} + \mathcal{F}^{\ast} - \mathcal{W} - \mathcal{W}^{\ast}](x,y) & = & 0 \, , \\
{[} \mathcal{Q}_{\mathcal{F}} + \mathcal{Q}^{\ast}_{\mathcal{F}} - \mathcal{Q}_{\mathcal{W}} - \mathcal{Q}^{\ast}_{\mathcal{W}} ](x,y) & = & 0 \, ,
\end{eqnarray}
the latter expression which is proved in Appendix \ref{app:ra_cancel}. Finally we also have used the expression $[ \Gamma_{\mathrm{u}} +  \Gamma^{\ast}_{\mathcal{U}} - 3\Gamma_{\mathcal{N}} - 3\Gamma^{\ast}_{\mathcal{N}}   ](x,y,z) \to 0$ which is true when symmetrised in the spacetime points. Note that all cancellations hold for any arbitrary Gaussian excited state.

\subsection{Specialisation to isotropic states}

The loops above can be written down very explicitly for an isotropic state, such that ({\it cf.}~Eq.~(\ref{matrixexcitedstate}) with $n(\mathbf{k}) \to n(|\mathbf{k}|)$)
\begin{eqnarray}
\mathfrak{n}(k) = \theta(k^0) n(|\mathbf{k}|) + \theta(- k^0) n(|\mathbf{k}|) = n(|\mathbf{k}|) \ .
\end{eqnarray}
Expressed as 
\begin{small}
\begin{eqnarray}
\Sigma_{\mathcal{F}}(k) & = & i \frac{M^2 (M^2-k^2)}{8 \pi ^2} \log \left(\frac{M^2}{\mu ^2}\right) \\
&& - i \int_{-1}^{+1} \exd y\; \frac{y^2 (5 y^2+3) (k^2)^2 -12 (2 y^2+1)k^2 M^2 +24 M^4}{256 \pi ^2} \log \bigg|\frac{(1-y^2)k^2 +4 M^2 }{\mu ^2}\bigg| \notag \\ 
&& - \frac{i k^2}{\pi^2} \int_{M}^\infty \exd \Omega \; \sqrt{\Omega^2 - M^2} \; n\big( \sqrt{\Omega^2 - M^2} \, \big) \notag \\
&& - \frac{ i (k^2 - 2M^2)^2 }{32 \pi^2 |\mathbf{k}| }\int_M^\infty \exd \Omega \;  n\big( \sqrt{ \Omega^2 - M^2 } \, \big) \; \log \bigg| \frac{(k^2 - 2 |\mathbf{k}| \sqrt{\Omega^2 - M^2})^2 - 4 k_0^2 \Omega^2 }{ (k^2 + 2 |\mathbf{k}| \sqrt{\Omega^2 - M^2})^2 - 4 k_0^2 \Omega^2  } \bigg| \notag \\
& & - \theta(-k^2 - 4M^2) \frac{ \, (k^2 - 2M^2)^2 }{64 \pi |\mathbf{k}| } \int_{\Omega_{-}}^{\Omega_{+}} \exd \Omega \; \Big[ 1 + 4 n\big( \sqrt{ \Omega^2 - M^2 } \, \big) \notag \\
&& \hspace{70mm} + 2 n\big( \sqrt{\Omega^2 - M^2} \big) \, n\big( \sqrt{ (\Omega-|k_0|)^2 - M^2} \big) \Big] \notag \\
&& - \theta(k^2) \frac{(k^2 - 2M^2)^2 }{32 \pi |\mathbf{k}| } \int_{\Omega_{+}}^\infty \exd \Omega \; \Big[ n\big( \sqrt{ \Omega^2 - M^2 } \, \big) + n\big( \sqrt{ (\Omega - |k_0|)^2 - M^2 } \, \big) \notag \\
&& \hspace{70mm} \, + 2 n\big( \sqrt{\Omega^2 - M^2} \big) \, n\big( \sqrt{ (\Omega-|k_0|)^2 - M^2} \big)  \Big] \notag \\
\Sigma_{\mathcal{W}}(k) & = & \theta(-k^2 - 4M^2) \frac{(k^2 -2M^2)^2}{32 \pi |\mathbf{k}|} \int_{ \Omega_{-} }^{\Omega_{+} } \exd \Omega \; \bigg[\theta(k^0) + 2 \theta(k^0) n\big( \sqrt{ \Omega^2 - M^2} \, \big) \\
&& \hspace{70mm} + n\big( \sqrt{ \Omega^2 - M^2} \, \big) n\big( \sqrt{ (\Omega-|k_0|)^2 - M^2} \, \big) \bigg] \notag \\
&&  + \theta(k^2) \frac{ (k^2 - 2M^2)^2 }{16\pi |\mathbf{k}|} \int_{\Omega_{+} }^{\infty} \exd \Omega\; \bigg[ \theta(-k^0) n\big( \sqrt{ \Omega^2 - M^2} \, \big) + \theta(k^0) n\big( \sqrt{ (\Omega-|k_0|)^2 - M^2} \, \big) \notag \\
&& \hspace{70mm} + n\big( \sqrt{\Omega^2 - M^2} \big) \, n\big( \sqrt{ (\Omega-|k_0|)^2 - M^2} \big) \bigg] \, ,\notag 
\end{eqnarray}
\end{small}\ignorespaces
where again $\Omega_{\pm} = \frac{|k_0|}{2} \pm \frac{|\mathbf{k}|}{2} \sqrt{ 1 + \frac{4M^2}{k^2} }$ from Eq.~(\ref{Omegapm}). We have used $\overline{\mathrm{MS}}$ renormalisation scheme. Notice that one very generally finds a new branch cut which develops for spacelike momenta with $k^2 < 0$. This is completely expected since the vacuum is not empty and reflects the presence of excited particles in the environment. 

\subsubsection{Thermal case: high-temperature limit}

We can specialize the above results further to reflect how the loops depend on an extra explicilt scale. An obvious choice is to pick a thermal state again, with $n$ the Bose-Einstein distribution. In the high temperature limit $\beta \to 0$, the thermal correction to each propagator dominate. As shown in Appendix \ref{App:SSBint}, the behaviour of the dominant part of the loops is
\begin{eqnarray}
 - \Sigma_{\mathcal{F}}(k) \simeq \Sigma_{\mathcal{W}}(k) & \simeq & \  \frac{\theta(-k^2 - 4 M^2) ( - k^2 + 2 M^2)^2}{ 32 \pi |\mathbf{k}|} \bigg[ \frac{ \log(\frac{\Omega_{+}}{\Omega_{-}}) - \log\left( \frac{|k_0| - \Omega_{+}}{|k_0| - \Omega_{-}} \right) }{|k_0|\beta^2} + \mathcal{O}(\beta^{-1}) \bigg] \label{thermalSigmas}  \qquad \\
&&  + \frac{\theta(k^2) ( - k^2 + 2 M^2)^2 }{16 \pi |\mathbf{k}|} \bigg[ \frac{ \log\left( \frac{\Omega_{+}}{\Omega_{+} - |k_0|} \right) }{|k_0|\beta^2} + \mathcal{O}(\beta^{-1}) \bigg] \, .\notag
\end{eqnarray}
Note that even in the high temperature limit these corrections remain non-local. This is in part because we are dealing with a weakly coupled system.

\section{SSB II: Abelian Higgs-Kibble Model \`a la Caldeira-Leggett.}
\label{SSBII}

In the previous section we considered SSB with the Higgs treated as the environment, to compare with the conventional low energy EFT obtained from integrating out the Higgs radial mode. This example is limited in one crucial aspect, since the Higgs radial mode has no charge there is no charge transfer with the heat bath. In this section we consider the Abelian Higgs-Kibble model coupled to a charged bath, following the gauged Caldeira-Leggett model considered in section~\ref{sec:CLmodel}. This leads to a more interesting EFT construction already at tree level and at quadratic order in the fields. The model we consider now has action
\begin{eqnarray} 
S & = & - \int \exd^4 x\; \bigg[ \frac{1}{4} F_{\mu\nu} F^{\mu\nu} +| D^\mu[A] \Phi|^2 + \lambda \(v^2 - \Phi^{\ast} \Phi \)^2 \nn \\
&+&  \sum_I \(- |D_0[A] \varphi_I|^2+\Gamma_I^2 |\varphi_I|^2-g_I (\Phi^* \varphi_I + \varphi_I^* \Phi) \) \bigg] \, ,\label{AbelianHiggsCD}
\end{eqnarray}
where we assume for simplicitly that all scalars have the same charge. This is clearly gauge invariant and we have sacrificed Lorentz invariance of the heat bath to simplify the analysis. Because of the quadratic mixing between the bath and Higgs field, the usual SSB vev is shifted and the bath fields acquire a vev
\be
\langle \Phi \rangle=v_0=\sqrt{v^2 + \frac{1}{2\lambda} \sum_I \frac{g_I^2}{\Gamma_I^2}} \,, \quad \langle \varphi_I \rangle= v_I=\frac{g_I}{\Gamma_I^2} v_0 \, .  \nn
\ee
In the SSB state we can consider the action in the unitary gauge by setting $\Phi=(v_0+\zeta/\sqrt{2})$ with $\zeta$ real and $\varphi_I=v_I+\frac{1}{\sqrt{2}}(\alpha_I+i\beta_I)$ with $\alpha_I$ and $\beta_I$ real. Note that $\varphi_I$ remains complex in the unitary gauge since there is only the gauge freedom to remove one phase.
The \stu field can then be reintroduced in the manner of a gauge transformation for each field, including those of the bath
\be
\Phi \rightarrow (v_0+ \frac{1}{\sqrt{2}}\zeta) e^{\tfrac{i \chi}{\sqrt{2} v_0}} \, , \quad \varphi_I \rightarrow ( v_I+\frac{1}{\sqrt{2}}(\alpha_I+i\beta_I)) \, e^{\tfrac{i \chi}{\sqrt{2} v_0}} \, ,
\ee
so that the gauge invariant action in \stu formulation is
\ba
S &=&\;
-\int \d^4 x \bigg( \frac{1}{4} F_{\mu\nu} F^{\mu\nu}
+ \frac{1}{2} (\partial_\mu \zeta)^2
+ \frac{(v_0 + \frac{1}{\sqrt{2}}\zeta)^2}{2 v_0^2}
   (\mathscr{D}_{\mu} \chi)^2 \\
&& + \sum_I \Bigg[-
  \frac{1}{2}(\partial_0 \alpha_I)^2-\frac{1}{2}(\partial_0 \beta_I)^2
  - \frac{(v_I + \frac{1}{\sqrt{2}}\alpha_I)^2+\frac{1}{2}\beta_I^2}{2v_0^2}
    (\mathscr{D}_0 \chi )^2
  + \Gamma_I^2 \(v_I^2 + \frac{1}{2}\alpha_I^2+\frac{1}{2}\beta_I^2 \)
  \nn \\
&& - \frac{1}{v_0} \left( v_I\partial_0 \beta_I +\frac{\alpha_I \partial_0 \beta_I -\beta_I \partial_0 \alpha_I}{\sqrt{2}} \right) {\mathscr{D}^{0} \chi}  - 2 g_I \left( v_0 v_I+\frac{1}{2} \zeta \alpha_I \right) \Bigg]\nn \\
&& +\lambda \bigg[ (v^2 - v_0^2)^2
+  (3v_0^2 - v^2)\zeta^2
+  \sqrt{2} v_0 \zeta^3
+ \frac{1}{4}\zeta^4 \bigg] \bigg) \, . \nn
\ea
where now $\mathscr{D}_\mu\chi \equiv \partial_\mu \chi - \sqrt{2} q v_0 A_{\mu}$. We see that the \stu field enters only with (covariant) derivative interactions as expected although now with $v$ replaced by $v_0$.
When $g_I=0$, $v_I=0$ and then this has a residual global symmetry $(\alpha_I+i \beta_I) \rightarrow e^{i \theta} (\alpha_I+i \beta_I)$. In general this is broken by the interaction between the system and environment with coupling constant $g_I$. For $g_I=0$, $v_I=0$ the residual global symmetry ensures that the influence functional obtained from integrating out the fields $ \alpha_I$ and $\beta_I$ is independent of $\chi$ (although not of $F_{\mu \nu}$). For finite $g_I$ and $v_I$ the breaking of the residual symmetry generates contributions to the influence functional from the \stu field already at quadratic order.
    
In the SSB case, the heat bath can be described by the fields $\alpha_I$ and $\beta_I$ which are themselves gauge invariant, being determined by the value of $\varphi$ in unitary gauge. Thus it is straightforward to specify a state for these fields in a gauge invariant manner, which as usual we will take to be in general a translation invariant Gaussian mixed state specified by its occupation number density $\mathfrak{n}_{I}(k)$ for each field label $I$. We will choose (for simplicity) not to integrate out the Higgs field $\zeta$. Dropping constant terms, the in-in action is then
\ba
 S_{\text{in-in}}&=& - \int \d^4 x \, \bigg[ \frac{1}{4} F^+_{\mu\nu} F_+^{\mu\nu}
+ \frac{1}{2} (\partial_\mu \zeta_+)^2+\lambda \(
(3v_0^2 - v^2)\zeta_+^2 +  \sqrt{2} v_0 \zeta_+^3 + \frac{1}{4}\zeta_+^4 \)  \\
&&  +\frac{(v_0 + \frac{1}{\sqrt{2}}\zeta_+)^2}{2v_0^2}
   (\mathscr{D}^+_{\mu} \chi_+)^2 -\sum_I \frac{v_I^2}{2v_0^2} (\mathscr{D}^+_{0} \chi_+ )^2 \bigg] \nn \\
&&+\int \d^4 x \, \bigg[ \frac{1}{4} F^-_{\mu\nu} F_-^{\mu\nu}
+ \frac{1}{2} (\partial_\mu \zeta_-)^2+\lambda \(
 (3v_0^2 - v^2)\zeta_-^2 +  \sqrt{2} v_0 \zeta_-^3 + \frac{1}{4}\zeta_-^4 \)  \nn \\
 &&  +\frac{(v_0 + \frac{1}{\sqrt{2}}\zeta_-)^2}{2v_0^2}
   (\mathscr{D}^-_{\mu} \chi_-)^2 -\sum_I \frac{v_I^2}{2v_0^2} (\mathscr{D}^-_{0} \chi_- )^2 \bigg] \nn \\
 &&+S_{\mathrm{IF}}[A_+,\zeta_+,\chi_+,A_-,\zeta_-,\chi_-] \, , \nn
\ea
with the influence functional now defined by the path integral
\ba
&& e^{i S_{\mathrm{IF}}[A_+,\zeta_+,\chi_+,A_-,\zeta_-,\chi_-]} =  \\ \nn
&& \int {\cal D}[\alpha_+,\beta_+]\int {\cal D}[\alpha_-,\beta_-] \, e^{i S_{\mathcal{E}}[\alpha_+,\beta_+]-i S_{\mathcal{E}}[\alpha_-,\beta_-] +i S_{\mathrm{int}}[\alpha_+,\beta_+,A_+,\zeta_{+},\chi_+]-i S_{\mathrm{int}}[\alpha_-,\beta_-,\chi_-,A_-,\zeta_{-},\chi_-] +i S_{i \epsilon}} \, ,
\ea
where now the environment is
\be
S_{\mathcal{E}}[\alpha,\beta]=\int \d^4 x \sum_I \[
  \frac{1}{2}(\partial_0 \alpha_I)^2+\frac{1}{2}(\partial_0 \beta_I)^2
  - \frac{1}{2}\Gamma_I^2 \( \alpha_I^2+\beta_I^2 \) \right] \, ,
\ee
and the interaction
\ba
&& S_{\mathrm{int}}[\alpha,\beta,A,\zeta,\chi]= \\ \nn
&& \int \d^4 x\sum_I \Bigg[ \frac{\sqrt{2} v_I\alpha_I+\frac{1}{2}(\alpha_I^2+\beta_I^2)}{2v_0^2}    (\mathscr{D}_0 \chi )^2+ \frac{1}{v_0} \(v_I\partial_0 \beta_I +\frac{\alpha_I \partial_0 \beta_I -\beta_I \partial_0 \alpha_I}{\sqrt{2}} \){\mathscr{D}^{0} \chi}  + g_I  \zeta \alpha_I  \Bigg] \, .
\ea
The $i \epsilon$ terms will be taken to be the standard ones for real scalars $\alpha_I$ and $\beta_I$ in a Gaussian mixed state which does not couple the different heat bath fields. Although formally the path integral may be performed exactly, we shall content ourselves with the tree level contribution that arises at quadratic order in $\mathscr{D}^{0} \chi$ as this already induces dissipation and noise for the free \stu fields. This arises from approximating the interaction as 
\be
 S_{\mathrm{int}}[\alpha,\beta,A,\zeta,\chi] \simeq   \int \d^4 x \sum_I g_I\Bigg[ \frac{ \alpha_I}{\sqrt{2} \Gamma_I^2 v_0}    (\mathscr{D}_0 \chi )^2- \frac{1}{ \Gamma_I^2}  \beta_I \partial_0 {\mathscr{D}^{0} \chi}  +   \zeta \alpha_I  \Bigg] \, , 
\ee
where we have used $v_{I}/v_0 = g_{I}/\Gamma_{I}^2$ and integrated by parts. The influence functional is then 
\ba
&& S_{\rm IF}[A_+,\chi_+,\zeta_+,A_-,\chi_-,\zeta_-] \; = \; \frac{1}{2} \int \d^4 x \int \d^4 y \sum_I i g_I^2 \begin{pmatrix} \tilde \zeta_I^+(x)
 &  -\tilde \zeta_I^-(x) \end{pmatrix} {\bf D}_I(x-y)\begin{pmatrix} \tilde \zeta_I^+(y)
 \\  -\tilde \zeta_I^-(y) \end{pmatrix} \qquad  \\
&& \hspace{24mm} + \frac{1}{2} \int \d^4 x \int \d^4 y \sum_I \frac{i g_I^2 }{\Gamma_I^4} \begin{pmatrix} \mathscr{D}_0^+ \chi^+(x) & -\mathscr{D}_0^- \chi^-(x) \end{pmatrix} \partial_0^2 {\bf D}_I(x-y) \begin{pmatrix} \mathscr{D}_0^+ \chi^+(y) \\ -\mathscr{D}_0^{-} \chi^-(y) \end{pmatrix} \, , \nn 
 \ea
 with
 \be
 \tilde \zeta_I^\pm(x)=\zeta^\pm(x) +\frac{1}{\sqrt{2} \Gamma_I^2 v_0} \(\mathscr{D}_0 \chi^\pm \)^2 \, ,
 \ee
and the propagator matrix written in momentum space $k=(\omega ,{\bf k})$,
\begin{small}
 \be
{\bf D}_I(k) = \begin{pmatrix}
-\frac{i}{-\omega^2+\Gamma_I^2-i \epsilon}+2 \pi \delta(-\omega^2+\Gamma_I^2) \mathfrak{n}_I(k) & \theta(-\omega) 2 \pi \delta(-\omega^2+\Gamma_I^2) +2 \pi \delta(-\omega^2+\Gamma_I^2) \mathfrak{n}_I(k)\\
\theta(\omega) 2 \pi \delta(-\omega^2+\Gamma_I^2) +2 \pi \delta(-\omega^2+\Gamma_I^2) \mathfrak{n}_I(k)& \frac{i}{-\omega^2+\Gamma_I^2+i \epsilon}+2 \pi \delta(-\omega^2+\Gamma_I^2) \mathfrak{n}_I(k) 
\end{pmatrix} \, .
\ee
\end{small}\ignorespaces
Given the Green's function equation
\be
\partial_0^2 {\bf D}_I(x-y) + \Gamma_I^2 {\bf D}_I(x-y) = -i  \delta^4(x-y) \begin{pmatrix} 1 & 0 \\ 0 & -1 \end{pmatrix} \, ,
\ee
 then the influence functional can be reorganised as
 \ba 
S_{\rm IF}[A_+,\chi_+,\zeta_+,A_-,\chi_-,\zeta_-]&=&-\frac{C}{2}  \int \d^4 x \[ (\mathscr{D}_0^+ \chi^+(x))^2-(\mathscr{D}_0^- \chi^-(x))^2 \] \label{IF100}  \\ 
&&-\frac{1}{2}  \int \d^4 x \int \d^4 y  \begin{pmatrix} \mathscr{D}_0^+ \chi^+(x) & -\mathscr{D}_0^- \chi^-(x) \end{pmatrix} {\bf d}(x-y) \begin{pmatrix} \mathscr{D}_0^+ \chi^+(y) \\ -\mathscr{D}_0^- \chi^-(y) \end{pmatrix} \nn  \\
&& + \frac{1}{2} \int \d^4 x \int \d^4 y \sum_I i g_I^2 \begin{pmatrix} \tilde \zeta_I^+(x)
 &  -\tilde \zeta_I^-(x) \end{pmatrix} {\bf D}_I(x-y)\begin{pmatrix} \tilde \zeta_I^+(y)
 \\  -\tilde \zeta_I^-(y) \end{pmatrix} \nn \, ,
 \ea
with 
\be
C=\sum_I \frac{ g_I^2 }{\Gamma_I^4}  \, , \quad \text{and} \quad 
{\bf d}(x-y)=\sum_I \frac{i g_I^2 }{ \Gamma_I^2} {\bf D}_I(x-y) \, . 
\ee
The first line in \eqref{IF100} describes a medium induced correction to the speed of propagation of the Goldstone/\stu modes. The second term encodes a gauge invariant form of dissipation and noise encapsulated by the non-local couplings ${\bf d}(x-y)$. The third line denotes dissipative and noise corrections to the Higgs together with interactions between the Higgs and Goldstone modes. If the Higgs fields $\zeta_{\pm}$ are further integrated out, the resulting contributions to the Goldstone/\stu EFT are quartic order at tree level and so focussing only on the quadratic terms (linear response) the in-in action is (ignoring the gauge fixing and ghost terms)
\ba
S_{\text{in-in}}[A_+,\chi_+,A_-,\chi_-] 
&=& \int d^4 x \left(
    -\frac{1}{4} F^+_{\mu\nu} F^{+\mu\nu}
    -\frac{1}{2} (\mathscr{D}^+_\mu \chi^+)^2+\frac{1}{4} F^-_{\mu\nu} F^{-\mu\nu}
    +\frac{1}{2} (\mathscr{D}^-_\mu \chi^-)^2
\right) \\
&& 
-C \int d^4 x\,
   \big[ (\mathscr{D}_0^+ \chi^+(x))^2
       -(\mathscr{D}_0^- \chi^-(x))^2 \big]
\nonumber\\
&& -\frac{1}{2} \int d^4 x \int d^4 y\;
   \begin{pmatrix}
      \mathscr{D}_0^+ \chi^+(x) & -\mathscr{D}_0^- \chi^-(x)
   \end{pmatrix}
   {\bf d}(x-y)
   \begin{pmatrix}
      \mathscr{D}_0^+ \chi^+(y) \\
      -\mathscr{D}_0^- \chi^-(y)
   \end{pmatrix}
 + \dots \, . \nonumber
\ea
As expected, the resulting open EFT has two copies of gauge invariance.  The full in-in action in Keldysh form is
\ba
S_{\text{in-in}}[A^\mathrm{r}, A^\mathrm{a}, \chi^\mathrm{r}, \chi^\mathrm{a} ] 
&=& \int d^4 x \bigg[
- \frac{1}{2} F^\mathrm{r}_{\mu\nu} F^{\mathrm{a}\mu\nu} 
- \mathscr{D}_\mu^\mathrm{r} \chi^{\mathrm{r}} \, \mathscr{D}^{\mathrm{a}\mu} \chi^{\mathrm{a}}
- 2 C  \, \mathscr{D}_0^\mathrm{r} \chi^{\mathrm{r}} \, \mathscr{D}_0^\mathrm{a} \chi^{\mathrm{a}}
\bigg] \\
&& - \int \frac{d^4 k}{(2\pi)^4} \Big[
\mathscr{D}_0^\mathrm{a} \chi^{\mathrm{a}}(-k) \,  d^D(\omega, \mathbf{k})  \, \mathscr{D}_0^\mathrm{r} \chi^{\mathrm{r}}(k)
+ \frac{i}{2} \mathscr{D}_0^\mathrm{a} \chi^{\mathrm{a}}(-k) \,  d^N(\omega, \mathbf{k}) \, \mathscr{D}_0^\mathrm{a} \chi^{\mathrm{a}}(k)
\Big] 
+ \dots \, ,\nonumber
\ea
with
\begin{align}
d^N(\omega, \mathbf{k}) 
&=\ \sum_I \frac{g_I^2 }{ \Gamma_I^2} \, 2\pi \, \delta( - \omega^2 + \Gamma_I^2 ) \, \big( \tfrac{1}{2} + \mathfrak{n}_I(\mathbf{k}) \big)  \, ,
\end{align}
and
\begin{align}
d^D(\omega, \mathbf{k}) 
= \sum_I \frac{g_I^2 }{ \Gamma_I^2} \frac{1}{-(\omega + i \epsilon)^2 + \Gamma_I^2} \,,
\end{align}
which determine the noise and dissipation terms in the \stu effective action. Note that the dissipative term never looks like a simple friction constant as in the usual Caldeira-Leggett models with Ohmic dissipation \cite{Caldeira:1982iu}. This is because of gauge invariance, forcing the dissipation to be built out of covariant derivatives of the \stu fields. It remains dissipative in the sense that it cannot be captured by an interaction in an in-out effective action or effective Hamiltonian and thus from the latter point of view implies a violation of energy conservation.

\section{Bottom-up open EFT constructions}
\label{BottomUp}

We are now in a position to address the challenge of constructing bottom-up open effective field theories (open EFTs).  
In the in-out formalism, the rules for constructing EFTs are well understood and the effective action is dictated by symmetries, and most importantly by locality. It is of great interest, particularly for cosmological and gravitational applications to where we are almost always working with EFTs to develop the equivalent rules for the in-in formalism. Implementing such a construction from the bottom up is highly non-trivial, in large part because locality is realised quite differently. In what follows, we outline the conditions that must be met and highlight the principal obstacles.

\subsection{Lorentz-covariant formulation of the influence functional}

To illustrate these challenges, it is useful to recall the linear response of a material medium to electromagnetic field. In vacuum, the field-strength tensor $F_{\mu\nu}$ fully encodes the properties of the electric and magnetic fields. In a medium, however, the fields induce polarisation and magnetisation, so the ``bare'' fields $E$ and $B$ are modified by the medium's response. A covariant formulation introduces the excitation tensor $H_{\mu\nu}$, which is related to $F^{\mu\nu}$ through constitutive relations. In the simplest setting these are linear and local, though in general they can be anisotropic, dispersive, dissipative, and non-local in both space and time.

A standard classical approach proceeds as follows: the charge/current density in Maxwell's equations can be separated into a contribution from the charged particles that constitute the medium, and a free part in the manner 
\be
\partial^{\mu}F_{\mu\nu} = - J_{\nu}^{\rm medium}- J_{\nu}^{\rm free} \, .
\ee
The split is made such that both terms are separately conserved $\partial^{\mu}J_{\nu}^{\rm medium}=0$, $\partial^{\mu}J_{\nu}^{\rm free}$=0.
The equations of motion for the medium can be solved in terms of the external electromagnetic field and $J_{\nu}^{\rm medium}$ then determined as a function thereof. Since this current must be conserved for an arbitrary external electromagnetic field, and the equations of motion of the medium have already been used, we infer that $J_{\nu}^{\rm medium}$ as a function of the electromagnetic field must be identically conserved. Such a current can be written in the form
\begin{equation}
 J_{\rm medium}^\mu  =  \partial_{\nu} P^{\mu\nu}\, ,
\end{equation}
with $P^{\mu\nu}$ an antisymmetric two form. The classical Maxwell equations may then be rewritten 
\be
\partial^{\mu}\( F_{\mu\nu}+ P_{\mu\nu} \) = - J_{\nu}^{\rm free} \, .
\ee
Comparing with Maxwell's equations in a medium in non-relativistic notation we recognize that $P_{0i}$ is the induced polarisation ${\bf P}$  and $\epsilon^{ijk} P_{ij}$ the induced Magnetisation ${\bf M}$, and so we may refer to $P_{\mu\nu}$ as the polarisation tensor.

The constitutive relations are the statement that the 6 functions $P_{\mu\nu}$ must be determined in terms of the 6 functions $F_{\mu\nu}$ specifying the electromagnetic field, $P_{\mu\nu}=P_{\mu\nu}[F_{\alpha\beta}]$. In complete generality these are potentially nonlinear and non-local, however they should respect causality in the sense
\be
\frac{\delta P_{\mu\nu}(x) }{\delta F_{\alpha \beta}(y)} = 0 \,, \quad \text{ unless $y$ lies in the past lightcone of $x$}\, .
\ee
Classically this is guaranteed, since the polarisation tensor is determined by solving the classical relativistic equations of motion causally (with retarded boundary conditions) in terms of the applied external field.\footnote{Note that most textbook treatments treat the charged medium as non-relativistic and so can only guarantee primitive causality which is the weaker statement that $\frac{\delta P_{\mu\nu}(x) }{\delta F_{\alpha \beta}(y)} = 0$ when $y^0>x^0$.}

It is straightforward from a bottom up point of view to write down an in-in effective action that reproduces these equations. The key is to recognize that the classical field can be identified with the retarded gauge field and the classical equation of motion is the one that follows from varying the action with respect to the advanced gauge field. Thus the required in-in action is\footnote{For simplicity of presentation we drop the gauge fixing and ghost terms needed for quantisation, these are easily added on.}
\be
S_{\text{in-in}} = \int \d^4 x \left[ -\frac{1}{2} \left(F_{\mu\nu}^\mathrm{r}+P_{\mu\nu}[F^\mathrm{r}_{\alpha\beta}] \right)F^{\mu\nu}_\mathrm{a}  + J_{\mu}^\mathrm{r} A^{\mu}_\mathrm{a} +J_{\mu}^\mathrm{a} A^{\mu}_\mathrm{r}  \right]\, ,
\ee
where we have included sources for the retarded and advanced fields associated with the `free' charges. This action clearly respects two copies of gauge invariance and leads to the classical equation of motion
\be
\frac{\delta S_{\text{in-in}}}{\delta A^\mathrm{a}_{\nu}(x)}=\partial^{\mu}\( F^\mathrm{r}_{\mu\nu}+ P_{\mu\nu}[F^\mathrm{r}_{\alpha\beta}] \) + J_{\nu}^\mathrm{r}=0 \, . 
\ee
More generally this action should be supplemented by noise terms which come from terms quadratic and higher in the advanced fields. Already to quadratic order in the Keldysh expansion we can write 
\ba
S_{\text{in-in}} &=& \int \d^4 x  \left[ -\frac{1}{2}  \left( F_{\mu\nu}^\mathrm{r}+P_{\mu\nu}[F^\mathrm{r}_{\alpha\beta}] \) F^{\mu\nu}_\mathrm{a} \right] + \int \d^4 x \int \d^4 y \, F^{\mu \nu}_\mathrm{a}(x) K_{\mu \nu ; \alpha \beta}[F^\mathrm{r}_{\alpha\beta}](x,y) F^{\alpha \beta}_\mathrm{a}(y) + \dots  \nn  \\
&&  + \int \d^4 x \[ J_{\mu}^\mathrm{r}(x) A^{\mu}_\mathrm{a}(x) +J_{\mu}^{\mathrm{a}}(x) A^{\mu}_\mathrm{r}(x)  \right] \, ,
\ea
where $K_{\mu \nu ; \alpha \beta}[F^\mathrm{r}_{\alpha\beta}](x,y)$ itself is in general a non-local and non-linear function of $F^\mathrm{r}_{\alpha\beta}$ with obvious antisymmetry properties on its first and second pair of indices. A bottom up construction of an open EFT for electromagnetism then amounts to specifying $P_{\mu\nu}[F^\mathrm{r}_{\alpha\beta}]$ and $K_{\mu \nu ; \alpha \beta}[F^\mathrm{r}_{\alpha\beta}]$ and ultimately the higher order terms in the Keldysh expansion. Although the symmetries are intact, unlike in the in-out formalism there is in general no expectation of locality of these functions nor of a simple power law expansion in energy scales. The exception of course is those contributions that do come from integrating out massive fields in the in-out formalism that are known to be captured by local operators. For example, the Euler-Heisenberg corrections $S_{\mathrm{EH}}$ derived in in-out formalism determine local corrections to the polarisation tensor
\be
P^{\mu\nu}[F^\mathrm{r}_{\alpha\beta}](x)= -\frac{\delta S_{\rm EH}[F^\mathrm{r}_{\alpha\beta}]}{\delta F_{\mu\nu}^\mathrm{r}(x)} + \dots \, .
\ee
In addition to that, we expect there'll be dissipative contributions coming from non-local operators in the open EFT.  To make progress we need to specify the constitutive relation. Defining
\be
H_{\mu\nu} = F_{\mu\nu}+ P_{\mu\nu}\, ,
\ee
we recognize $H_{0i}$ as ${\bf D}$ and $\epsilon^{ijk} H_{jk}$ as ${\bf H}$. 
the simplest approximation is that the constitutive relation is linear 
\begin{equation}
\label{ConstitutiveRelation1}
H_{\rho\sigma}(x) \equiv 
\int \d^4 y \,  \chi_{\alpha\beta\rho\sigma}(x,y) F^{\alpha\beta}(y) \, ,
\end{equation}
with causality requiring $\chi_{\alpha\beta\rho\sigma}(x,y)$ only has support for $y$ in the past lightcone of $x$. The rank-4 tensor density $\chi^{\mu\nu\alpha\beta}$  satisfies, the following symmetries
\begin{equation}
\label{constSymmetries}
\chi^{\mu\nu\alpha\beta} = -\chi^{\nu\mu\alpha\beta} = -\chi^{\mu\nu\beta\alpha}\, .
\end{equation}
Truncating at quadratic order in fields, we obtain the influence functional that describes the linear response
\ba
\label{IFconst}
S_{\text{in-in}}  & \equiv & -\frac{1}{2}
\int \d^4 x \int \d^4 y \,  F_\mathrm{r}^{\alpha\beta}(x) \chi_{\alpha\beta\rho\sigma}(x,y) F_\mathrm{a}^{\rho \sigma}(y)
+ \int \d^4 x \int \d^4 y \, F^{\mu \nu}_\mathrm{a}(x) K_{\mu \nu ; \alpha \beta}[F^\mathrm{r}_{\alpha\beta}](x,y) F^{\alpha \beta}_\mathrm{a}(y)  \nn  \\
&&  + \int \d^4 x \left[ J_{\mu}^\mathrm{r}(x) A^{\mu}_\mathrm{a}(x) +J_{\mu}^\mathrm{a}(x) A^{\mu}_\mathrm{r}(x)  \right] \, .
\ea
This object can exhibit a highly intricate tensorial and non-local structure. Nevertheless, under symmetry constraints or locality considerations, it is possible to parameterize the allowed terms in the open EFT.

To be concrete we now consider a system which exhibits space-time translation invariance, for which the medium is isotropic. The effects of the medium are then encoded in a background frame field $n_\mu$ which breaks Lorentz boosts. We write all possible terms consistent with the symmetries and the available degrees of freedom parametrised by arbitrary, generally non-local functions $f_{n}\equiv f_n(x-y)$ and $g_{n}\equiv g_n(x-y)$, with $n = 1, 2, \ldots$ labelling the couplings. 
For completion, we include all possible parity-odd operators consistent with these ingredients, where we use the definition for the dual $\tilde F_{\mu\nu} \equiv \frac{1}{2} \epsilon_{\mu\nu\rho\sigma} F^{\rho\sigma}$ for both retarded and advanced fields.
In addition we should allow for Chern-Simons like interactions.
Using this general setup, we obtain an effective action of the form
\begin{align}
S_{\text{in-in}}=   \int \exd^4 x \, \bigg[- \frac{1}{2}F^\mathrm{r}_{\alpha\beta} F_\mathrm{a}^{\alpha\beta} + \theta\,F_\mathrm{r}^{\alpha\beta} \tilde{F}^\mathrm{a}_{\alpha\beta}  \bigg] +S_{\rm IF} \, ,
\end{align}
where
\begin{equation}
\label{IFdef}
 S_{\rm IF} \equiv    S_{\rm diss} + S_{\rm noise} \, .
\end{equation}
We consider all possible terms that could contribute to the dissipative and noise effective action up to quadratic order in the fields, and with an infinite amount of derivatives. Due to isotropy and translation invariance, time-derivatives and spatial Laplacian operators can be absorbed into a redefinition of the non-local coupling functions, so what really matters is the different way of contracting vector indices to construct scalar operators.
The number of possibilities is significantly reduced by using the Bianchi identity 
\begin{equation}
\partial_{[\alpha}F_{\beta\gamma]} = 0 \, ,
\end{equation}
or in terms of components
\begin{equation}
\pd_i B_i =0, \quad \text{and} \quad  \pd_0 B_k = - \epsilon_{ijk} \pd_i E_j \, ,
\end{equation} 
which applies to both advanced and retarded fields. Given this, the remaining independent contributions to the dissipative part of the action, up  to quadratic order in the fields, written covariantly are 
\ba
S_{\rm diss} & = &  \int \exd^4 x \int \exd^4 y \,  \bigg\{ f_1(x-y) \, F^\mathrm{r}_{\alpha\beta}(x) F_\mathrm{a}^{\alpha\beta}(y) +  f_2(x-y) \, F_\alpha^{\mathrm{r}\gamma}(x) F^\mathrm{a}_{\gamma\beta}(y) \,  n^{\alpha} n^{\beta} \label{IFdiss2} \\
&& + f_3(x-y) \,F^\mathrm{r}_{\alpha\beta}(x) \tilde{F}_\mathrm{a}^{\alpha\beta}(y) + f_4(x-y) \, F_\alpha^{\mathrm{r}\gamma} (x) \tilde{F}^\mathrm{a}_{\beta\gamma}(y) \, n^\alpha n^\beta
 + f_5(x-y) \,  \tilde F^\mathrm{r}_{\alpha\beta}(x) A_{\mathrm{a}}^\beta(y) \, n^\alpha \nn \\
&&  + f_6(x-y) \, F_\mathrm{r}^{\gamma\alpha}(x) \pd_\gamma \tilde{F}^\mathrm{a}_{\beta\alpha}(y) \, n^\beta  
 +f_7(x-y) \, \tilde{F}^{\mathrm{r}}_{\alpha\beta}(x) \pd_\gamma F_\mathrm{a}^{\gamma\alpha}(y) \, n^\beta  \nn \\ 
 && + f_8(x-y) \, \pd_\alpha F_\mathrm{r}^{\alpha\beta}(x) \pd_\gamma F_\beta^{\mathrm{a}\gamma}(y)
 + f_9(x-y) \, \pd_\alpha F^\alpha_{\mathrm{r}\sigma}(x) \pd_\nu F^\nu_{\mathrm{a}\rho}(y) \, n^\sigma n^\rho \bigg\} \, , \nn
\ea
and similarly for the noise part of the action 
\ba
S_{\rm noise} & = & \frac{i}{2} \int \exd^4 x \int \exd^4 y \, \bigg\{ g_1(x-y) \, F^\mathrm{a}_{\alpha\beta}(x) F_\mathrm{a}^{\alpha\beta}(y) +   g_2(x-y) \, F^\mathrm{a}_{\alpha\gamma}(x) F_\beta^{ \mathrm{a}\gamma}(y)\,n^{\alpha} n^{\beta} \label{IFnoise2} \\
&& +  g_3(x-y) \,F^\mathrm{a}_{\alpha\beta}(x) \tilde{F}_\mathrm{a}^{\alpha\beta}(y) + g_4(x-y) \, F_\alpha^{\mathrm{a}\gamma}(x) \tilde{F}^\mathrm{a}_{\beta\gamma}(y) \,  n^\alpha n^\beta
 + g_5(x-y) \,  \tilde{F}^\mathrm{a}_{\alpha\beta}(x) A_{\mathrm{a}}^\alpha(y)  \, n^\beta \nn \\
&& +  g_6(x-y)\, F_\mathrm{a}^{\gamma\alpha}  \pd_\gamma \tilde{F}^\mathrm{a}_{\beta\alpha} \,  n^\beta +   g_7(x-y) \, \pd_\alpha F_\mathrm{a}^{\alpha\beta}(x) \pd_\gamma F_\beta^{\mathrm{a}\gamma}(y)
 + g_{8}(x-y) \, \pd_\alpha F^\alpha_{\mathrm{a}\sigma}(x) \pd_\nu F^\nu_{\mathrm{a}\rho} \, n^\sigma n^\rho \bigg\}\, . \nn
\ea
We have included here higher derivative terms such as $\pd_\nu F^\nu_{ \ \rho}(y)$ which in S-matrix discussions are usually removed by field redefinitions. There are two reasons for this, one is that performing a field redefinition will induce corrections in the charged matter action which we have not specified. Secondly performing the field redefinition in the open EFT changes the connection between the open EFT gauge field and the original closed system one. Since correlation functions are not invariant under field redefinitions, and in the UV/closed system there is a notion of minimal coupling which clearly identifies a specific definition of the gauge field, it is undesirable to perform the field redefinition in the open EFT.\footnote{One issue for example is that field redefinitions that are perturbatively local in an EFT expansion may be non-local in the full UV/closed theory.} 
We should allow for all terms that can be obtained by tracing over the environment in the open EFT. 
We have also included parity violating terms in the EFT still to accommodate for an intrinsically chiral medium. These include the Chern-Simons terms with couplings $f_5$, and $g_5$ as well as the parity-violating terms with couplings $f_6, f_7$ and $g_6$ lead to a helicity-dependent propagation of photons traversing the medium. 

The structure of the effective action becomes more transparent if we work in terms of observables in the frame of the medium. Denoting the timelike vector $n^\alpha = (1, \mathbf 0)$, ($n_\lambda n^\lambda = -1$) then the retarded and advanced field strength tensor can be decomposed as $F_{0i} = - E_i$, $F_{ij} = \epsilon_{ijk} B^k$, and using that $\epsilon^{ijk} \epsilon_{jkl} = 2 \delta^i_l$ and $\epsilon_{0ijk} = - \epsilon_{ijk}$, then  (\ref{IFdiss2}) and (\ref{IFnoise2}) become
\ba
S_{\rm diss} & = & \int \exd^4 x \int \exd^4 y \, \bigg\{ 
 f_1(x-y) \, \Big[ E^\mathrm{r}_i(x)\,E^\mathrm{a}_{i}(y) - B^\mathrm{r}_i(x)\,B^\mathrm{a}_{i}(y)
   \Big] + f_2(x-y)\, E^\mathrm{r}_i(x) \, E^\mathrm{a}_i(y)  \\
  && + f_3(x-y) \, \Big[ E^\mathrm{r}_i(x)\,B^\mathrm{a}_{i}(y) + B^\mathrm{r}_i(x)\,E^\mathrm{a}_{i}(y)
   \Big] + f_4(x-y)\, E^\mathrm{r}_i(x) \, B^\mathrm{a}_i(y) 
- f_5(x-y)\, B^\mathrm{r}_i(x) \, A^\mathrm{a}_i(y) \nn \\
&& - f_6(x-y) \,  \epsilon_{ijk} B^\mathrm{r}_k(x) \, \pd_{i}  B^\mathrm{a}_j(y)   + f_8 \, (x-y) \, \pd_i E^\mathrm{r}_i(x) \, \pd_j E^\mathrm{a}_j(y)\bigg\} \, , \nn
\ea
and
\ba
S_{\rm noise} & = & \frac{i}{2} \int \exd^4 x \int \exd^4 y \, \bigg\{ \, g_1(x-y)\Big[ E^\mathrm{a}_i(x)\,E^\mathrm{a}_{i}(y) - B^\mathrm{a}_i(x)\,B^\mathrm{a}_{i}(y)
   \Big] + g_2(x-y)\, E^\mathrm{a}_i(x) E^\mathrm{a}_i(y) \\
&& + \,g_3(x-y)\Big[ E^\mathrm{a}_i(x)\,B^\mathrm{a}_{i}(y) + B^\mathrm{a}_i(x)\,E^\mathrm{a}_{i}(y)
   \Big] +   g_4(x-y)\,  B^\mathrm{a}_i(x)  E^\mathrm{a}_i(y) - g_5(x-y)\, B^\mathrm{a}_i(x) A^\mathrm{a}_i(y) \nonumber \\
&& - g_6(x-y) \,  \epsilon_{ijk} B^\mathrm{a}_k(x) \pd_{i} B^\mathrm{a}_j(y) +   g_7(x-y) \, \pd_i E^\mathrm{a}_i(x) \pd_j E^\mathrm{a}_j(y)
\bigg\} \, , \nn
\ea
where we have absorbed a factor of $(-2)$ inside the couplings $f_1, f_3, g_1$, and $g_3$, and have combined several as they simply give the same contribution in the matter frame. Note, a term of the form $E_i A_i$ would violate gauge invariance of the effective action and therefore, is not permitted. As a concrete example the action recently considered in \cite{Yoshimura:2026vil} is a special case of this since the non-local kernels can accomodate local time derivatives of the retarded fields to any order in derivatives in the manner
\be
f_n(x-y) = \sum_{r=0}^{p} \frac{1}{r!}c_{nr} \partial_{x^0}^r \delta(x-y) \, , \quad g_n(x-y) = \sum_{r=0}^{p} \frac{1}{r!}g_{nr} \partial_{x^0}^r \delta(x-y)\, .
\ee
We can compare the form of the effective action in  (\ref{IFdiss2}) and (\ref{IFnoise2}) to the one we derived earlier in scalar QED (\ref{IFscalarQEDra}). For this it is convenient to go to momentum space, where 
\ba
S_{\rm diss} & = & \int \frac{\d^4 k}  {(2\pi)^4} \, \bigg\{  2\,f_{1}(-k)\,    A_\mathrm{r}^{\beta}(k) \left( k^2 \eta_{\alpha\beta} - k_{\alpha}k_{\beta} \right) A_\mathrm{a}^{\alpha}(-k) + f_{2}(-k) \Big[ A^{\mathrm{r}}_{\gamma}(k) (k \cdot n)^2 A_\mathrm{a}^{\gamma}(-k)  \\
&& +
   A_\mathrm{r}^{\beta}(k) \big( k^2 n_{\alpha} n_{\beta}  - (k \cdot n) [
k_{\alpha}  n_{\beta} + k_{\beta} n_{\alpha} ]
\big) \,  A_\mathrm{a}^{\alpha}(-k) \Big] \nonumber
\\
&& - 2 f_4 (-k) \,  \epsilon_{\alpha\beta\delta\sigma} \, A_\mathrm{r}^{\beta}(k) \, (k\cdot n) \, k^{\delta} n^{\sigma}  A_\mathrm{a}^{\alpha}(-k) - i \, 2 f_5(-k) \, \epsilon_{\alpha\beta\gamma\delta}  \, A_\mathrm{r}^\beta(k) \, k^\gamma n^\delta \,  A_\mathrm{a}^\alpha(-k) \nonumber 
\\
&& + 2 f_6(-k) \epsilon_{\beta\gamma\alpha\delta} \,  A_\mathrm{r}^\gamma(k) k^2 k^\alpha n^\delta A_\mathrm{a}^\beta(-k) +   A_\mathrm{r}^\beta(k) \big[f_8(-k) k^2 + f_9(-k) (k \cdot n)^2 \big]  k_\alpha k_\beta A_\mathrm{a}^\alpha(-k)  \bigg\} \, , \nn \qquad
\ea
and
\ba
S_{\rm noise} & = & \frac{i}{2} \int \frac{\d^4 k}  {(2\pi)^4} \,\bigg\{ 2 g_{1}(-k)\,    A_\mathrm{a}^{\beta}(k) \big( k^2 \eta_{\alpha\beta} - k_{\alpha}k_{\beta} \big) A_\mathrm{a}^{\alpha}(-k) + g_{2} \Big[(-k) \, A^\mathrm{a}_{\gamma}(k) (k \cdot n)^2  A_\mathrm{a}^{\gamma}(-k) \\
&&  +
A_\mathrm{a}^{\beta}(k)   \big( k^2 n_{\alpha} n_{\beta}  - (k \cdot n) [
k_{\alpha}  n_{\beta} + k_{\beta} n_{\alpha} ]
\big) A_\mathrm{a}^{\alpha}(-k) \Big] \nonumber \\
&& - 2 g_{4}(-k) \,  \epsilon_{\alpha\beta\delta\sigma} \, A_\mathrm{a}^{\beta}(k) \, (k\cdot n) \, k^{\delta} n^{\sigma}  A_\mathrm{a}^{\alpha}(-k) - i \, 2 g_5(-k) \, \epsilon_{\alpha\beta\gamma\delta}  \, A_\mathrm{a}^\beta(k) \, k^\gamma n^\delta \,  A_\mathrm{a}^\alpha(-k) \nn \\
&& +  2 g_6(-k) \, \epsilon_{\beta\gamma\alpha\delta} \,  A_\mathrm{a}^\gamma(k) k^2 k^\alpha n^\delta \,  A_\mathrm{a}^\beta(-k)  +   A_\mathrm{a}^\beta(k) \big[g_7(-k) k^2 + g_8(-k) (k \cdot n)^2 \big]  k_\alpha k_\beta A_\mathrm{a}^\alpha(-k)  \bigg\}\, . \nn
\ea
To match with the $\mathcal{O}(q^2)$ influence functional in (\ref{ScalarQEDra}) from the earlier thermal scalar QED top-down derivation, it turns out that one needs to set the chiral and higher-derivative contributions to zero. Combining the four remaining non-local couplings into two kernels $\mathcal{S}_{\alpha\beta}$ and $\mathcal{D}_{\alpha\beta}$ the influence functional is
\begin{align}
S_{\rm IF} & = q^2 \int \frac{\d^4 k} {(2\pi)^4} \bigg[ -  A_\mathrm{a}^\alpha(-k)\mathcal{D}_{\alpha\beta}(k) A_\mathrm{r}^\beta(k)  + \frac{i}{2} A_\mathrm{a}^\alpha(-k)  \mathcal{S}_{\alpha\beta}(k) A_\mathrm{a}^\beta(k) \bigg] \, ,
\end{align}
with 
\ba
- q^2 \mathcal{D}_{\alpha\beta}(k) & = &  2 f_{1}(-k)    (k^2 \eta_{\alpha\beta} - k_{\alpha} k_{\beta} ) + f_{2}(-k) \Big( (n \cdot k)^2  \eta_{\alpha\beta} + k^2 n_{\alpha} n_{\beta}  - (k \cdot n) [
k_{\alpha}  n_{\beta} + k_{\beta} n_{\alpha} ] \Big) \, , \qquad \quad \\
q^2 \mathcal{S}_{\alpha\beta}(k)  & = & 2 g_{1}(-k) (k^2 \eta_{\alpha\beta} - k_{\alpha}k_{\beta} ) + g_{2}(-k) \Big( (n \cdot k)^2  \eta_{\alpha\beta} + k^2 n_{\alpha} n_{\beta}  - (k \cdot n) [
k_{\alpha}  n_{\beta} + k_{\beta} n_{\alpha} ] \Big) \, ,
\ea
being the dissipation and noise kernels, respectively.
One can check explicitly that $ k^\mu \mathcal{D}_{\mu\nu} = k^\mu \mathcal{S}_{\mu\nu} =0$ consistent with gauge invariance. Next, we can decompose into longitudinal and transverse projectors from Eq.~(\ref{PLPT_sum}), as
\begin{align}
 \mathcal{D}_{\alpha\beta} & = \mathcal{D}_\mathrm{L}(k) \, \mathcal{P}^\mathrm{L}_{\alpha\beta} + \mathcal{D}_\mathrm{T}(k) \, \mathcal{P}^\mathrm{T}_{\alpha\beta}  \, ,
\\
 \mathcal{S}_{\alpha\beta} & = \mathcal{S}_\mathrm{L}(k) \, \mathcal{P}^\mathrm{L}_{\alpha\beta} + \mathcal{S}_\mathrm{T}(k) \, \mathcal{P}^\mathrm{T}_{\alpha\beta}  \, ,
\end{align}
for some scalar kernels $\mathcal{D}_{L,T}(k)$ and $\mathcal{S}_{L,T}(k)$. Matching carefully to the above expressions, we get: 
\begin{equation}
  \begin{split}
    - q^2 \mathcal{D}_\mathrm{L}(k) & = 2 f_1(-k) - f_2(-k) \, ,\\
    - q^2 \mathcal{D}_\mathrm{T}(k) & = \frac{2 k^2}{|\mathbf k|^2} f_1(-k) + \frac{k_0^2}{|\mathbf k|^2} f_2(-k) \, ,
  \end{split}
\qquad \qquad
  \begin{split}
    q^2 \mathcal{S}_\mathrm{L}(k) & = 2 g_1(-k) - g_2(-k) \, ,\\
    q^2 \mathcal{S}_\mathrm{T}(k) & =  \frac{2k^2}{|\mathbf k|^2} g_1(-k) + \frac{k_0^2}{|\mathbf k|^2} g_2(-k) \, .
  \end{split}
\end{equation}
This gives precisely the influence functional  of the form of Eq.~(\ref{ScalarQEDra}) derived from  top-down.

\subsection{Microcausality Constraints on Influence Kernels}

The functions $f_n(x-y)$, or influence kernels, that arise in the dissipative part of the action are not arbitrary, they must crucially satisfy the relativistic microcausality condition (one of the Wightman axioms)
\be
f_n(x-y) = 0 \, \quad \text{unless } \, (x-y)^2 \le 0 \, \text{ and } \, x^0>y^0 \,.
\ee
In explicit examples this arises automatically since these contributions arise directly from the retarded propagators of the states which are integrated out. This condition ensures that the equations of motion for the retarded gauge fields only depend on the causal past of the time at which the leading two time derivative terms are evaluated. A recent discussion of microcausality in this context, albeit for the 1PI effective action rather than the influence functional is given in \cite{Creminelli:2024lhd}. See also \cite{Hui:2025aja,Creminelli:2025rxj}.

The influence kernels must also be consistent with the properties of a density matrix. In particular reality of the influence functional (equivalent to hermicity of the reduced density matrix or that $S_{\mathrm{IF}}[A_\mathrm{r},A_\mathrm{a}]^{\ast} = -S_{\mathrm{IF}}[A_\mathrm{r},-A_\mathrm{a}]$) determines that $f_{n},g_{n}$ are real functions.

\subsection{Symmetry breaking case} 

In the previous section, we worked out the effective action for linear response theory. In the presence of SSB of the gauge symmetry it is sufficient to consider the addition of operators that depend on the Goldstone/stu fields that are consistent with the assumed symmetries.
We are interested in an in-in action of the form\footnote{Again for simplicity of presentation we drop the gauge fixing and ghost terms.}
\begin{align}
S_{\text{in-in}} =   \int \d^4 x \,\bigg[- \frac{1}{2}F^\mathrm{r}_{\alpha\beta} F_\mathrm{a}^{\alpha\beta} + \theta\,F_\mathrm{r}^{\alpha\beta} \tilde{F}_\mathrm{a}^{\alpha\beta}  - \mathscr{D}^\mathrm{r}_{\mu}\chi_\mathrm{r}(x) \mathscr{D}^{\mu}_{\mathrm{a}} \chi_\mathrm{a}(x) \bigg] +S_{\rm IF} \, ,
\end{align}
where $S_{\rm IF}$ was defined in (\ref{IFdef}).
Using the same symmetry assumptions (isotropy and spacetime translations) the additional contributions to the dissipative action in this case will look like 
\begin{align}
\Delta S^{\rm diss} & = \int \d^4 x \int \d^4 y \, \bigg\{ f_{10}(x-y) \, \mathscr{D}^\mathrm{r}_{0} \chi_\mathrm{r}(x)  \mathscr{D}^\mathrm{a}_{0}\chi_\mathrm{a}(y) +  f_{11}(x-y) \, \eta^{ij} \mathscr{D}_i^\mathrm{r}  \chi_\mathrm{r}(x) \mathscr{D}_j^\mathrm{a} \chi_\mathrm{a}(y) \nonumber
\\&+ h_1(x-y) \, F^\mathrm{r}_{0i}(x) \mathscr{D}_i^\mathrm{a} \chi_\mathrm{a}(y)  + h_2(x-y) \, \mathscr{D}_i^\mathrm{r} \chi_\mathrm{r}(x) F^\mathrm{a}_{0i}(y) 
+  h_3(x-y) \, \epsilon_{ijk} F^\mathrm{r}_{ij}(x) \mathscr{D}_k^\mathrm{a} \chi_\mathrm{a}(y) \nonumber
\\& +  h_4(x-y) \, \epsilon_{ijk} \mathscr{D}^\mathrm{r}_{k} \chi_\mathrm{r}(x) F^\mathrm{a}_{ij}(y) \bigg\}
\end{align}
and the noise contributions will be
\begin{align}
\Delta S^{\rm noise} & = \int \d^4 x \int \d^4 y \, \bigg\{ g_9(x-y) \, \mathscr{D}_0^\mathrm{a} \chi_\mathrm{a}(x)  \mathscr{D}^\mathrm{a}_{0}\chi_\mathrm{a}(y) +  g_{10}(x-y) \, \eta^{ij} \mathscr{D}_i^\mathrm{a}  \chi_\mathrm{a}(x) \mathscr{D}_j^\mathrm{a} \chi_\mathrm{a}(y) \nonumber
\\&+ c_1(x-y) \,  F^\mathrm{a}_{0i}(x) \mathscr{D}_i^\mathrm{a} \chi_\mathrm{a}(y)  
 + c_2(x-y) \, \epsilon_{ijk}  F^\mathrm{a}_{ij}(x) \mathscr{D}_k^\mathrm{a} \chi_\mathrm{a}(y) \bigg\} \, .
\end{align}
The example influence functional \eqref{IF100} is a particularly simple example of this general structure.

\subsection{Local limit} 

As we have already discussed in Sec.~\ref{nonlocality}, the  influence functional is parametrised by complex couplings that are generically non-local in space and time. Let us reiterate that ``non-local in time'' means the equations of motion contain memory integrals, convolutions over past times, while ``non-local in space'' refers to spatial dispersion, where the system’s response depends on gradients in a non-perturbative way, or equivalently, on spatial momenta in forms that cannot be captured by a finite gradient series. 

When there is a sufficient separation of scales, a local limit can sometimes be identified. This might occur in certain parameter regimes in the construction described above, but it should be regarded as an exception rather than the norm. Likewise, a short correlation length together with a sufficiently regular kernel near $\mathbf k=0 $ ({\it eg.}, screening, corresponding to  the absence of gapless modes)  allows for a spatial gradient expansion. In this case, the kernels correspond to response functions that are analytic in $k$, such as polynomial expansions capturing local derivative corrections.

In general, a pronounced separation of scales typically suppresses genuine open system effects. In particular, as we discussed in Sec.~\ref{MassLimit}, if all environmental excitations that couple to the photon are heavy $m \gg \{|k|, T\}$, Boltzmann loops are mass suppressed as $\sim e^{-m/T}$ (already at $n=1)$. In this case, non-local tails are exponentially small ({\it eg.}, finite temperature effects or multi-particle interactions). In this regime the open EFT reduces to a Wilsonian expansion. Stated differently: the very conditions that justify a gradient expansion are typically the same conditions that suppress environmental effects.  

In cosmology we expect open system effects become relevant whenever at least one of the following is present: (i) light ({\it eg.}, gapless) degrees of freedom interact with the system generating memory effects and damping \cite{Weinberg:2003ur}, (ii) strongly coupled or otherwise non-perturbative regimes ({\it eg.}, reheating/preheating \cite{Kofman:1997yn}, electroweak phase transitions \cite{Morrissey:2012db}), (iii) horizons, ({\it eg.}, cosmological/black-hole horizons \cite{gibbons1977cosmological, goldberger2006dissipative, Danielson:2022tdw}), iv) stochastic backreaction \cite{starobinsky2005stochastic,  baumann2012cosmological}.

\section{Conclusions}

In this work, we clarified the description of gauge theories for open systems and how it differs from the familiar case of open systems with global symmetries. We have emphasised the power of the BRST symmetry in the Schwinger-Keldysh formalism. As in the case of ordinary global symmetries, the in-in boundary conditions explicitly break the naive doubled BRST symmetry down to a single diagonal one. This is equally true both for the full UV/closed system, and the open system that arises from tracing out degrees of freedom. In particular the resulting Feynman-Vernon influence functional is invariant under the diagonal BRST symmetry. In the case of an Abelian theory considered here this is sufficient to guarantee that the influence functional remains exactly invariant under two copies of the gauge symmetry, retarded and advanced. This holds independently of the choice of state, or the presence of symmetry breaking.

The precise way in which gauge invariance of the influence functional is maintained is different depending on whether the gauge symmetry is spontaneously broken or not. When there is no symmetry breaking, open-system effects then appear through non-local, state-dependent response and noise kernels. The non-local functions maintain gauge invariance via Wilson lines built into the propagators of charged states. In the broken case, the existence of a preferred gauge (unitary gauge) simplifies the analysis, and the open system is in effect unambiguously defined in unitary gauge. The generic open EFT is then captured by an EFT for the gauge fields and \stu degrees of freedom. 

We have illustrated these general statements through a range of explicit top-down constructions. These included gauged Caldeira-Leggett models, open formulations of spinor and scalar QED (both at zero temperature and in thermal states), and extensions to spontaneously broken phases in the Abelian-Higgs-Kibble model. In all cases, integrating out environmental degrees of freedom produces dissipative and stochastic effects that mix the two CTP branches, yet the influence functional continues to respect retarded and advanced gauge invariance once the required Wilson-line dressings are accounted for. In thermal relativistic examples, we explicitly demonstrated that genuine openness is generically non-local in space and time, with local effective descriptions emerging only in appropriate Wilsonian limits.

We also developed a complementary bottom-up framework for open electromagnetic EFTs formulated directly at the level of the in-in action. By imposing diagonal BRST invariance and spacetime symmetries, we derived the most general non-local influence functional at quadratic order consistent with retarded and advanced gauge invariance, both in unbroken and symmetry-broken phases, for the case of an isotropic and translation-invariant medium. This provides a systematic parameterisation of open gauge dynamics that can arise from sensible UV completions.

Our analysis further clarifies recent claims in the literature that associate openness in gauge theories with a breaking or deformation of advanced gauge symmetry. For theories with global symmetries, the global $G \times G$ symmetries are generically broken to the diagonal subgroup $G_{\rm diag}$ by the in-in boundary conditions (both at the final and initial times). In other words, the advanced global symmetry is broken. By contrast for gauge theories the advanced local symmetry remains intact (up to ghost interactions in the non-Abelian case). The apparent contradiction is resolved by analyzing the decoupling limit of the gauge case, which naturally reproduces the breaking of the global diagonal symmetry. At an elementary level this can be understood because the decoupling limit of the diagonal BRST transformation reduces to a diagonal global symmetry alone.

Our analysis has focussed primarily on Gaussian initial states for simplicity, and we have expanded on the way these can be easily described through the $i \epsilon$ prescription, and how the latter can be ensured to be BRST invariant. However, generic states defined at finite time can easily be described. 
The BRST symmetry provides a simple definition of when states are physical, which can easily be incorporated into the Schwinger-Keldysh path integral, and in particular the $i \epsilon$ terms when used should be BRST invariant.

Whilst our specific examples have focused on Abelian theories, the underlying mechanisms of BRST doubling on the Schwinger-Keldysh contour, the essential role of Wilson lines and \stu fields, the (diagonal) BRST invariance of the influence functional, and the emergence of non-local influence functionals-apply equally to non-Abelian gauge theories \cite{Kaplanek:2026kpp} and, by extension, to gravity. The main additional complication is the interacting and state-dependent ghost sector, a feature already familiar from real-time treatments of QCD. The interactions of the ghosts with the environment degrees of freedom will in general lead to non-trivial ghost interactions in the Feynman-Vernon influence functional. These will spoil the naive doubled gauge invariance of the influence functional, but will do so in such a way that the diagonal BRST invariance is intact \cite{Kaplanek:2026kpp}. Nevertheless, interactions that do not arise with accompanying ghost interactions must be gauge invariant under both advanced and retarded non-Abelian gauge transformations. Thus while technically slightly more subtle, there is no significant conceptual change from the Abelian treatment.
We therefore expect our conclusions to carry over directly to non-Abelian gauge theories and to gravitational open EFTs, with potential applications ranging from plasma physics to cosmology.
In summary, openness in gauge theories does not require breaking the advanced gauge symmetry. Instead, it is naturally and consistently realised through BRST-invariant, non-local influence functionals that preserve retarded and advanced Ward identities while encoding dissipation, noise, and memory effects in a state-dependent manner.

\section*{Acknowledgements}

We thank Daniel Baumann, Cliff Burgess, Perseas Christodoulidis, Thomas Colas, Claudia de Rham, Matthew Johnson, Amaury Micheli, Ryo Namba, Toshifumi Noumi, Antonio Padilla, Misao Sasaki, Sergey Sibiryakov for useful discussions. AJT is supported by the STFC Consolidated Grant ST/X000575/1. MM is supported Kavli IPMU which was established by
the World Premier International Research Center Initiative (WPI), MEXT, Japan. MM is also grateful for the hospitality of Perimeter
Institute where part of this work was carried out. Her visit to Perimeter Institute was supported by a
grant from the Simons Foundation (1034867, Dittrich). Research at Perimeter Institute is supported
in part by the Government of Canada through the Department of Innovation, Science and Economic
Development and by the Province of Ontario through the Ministry of Colleges and Universities.

\appendix

\section{Path Integral Measure and Matthews' theorem}

\label{app:measure}

Throughout this paper, we will work with the covariant (configuration space) form of the path integral. In doing so, we rely on Matthews' theorem \cite{PhysRev.76.684.2} which can in effect be stated as the operator statement
\be
{\cal T} e^{-i \int \d^4 x \, \hat {\cal H}_{\rm int}(x)} \equiv {\cal T}^* e^{i \int \d^4 x \, \hat {\cal L}_{\rm int}(x)} \, ,
\ee
with ${\cal T}$ the Dyson time ordering operator and ${\cal T}^*$ the covariant time ordering operator of  Nishijima \cite{10.1143/ptp/5.3.405} or equivalently as the path integral statement that 
\be \label{eq:pathintegral}
Z=\int {\cal D}[\phi_I] \int {\cal D}[\pi_I] \;  e^{i \int \d^4 x \sum_I \pi_I \partial_t  \phi_I -{\cal H}[\phi_I,\pi_I]} = \int {\cal D}[\phi_I] \; e^{i S[\phi_I]} \, ,
\ee
for generic fields $\phi_I$ of any spin. This holds even when 
\be
\hat {\cal H}_{\rm int}(x) \neq  - \hat {\cal L}_{\rm int}(x) \, .
\ee
The content of Matthews' theorem is that even in theories with derivative interactions for which the interaction Hamiltonian and Lagrangian differ by more than a sign, the difference between the two is entirely captured by distinction between the ordinary Dyson time ordering and the covariant time ordering.  At loop level, Matthews' theorem is known to hold for derivative interactions provided a covariant regularisation scheme such as dimensional regularisation is used \cite{PhysRevD.11.848}. 

The path integral explanation of Matthews' theorem is contained in the relationship between the canonical/phase-space path integral for which the measure is trivial (unity), and the covariant path integral for which the measure is not necessarily unity. For theories in which the phase space action is Gaussian in field momenta, it is straightforward to compute the covariant path integral measure exactly and there are well known examples where this measure is field dependent \cite{Salam:1970fso}. In general we should identify 
\be
 \int {\cal D}[\pi_I]\; e^{i \int \d^4 x \sum_I \pi_I \partial_t  \phi_I -{\cal H[\phi_I,\pi_I]}} = \mu[\phi_I]  e^{i S[\phi_I]} \, ,
\ee
to define the measure $\mu[\phi_I] $ and insert this measure in the path integral on the RHS of \eqref{eq:pathintegral}. DeWitt has given an exact expression for the covariant path integral measure at one-loop by careful consideration of the time-ordering operation \cite{DeWitt:1967ub,DeWitt:2003pm}. This has a compact expression in terms of the advanced Green's functions. For example, for a scalar field theory, the correct covariant path integral to one-loop order is
\be
Z=  \int {\cal D} [\phi]\; \sqrt{\det [G_A[\phi]]}e^{i S[\phi]} \, ,
\ee
where $G_A[\phi](x,y)$ is the advanced Green's function which satisfies\footnote{Contrary to the usual classical convention, we include an $i$ in the Green's function equation so that $G_A$ satisfies the same equation as the Feynman propagator around a background.}
\be \label{GAequation}
\int \d^4 z \; \frac{\delta^2 S[\phi]}{\delta \phi(x) \delta \phi(z)}G_A[\phi](z,y)=i \delta^4(x,y) \, .
\ee
With this choice the 1PI effective action at one-loop is
\be
\Gamma[\phi] = S[\phi]-\frac{i}{2} \Tr \log G_F[\phi]+\frac{i}{2} \Tr \log G_A[\phi] \, ,
\ee
with $G_F[\phi]$ the equivalent solution of \eqref{GAequation} with Feynman boundary conditions. Almost all discussions of the 1PI effective action drop the contribution from $\Tr \log G_A[\phi]$. The reason this is allowed is, if $\Tr \log G_A[\phi]$ is computed perturbatively in dimensional regularisation then its result is trivial. The reason being is that $\Tr \log G_A[\phi]$ leads only to power law divergent contributions that vanish in dimensional regularisation. This is best understood in real space. When computing perturbative corrections to the advanced Green's functions we will naturally get closed cycles in the sense of factors such as
\be
G_A(x,y) G_A(y,z) G_A(z,x)\, .
\ee
Since $G_A(x,y)$ vanishes unless $x$ is in the past of $y$ the only support this expression has is for coincident singularities when $x=y=z$, which is equivalent in momentum space to only receiving contributions from the arcs at infinity.
More precisely, as shown by DeWitt, the contribution of $\Tr \log G_A[\phi] $ serves to justify the evaluation of integrals in dimensional regularisation via Wick rotation to the Euclidean \cite{DeWitt:2003pm} since it serves to cancel the arcs at infinity usually neglected in arguments related to Wick rotation in dimensional regularisation.
Thus, whilst technically $\mu \neq 1$, it is for all intents and purposes unity provided loop corrections are computed in dimensional regularisation via Wick rotation. Hence, the covariant path integral, which is computing matrix elements of ${\cal T}^* e^{i \int \d^4 x \, \hat {\cal L}_{\rm int}(x) }$ is equivalent to the canonical path integral, which is computing matrix elements of ${\cal T} e^{-i \int \d^4 x \, \hat {\cal H}_{\rm int}(x)} $ which is Matthews' theorem.

Despite this, there are situations where it is helpful to include the measure. Notably, the one-loop correction can be written as a ratio of Fredholm determinants 
\be
\Gamma[\phi] = S[\phi]-\frac{i}{2} \log \( \frac{\det G_F[\phi]}{\det  G_A[\phi]} \)\, ,
\ee
which can be useful for analyzing its properties since such ratios are well known to be better defined than the individual determinants. In the context of the in-in formalism however, the path integral measure plays a crucial role. The in-in version of the 1PI effective action is to one-loop order
\be
\Gamma[\phi_+,\phi_-]=S[\phi_+]-S[\phi_-]-\frac{i}{2} \Tr \log {\bf G}[\phi_+,\phi_-]+\frac{i}{2} \Tr \log {\bf G}_A[\phi_+,\phi_-]
\ee
Here ${\bf G}[\phi_+,\phi_-]$ is now the matrix of propagators for fluctuations in $\phi$ around a background value, with Feynman boundary conditions of the CTP in a background which may be different on each branch. Similarly ${\bf G}_A[\phi_+,\phi_-]$ is the equivalent matrix of propagators with advanced boundary conditions relative to the CTP (this is defined in \eqref{advancedpropagator}). It is straightforward to show that 
\be
\det {\bf G}_A[\phi_+,\phi_-]=\det G_A[\phi_+] \det G_R[\phi_-]  \, .
\ee
Now if the background advanced field is set to zero, $\phi_{\pm}=\phi_\mathrm{r}$, then it may be easily shown in the Keldysh basis that ${\bf G}_A[\phi_\mathrm{r},\phi_\mathrm{r}]$ takes the form of a block matrix with a zero on the off-diagonal. Hence the determinant reduces to 
\be
\det {\bf G}[\phi_\mathrm{r},\phi_\mathrm{r}]=\det G_A[\phi_\mathrm{r}] \det G_R[\phi_\mathrm{r}] \, .
\ee
Putting this together
\be
\Gamma[\phi_\mathrm{r},\phi_\mathrm{r}]=0 \, .
\ee
This relation holds as an identity, only because we have included the measure factor. 

A similar argument applies to the influence functional. The influence functional should satisfy
\be
S_{\rm IF}[\phi_\mathrm{r},\phi_\mathrm{r}]=0\, ,
\ee
by virtue of unitarity, since the interactions induced by the background field on each branch should cancel. This is manifest in the canonical/phase-space formalism, but is not manifest in the covariant one. 
Let us now consider a scalar $\phi$ coupled to a second field $H$ which we integrate out in the in-in path integral
\be
\tilde \mu[\phi_+,\phi_-]\; e^{i S_{\rm IF}[\phi_+,\phi_-]} = \int {\mathcal D}[H_+,H_-] \; \mu[\phi_+,\phi_-,H_+,H_-]\; e^{i S[\phi_+,\phi_-,H_+,H_-]+i S_{i \epsilon}}
\ee
where we have now been careful to include a possible covariant measure factor $\mu[\phi_+,\phi_-,H_+,H_-]$. in the full path integral, and have similarly maintained an effective measure $\tilde \mu[\phi_+,\phi_-] $ for the reduced system.

Following DeWitt, at one-loop level $\mu[\phi_+,\phi_-,H_+,H_-]$ will be given by square root of the determinant of now a 4 by 4 matrix of propagators which satisfy the matrix Green's function equation
\be \label{GAequation}
\sum_{\beta = \pm}\int \d^4 z \begin{pmatrix} \frac{\delta^2 S[\phi]}{\delta \phi_{\alpha}(x) \delta \phi_{\beta}(z)} & \frac{\delta^2 S[\phi]}{\delta \phi_{\alpha}(x) \delta H_{\beta}(z)} \\ \frac{\delta^2 S[\phi]}{\delta H_{\alpha}(x) \delta \phi_{\beta}(z)}  & \frac{\delta^2 S[\phi]}{\delta H_{\alpha}(x) \delta H_{\beta}(z)} \end{pmatrix}{\bf G}_{A \beta \gamma}[\phi_+,\phi_-,H_+,H_{-}](x,y) =i \delta^4(x,y) \delta_{\alpha \gamma}\begin{pmatrix} 1 & 0 \\ 0 & 1 \end{pmatrix}  \, ,
\ee
with $\alpha,\beta,\gamma$ taking values $\pm$ and with advanced boundary conditions on the CTP used, so that 
\be
\mu[\phi_+,\phi_-,H_+,H_-]=\sqrt{\det {\bf G}_{A \beta \gamma}[\phi_+,\phi_-,H_+,H_{-}]} \, .
\ee
However, if $S[\phi_+,\phi_-,H_+,H_-]$ is organised so that it has no linear terms in $H_{\pm}$, then the loop expansion is based around $H_{\pm}=0$ which means to one-loop level it is sufficient to evaluate $\mu[\phi_+,\phi_-,0,0]$. With this choice, the cross terms such as $ \frac{\delta^2 S[\phi]}{\delta H_{\alpha}(x) \delta \phi_{\beta}(z)} =0$ vanish and so the determinant factorizes into that for the $\phi$ and $H$ in a background $\phi_{\pm}$ separately
\be
\mu[\phi_+,\phi_-,0,0] = \sqrt{\det {\bf G}_{\phi A}[\phi_+,\phi_-] } \; \sqrt{\det {\bf G}_{H A}[\phi_+,\phi_-] } \, .
\ee
We can now identify 
\be
\tilde \mu[\phi_+,\phi_-] =\sqrt{\det {\bf G}_{\phi A}[\phi_+,\phi_-]} \, ,
\ee
as the correct measure for the subsequent $\phi$
 path integrals to one-loop.
Hence we infer that to one-loop order 
\be
S_{\rm IF}[\phi_+,\phi_-]=-\frac{i}{2} \Tr \log {\bf G}_H[\phi_+,\phi_-]+\frac{i}{2} \Tr \log {\bf G}_{H A}[\phi_+,\phi_-]
\ee
with ${\bf G}_H$ now the matrix of propagators of the $H$ field in a background of $\phi$.
Hence 
\be
S_{\rm IF}[\phi_\mathrm{r},\phi_\mathrm{r}]=-\frac{i}{2} \Tr \log {\bf G}_H[\phi_\mathrm{r},\phi_\mathrm{r}]+\frac{i}{2} \Tr \log {\bf G}_{H A}[\phi_\mathrm{r},\phi_\mathrm{r}]=0 \, .
\ee
In summary, in the covariant formalism, the unitarity of the influence functional is guaranteed only when the appropriate covariant path integral measure is included. By contrast in the phase space path integral formalism, no measure is needed and unitarity holds straightforwardly. Matthews' theorem is contained in the result that when $\Tr \log {\bf G}_{\phi A}[\phi_+,\phi_-]$ or $\Tr \log {\bf G}_{H A}[\phi_+,\phi_-]$ is computed in dimensional regularisation, it gives no new contribution and serves only to justify the computation of loop integrals via Wick rotation \cite{DeWitt:2003pm} which explains its near universal neglect in most discussions.

Note that several recent papers have proposed a definition of the measure to all loop orders in the in-in path integral (with no clear derivation from in-out) by utilizing a novel BRST-like symmetry \cite{crossley2017effective,Liu:2018kfw,Haehl:2015foa,Haehl:2016pec} (which has no relation to the usual BRST of gauge theories). This is designed to ensure that $S_{\rm IF}[\phi_\mathrm{r},\phi_\mathrm{r}]=0$ to all loop orders. It would be interesting to explore the relation of this conjecture to the well established measure of DeWitt and the true measure of the phase space path integral (see also \cite{Gao:2018bxz} for a related discussion).

\section{In-in propagators and Wilson lines}

\label{app:Wilson}

To illustrate the subtleties of in-in quantisation for gauge theories let us consider a simple gauged complex scalar quantum mechanics theory with harmonic oscillator action
\be
S[\Phi,\Phi^*,A] = \int_{\ti}^{\tf} \d t \left( | D_0[A]\Phi|^2-\Gamma^2 |\Phi|^2 \right) \, ,
\ee
with $D_0[A]=\partial_t - i q A_0(t)$.
We define the in-in connected generating functional ({\it cf.}~Eq.~(\ref{Zinin})) via
\ba
&& e^{i W[J_+,J_-,A_+,A_-] }  \\
&& \quad = \int \d[ \Upphi_\mathrm{f} ,\Upphi^*_\mathrm{f},\Upphi_{\pm \mathrm{i}},\Upphi^*_{\pm \mathrm{i}}] \; \langle \Upphi_{+}\Upphi_+^*| \rho | \Upphi_{-}\Upphi_-^* \rangle \int_{\Upphi_{+\mathrm{i}}\Upphi^*_{+ \mathrm{i}}}^{\Upphi_\mathrm{f}\Upphi^*_\mathrm{f}}  {\cal D}[\Phi_+,\Phi^{\ast}_+] \int_{\Upphi_{-\mathrm{i}}\Upphi^*_{- \mathrm{i}}}^{\Upphi_\mathrm{f}\Upphi^*_\mathrm{f}}  {\cal D}[\Phi_-,\Phi^{\ast}_-] \nn \\
&& \qquad \times \; e^{i S[\Phi_+,\Phi_+^*,A_+]-iS[\Phi_-,\Phi_-^*,A_-]+i \int_{\ti}^{\tf} \d t \( J_+^*(t) \Phi_+(t)+ \Phi_+^*(t) J_+(t)- J_-^*(t) \Phi_-(t)- \Phi_-^*(t) J_-(t)\)} \, , \nn
\ea
accounting for the different gauge fields on each branch. The initial state should be invariant under gauge transformations at the initial time. To keep the problem exactly solvable we will consider a Gaussian density matrix
\be
\langle \Upphi_{+}\Upphi_+^*| \rho | \Upphi_{-}\Upphi_-^* \rangle =A e^{-\kappa | \Upphi_+|^2-\kappa | \Upphi_-|^2} \, .
\ee
Since we are considering a free theory, with Gaussian initial state, the generating functional at tree level can be determined by solving the classical equations of motion with sources, accounting for the in-in boundary conditions, and substituting the solution back in. 
The classical problem including its boundary conditions can be summarised by the combined in-in action
\ba
S_{\text{in-in}}&=&S[\Phi_+,\Phi_+^*,A_+]-S[\Phi_-,\Phi_-^*,A_-]+\int_{\ti}^{\tf} \d t \( J_+^* \Phi_+(t)+ \Phi_+^* J_+(t)- J_-^* \Phi_-(t)- \Phi_-^* J_-(t)\) \nn  \\
&&+  \lambda^*(\Phi_+(\tf)-\Phi_-(\tf))+(\Phi_+^*(\tf)-\Phi^*_-(\tf))\lambda+i \kappa | \Phi_+(\ti)|^2+i\kappa | \Phi_-(\ti)|^2 \, .
\ea
Where we have introduced a complex Lagrange multiplier $\lambda$ to impose the final time constraint. Extremizing the action $\delta S_{\text{in-in}}=0$ the equations of motion are
\ba
&& D_0[A_+]^2 \Phi_+ + \Gamma^2 \Phi_+ = J_+(t) \, , \\
&& D_0[A_-]^2 \Phi_- + \Gamma^2 \Phi_- = J_-(t) \, , 
\ea
with boundary conditions
\begin{equation} \label{QMBC}
  \begin{split}
    D_0[A_+]\Phi_+(\ti) & = i \kappa \Phi_+(\ti) \, ,\\
    \Phi_+(\tf) &= \Phi_-(\tf)\, ,
  \end{split}
\qquad \qquad
  \begin{split}
    D_0[A_-]\Phi_-(\ti) & = -i \kappa \Phi_-(\ti) \, , \\
    \dot \Phi_+(\tf)&=\dot \Phi_-(\tf) \, ,
  \end{split}
\end{equation}
remembering that $A_0^+(\tf)=A^-_0(\tf)$. There are similar boundary conditions for $\Phi_{\pm}^*$ with $\kappa \rightarrow -\kappa$.
Consider now a redefinition of both branches of fields and sources via a Wilson line connected to the initial time,
\be
\Phi_{\pm}(t) = e^{iq \int_{\ti}^t \d \tau A^{\pm}_0(\tau)}\sigma_{\pm}(t)\, , \qquad \mathrm{and} \qquad J_\pm(t)=e^{iq \int_{\ti}^t \d \tau A^{\pm}_0(\tau)}j_{\pm}(t)  \, .
\ee
The renders the upper line of boundary conditions (\ref{QMBC}) standard
\be
\partial_t \sigma_+(\ti) = i \kappa \sigma_+(\ti)\, , \quad \partial_t \sigma_-(\ti) = -i \kappa \sigma_-(\ti)\, , \\
\ee
together with the equations of motion 
\ba
&& \partial_t^2 \sigma_+ + \Gamma^2 \sigma_+ = j_+(t) \, ,\\
&& \partial_t^2 \sigma_- + \Gamma^2 \sigma_- = j_-(t) \, ,
\ea
but the lower line of the boundary conditions in (\ref{QMBC}) now contain explicit Wilson lines for the advanced gauge field $A^\mathrm{a}_0=A^+_0-A^-_0$
\be
\sigma_{+}(\tf) = e^{-i q \int_{\ti}^{\tf} \d \tau A^{\mathrm{a}}_0(\tau)} \sigma_{-}(\tf) \, , \quad 
\dot \sigma_{+}(\tf) = e^{-i q \int_{\ti}^{\tf} \d \tau A^{\mathrm{a}}_0(\tau)} \dot \sigma_{-}(\tf) \, .
\ee
We can further rescale
\be
\sigma_{\pm}(t) =  e^{\mp \tfrac{i}{2} q \int_{\ti}^{\tf} \d \tau A^{\mathrm{a}}_0(\tau)} \tilde \sigma_{\pm}(t) \, , \quad j_{\pm}(t) =  e^{\mp \tfrac{i}{2} q \int_{\ti}^{\tf} \d \tau A^{\mathrm{a}}_0(\tau)} \tilde j_{\pm}(t) \, ,
\ee
then the tilde variable satisfy the standard equations in the absence of gauge fields
\begin{equation}
  \begin{split}
    \partial_t \tilde \sigma_+(\ti) & = i \kappa \tilde \sigma_+(\ti) \, ,\\
    \partial_t^2 \tilde \sigma_+ + \Gamma^2 \tilde \sigma_+ & = \tilde j_+(t) \, , \\
    \tilde \sigma_{+}(\tf) & = \tilde \sigma_{-}(\tf) \, ,
  \end{split}
\qquad \qquad 
  \begin{split}
    \partial_t \tilde  \sigma_-(\ti) & = -i \kappa \tilde \sigma_-(\ti) \, ,\\
    \partial_t^2 \tilde \sigma_- + \Gamma^2 \tilde \sigma_- & = \tilde j_-(t) \, , \\
    \dot{\tilde \sigma}_{+}(\tf) & = \dot{\tilde \sigma}_{-}(\tf) \, .
  \end{split}
\end{equation}
Consequently the generating function will take the standard form in terms of $\tilde j_{\pm}$. In other words 
\be
 W[J_+,J_-,A_+,A_-]= 
 i \int_{\ti}^{\tf} \d t  \int_{\ti}^{\tf} \exd t'  \begin{pmatrix} \tilde j_+^*(t) & -\tilde j_-^*(t) \end{pmatrix} {\bf D}(t,t') \begin{pmatrix}  \tilde j_+(t') \\ -\tilde j_-(t') \end{pmatrix}  \, ,
\ee
with $\bf D$ the standard matrix of propagators for a single complex oscillator. Converting back to the original source variables we have
\be
 W[J_+,J_-,A_+,A_-]= 
 i \int_{\ti}^{\tf} \d t \int_{\ti}^{\tf} \exd t'  \begin{pmatrix} J_+^*(t) & -J_-^*(t) \end{pmatrix} V(t) {\bf U}(t)  {\bf D}(t,t') {\bf U}^{\dagger}(t') V^*(t') \begin{pmatrix}  J_+(t') \\  -J_-(t') \end{pmatrix} \, ,
\ee
with Wilson line 
\be
V(t)=e^{ iq  \int^{t}_{\ti} \d t' A_0^{\mathrm{r}}(t') } \,, 
\ee
and
\be
{\bf U}(t) = \begin{pmatrix} U_+(t) &0 \\ 0 & U_{-}(t)\end{pmatrix} \, ,
\ee
with 
\be
U_{\pm}(t)=e^{\mp \frac{i}{2}q \int^{\tf}_{t} \d t' A_0^{\mathrm{a}}(t') } \, .
\ee
In short, the retarded gauge fields are naturally tethered by Wilson lines to the initial time, and the advanced ones to the final time. This is the $0+1$ dimensional version of the discussion in Sec.~\ref{sec:CLmodel}. We stress that this emerges simply from solving the classical equations of motion with in-in boundary conditions.

\section{Noether's theorem in the Influence Functional}
\label{App:Noether}

In general an influence functional is defined by a partial path integral, {\it i.e.} a subset of degrees of freedom are integrated out.  In the present work we are mainly interesting in the influence functional for the photon obtained by integrating out charged matter. For example in spinor QED we have (ignoring the measure)
\ba
e^{i S_{\rm IF}[A_+,A_-]} &=&\int {\mathcal D} [\Psi_+,\bar \Psi_+,\Psi_-,\bar \Psi_- ] \\
&&e^{i \int \d^4 x \( i\overline{\Psi}_+ \slashed{\partial} \Psi_+ - m \overline{\Psi}_+ \Psi_+-i \overline{\Psi}_- \slashed{\partial} \Psi_- + m \overline{\Psi}_- \Psi_- +\overline{\Psi}_+ q\slashed{A}_+ \Psi_+-  \overline{\Psi}_- q\slashed{A}_- \Psi_-\)} \nn
\ea
where we have suppressed the final time integrals and initial state specification.
Within this path integral $A_{\pm}$ should be regarded as fixed background fields. From this point of view, the action that defines this path integral is not gauge or BRST invariant, since we should not transform $A_{\pm}$. It is only once we choose to integrate over the photon that BRST invariance is recovered.

If we ignore the initial/final conditions, it is, however, invariant under two copies of global $U(1)$ transformations. The final time conditions break this down to the diagonal group but locally away from $\ti$ we may still apply the path integral version of Noether's theorem. Specifically, let us consider an infinitesimal field redefinition of the form
\be
\Psi_{\pm}(x) \rightarrow \Psi_{\pm}(x)+i \lambda_{\pm}(x) \Psi_{\pm}(x)
\ee
where $ \lambda_{\pm}(x)$ vanishes at $\ti$ so that the initial/final conditions are invariant. Invariance of the path integral under field redefinitions then implies 
\ba
&&  \int {\mathcal D} [\Psi_+,\bar \Psi_+,\Psi_-,\bar \Psi_- ]  \, \int \d^4 x \; \partial_{\mu} \lambda_{\pm}(x) \Psi_{\pm}(x) \gamma^{\mu} \Psi_{\pm}(x) e^{i \int \d^4 y \dots} = \nn \\
&&  \int {\mathcal D} [\Psi_+,\bar \Psi_+,\Psi_-,\bar \Psi_- ] \(  -  \int \d^4 x \; \lambda_{\pm}(x)  \partial_{\mu} (\Psi_{\pm}(x) \gamma^{\mu} \Psi_{\pm}(x) )\) e^{i \int \d^4 y \dots} \ = \ 0 \, .
\ea
Since inserting the divergence of a current in the influence functional path integral vanishes, it follows that 
\be
\partial_{\mu } \frac{\delta S_{\rm IF}[A_+,A_-]}{\delta A^{\pm}_{\mu}(x)} =0 \, ,
\ee
which is a statement that $S_{\rm IF}[A_+,A_-]$ is invariant under two copies of gauge transformations. This is of course consistent with the necessary diagonal BRST invariance recovered once the photon (and ghosts) are integrated over.

Note that despite familiarity, this argument is slightly different from the conventional arguments for Ward/Slavnov-Taylor identities because we are dealing with a partial path integral. The `current' above is not the full quantum current since it does not include any contributions from photon loops. It is the partial current obtained assuming that the photon fields are background values.

\section{Thermal Scalar QED details}
\label{App:thermalQED}

In this appendix we present the details that go into the results presented in \S\ref{sec:thermal}. 
The contributing correlators are the finite temperature free Feynman propagator from Eq.~(\ref{Feynman_def}) and free Wightman function from Eq.~(\ref{Wightman_def}). After a straightforward calculation, one gets explicitly in position space:
\begin{eqnarray}
S_{\mathrm{IF}}[A_{+},A_{-}] &\simeq& q^2 \int \mathrm{d}^4 x \;\bigg( - A_{+\mu}(x) A_{+}^{\mu}(x) + A_{-\mu}(x) A_{-}^{\mu}(x) \bigg) \; \mathcal{F}(x,x) \label{SIF_infty1}  \\
&  & + i q^2  \int  \mathrm{d}^4 x  \int  \mathrm{d}^4 y \; \bigg( A_{+}^\mu(x) A_{+}^\nu(y) \bigg[ \mathcal{F}(x,y) \frac{\partial^2 \mathcal{F}(x,y)  }{\partial x^{\mu} \partial y^{\nu}} - \frac{\partial \mathcal{F}(x,y) }{\partial x^{\mu}} \frac{\partial \mathcal{F}(x,y) }{\partial y^{\mu}}  \bigg]  \notag \\
&& \hspace{23mm} - A_{+}^\mu(x) A_{-}^\nu(y)  \bigg[ \mathcal{W}^{\ast}(x,y) \frac{\partial^2 \mathcal{W}^{\ast}(x,y)  }{\partial x^{\mu} \partial y^{\nu}} - \frac{\partial \mathcal{W}^{\ast}(x,y) }{\partial x^{\mu}} \frac{\partial \mathcal{W}^{\ast}(x,y) }{\partial y^{\mu}}  \bigg] \notag  \\
&& \hspace{23mm} - A_{-}^\mu(x) A_{+}^\nu(y)  \bigg[ \mathcal{W}(x,y) \frac{\partial^2 \mathcal{W}(x,y)  }{\partial x^{\mu} \partial y^{\nu}} - \frac{\partial \mathcal{W}(x,y) }{\partial x^{\mu}} \frac{\partial \mathcal{W}(x,y) }{\partial y^{\mu}}  \bigg] \notag  \\
&& \hspace{21.5mm}+ A_{-}^\mu(x) A_{-}^\nu(y) \bigg[ \mathcal{F}^{\ast}(x,y) \frac{\partial^2 \mathcal{F}^{\ast}(x,y) }{\partial x^{\mu} \partial y^{\nu}} - \frac{\partial \mathcal{F}^{\ast}(x,y) }{\partial x^{\mu}} \frac{\partial \mathcal{F}^{\ast}(x,y) }{\partial y^{\mu}}  \bigg] \bigg) + \mathcal{O}(q^4) \notag
\end{eqnarray}
Going to momentum space results in Eq.~(\ref{thermalIF_loops}) from the main text.

\subsection{Vacuum contribution}
\label{App:thermalVac}

Because the influence functional (\ref{thermalIF_loops}) contains UV divergences, it is convenient to begin with the vacuum calculation. The key point is that the UV behaviour of thermal states is the same as that of the vacuum, since both share the same universal short-distance singularities. This makes renormalisation simpler to handle in the vacuum case, with thermal effects included afterward.

In practice, this means taking the zero-temperature limit $\beta \to \infty$, so that the loop integrals in (\ref{Pibeta}) and (\ref{Nbeta}) get replaced by
\begin{equation}
\Pi^{\beta}_{\mu\nu} \to \Pi_{\mu\nu} \qquad \mathrm{and} \qquad  \mathcal{N}^{\beta}_{\mu\nu} \to \mathcal{N}_{\mu\nu} \ , 
\end{equation}
with the free thermal propagators are replaced with their vacuum counterparts (see Eq.~(\ref{freeprop_thermal}))
\begin{equation}  \label{freeprop_vac}
\mathcal{F}^{\rm vac}(k) = \frac{- i }{k^2  + m^2 - i \epsilon} \qquad \mathrm{and} \qquad \Wvac(k) = 2 \pi \delta(k^2 + m^2) \theta(k^0)  \ . 
\end{equation}

\subsubsection{Computing $\Pi_{\mu\nu}$}

The loop integral $\Pi_{\mu\nu}$ is nothing but the familiar photon self-energy. Since its computation is standard textbook material, we only sketch the derivation here for completeness. We compute
\begin{eqnarray}
\Pi_{\mu\nu}(k) & \equiv &  2 \eta_{\mu\nu} \int \frac{\exd^4 \ell}{(2\pi)^4}   \mathcal{F}^{\rm vac}(\ell)  - 2 i \int \frac{\exd^4 \ell}{(2\pi)^4}  \mathcal{F}^{\rm vac}(\ell)   \mathcal{F}^{\rm vac}(\ell-k)  ( 2 \ell_{\mu} - k_\mu) \ell_{\nu} \qquad \\
& = & - 2 i \int \frac{\exd^4 \ell}{(2\pi)^4} \; \bigg[ \frac{ \eta_{\mu\nu} }{\ell^2 +m^2 - i \epsilon} - \frac{ ( 2 \ell_{\mu} - k_\mu) \ell_{\nu}  }{\big[ \ell^2 +m^2 - i \epsilon \big]\big[ (\ell - k)^2 +m^2 - i \epsilon \big] }  \bigg] \\
& = & - 2 i \int_{-1}^{+1} \exd y \int \frac{\exd^4 \ell}{(2\pi)^4} \; \frac{  \eta_{\mu\nu} \big[ \ell^2 - 2 \ell \cdot k + k^2 + m^2 \big] - ( 2 \ell_{\mu} - k_\mu) \ell_{\nu}  }{ 2 \big( \ell^2+  (y-1) \ell \cdot k -\frac{1}{2} (y-1) k^2+m^2-i \epsilon \big)^2 } 
\end{eqnarray}
where in the last step we introduced a Feynman parameter using
\begin{equation}
\frac{1}{AB} = 2 \int_{-1}^{+1} \frac{\mathrm{d}y}{[(1+y) A + (1- y) B ]^{2}} \ .
\end{equation}
Shifting the loop momentum to $\ell \to p = \ell + \tfrac{1}{2}(y-1) k$ completes the square in the denominator, giving
\begin{equation}
\Pi_{\mu\nu}(k) = - i \int_{-1}^{+1} \exd y \int \frac{\exd^4 p}{(2\pi)^4} \; \frac{ \eta_{\mu\nu} \big[ p^2 + \frac{1}{4} (y^2 +1) k^2 + m^2 \big] -  2 p_{\mu} p_{\nu} - \frac{1}{2} y^2  k_{\mu} k_{\nu} }{ ( p^2 + \frac{1}{4}\Sigma  - i \epsilon )^2 } \label{loopPi4}
\end{equation}
with $\Sigma \equiv (1 - y^2) k^2+4 m^2$ (where we've dropped terms linear in $y$ or $k$ which vanish by symmetry of the integrals). Our goal is to evaluate the loop integral using dimensional regularisation. By Lorentz invariance, any term proportional to $p_\mu p_\nu$ must reduce to $\eta_{\mu\nu}$ times a scalar function, so we use the replacement
\begin{equation} \label{pmupnu_loop}
\int \frac{\exd^D p}{(2\pi)^D} \; \frac{ p_{\mu} p_{\nu} }{ ( p^2 + \frac{1}{4} \Sigma - i \epsilon )^2 }  \to \frac{ \eta_{\mu\nu} }{ D } \int \frac{\exd^D p}{(2\pi)^D} \; \frac{ p^2 }{ ( p^2 + \frac{1}{4} \Sigma - i \epsilon )^2 }  \ .
\end{equation}
Applying this to Eq.~(\ref{loopPi4}) and promoting the integral to $D$ dimensions gives
\begin{equation}
\Pi_{\mu\nu}(k) =  \int_{-1}^{+1} \exd y  \; \bigg( \Big[ \tfrac{1}{4} (y^2 + 1) \eta_{\mu\nu} k^2 + \eta_{\mu\nu} m^2 - \frac{1}{2} y^2 k_\mu k_\nu \Big] \mathscr{L}_{(0,2)}(\tfrac{1}{4}\Sigma) + \eta_{\mu\nu} \left( 1 - \tfrac{2}{D} \right) \mathscr{L}_{(1,2)}(\tfrac{1}{4}\Sigma) \bigg) \label{loopPi_12}
\end{equation}
where we define the elementary loop integral for integers $n$ and $N$,
\begin{equation} \label{curlyLnN}
\mathscr{L}_{(n,N)}( m^2 ) \equiv \mu^{4-D}  \int \frac{\exd^D p}{(2\pi)^D} \frac{-i  (p^2)^{n} }{( p^2 + m^2 - i \epsilon )^N}
\end{equation}
with $\mu > 0$ is a renormalisation mass scale. Evaluating the loop integral follows standard techniques, taking care to Wick rotate $p^0$ in the correct direction due to the poles located at $p_0 \simeq \pm \sqrt{ |\mathbf{p}|^2 + m^2 } \mp i \epsilon$, with the result
\begin{equation} \label{curlyLnN_ans}
\mathscr{L}_{(n,N)}( m^2 ) = \mu^{4 + 2 n -2 N} \frac{ \Gamma \left(\frac{D}{2}+n\right) \Gamma \left(-\frac{D}{2}-n+N\right) }{ (4 \pi)^{D/2}  \Gamma \left(\frac{D}{2}\right) \Gamma(N)} \left(\frac{m^2-i \epsilon }{\mu ^2}\right)^{\tfrac{D}{2}+n-N}
\end{equation}
where $\gamma$ is the Euler-Mascheroni constant. Note that we must keep the $i \epsilon$ explicit in the above formula for when the argument has an undecided sign. Inserting this result into (\ref{loopPi_12}) along with $\Sigma(y) = (1 - y^2) k^2 + 4m^2$ naturally factors out a $k^2 \eta_{\mu\nu} - k_\mu k_\nu$ structure, yielding
\begin{equation}
\Pi_{\mu\nu}(k) = \Big[ k^2 \eta_{\mu\nu} - k_\mu k_\nu \Big] \int_{-1}^{+1} \exd y \; \frac{y^2}{2} \cdot \frac{\Gamma(\frac{4-D}{2})}{(4\pi)^{D/2}} \left( \frac{\mu^2}{\Sigma(y)/4 - i \epsilon} \right)^{\tfrac{4-D}{2}} \; 
\end{equation}
Expanding for $4 - D \ll 1$ gives 
\begin{equation}
\Pi_{\mu\nu}(k) \simeq \Big( k^2 \eta_{\mu\nu} - k_\mu k_\nu \Big) \bigg[ \frac{1}{24\pi^2 (4 - D)}  - \frac{1}{32\pi^2} \int_{-1}^{+1} \exd y \; y^2\; \log \bigg(\frac{ (1 - y^2) k^2 + 4m^2 -i \epsilon }{16\pi e^{-\gamma}\mu ^2} \bigg) \bigg] \label{Pimunu_result}
\end{equation}
where $\gamma$ is the Euler-Mascheroni constant.

\subsubsection{Computing $\mathcal{N}_{\mu\nu}$}

We now focus on the loop $\mathcal{N}_{\mu\nu}$ which causes the two branches of the in-in contour in the influence functional to interact. We compute:
\begin{equation}
\mathcal{N}_{\mu\nu}(k) = 2 \int \frac{\exd^4 \ell}{(2\pi)^4} \Wvac(\ell) \Wvac(-k-\ell) ( 2 \ell_{\mu} + k_\mu ) \ell_{\nu} \ .
\end{equation}
Note that we have two copies of the gauge symmetry, which means the above satisfies the Ward identity $k^\mu \mathcal{N}_{\mu\nu}(k) = 0$ and in addition is also Lorentz invariant. This means that one can express
\begin{equation}
\mathcal{N}_{\mu\nu}(k) = \big(  k^2 \eta_{\mu\nu} - k_\mu k_\nu \big) \mathcal{N}(k) \quad \mathrm{with}\ \mathcal{N}(k) \equiv \frac{ 2 }{ 3 k^2 } \int \frac{\exd^4 \ell}{(2\pi)^4} \Wvac(\ell)\Wvac(-k-\ell) ( 2 \ell^2 + k \cdot \ell )
\end{equation}
similar to Eq.~(\ref{Pimunu_result}). Let us evaluate this now using the explicit form of the Wightman function (\ref{freeprop_vac}), where:
\begin{eqnarray}
\mathcal{N}(k)  & =  & \frac{ 1 }{ 6 \pi^2 k^2 } \int \exd^4 \ell\; \theta(\ell_0) \theta(-k_0 - \ell_0 )  \delta(\ell^2 + m^2) \delta\big(\ell^2 + 2 \ell \cdot k + k^2 + m^2\big)\; ( 2 \ell^2 + \ell \cdot k ) \qquad \\
& = & \frac{  \theta(-k^0) }{ 6 \pi^2 k^2 } \int_{0}^{-k_0} \exd \ell_0 \int \exd^3 \boldsymbol{\ell} \; \delta(\ell^2 + m^2) \delta\big(\ell^2 + 2 \ell \cdot k + k^2 + m^2\big)\; ( 2 \ell^2 + \ell \cdot k ) \\
& = &  \frac{  \theta(-k^0) }{ 3 \pi k^2 } \int_{0}^{-k_0} \exd \ell_0 \int_{m}^{\infty} \exd \Omega \int_{-1}^{+1} \exd \mu \;  \; \delta( - \ell_0^2 + \Omega^2) \delta\big( - \ell_0^2 + \Omega^2 - 2 \ell_0 k_0 + 2 \mu \sqrt{\Omega^2 - m^2} |\mathbf{k}| + k^2 \big) \notag \\
&& \hspace{20mm} \times \; \Omega \sqrt{\Omega^2 - m^2} ( - 2 \ell_0^2 + 2 \Omega^2 - 2m^2 - \ell_0 k_0 + \mu L |\mathbf{k}| )
\end{eqnarray}
In the last equality we used rotational symmetry, used spherical coordinates $(|\boldsymbol{\ell}|, \theta, \varphi)$, integrated over $\varphi$ and changed variables $|\boldsymbol{\ell}| \to \Omega  = \sqrt{|\boldsymbol{\ell}|+ m^2}$ and $\theta \to \mu = \cos \theta$. Integrating over the first $\delta$-function restricts $m < \Omega < - k_0$ and one replaces $\theta(-k^0) \to \theta(-k_0 - m) $ and one simplifies to find 
\begin{equation}
\mathcal{N}(k) = \sfrac{\theta(-k_0 - m) }{ 12 \pi k^2 |\mathbf{k}| } \int_{m}^{-k_0} \exd \Omega \int_{-1}^{+1} \exd \mu \; ( - 2 m^2 - \Omega k_0 + \mu \sqrt{\Omega^2 - m^2} |\mathbf{k}| ) \; \delta\big( \mu - \tfrac{2 \Omega k_0 - k^2}{2 \sqrt{\Omega^2 - m^2} |\mathbf{k}|} \big)\; .
\end{equation}
Now integrating over $\mu$ in the $\delta$-function restricts the parameters to the region
\begin{eqnarray}
-1 <  \frac{2 \Omega k_0 - k^2}{2 \sqrt{\Omega^2 - m^2} |\mathbf{k}|} < 1 \ .
\end{eqnarray}
This restriction along with $|\mathbf{k}| > 0$ and $-k_0 > \Omega > m > 0$ reduces to the conditions
\begin{eqnarray}
0 < 4m^2 < - k^2 \quad \mathrm{and} \quad - \sfrac{k_0}{2} -\sfrac{|\mathbf{k}|}{2} \sqrt{1 + \sfrac{4 m^2}{k^2} } < \Omega < - \sfrac{k_0}{2} + \sfrac{|\mathbf{k}|}{2} \sqrt{1 + \sfrac{4 m^2}{k^2} } \ .
\end{eqnarray}
This clearly restricts the range on the $\Omega$, while also fixing timelike $k^2$ above the particle production threshold, and also enforcing that the sign of $k_0$ is negative, giving
\begin{eqnarray}
\mathcal{N}(k) & = & \frac{  \theta(-k^0) \theta( - k^2 - 4m^2 ) }{ 12 \pi k^2 |\mathbf{k}| } \int_{-\frac{k_0}{2} - \frac{|\mathbf{k}|}{2} \sqrt{1 + \frac{4 m^2}{k^2} }}^{-\frac{k_0}{2} + \frac{|\mathbf{k}|}{2} \sqrt{1 +\frac{4 m^2}{k^2} }} \exd \Omega \; \frac{ - k^2 - 4m^2 }{2} \\
& = & - \frac{  \theta(-k^0) \theta(- k^2 - 4m^2 ) }{ 24 \pi } \; \left( 1 + \frac{4m^2}{k^2} \right)^{3/2} \label{Nmunu_result}
\end{eqnarray}
Notice how this answer is finite and is purely real.

\subsubsection{Renormalisation}

We have now established that $\Pi_{\mu\nu}$ is divergent and $\mathcal{N}_{\mu\nu}$ is finite. Consequently, only the $++$ and $--$ components of the influence functional exhibit divergences. Moreover, using the result (\ref{Pimunu_result}) these divergences take the form
\begin{equation}
\mp \frac{q^2}{2} \int \frac{\exd^4 k}{(2\pi)^4} \; A_{\pm}^\mu(k) \Pi_{\mu\nu}(k) A_{\pm}^\nu(-k) \  \supset \  \mp \frac{q^2}{2} \int \frac{\exd^4 k}{(2\pi)^4} A_{\pm}^\mu(k) \frac{k^2 \eta_{\mu\nu} - k_\mu k_\nu}{24\pi^2 (4 - D)} A_{\pm}^\nu(-k) \ .
\end{equation}
in the vacuum version of the influence functional (\ref{thermalIF_loops}). The momentum structure of the divergence matches precisely that of the $F^2$ term in the free action (\ref{SS_sec4}). This means that the divergence must be absorbed through a field-strength renormalisation, implemented by augmenting the interaction (\ref{QEDactionINT}) with an $\mathcal{O}(q^2)$ counterterm
\begin{equation}
- \frac{q^2 c}{4} \int \exd^4 x\; F_{\mu\nu} F^{\mu\nu} = - \frac{q^2 c}{2} \int \exd^4 x\; A^{\mu}(k) (k^2 \eta_{\mu\nu} - k_\mu k_\nu) A^{\nu}(-k) \ .
\end{equation}
This counterterm involves only the gauge field, and therefore modifies exclusively the $++$ and $--$ terms in the influence functional. The choice
\begin{equation} \label{selfenergy_c}
c = - \frac{1}{24\pi^2 (4 - D)} + \frac{1}{48\pi^2} \log\left( \frac{m^2}{4 \pi e^{-\gamma} \mu^2} \right)
\end{equation}
then gives rise to the vacuum influence functional 
\begin{small}
\begin{eqnarray}
\lim_{\beta \to\infty} S_{\mathrm{IF}}[ A_{+}, A_{-} ] & \simeq & \frac{q^2}{2} \int \frac{ \mathrm{d}^4k}{(2\pi)^4} \; (k^2 \eta_{\mu\nu} - k_\mu k_\nu ) \bigg[ - A_{+}^\mu(k)  \Pi(k) A_{+}^\nu(-k) - i A_{+}^\mu(k) \mathcal{N}(-k) A_{-}^\nu(-k) \label{IF_vac}  \\
&& \hspace{41mm} - i A_{-}^\mu(k) \mathcal{N}(k) A_{+}^\nu(-k) + A_{-}^\mu(k) \Pi^{\ast}(k) A_{-}^\nu(-k) \bigg] + \mathcal{O}(q^4)  \notag
\end{eqnarray}
\end{small}\ignorespaces
with $\mathcal{N}$ given in Eq.~(\ref{Nmunu_result}), and where
\begin{equation} \label{Pi_answer}
\Pi(k) = \frac{1}{32\pi^2} \int_{-1}^{+1} \exd y \; y^2\; \log \bigg(\frac{4m^2}{ (1 - y^2) k^2 + 4m^2 -i \epsilon } \bigg) \ .
\end{equation}
the coefficient in the counterterm (\ref{selfenergy_c}) was chosen to impose the condition $\Pi(0) =0$. This is the same renormalisation prescription used in the ordinary in-out formalism, ensuring that the photon self-energy remains regular at $k^2 =0$. In particular, it guarantees that the resummed photon propagator always has unit residue at the pole $k^2 =0$, which physically enforces that the photon remains exactly massless to all orders in perturbation theory.

To evaluate the limit $\epsilon \to 0{+}$ in Eq.~(\ref{Pi_answer}), one uses the identity $\log( \frac{1}{z - i\epsilon} ) = \log\left| \frac{1}{z} \right| + i \pi \theta(-z)$ for real $z$, which yields the real and imaginary parts:
\begin{eqnarray} 
\mathrm{Re}[\Pi(k)] & = & \frac{1}{32\pi^2} \int_{-1}^{+1} \exd y \; y^2\; \log \Big| \frac{4m^2}{ (1 - y^2) k^2 + 4m^2} \Big| \label{RePi_answer} \\
\mathrm{Im}[\Pi(k)] & = & \frac{\theta( - k^2 - 4m^2 )}{48\pi} \left( 1 + \frac{4m^2}{k^2}  \right)^{3/2} \label{ImPi_answer}
\end{eqnarray}
the $y$-integral for the imaginary part is straightforward, while the real part can also be evaluated explicitly but leads to cumbersome expressions. It is therefore standard to leave the real part in the compact Feynman-parameter form above.

Finally, by combining (\ref{Nmunu_result}) with (\ref{RePi_answer})-(\ref{ImPi_answer}), one can verify explicitly the identity
\begin{equation} \label{magic}
2\mathrm{Im}[ \Pi(k^2) ] +\mathcal{N}(k) + \mathcal{N}(-k) = 0 
\end{equation}
which ensures that $S_{\mathrm{IF}}[\varphi,\varphi] =0$ (the condition that preserves the trace of the reduced density matrix).

\subsection{Thermal corrections}
\label{App:thermal_corr}

With the renormalisation of the vacuum influence functional (\ref{IF_vac}) already complete, we can now proceed to compute the thermal corrections appearing in the influence functional (\ref{thermalIF_loops}) in the main text. The thermal loops $\Pi_{\mu\nu}^{\beta}$ and $\mathcal{N}_{\mu\nu}^{\beta}$ are built from the free thermal propagators (\ref{freeprop_thermal}). It is convenient to decompose these into the sum of the vacuum propagators from Eq.~(\ref{freeprop_vac}) and a thermal correction:
\begin{equation} \label{freeprop_thermal_APP}
\mathcal{F}^{\beta}(k) = \mathcal{F}^{\rm vac}(k)  + \mathcal{C}^{\beta}(k) \qquad \mathrm{and} \qquad \mathcal{W}^{\beta}(k) = \Wvac(k) +  \mathcal{C}^{\beta}(k) \ .
\end{equation}
where both propagators share the same correction
\begin{eqnarray} \label{thermalprop_corr}
\mathcal{C}^{\beta}(k) \equiv  \frac{2 \pi \delta(k^2 + m^2)}{ e^{\beta |k_0|}  - 1 } \ .
\end{eqnarray}
We now take the thermal loops from the main text and separate out the vacuum contributions $\Pi_{\mu\nu}$ and $\mathcal{N}_{\mu\nu}$ from Appendix \ref{App:thermalVac}, writing
\begin{eqnarray}
\Pi^{\beta}_{\mu\nu}(k) = \Pi_{\mu\nu}(k) + \overline{\Pi}^{\beta}_{\mu\nu}(k) \qquad \mathrm{and} \qquad \mathcal{N}^{\beta}_{\mu\nu}(k) = \mathcal{N}_{\mu\nu} + \overline{\mathcal{N}}^{\beta}_{\mu\nu}(k) \ .
\end{eqnarray}
Using the definitions (\ref{Pibeta}) and (\ref{Nbeta}) together with the propagator decomposition above, the thermal loop corrections can then be written as
\begin{footnotesize}
\begin{eqnarray}
\overline{\Pi}^{\beta}_{\mu\nu}(k) & = & \eta_{\mu\nu} \mathcal{T}^{\beta} - 2 i \int \frac{\exd^4 \ell}{(2\pi)^4} \bigg( \mathcal{C}^{\beta}(\ell)  \mathcal{F}^{\rm vac}(\ell-k)  ( 2 \ell_{\mu} - k_\mu) ( 2 \ell_{\nu} - k_\nu)+ \mathcal{C}^{\beta}(\ell) \mathcal{C}^{\beta}(\ell-k) ( 2 \ell_{\mu} - k_\mu) \ell_{\nu} \bigg) \qquad \label{Picorr_def}  \\
\overline{\mathcal{N}}^{\beta}_{\mu\nu}(k) & = & 2 \int \frac{\exd^4 \ell}{(2\pi)^4} \bigg( \mathcal{C}^{\beta}(\ell) \Wvac(\ell-k) ( 2 \ell_{\mu} - k_\mu) ( 2 \ell_{\nu} - k_\nu )+ \mathcal{C}^{\beta}(\ell) \mathcal{C}^{\beta}(\ell-k) ( 2 \ell_{\mu} - k_\mu) \ell_{\nu} \bigg) \label{Ncorr_def} 
\end{eqnarray}
\end{footnotesize}\ignorespaces
with the definition
\begin{eqnarray} \label{bubbledef}
\mathcal{T}^{\beta} & : = & 2 \int \frac{\exd^4 \ell}{(2\pi)^4} \; \mathcal{C}^{\beta}(\ell) \ .
\end{eqnarray}
the tensorial structure is cumbersome to handle directly, so it is useful to reduce it to scalars by contracting with the available tensors. The only nontrivial options are $\eta^{\mu\nu}$ and $n^{\mu}$, since any contraction with $k^\mu$ vanishes due to gauge invariance. Carrying out these contractions gives:
\begin{eqnarray}
 \eta^{\mu\nu} \overline{\Pi}^{\beta}_{\mu\nu}(k) & = & 4 \mathcal{T}^{\beta} - \mathcal{E}^{\beta\mathcal{F}}(k) + i \mathcal{E}^{\beta\beta}(k) +  \mathcal{K}^{\beta\mathcal{F}}(k) - i \mathcal{K}^{\beta \beta}(k)  \label{etaXmunu2} \\
n^{\mu} n^{\nu} \overline{\Pi}^{\beta}_{\mu\nu}(k) & = & - \mathcal{T}^{\beta} + \mathcal{E}^{\beta\mathcal{F}}(k) - i \mathcal{E}^{\beta\beta}(k) \label{nnXmunu2} \\
 \eta^{\mu\nu} \overline{\mathcal{N}}^{\beta}_{\mu\nu}(k) & = & -  \mathcal{E}^{\beta\mathcal{W}}(k) - \mathcal{E}^{\beta\beta}(k)  + \mathcal{K}^{\beta\mathcal{W}}(k) + \mathcal{K}^{\beta \beta}(k) \label{etaYmunu} \\
n^{\mu} n^{\nu} \overline{\mathcal{N}}^{\beta}_{\mu\nu}(k) & = & \mathcal{E}^{\beta\mathcal{W}}(k) + \mathcal{E}^{\beta\beta}(k) \label{nnYmunu}
\end{eqnarray}
Along with the definitions:
%
\ba \label{KE_integrals}
&& \mathcal{K}^{\beta\mathcal{F}}(k)  \equiv  - 2 i \int \frac{\exd^4 \ell}{(2\pi)^4} \;   \mathcal{C}^{\beta}(\ell)   \mathcal{F}^{\rm vac}(\ell-k)  \; | 2 \boldsymbol{\ell} - \mathbf{k} |^2 \\
 &&  \mathcal{E}^{\beta\mathcal{F}}(k)  \equiv  -2  i \int \frac{\exd^4 \ell}{(2\pi)^4} \mathcal{C}^{\beta}(\ell)  \mathcal{F}^{\rm vac}(\ell-k)  ( 2 \ell_{0} - k_0)^2 \\
 &&     \mathcal{K}^{\beta\mathcal{W}}(k)  \equiv 2  \int \frac{\exd^4 \ell}{(2\pi)^4} \; \mathcal{C}^{\beta}(\ell) \Wvac(\ell-k) \; | 2 \boldsymbol{\ell} - \mathbf{k} |^2 \\
  && \mathcal{E}^{\beta\mathcal{W}}(k)  \equiv  2 \int \frac{\exd^4 \ell}{(2\pi)^4} \; \mathcal{C}^{\beta}(\ell) \Wvac(\ell-k) ( 2 \ell_{0} - k_0)^2  \\
&&  \mathcal{K}^{\beta \beta}(k)  \equiv  2 \int \frac{\exd^4 \ell}{(2\pi)^4} \; \mathcal{C}^{\beta}(\ell)  \mathcal{C}^{\beta}(\ell-k) ( 2 |\boldsymbol{\ell}|^2 - \mathbf{k} \cdot \boldsymbol{\ell} ) \\
&& \mathcal{E}^{\beta\beta}(k)  \equiv  2 \int \frac{\exd^4 \ell}{(2\pi)^4} \; \mathcal{C}^{\beta}(\ell)  \mathcal{C}^{\beta}(\ell-k)  ( 2 \ell_{0}^2 - k_0\ell_0)
\ea
%
these integrals are ultimately what get related to the functions $\Pi_{\mathrm{L},\mathrm{T}}^{\beta}$ and $\mathcal{N}_{\mathrm{L},\mathrm{T}}^{\beta}$ in the main text.

\subsubsection{Simplification of integrals}
\label{App:simplify}

The integrals defined in (\ref{KE_integrals}) are complicated, and it is difficult to obtain closed-form expressions. For this reason, we simplify the four-dimensional loop integrals to single energy integrals, which are easier to analyze, especially when taking limits.

We begin with the integral from Eq.~(\ref{bubbledef}). Performing the $\ell_0$ integration using the $\delta$-function in (\ref{thermalprop_corr}), and then integrating over the angular variables yields
\begin{equation} \label{bubble_ans}
\mathcal{T}^{\beta} = 2 \int \frac{\exd^4 \ell}{(2\pi)^4}   \frac{2 \pi \delta(\ell^2 + m^2)}{ e^{\beta |\ell_0|}  - 1 } = \frac{1}{\pi^2} \int_{m}^\infty \exd \Omega \; \frac{\sqrt{\Omega^2 - m^2}}{ e^{\beta \Omega }  - 1 }
\end{equation}
where we have defined the energy variable $\Omega = \sqrt{|\boldsymbol{\ell}|^2 + m^2}$, which will be used extensively below. Even this simplest integral does not admit a closed-form solution. However, its high-temperature limit can be shown to be
\begin{equation}
\mathcal{T}^{\beta} \simeq \frac{1}{6\beta^2} \qquad (m\beta \ll 1) \ .
\end{equation}

\subsubsection*{Simplification of integrals involving $\mathcal{F}$}

The integrals involving $F^{\mathrm{vac}}$ in Eq.~(\ref{KE_integrals}) take more care to simplify. For example, the first integral $ \mathcal{K}^{\beta\mathcal{F}}(k)$ becomes:
\begin{footnotesize}
\begin{eqnarray}
 \mathcal{K}^{\beta\mathcal{F}}(k) & = & - 2 i \int \frac{\exd^4 \ell}{(2\pi)^4} \; \frac{ 2 \pi \delta(\ell^2 + m^2) }{ e^{\beta |\ell_0|}  - 1  } \cdot  \frac{- i }{(\ell- k)^2  + m^2 - i \epsilon} \cdot ( |\mathbf{k}|^2 - 4 \mathbf{k} \cdot \boldsymbol{\ell} + 4 | \boldsymbol{\ell}| ^2 ) \\
& = & \frac{1}{8\pi^2 | \mathbf{k} |} \int_m^\infty \; \sfrac{  \exd \Omega }{ e^{\beta \Omega }  - 1 } \int_{-1}^{+1} \exd \mu \; \bigg[ \sfrac{ |\mathbf{k}|^2 - 4 |\mathbf{k}| \sqrt{ \Omega^2 - m^2 } \mu +  4 \Omega^2 - 4 m^2 }{  \mu - \frac{ k^2 + 2 k_0 \Omega }{2 | \mathbf{k} | \sqrt{ \Omega^2 - m^2 }} +  i \epsilon } +  \sfrac{ |\mathbf{k}|^2 - 4 |\mathbf{k}| \sqrt{ \Omega^2 - m^2 } \mu +  4 \Omega^2 - 4 m^2 }{ \mu -  \frac{ k^2 - 2 k_0 \Omega }{ 2 | \mathbf{k} | \sqrt{ \Omega^2 - m^2 } }  + i \epsilon } \bigg] \qquad
\end{eqnarray}
\end{footnotesize}\ignorespaces
In the last equality we've integrated over $\ell_0$ in the $\delta$-function, switched to spherical coordinates $(|\boldsymbol{\ell}|, \theta, \varphi)$, integrated over $\varphi$ and swapped $|\boldsymbol{\ell}| \to \Omega = \sqrt{ |\boldsymbol{\ell}|^2 + m^2 }$ and $\theta \to \mu = \cos \theta$. One can simplify this by using the identity $\frac{a-b \mu }{\mu -x}+\frac{a-b \mu }{\mu -y} = - 2 b + \frac{a-b x}{\mu -x}+\frac{a-b y}{\mu -y}$ which writes the above as
\begin{small}
\begin{eqnarray}
 \mathcal{K}^{\beta\mathcal{F}}(k) & = &  \frac{1}{8 \pi^2 |\mathbf{k}|} \int_m^\infty \sfrac{ \exd \Omega  }{ e^{\beta \Omega }  - 1 } \bigg[ - 8 |\mathbf{k}| \sqrt{\Omega^2 - m^2} + (2k_0^2 - |\mathbf{k}|^2 - 4 k_0 \Omega + 4 \Omega^2 - 4 m^2) \; g\Big( \tfrac{ k^2 + 2 k_0 \Omega }{ 2 | \mathbf{k} | \sqrt{ \Omega^2 - m^2 } } \Big) \notag \\
 && \hspace{25mm} + (2k_0^2 - |\mathbf{k}|^2 + 4 k_0 \Omega + 4 \Omega^2 - 4 m^2) \; g\Big( \tfrac{ k^2 - 2 k_0 \Omega }{ 2 | \mathbf{k} | \sqrt{ \Omega^2 - m^2 } } \Big) \bigg] \ .
\end{eqnarray}
\end{small}\ignorespaces
where for convenience we've defined the function 
\begin{eqnarray} \label{f_def_int}
g(x) \equiv \int_{-1}^{+1} \frac{\exd \mu }{\mu - x + i \epsilon} \ = \ \log \left| \frac{x - 1}{x+1} \right| - i \pi \theta( 1 - x^2  ) \qquad \mathrm{for \ }x \in \mathbb{R} \ .
\end{eqnarray}
the real part is easily seen to simplify to
\begin{small}
\begin{eqnarray}
\mathrm{Re}\left[ \mathcal{K}^{\beta\mathcal{F}}(k) \right] & = & - \mathcal{T}^{\beta} +   \frac{1}{8 \pi^2 |\mathbf{k}|} \int_m^\infty \sfrac{ \exd \Omega  }{ e^{\beta \Omega }  - 1 } (2k_0^2 - |\mathbf{k}|^2 + 4 \Omega^2 - 4 m^2) \; \log \left| \frac{ (k^2- 2 |\mathbf{k}| \sqrt{\Omega^2 - m^2})^2 - 4 k_0^2 \Omega^2 }{  (k^2 + 2 |\mathbf{k}| \sqrt{\Omega^2 - m^2})^2 - 4 k_0^2 \Omega^2  } \right| \notag \\
&& \qquad \quad + \frac{k_0}{2 \pi^2 |\mathbf{k}|} \int_m^\infty \exd \Omega \;  \sfrac{ \Omega }{ e^{\beta \Omega }  - 1 } \; \log \left| \frac{ (k^2)^2 - 4 ( k_0 \Omega + |\mathbf{k}| \sqrt{\Omega^2-m^2} )^2 }{ (k^2)^2 - 4 ( k_0 \Omega - |\mathbf{k}| \sqrt{\Omega^2-m^2} )^2 } \right|  
\end{eqnarray}
\end{small}\ignorespaces
where we've also used the result (\ref{bubble_ans}). Meanwhile, the imaginary part has the form:
\begin{small}
\begin{eqnarray}
\mathrm{Im}\left[ \mathcal{K}^{\beta\mathcal{F}}(k) \right]& = & - \frac{1}{8 \pi |\mathbf{k}|} \int_m^\infty \sfrac{ \exd \Omega  }{ e^{\beta \Omega }  - 1 } \bigg[ (2k_0^2 - |\mathbf{k}|^2 - 4 k_0 \Omega + 4 \Omega^2 - 4 m^2) \; \theta\Big(1 -  \tfrac{ ( k^2 + 2 k_0 \Omega )^2 }{ 4 | \mathbf{k} |^2 ( \Omega^2 - m^2 ) } \Big) \\
 && \hspace{40mm} + (2k_0^2 - |\mathbf{k}|^2 + 4 k_0 \Omega + 4 \Omega^2 - 4 m^2) \; \theta\Big(1 -  \tfrac{ ( k^2 - 2 k_0 \Omega )^2 }{ 4 | \mathbf{k} |^2 ( \Omega^2 - m^2 ) } \Big)  \bigg] \ . \notag
\end{eqnarray}
\end{small}\ignorespaces
Carefully tracking where the arguments of the Heaviside functions are positive turns the imaginary part into
\begin{small}
\begin{eqnarray}
\mathrm{Im}\left[ \mathcal{K}^{\beta\mathcal{F}}(k) \right] & = & - \frac{ \theta( -k^2 - 4m^2 ) }{8 \pi |\mathbf{k}|} \int_{\Omega_{-}}^{\Omega_{+}} \sfrac{ \exd \Omega \; ( 4 \Omega^2 - 4 m^2 + 2k_0^2 - |\mathbf{k}|^2 - 4 |k_0| \Omega  ) }{ e^{\beta \Omega }  - 1 }  \\
&& - \frac{\theta(k^2)}{8 \pi |\mathbf{k}|} \bigg[ \; \int_{\Omega_{+}}^\infty \sfrac{ \exd \Omega \; ( 4 \Omega^2 - 4 m^2 + 2k_0^2 - |\mathbf{k}|^2 - 4 |k_0| \Omega )  }{ e^{\beta \Omega }  - 1 }  +\int_{-\Omega_{-} }^\infty \sfrac{ \exd \Omega \; (4 \Omega^2 - 4 m^2 + 2k_0^2 - |\mathbf{k}|^2 + 4 |k_0| \Omega ) }{ e^{\beta \Omega }  - 1 }   \bigg] \notag \, ,
\end{eqnarray}
\end{small}\ignorespaces
where we have used the definition of $\Omega_{\pm}$ from (\ref{Omegapm}), where we note that $- \Omega_{-} = |\Omega_{-}|$ is positive in the spacelike case when $k^2 >0$. Notice that $- \Omega_{-} + |k_0| = \Omega_{+}$ which means that if we take $\Omega \to \Omega + |k_0| $ in the very last integral, then we can write the $\theta(k^2)$ term as a single integral over the range $\Omega > \Omega_{+}$ such that
\begin{eqnarray}
\mathrm{Im}\left[ \mathcal{K}^{\beta\mathcal{F}}(k) \right] & = & - \frac{ \theta( -k^2 - 4m^2 ) }{8 \pi |\mathbf{k}|} \int_{\Omega_{-}}^{\Omega_{+}} \sfrac{ \exd \Omega \; ( 4 \Omega^2 - 4 m^2 + 2k_0^2 - |\mathbf{k}|^2 - 4 |k_0| \Omega  ) }{ e^{\beta \Omega }  - 1 }  \\
&& - \frac{\theta(k^2)}{8 \pi |\mathbf{k}|} \int_{\Omega_{+}}^\infty \exd \Omega \; \bigg[ \frac{1}{e^{\beta \Omega }  - 1} + \frac{1}{ e^{\beta ( \Omega - |k_0| )  }  - 1 } \bigg] ( 4 \Omega^2 - 4 m^2 + 2k_0^2 - |\mathbf{k}|^2 - 4 |k_0| \Omega ) \ . \notag
\end{eqnarray}
In an extremely similar computation one finds that the integral $\mathcal{E}^{\beta\mathcal{F}}(k)$ has real and imaginary parts:
\begin{eqnarray}
\mathrm{Re}\left[ \mathcal{E}^{\beta\mathcal{F}}(k) \right] & = & \frac{1}{8 \pi^2 | \mathbf{k} |}  \int_m^\infty \frac{\exd \Omega }{e^{\beta \Omega }  - 1 } (k_0^2 +  4 \Omega^2) \log \left| \frac{ (k^2- 2 |\mathbf{k}| \sqrt{\Omega^2 - m^2})^2 - 4 k_0^2 \Omega^2 }{  (k^2 + 2 |\mathbf{k}| \sqrt{\Omega^2 - m^2})^2 - 4 k_0^2 \Omega^2  } \right| \\
& &+ \frac{k_0}{2 \pi^2 | \mathbf{k} |}  \int_m^\infty \exd \Omega  \; \frac{  \Omega}{e^{\beta \Omega }  - 1 } \; \log \left| \frac{ (k^2)^2 - 4 ( k_0 \Omega + |\mathbf{k}| \sqrt{\Omega^2-m^2} )^2 }{ (k^2)^2 - 4 ( k_0 \Omega - |\mathbf{k}| \sqrt{\Omega^2-m^2} )^2 } \right| \notag \\
\mathrm{Im}\left[ \mathcal{E}^{\beta\mathcal{F}}(k) \right] & = &  -  \frac{\theta(- k^2 - 4m^2) }{8 \pi | \mathbf{k} |}  \int_{\Omega_{-} }^{ \Omega_{+} } \frac{\exd \Omega }{e^{\beta \Omega }  - 1 } (k_0^2 - 4 |k_0|  \Omega +  4 \Omega^2)  \\
&\  & -  \frac{\theta(k^2)}{8 \pi | \mathbf{k} |} \int_{ \Omega_{+} }^\infty \exd \Omega \; \bigg[ \frac{1}{e^{\beta \Omega }  - 1 } + \frac{1}{e^{\beta (\Omega - |k_0| ) }  - 1 } \bigg]  (k_0^2 - 4 | k_0|  \Omega + 4 \Omega^2)  \notag \, .
\end{eqnarray}

\subsubsection*{Simplification of integrals with two $\delta$-functions}

The integrals in the rightmost column of Eq.~(\ref{KE_integrals}) each contain two $\delta$-functions and are purely real. they take the general form
\begin{eqnarray} \label{loopLf_def}
\mathscr{L}_{f}(k) & = & 2 \int \frac{\exd^4 \ell}{(2\pi)^4} \;  2 \pi \delta\big(\ell^2 + m^2\big) \cdot 2 \pi \delta\big( (k-\ell)^2 + m^2\big) \cdot f\big( \ell_0, |\boldsymbol{\ell}|, \mathbf{k} \cdot \boldsymbol{\ell} , k_0 \big) \, ,
\end{eqnarray}
where $f$ is a shorthand for the accompanying functions in the relevant propagator (and $f$ always has mass dimension 2). Rotational invariance is built into $f$, since it depends only on scalar combinations of $\ell$ and $k$. It is also important to keep track of $k_0$ explicitly, as later expressions depend on its sign. Integrating over $\ell_0$ in the first $\delta$-function, using spherical coordinates $(|\boldsymbol{\ell}|, \theta, \varphi)$, integrating over $\varphi$ and swapping $|\boldsymbol{\ell}| \to \Omega = \sqrt{ |\boldsymbol{\ell}|^2 + m^2 }$ and $\theta \to \mu = \cos \theta$ leads to:
\begin{eqnarray}
\mathscr{L}_{f}(k) & = & \frac{1}{4\pi |\mathbf{k}|} \int_{m}^{\infty} \exd \Omega \int_{-1}^{+1} \exd \mu \; \delta\big( \mu - \tfrac{ k^2 + 2 k_0 \Omega }{ 2 | \mathbf{k} | \sqrt{ \Omega^2 - m^2 } } \big) \; f\big( \Omega ,  \sqrt{ \Omega^2 - m^2 }, | \mathbf{k} |  \sqrt{ \Omega^2 - m^2 } \mu, k_0\big)  \\
&& \qquad + \frac{1}{4\pi |\mathbf{k}|} \int_{m}^{\infty} \exd \Omega  \int_{-1}^{+1} \exd \mu \; \delta\big( \mu - \tfrac{ k^2 - 2 k_0 \Omega }{ 2 | \mathbf{k} | \sqrt{ \Omega^2 - m^2 } } \big) \; f\big( - \Omega ,  \sqrt{ \Omega^2 - m^2 }, | \mathbf{k} |  \sqrt{ \Omega^2 - m^2 } \mu, k_0\big) \notag \, .
\end{eqnarray}
Integrating over $\mu$ now in the remaining $\delta$-functions enforces $-1 < \tfrac{ k^2 \pm 2 k_0 \Omega }{ 2 | \mathbf{k} | \sqrt{ \Omega^2 - m^2 } } < +1$, which eventually turns the above into
\begin{footnotesize}
\begin{eqnarray}
 \mathscr{L}_{f}(k) & = & \sfrac{\theta(-k^2 - 4m^2)}{4\pi |\mathbf{k}|} \int_{ \Omega_{-} }^{\Omega_{+} } \exd \Omega \; \Scale[0.90]{ \bigg[ \theta(k^0)  f\big( \Omega ,  \sqrt{ \Omega^2 - m^2 }, \tfrac{k^2+2|k_0| \Omega}{2}, |k_0| \big) +   \theta(-k^0) f\big( - \Omega ,  \sqrt{ \Omega^2 - m^2 }, \tfrac{k^2+2|k_0| \Omega}{2}  , - |k_0| \big) \bigg] } \notag \\
&&  +\sfrac{\theta(k^2)}{4\pi |\mathbf{k}|} \int_{\Omega_{+} }^{\infty} \exd \Omega\;  \Scale[0.90]{ \bigg[ \theta(k^0) f\big( \Omega ,  \sqrt{ \Omega^2 - m^2 }, \tfrac{k^2+ 2|k_0| \Omega}{2} , |k_0| \big) + \theta(-k^0) f\big(- \Omega ,  \sqrt{ \Omega^2 - m^2 }, \tfrac{k^2+ 2|k_0| \Omega}{2} , - |k_0| \big) \bigg] } \label{loopLf_ans}  \\
&& + \sfrac{\theta(k^2)}{4\pi |\mathbf{k}|} \int_{- \Omega_{-} }^{\infty} \exd \Omega\;  \Scale[0.90]{ \bigg[ \theta(k^0) f\big(- \Omega ,  \sqrt{ \Omega^2 - m^2 }, \tfrac{k^2- 2 |k_0| \Omega}{2} , |k_0| \big) + \theta(-k^0) f\big( \Omega ,  \sqrt{ \Omega^2 - m^2 }, \tfrac{k^2-2|k_0| \Omega}{2} , -|k_0| \big) \bigg] } \notag \, , 
\end{eqnarray}
\end{footnotesize}\ignorespaces
where again we've used $\Omega_{\pm}$ from Eq.~(\ref{Omegapm}). Note that we also make use of this formula in Appendix \ref{App:SSBint}. One can use this formula for example to compute $\mathcal{K}^{\beta \beta}(k)$ where one has
\begin{equation}
\mathcal{K}^{\beta \beta}(k) = 2 \int \frac{\exd^4 \ell}{(2\pi)^4}   \frac{ 2 \pi \delta(\ell^2 + m^2) }{ e^{\beta |\ell_0|}  - 1  }   \frac{ 2 \pi \delta( (k-\ell)^2 + m^2) }{ e^{\beta |k_0-\ell_0|}  - 1  } ( 2 |\boldsymbol{\ell}|^2 - \mathbf{k} \cdot \boldsymbol{\ell} ) \, .
\end{equation}
Comparing this with the definition of  in Eq.~(\ref{loopLf_def}), we see that this integral  clearly corresponds to a function
\begin{equation}
f\big( \ell_0, |\boldsymbol{\ell}|, \mathbf{k} \cdot \boldsymbol{\ell} , k_0 \big) \ \to \ \frac{ 1 }{ e^{\beta |\ell_0|}  - 1  } \cdot  \frac{ 2 |\boldsymbol{\ell}|^2 - \mathbf{k} \cdot \boldsymbol{\ell} }{ e^{\beta |k_0-\ell_0|}  - 1  }  \ .
\end{equation}
Plugging this function into Eq.~(\ref{loopLf_ans}) and simplifying gives:
\begin{eqnarray}
- i \mathcal{K}^{\beta \beta}(k)  & = & - \frac{i \theta(-k^2 - 4m^2)}{8 \pi |\mathbf{k}|} \int_{\Omega_{-}}^{\Omega_{+}} \frac{ \exd \Omega }{e^{\beta \Omega}  - 1 } \cdot \frac{ 4 \Omega^2 - 4 m^2 + k_0^2 - |\mathbf{k}|^2  - 2 |k_0| \Omega  }{ e^{\beta ( |k_0|-\Omega)}  - 1 }  \\
&& - \frac{i \theta(k^2) }{8 \pi |\mathbf{k}|} \int_{\Omega_{+}}^{\infty} \frac{ \exd \Omega }{ e^{\beta \Omega}  - 1 }  \cdot \frac{ 2 \big( 4 \Omega^2 - 4 m^2 + 2 k_0^2 - |\mathbf{k}|^2 - 4 |k_0| \Omega \big)  }{ e^{\beta ( \Omega - |k_0| ) }  - 1 } \, . \notag
\end{eqnarray}
One computes similarly the remaining integrals:
\begin{small}
\begin{eqnarray}
\mathcal{E}^{\beta\beta}(k) & = & \frac{\theta(-k^2 - 4m^2)}{8 \pi |\mathbf{k}|} \int_{\Omega_{-}}^{\Omega_{+}} \frac{ \exd \Omega }{ e^{\beta \Omega}  - 1 } \cdot \frac{ 4 \Omega^2 - 2 |k_0| \Omega }{ e^{\beta ( |k_0|-\Omega) }  - 1  }  + \frac{ \theta(k^2) }{16\pi |\mathbf{k}|} \int_{\Omega_{+}}^{\infty} \frac{ \exd \Omega }{ e^{\beta \Omega}  - 1 } \cdot \frac{ 2 ( 2 \Omega - |k_0| )^2}{ e^{\beta ( \Omega - |k_0|)}  - 1  } \qquad  \\
\mathcal{K}^{\beta\mathcal{W}}(k) & = & \frac{\theta(-k^2 - 4m^2)}{4\pi |\mathbf{k}|} \int_{\Omega_{-}}^{\Omega_{+}} \frac{\exd \Omega}{ e^{\beta \Omega} - 1 } \; \theta(-k^0) \big[  (2\Omega - |k_0|)^2 - k^2 - 4m^2 \big]  \\
&& + \frac{\theta(k^2) }{4\pi |\mathbf{k}|} \int_{\Omega_{+}}^{\infty} \exd \Omega \; \bigg[ \frac{\theta(k^0)}{e^{\beta \Omega} - 1 } + \frac{\theta(-k^0)}{e^{\beta ( \Omega - |k_0| ) } - 1 } \bigg] \big[  (2\Omega - |k_0|)^2 - k^2 - 4m^2 \big] \notag \\
\mathcal{E}^{\beta\mathcal{W}}(k) & = & \frac{\theta(-k^2 - 4m^2)}{4\pi |\mathbf{k}|} \int_{\Omega_{-}}^{\Omega_{+}} \frac{\exd \Omega}{ e^{\beta \Omega} - 1 } \; \theta(-k^0)  \; (2\Omega - |k_0|)^2 \\
&& + \frac{\theta(k^2) }{4\pi |\mathbf{k}|} \bigg[ \; \int_{\Omega_{+}}^{\infty} \exd \Omega \; \bigg[  \frac{\theta(k^0)}{ e^{\beta \Omega} - 1 } + \frac{\theta(-k^0)}{e^{\beta ( \Omega - |k_0| ) } - 1 } \bigg] (2\Omega - |k_0|)^2 \notag \, .
\end{eqnarray}
\end{small}\ignorespaces

\subsubsection{Results}
\label{App:thermalresults}

Using the results above, we now express $\Pi_{\mathrm{L},\mathrm{T}}^{\beta}$ and $\mathcal{N}_{\mathrm{L},\mathrm{T}}^{\beta}$ in terms of the previously defined integrals, so that they can be applied directly in the influence functional (\ref{thermalIF_ans}). After performing the same field strength renormalisation as in the vacuum case (with $c$ from Eq.~(\ref{selfenergy_c})), we obtain
\begin{eqnarray}
\Pi^{\beta}_{\mu\nu}(k) + c (k^2 \eta_{\mu \nu} - k_\mu k_\nu ) & = & (k^2 \eta_{\mu\nu} - k_\mu k_\nu ) \Pi(k) + \overline{\Pi}^{\beta}_{\mu\nu}(k) \\
\mathcal{N}^{\beta}_{\mu\nu}(k) & = & (k^2 \eta_{\mu\nu} - k_\mu k_\nu ) \mathcal{N}(k) + \overline{\mathcal{N}}^{\beta}_{\mu\nu}(k) \, ,
\end{eqnarray}
where $\Pi$ and $\mathcal{N}$ are the vacuum results from Appendix \ref{App:thermalVac}. These are precisely the renormalised contributions appearing in the influence functional (\ref{thermalIF_ans}). We now decompose these finite terms as
\begin{eqnarray}
(k^2 \eta_{\mu\nu} - k_\mu k_\nu ) \Pi(k) + \overline{\Pi}^{\beta}_{\mu\nu}(k) & = & \Pi^{\beta}_{\mathrm{L}}(k) \mathcal{P}^{\mathrm{L}}_{\mu\nu}(k)  \; +  \Pi_{\mathrm{T}}^{\beta}(k)  \mathcal{P}^{\mathrm{T}}_{\mu\nu}(k) \label{Pi_LT} \\
(k^2 \eta_{\mu\nu} - k_\mu k_\nu ) \mathcal{N}(k) + \overline{\mathcal{N}}^{\beta}_{\mu\nu}(k)  & = & \mathcal{N}^{\beta}_{\mathrm{L}}(k) \mathcal{P}^{\mathrm{L}}_{\mu\nu}(k)\; + \mathcal{N}_{\mathrm{T}}^{\beta}(k)  \mathcal{P}^{\mathrm{T}}_{\mu\nu}(k) 
\end{eqnarray}
with $\mathcal{P}^{\mathrm{L}, \mathrm{T}}_{\mu\nu}$ given in Eq.~ (\ref{PTPL_matrices}).  This system of equations can be solved by contracting both sides with $\eta^{\mu\nu}$ and $n^{\mu} n^{\nu}$, making use of the relations:
\begin{equation}
  \begin{split}
\eta^{\mu\nu} \mathcal{P}^{\mathrm{L}}_{\mu\nu}(k) & = k^2 \\
n_{\mu} n_{\nu} \mathcal{P}^{\mathrm{L}}_{\mu\nu}(k) & = - |\mathbf{k}|^2
  \end{split}
\hspace{30mm}
  \begin{split}
    \eta^{\mu\nu} \mathcal{P}^{\mathrm{T}}_{\mu\nu}(k) & = 2 |\mathbf{k}|^2 \\
   n_{\mu}n_{\nu} \mathcal{P}^{\mathrm{T}}_{\mu\nu}(k) & =  0 \, ,
  \end{split}
\end{equation}
these expressions along with Eqs.~(\ref{etaXmunu2})-(\ref{nnYmunu}) give
\begin{eqnarray}
\label{PiLbeta}
\Pi^{\beta}_{\mathrm{L}}(k) & = & \Pi(k) + \sfrac{1}{|\mathbf{k}|^2} \Big( \mathcal{T}^{\beta} - \mathcal{E}^{\beta\mathcal{F}}(k) + i \mathcal{E}^{\beta\beta}(k) \Big) \\
\Pi^{\beta}_{\mathrm{T}}(k) & = & \frac{k^2}{|\mathbf{k}|^2} \Pi(k) + \sfrac{ 1 }{2|\mathbf{k}|^4} \Big( (k_0^2 + 3 |\mathbf{k}|^2) \mathcal{T}^{\beta} - k_0^2 [ \mathcal{E}^{\beta\mathcal{F}}(k) - i \mathcal{E}^{\beta\beta}(k) ] + |\mathbf{k}|^2 [  \mathcal{K}^{\beta\mathcal{F}}  - i \mathcal{K}^{\beta \beta}  ]  \Big) \qquad \\
\mathcal{N}^{\beta}_{\mathrm{L}}(k) & = & \mathcal{N}(k) - \sfrac{1}{|\mathbf{k}|^2} \Big(  \mathcal{E}^{\beta\mathcal{W}}(k) + \mathcal{E}^{\beta\beta}(k) \Big) \\
\mathcal{N}^{\beta}_{\mathrm{T}}(k) & = & \frac{k^2}{|\mathbf{k}|^2} \mathcal{N}(k) + \sfrac{ 1 }{2|\mathbf{k}|^4} \Big( - k_0^2 [ \mathcal{E}^{\beta\mathcal{W}}(k) + \mathcal{E}^{\beta\beta}(k) ] + |\mathbf{k}|^2 [  \mathcal{K}^{\beta\mathcal{W}} + \mathcal{K}^{\beta \beta}  ]  \Big) \qquad \, ,
\label{NTbeta}
\end{eqnarray}
where we can finally plug in the expressions from Appendix \ref{App:simplify} to get:
\begin{footnotesize}
\begin{eqnarray}
\mathrm{Re}\big[ \Pi^{\beta}_{\mathrm{L}}(k) \big] & = & \frac{1}{32\pi^2} \int_{-1}^{+1} \exd y \; y^2\; \log \Big| \frac{4m^2}{ (1 - y^2) k^2 + 4m^2} \Big| \qquad + \frac{1}{\pi^2 |\mathbf{k}|^2} \int_{m}^\infty \exd \Omega \; \frac{\sqrt{\Omega^2 - m^2}}{ e^{\beta \Omega }  - 1 } \\
&& - \frac{1}{8 \pi^2 | \mathbf{k} |^3}  \int_m^\infty \frac{\exd \Omega }{e^{\beta \Omega }  - 1 } (k_0^2 +  4 \Omega^2) \log \left| \frac{ (k^2- 2 |\mathbf{k}| \sqrt{\Omega^2 - m^2})^2 - 4 k_0^2 \Omega^2 }{  (k^2 + 2 |\mathbf{k}| \sqrt{\Omega^2 - m^2})^2 - 4 k_0^2 \Omega^2  } \right| \notag \\
& & - \frac{k_0}{2 \pi^2 | \mathbf{k} |^3}  \int_m^\infty \exd \Omega  \; \frac{  \Omega}{e^{\beta \Omega }  - 1 } \; \log \left| \frac{ (k^2)^2 - 4 ( k_0 \Omega + |\mathbf{k}| \sqrt{\Omega^2-m^2} )^2 }{ (k^2)^2 - 4 ( k_0 \Omega - |\mathbf{k}| \sqrt{\Omega^2-m^2} )^2 } \right| \notag \\
&& \notag \\
\mathrm{Im}\big[ \Pi^{\beta}_{\mathrm{L}}(k) \big] & = & \frac{\theta( -k^2- 4m^2 )}{8 \pi |\mathbf{k}|^3 } \int_{\Omega_{-}}^{\Omega_{+}} \exd \Omega \; \bigg( \sfrac{1}{2} \coth\left( \sfrac{\beta \Omega}{2} \right) (2\Omega - |k_0|)^2 + \frac{ 4 \Omega^2 - 2 |k_0| \Omega }{ [ e^{\beta \Omega}  - 1 ] [ e^{\beta ( |k_0|-\Omega) }  - 1 ] } \bigg)  \notag \\
& & + \frac{ \theta(k^2) }{8 \pi |\mathbf{k}|^3} \int_{\Omega_{+}}^{\infty} \exd \Omega \; \frac{1 + e^{\beta |k_0|}  }{[ e^{\beta \Omega}  - 1 ][ 1 - e^{-\beta ( \Omega - |k_0|)} ]} (2\Omega - |k_0|)^2  \qquad \\
&& \notag \\
\mathrm{Re}\left[ \Pi^{\beta}_{\mathrm{T}}(k) \right] & = &  \frac{k^2}{32\pi^2|\mathbf{k}|^2} \int_{-1}^{+1} \exd y \; y^2\; \log \Big| \frac{4m^2}{ (1 - y^2) k^2 + 4m^2} \Big| + \  \frac{k_0^2 + 2 |\mathbf{k}|^2}{2 \pi^2|\mathbf{k}|^4 }  \int_{m}^\infty \exd \Omega \;  \frac{\sqrt{\Omega^2 - m^2}}{ e^{\beta \Omega }  - 1 }  \\
&& + \frac{ 1 }{16 \pi^2 | \mathbf{k} |^5}  \int_m^\infty \frac{\exd \Omega }{e^{\beta \Omega }  - 1 } \big[ k^2 ( 4\Omega^2 - k^2 ) - 4 |\mathbf{k}|^2 m^2 \big] \log \left| \frac{ (k^2- 2 |\mathbf{k}| \sqrt{\Omega^2 - m^2})^2 - 4 k_0^2 \Omega^2 }{  (k^2 + 2 |\mathbf{k}| \sqrt{\Omega^2 - m^2})^2 - 4 k_0^2 \Omega^2  } \right| \notag \\
& & + \frac{ k^2 \cdot k_0 }{4 \pi^2 | \mathbf{k} |^5}  \int_m^\infty \exd \Omega  \; \frac{  \Omega}{e^{\beta \Omega }  - 1 } \; \log \left| \frac{ (k^2)^2 - 4 ( k_0 \Omega + |\mathbf{k}| \sqrt{\Omega^2-m^2} )^2 }{ (k^2)^2 - 4 ( k_0 \Omega - |\mathbf{k}| \sqrt{\Omega^2-m^2} )^2 } \right| \notag \\
&& \notag \\
\mathrm{Im}\left[\Pi^{\beta}_{\mathrm{T}}(k) \right] & = & - \frac{\theta( -k^2- 4m^2 )}{16\pi |\mathbf{k}|^5 } \int_{\Omega_{-}}^{\Omega_{+}} \exd \Omega \; \Scale[0.92]{  \bigg( \frac{1}{2} \coth\left( \sfrac{\beta\Omega}{2}\right) \Big( k^2 (2\Omega - |k_0|)^2  - (k^2 + 4m^2) |\mathbf{k}|^2 \Big) + \frac{ k^2 (4\Omega^2 - 2 |k_0|\Omega)  - (k^2 + 4m^2) |\mathbf{k}|^2 }{[ e^{\beta \Omega}  - 1 ] [ e^{\beta ( |k_0|-\Omega) }  - 1 ]} \bigg) } \notag  \\
&& - \frac{  \theta(k^2)  }{16\pi |\mathbf{k}|^5} \int_{\Omega_{+}}^{\infty} \exd \Omega \; \frac{ 1 + e^{\beta |k_0|} }{[e^{\beta \Omega} - 1] [ 1 - e^{-\beta (\Omega -  |k_0| ) } ] }  \Big( k^2 (2\Omega - |k_0|)^2  - (k^2 + 4m^2) |\mathbf{k}|^2 \Big) \, .
\end{eqnarray}
\end{footnotesize}\ignorespaces
We also get the purely real functions:
\begin{footnotesize}
\begin{eqnarray}
\mathcal{N}^{\beta}_{\mathrm{L}}(k) & = & - \frac{\theta(-k^2 - 4m^2)}{8\pi |\mathbf{k}|^3} \int_{\Omega_{-}}^{\Omega_{+}}\exd \Omega \; \bigg( \theta(-k^0)  \coth\left( \sfrac{\beta\Omega}{2} \right) (2\Omega - |k_0|)^2 +\frac{ 4 \Omega^2 - 2 |k_0| \Omega }{ [ e^{\beta \Omega} - 1 ] [ e^{\beta ( |k_0|-\Omega) }  - 1 ]  } \bigg) \\
&&- \frac{\theta(k^2) }{4\pi |\mathbf{k}|^3} \int_{\Omega_{+}}^{\infty} \exd \Omega \; \bigg( \frac{\theta(k^0)}{ e^{\beta \Omega} - 1 } + \frac{ \theta(-k^0)}{e^{\beta ( \Omega - |k_0| ) } - 1 } +  \frac{1}{[ e^{\beta \Omega}  - 1 ][ e^{\beta ( \Omega - |k_0|)}  - 1 ] } \bigg) (2\Omega - |k_0|)^2 \notag \\
&& \notag \\
 \mathcal{N}^{\beta}_{\mathrm{T}}(k) & = & \frac{ \theta(-k^2 - 4m^2)}{16\pi |\mathbf{k}|^5} \int_{\Omega_{-}}^{\Omega_{+}} \exd \Omega \; \Scale[0.92]{ \bigg( \theta(-k^0) \coth\left( \tfrac{\beta \Omega}{2} \right) \Big( k^2 (2\Omega - |k_0|)^2  - (k^2 + 4m^2) |\mathbf{k}|^2 \Big) + \frac{  k^2 ( 4 \Omega^2 - 2 |k_0| \Omega ) - (k^2 +4m^2) |\mathbf{k}|^2 }{ [ e^{\beta \Omega} - 1][e^{\beta ( |k_0|-\Omega) }  - 1 ] } \bigg) } \notag \\
&+&  \frac{ \theta(k^2)  }{8 \pi |\mathbf{k}|^5} \int_{\Omega_{+}}^{\infty} \exd \Omega \; \bigg( \sfrac{\theta(k^0)}{ e^{\beta \Omega} - 1 } + \sfrac{ \theta(-k^0)}{e^{\beta ( \Omega - |k_0| ) } - 1 } + \sfrac{1}{[ e^{\beta \Omega}  - 1 ][ e^{\beta ( \Omega - |k_0|)}  - 1 ] } \bigg) \Big( k^2 (2\Omega - |k_0|)^2  - (k^2 + 4m^2) |\mathbf{k}|^2 \Big) \, .
\end{eqnarray}
\end{footnotesize}\ignorespaces

\newpage

\section{SSB details}
\label{App:SSBint}

We here provide some extra details underlying the calculation of the influence functional in Eq.~(\ref{SSB_SIF_v4}). We begin with the tadpole contributions $\langle H_{\pm}(x) \rangle_{\mathcal{E}}$, which must be addressed even in the $\mathcal{O}(v^{-2})$ influence functional. Using the definition of the environmental average in Eq.~(\ref{SSB_average}) and performing the necessary Wick contractions, one finds
\begin{equation} \label{App_tadpolecancel} \langle H_{+}(x) \rangle_{\mathcal{E}} \simeq \frac{1}{\sqrt{2}v M^2} \int \frac{\exd^4 \ell}{(2\pi)^4} \; \ell^2  \mathcal{F}(\ell)  - \frac{a_1}{\sqrt{2}v M^2} + \mathcal{O}(v^{-3}) \, , \end{equation} 
which explicitly involves the counterterm $a_{1}$ introduced in Eq.~(\ref{SSB_SE}). In deriving this we have used
\begin{equation} \label{zeromom_FW}
\mathcal{F}(0) = - i M^{-2} \qquad \mathrm{and} \qquad \mathcal{W}(0) =0 \, ,
\end{equation}
which hold for any excited state (and follow from Eq.~(\ref{heavy_FW})). The above is always real-valued, and one has  $\langle H_{-}(x) \rangle_{\mathcal{E}} = \langle H_{+}(x) \rangle_{\mathcal{E}}$. Imposing the condition that the tadpole vanish fixes
\begin{equation} 
\label{tadpole_renorm_cond}
a_{1} \simeq \int \frac{\exd^4 \ell}{(2\pi)^4} \; \ell^2  \mathcal{F}(\ell)  +\mathcal{O}(v^{-2}) \ .
\end{equation}
a result that also reappears in the renormalisation of the two-point functions.

\begin{figure}[h]
\begin{center}
\includegraphics[width=135mm]{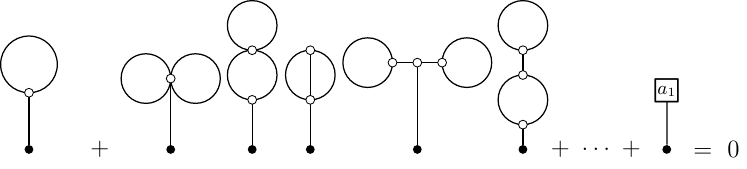}
\caption{\small The $\mathcal{O}(v^{-1})$ and $\mathcal{O}(v^{-3})$ contributions to $\langle H_{\pm}(x) \rangle_{\mathcal{E}}$.  Cubic, quartic, and quintic interactions from Eq.~(\ref{SSB_SE}) contribute to the subleading two-loop diagrams shown, and the $a_{1}$ counterterm is required to cancel the resulting tadpoles order by order in the $1/v$ expansion.} \label{Fig:SSBTadpoles}
\end{center}
\end{figure}

At $\mathcal{O}(v^{-4})$ the influence functional contains additional $\mathcal{O}(v^{-3})$ two-loop tadpole contributions, shown in Fig.~\ref{Fig:SSBTadpoles}. These two-loop corrections are $x$-independent constants and, although their explicit evaluation is technically involved, it is unnecessary: since they merely shift the tadpole by a constant amount, one may simply impose that the linear counterterm proportional to $a_{1}$ cancels them at the corresponding order in the $1/v$ expansion. For this reason, we do not compute them explicitly.

For the $\mathcal{O}(v^{-4})$ influence functional, one must compute the $\mathcal{O}(v^{-2})$ loops corrections to the two-point function appearing in Eq.~(\ref{SSB_SIF_3pt_average}).  Performing the required Wick contractions, one finds that the $\mathcal{O}(v^{-2})$ corrections $\QFe$ and $\QWe$ appearing in Eq.~(\ref{SSB_SIF_v4}) are
\begin{small}
\begin{eqnarray}
\QFe(k) & = & - 2 i  \Big( \mathcal{F}(k)^2 - \mathcal{W}(k) \mathcal{W}(-k) \Big)  \int \frac{\exd^4 \ell}{(2 \pi)^4} \big( k^2 + \ell^2 \big)  \mathcal{F}(\ell)  \\
&& + i \Big( \mathcal{F}(k)^2 - \mathcal{W}(k) \mathcal{W}(-k) \Big) \cdot \frac{ k^2 }{ M^2 }  \int \frac{\exd^4 \ell}{(2\pi)^4} \; \ell^2  \mathcal{F}(\ell)  \notag \\
&& - \frac{1}{4} \mathcal{F}(k)^2 \int \frac{\exd^4 \ell}{(2\pi)^4}  \; \Big( k^2 + (k -\ell)^2 + \ell^2 \Big)^2  \mathcal{F}(\ell)   \mathcal{F}(k - \ell )  \notag \\
&& + \frac{1}{4} \mathcal{F}(k) \mathcal{W}(k) \int \frac{\exd^4 \ell}{(2\pi)^4} \Big( k^2 + (k + \ell)^2 + \ell^2 \Big) \mathcal{W}(\ell)  \mathcal{W}( - k - \ell ) \notag \\
&& + \frac{1}{4} \mathcal{F}(k) \mathcal{W}(-k) \int \frac{\exd^4 \ell}{(2\pi)^4} \Big( k^2 + (k- \ell)^2 + \ell^2 \Big) \mathcal{W}(\ell) \mathcal{W}( k - \ell ) \notag \\
& & - \frac{1}{4} \mathcal{W}(k) \mathcal{W}(-k) \int \frac{\exd^4 \ell}{(2\pi)^4} \; \Big( k^2 + \ell^2 + (k -\ell)^2 \Big)^2 \mathcal{F}^*(\ell)   \mathcal{F}^*(k - \ell )  \notag \\
& & - i \Big( a_2 + \left[ b_2 + \frac{a_1}{M^2} \right] k^2\Big) \Big( \mathcal{F}(k)^2 - \mathcal{W}(k)  \mathcal{W}(-k) \Big) \, ,  \notag
\end{eqnarray} 
\end{small}\ignorespaces
as well as
\begin{small}
\begin{eqnarray}
\QWe(k) & = & - 2 i \mathcal{W}(k) \Big( \mathcal{F}(k) -   \mathcal{F}^*(k)   \Big) \int \frac{\exd^4 \ell}{(2 \pi)^4} \big( k^2 + \ell^2 \big)  \mathcal{F}(\ell)  \label{QWE_exp} \\
&& + i \mathcal{W}(k) \Big(  \mathcal{F}(k)  -  \mathcal{F}^*(k)    \Big) \cdot \frac{k^2}{M^2} \int \frac{\exd^4 \ell}{(2\pi)^4} \; \ell^2  \mathcal{F}(\ell)  \notag \\
&& - \frac{1}{4} \mathcal{W}(k)  \mathcal{F}(k) \int \frac{\exd^4 \ell}{(2\pi)^4}  \; \Big( k^2 + (k -\ell)^2 + \ell^2 \Big)^2  \mathcal{F}(\ell)   \mathcal{F}(k - \ell )  \notag \\
&& + \frac{1}{4}  \mathcal{F}^*(k)   \mathcal{F}(k) \int \frac{\exd^4 \ell}{(2\pi)^4}  \; \Big( k^2 + (k -\ell)^2 + \ell^2 \Big)^2 \mathcal{W}(\ell) \mathcal{W}(k - \ell ) \notag \\
&& + \frac{1}{4} \mathcal{W}(k) \mathcal{W}(k) \int \frac{\exd^4 \ell}{(2\pi)^4}  \; \Big( k^2 + (k + \ell)^2 + \ell^2 \Big)^2 \mathcal{W}(\ell) \mathcal{W}(- k - \ell ) \notag \\
&& - \frac{1}{4} \mathcal{W}(k)  \mathcal{F}^*(k)   \int \frac{\exd^4 \ell}{(2\pi)^4}  \; \Big( k^2 + (k -\ell)^2 + \ell^2 \Big)^2 \mathcal{F}^*(\ell)   \mathcal{F}^*(k - \ell )  \notag \\
& & - i \Big( a_2 + \left[ b_2 + \frac{a_1}{M^2} \right] k^2\Big) \mathcal{W}(k) \Big(  \mathcal{F}(k)  -  \mathcal{F}^*(k)   \Big) \notag \, , 
\end{eqnarray}
\end{small}\ignorespaces
where we have again used (\ref{zeromom_FW}) for the connected tadpole terms in the second line of each expression, and we note that $\mathcal{F}(-k) = \mathcal{F}(k)$, however $\mathcal{W}(-k) \neq \mathcal{W}(k)$. Notice that the tadpole cancellation from Eq.~(\ref{tadpole_renorm_cond}) means that $a_1$ exactly cancels off the second line in each expression. The remaining terms can be simplified by defining the three independent integrals
\begin{eqnarray}
I(k) & : = & - 2 \int \frac{\exd^4 \ell}{(2 \pi)^4} \big( k^2 + \ell^2 \big)  \mathcal{F}(\ell)  \label{Iloopdef}\\
L_{\mathcal{F}(k)}& \equiv & \int \frac{\exd^4 \ell}{(2\pi)^4}  \; \Big( k^2 + (k -\ell)^2 + \ell^2 \Big)^2  \mathcal{F}(\ell)   \mathcal{F}(k - \ell )  \label{LFloopdef} \\
L_{\mathcal{W}}(k) & \equiv & \int \frac{\exd^4 \ell}{(2\pi)^4}  \; \Big( k^2 + (k -\ell)^2 + \ell^2 \Big)^2 \mathcal{W}(\ell) \mathcal{W}(k - \ell ) \label{LWloopdef} \, ,
\end{eqnarray}
which rewrite the above loop corrections as
\begin{eqnarray}
\QFe(k) & = & i \Big(  \mathcal{F}(k)^2 - \mathcal{W}(k) \mathcal{W}(-k) \Big) \Big( I(k) - \frac{1}{4} \mathrm{Im}[ L_{\mathcal{F}}(k) ] - a_2 - b_2 k^2 \Big) \label{QFE_div} \qquad \\
&& - \frac{1}{4} \Big( \mathcal{F}(k)^2 + \mathcal{W}(k) \mathcal{W}(-k) \Big) \mathrm{Re}[ L_{\mathcal{F}}(k) ]  \notag \\
&& + \frac{1}{4} \mathcal{F}(k) \Big( \mathcal{W}(k) L_{\mathcal{W}}(-k) + \mathcal{W}(-k) L_{\mathcal{W}}(k) \Big) \notag \, , 
\end{eqnarray}
and
\begin{eqnarray}
\QWe(k) & = & i \mathcal{W}(k) \Big( \mathcal{F}(k) -   \mathcal{F}^*(k)   \Big) \Big( I(k) - \frac{1}{4} \mathrm{Im}[L_{\mathcal{F}}(k)] - a_2 - b_2 k^2 \Big) \qquad  \label{QWE_div} \\
&& - \frac{1}{4} \mathcal{W}(k) \Big( \mathcal{F}(k) +  \mathcal{F}^*(k)   \Big) \mathrm{Re}[ L_{\mathcal{F}}(k) ] \notag \\
&& + \frac{1}{4}  \mathcal{F}^*(k)   \mathcal{F}(k) L_{\mathcal{W}}(k) + \frac{1}{4} \mathcal{W}(k) \mathcal{W}(k) L_{\mathcal{W}}(-k) \ . \notag
\end{eqnarray}
All divergences are contained in the combination 
$I(k) - \tfrac{1}{4}\,\mathrm{Im}[L_{\mathcal{F}}(k)]$ appearing in both expressions, which is why this structure is accompanied by counterterms. Our next task is to evaluate these loop integrals and then renormalize them appropriately, noting that they represent the in-in self-energies of the Higgs field.

\subsection{Divergences and renormalisation}

Let us split apart the computation of the integrals from Eqs.~(\ref{Iloopdef})-(\ref{LWloopdef}) into vacuum and non-vacuum contributions such that
\begin{eqnarray} \label{deltasplit_ints}
I(k) & = & I^{\mathrm{vac}}(k) + \delta I(k) \qquad \mathrm{and} \qquad L_{\mathcal{F},\mathcal{W}}(k) = L^{\mathrm{vac}}_{\mathcal{F},\mathcal{W}}(k) + \delta L_{\mathcal{F},\mathcal{W}}(k) \, ,
\end{eqnarray}
where the vacuum contributions are given by
\begin{eqnarray}
\Ivac(k) & : = & - 2 \int \frac{\exd^4 \ell}{(2 \pi)^4} \big( k^2 + \ell^2 \big)  \mathcal{F}^{\rm vac}(\ell)  \\
\LFvac(k) & \equiv & \int \frac{\exd^4 \ell}{(2\pi)^4}  \; \Big( k^2 + (k -\ell)^2 + \ell^2 \Big)^2  \mathcal{F}^{\rm vac}(\ell)   \mathcal{F}^{\rm vac}(k - \ell )  \\
\LWvac(k) & \equiv & \int \frac{\exd^4 \ell}{(2\pi)^4}  \; \Big( k^2 + (k -\ell)^2 + \ell^2 \Big)^2 \Wvac(\ell) \Wvac(k - \ell ) \, ,
\end{eqnarray}
with the free vacuum propagators $\Delta_{\mathcal{F},\mathcal{W}}^{\mathrm{vac}}$ given by the first terms in Eq.~(\ref{FWwithn_MAIN}) (analogous to Eq.~(\ref{freeprop_vac})). The vacuum contributions are evaluated similar to those in Appendix \ref{App:thermalVac} using dimensional regularisation. We first compute
\begin{eqnarray}
\Ivac(k) & = & - 2 k^2 \mathscr{L}_{(0,1)}(M^2) - 2 \mathscr{L}_{(1,1)}(M^2) \\
& = & - \frac{2 M^2 \big( k^2 - M^2 \big) \Gamma \left(1-\frac{D}{2}\right)}{(4 \pi )^{D/2}} \bigg(\frac{M^2}{\mu ^2}\bigg)^{\tfrac{D-4}{2}} \\
& \simeq & \frac{M^2 (k^2 - M^2)}{4 \pi ^2 (4-D)} - \frac{M^2 (k^2-M^2)}{8 \pi ^2} \log \left(\frac{M^2}{4 \pi  e^{1-\gamma } \mu ^2}\right)  +\mathcal{O}(4-D) \, , 
\end{eqnarray}
which is expressed in terms of the elementary loop integrals $\mathcal{L}_{(n,N)}$ defined in Eq.~(\ref{curlyLnN}) (and  evaluated in Eq.~(\ref{curlyLnN_ans})). Similarly, standard manipulations give
\begin{eqnarray}
\LFvac(k) & = & - \frac{1}{2} \int \frac{\exd^4 \ell}{(2\pi)^4}  \; \Big( k^2 + (k -\ell)^2 + \ell^2 \Big)^2 \int_{-1}^{+1} \frac{\exd y}{\big[ \ell^2 + M^2 +(y-1) ( k \cdot \ell - \frac{1}{2} k^2 ) - i \epsilon \big]^2} \qquad \\
& = & - 2 \int_{-1}^{+1} \exd y \int \frac{\exd^4 p}{(2\pi)^4}  \; \frac{ [ p^2 - y p \cdot k + \frac{1}{4} ( 3 + y^2 ) k^2 ]^2}{(p^2 + \frac{1}{4}\Sigma(y) - i \epsilon )^2}
\end{eqnarray}
where $\Sigma(y) = (1 - y^2) k^2 + 4M^2$. Expanding this out, the terms odd in $p$ vanish, and there is also a term $\propto (p \cdot k)^2  = k^\mu k^\nu p_{\mu} p_\nu$ on which we use (\ref{pmupnu_loop}), and so
\begin{eqnarray}
\LFvac(k) & = & - 2 \int_{-1}^{+1} \exd y \int \frac{\exd^D p}{(2\pi)^D}  \; \frac{ (p^2)^2 + [ (\frac{1}{D} + \frac{1}{2}) y^2 + \frac{3}{2} ] k^2 p^2 + \frac{1}{16}(3+y^2)^2 (k^2)^2 }{(p^2 + \frac{1}{4}\Sigma(y) - i \epsilon )^2} \\
& = & - 2 i \int_{-1}^{+1} \exd y \;  \bigg[ \mathscr{L}_{(2,2)}(\tfrac{1}{4}\Sigma) + \frac{(\frac{2}{D} + 1) y^2 +3}{2} k^2 \mathscr{L}_{(1,2)}(\tfrac{1}{4}\Sigma) + \frac{(3+y^2)^2 (k^2)^2}{16} \mathscr{L}_{(0,2)}(\tfrac{1}{4}\Sigma) \bigg] \notag \, .
\end{eqnarray}
Evaluating this using Eq.~(\ref{curlyLnN_ans}) gives
\begin{small}
\begin{eqnarray}
\LFvac(k) & = & 4 i \int_{-1}^{+1} \exd y \; \frac{\Gamma \left(1-\frac{D}{2}\right) }{(16 \pi )^{D/2}} \bigg( \frac{(1-y^2)k^2+4 M^2 - i \epsilon }{\mu ^2}\bigg)^{\tfrac{D-4}{2}}  \\
&& \quad \times\; \bigg[ (k^2)^2 \left((D+1) y^4+(2 D-5) y^2+D-4\right)  - 4k^2 M^2 \left((D+2) y^2+D-1\right)  + 4 (D+2) M^4 \bigg] \notag  \\
& \simeq & - i \; \frac{(k^2)^2 - 10 k^2 M^2 + 12 M^4}{8 \pi^2 (4-D)} \ + i \; \frac{11 (k^2)^2-10 k^2 M^2-6 M^4}{120 \pi ^2} \\
&&+ i \int_{-1}^{+1} \exd y\; \frac{y^2 (5 y^2+3) (k^2)^2 -12 (2 y^2+1)k^2 M^2 +24 M^4}{64 \pi ^2} \log \bigg(\frac{(1-y^2)k^2 +4 M^2-i \epsilon }{16 \pi  e^{{3}/{5}-\gamma } \mu ^2}\bigg) \notag \, .
\end{eqnarray}
\end{small}\ignorespaces
Let us use $\log(x - i \epsilon) = \log|x| - i \pi \theta(-x)$, to find
\begin{eqnarray}
\mathrm{Re}[\LFvac(k)] & = & \frac{\theta(-k^2 - 4M^2)}{16\pi} (k^2 - 2M^2)^2 \sqrt{ 1 + \frac{4m^2}{k^2} } \\
\mathrm{Im}[\LFvac(k)] & = & - \frac{(k^2)^2 - 10 k^2 M^2 + 12 M^4}{8 \pi^2 (4-D)} \ + \frac{11 (k^2)^2-10 k^2 M^2-6 M^4}{120 \pi ^2} \\
&& + \int_{-1}^{+1} \exd y\; \frac{y^2 (5 y^2+3) (k^2)^2 -12 (2 y^2+1)k^2 M^2 +24 M^4}{64 \pi ^2} \log \Big| \frac{(1-y^2)k^2 +4 M^2}{16 \pi  e^{{3}/{5}-\gamma } \mu ^2} \Big| \notag \, .
\end{eqnarray}
Finally we trivially compute
\begin{eqnarray}
\LWvac(k) & = & \frac{\theta(k^0) \theta(-k^2 - 4M^2)}{8\pi} (k^2 - 2M^2)^2 \sqrt{ 1 + \frac{4m^2}{k^2} } \, , 
\end{eqnarray}
which is finite. Note that the non-vacuum contributions to the loops $\delta I$ and $\delta L_{\mathcal{F},\mathcal{W}}$ do not diverge, which means that the only divergent contributions to the corrections (\ref{QFE_div}) and (\ref{QWE_div}) appear in through the combination:
\begin{equation} \label{divergence_structure}
\Big( I(k) - \frac{1}{4}\mathrm{Im}[L_{\mathcal{F}}(k)] \Big)_{\mathrm{divergence}} \simeq \frac{(k^2+M^2)^2 - 4 k^2 M^2 + 3 M^4}{32\pi^2(4-D)} + \ldots 
\end{equation}

\subsubsection{Counterterm for composite operator divergences}

Eqs.~(\ref{QFE_div}) and (\ref{QWE_div}) show that the divergence structure appears along with counterterms $- a_2 -  b_2 k^2$. Examination of Eq.~(\ref{divergence_structure}) shows however that there is a divergence proportional to $(k^2)^2$ divergence that appears, which cannot be absorbed by these counterterms alone.

The source of this unusual divergence proportional to $(k^2)^2$ is the fact that $H$ is a composite operator, as explained in the text surrounding Eq.~(\ref{OPE}). The remedy is to absorb this divergence into the operator $\propto (\mathscr{D}\chi)^4$ in the system action (\ref{SSB_SS}). One finds that
\begin{small}
\begin{eqnarray}
&& S_{\mathcal{S}}[A_{+},\chi_{+},c_{+},\bar{c}_{+}]-  S_{\mathcal{S}}[A_{-},\chi_{-},c_{-},\bar{c}_{-}] + S_{\mathrm{IF}}[A_{+}, \chi_{+}, A_{-}, \chi_{-} ] \notag \\
&& \hspace{2mm} \supset - \frac{b_4}{4v^4} \int \exd^4 x\; \bigg[ \big(\mathscr{D}_{+}\chi_{+}(x)\big)^4 - \big(\mathscr{D}_{-}\chi_{-}(x)\big)^4  \bigg] \\
&& \hspace{6mm} + \frac{i}{4v^4} \int \exd^4 x \int \exd^4 y\; \bigg[ \big( \mathscr{D}_{+}\chi_{+}(x) \big)^2 \mathcal{Q}_{\mathcal{F}}(x,y) \big( \mathscr{D}_{+}\chi_{+}(y) \big)^2 +  \big( \mathscr{D}_{-}\chi_{-}(x) \big)^2 \mathcal{Q}^{\ast}_{\mathcal{F}}(x,y) \big( \mathscr{D}_{-}\chi_{-}(y) \big)^2 \bigg] \notag \\
&& \hspace{2mm} \supset  - \frac{1}{4v^4} \int \exd^4 x \int \exd^4 y\; \bigg[ \big( \mathscr{D}_{+}\chi_{+}(x) \big)^2 \big( \mathscr{D}_{+}\chi_{+}(y) \big)^2 -  \big( \mathscr{D}_{-}\chi_{-}(x) \big)^2 \big( \mathscr{D}_{-}\chi_{-}(y) \big)^2 \bigg]  \bigg[ b_{4} \delta(x-y) +  \mathrm{Im}[\mathcal{Q}_{F}(x,y)] \bigg] \notag \, , 
\end{eqnarray}
\end{small}\ignorespaces
where we focus only on the terms on the same branch which can be affected by this local counterterm. The coefficient $b_{4}\,\delta(x-y)$ appears exclusively in combination with $\mathrm{Im}[\mathcal{Q}_{\mathcal{F}}(x,y)]$, so absorbing $b_{4}$ into the definition of $\mathcal{Q}_{\mathcal{F}}(x,y)$ implies in momentum space that
\begin{eqnarray}
\QFe(k) & \to & \QFe(k) + i b_4 \\
& = & i \Big(  \mathcal{F}(k)^2 - \mathcal{W}(k) \mathcal{W}(-k) \Big) \Big( I(k) - \tfrac{1}{4} \mathrm{Im}[ L_{\mathcal{F}}(k) ] - a_2 - b_2 k^2 \Big) \label{QFE_b4} \\
&& - \tfrac{1}{4} \Big( \mathcal{F}(k)^2 + \mathcal{W}(k) \mathcal{W}(-k) \Big) \mathrm{Re}[ L_{\mathcal{F}}(k) ]  \notag \\
&& + \tfrac{1}{4} \mathcal{F}(k) \Big( \mathcal{W}(k) L_{\mathcal{W}}(-k) + \mathcal{W}(-k) L_{\mathcal{W}}(k) \Big) +  i b_4 \notag \, .
\end{eqnarray}
It is not immediately obvious, but the parameter $b_{4}$ enters the same bracket as the other counterterms. To make this clear, we rewrite the propagator structure multiplying the divergences in a more convenient form. As a first step, we take the $\epsilon \to 0^{+}$ limit of the free propagators in Eq.~(\ref{FWwithn_MAIN}), yielding
\begin{equation}
\mathcal{F}(k) = \frac{ - i}{k^2 + M^2} + 2 \pi \left( \tfrac{1}{2} + \mathfrak{n}(k) \right) \delta(k^2 + M^2) \quad \mathrm{and} \quad \mathcal{W}(k) =  2 \pi \left( \theta(k_0) + \mathfrak{n}(k) \right) \delta(k^2 + M^2) \label{FWwithn_APP} \, ,
\end{equation}
where we used the Sokhotski-Plemelj relations (\ref{SP_relations}). Using these expressions one finds that
\begin{eqnarray} \label{regulatedprops}
i \Big(  \mathcal{F}(k)^2 - \mathcal{W}(k) \mathcal{W}(-k) \Big) & = & \frac{-i}{(k^2 +M^2)^2} - 2 \pi \big( \tfrac{1}{2} + \mathfrak{n}(k) \big) \delta'(k^2 + M^2) \, , 
\end{eqnarray}
where we have used the regularisation \cite{Bellac:2011kqa}
\begin{equation} 
\frac{\delta(x)}{x+i \epsilon} = - \frac{1}{2} \delta'(x) - i \pi \big[ \delta(x) \big]^2 \, , 
\end{equation}
as well as the Sokhotski-Plemelj relations (\ref{SP_relations}) again which implies $(x - i \epsilon)^{-2} = x^{-2} - i \pi \delta'(x)$ (where formally $x^{-2} \to \mathcal{H}(x^{-2})$ with $\mathcal{H}$ the Hadamard finite part). Now using the identity $x^2 \delta'(x) = 0$ one can use the above to also show that
\begin{eqnarray}
(k^2 + M^2)^2 \cdot i \Big(  \mathcal{F}(k)^2 - \mathcal{W}(k) \mathcal{W}(-k) \Big) & = & - i  \, , 
\end{eqnarray}
which means that (\ref{QFE_b4}) can be written more usefully as
\begin{eqnarray}
\QFe(k) & = & i \Big(  \mathcal{F}(k)^2 - \mathcal{W}(k) \mathcal{W}(-k) \Big) \Big( I(k) - \tfrac{1}{4} \mathrm{Im}[ L_{\mathcal{F}}(k) ] - a_2 - b_2 k^2 - b_4 (k^2+M^2)^2 \Big) \notag \\
&& - \tfrac{1}{4} \Big( \mathcal{F}(k)^2 + \mathcal{W}(k) \mathcal{W}(-k) \Big) \mathrm{Re}[ L_{\mathcal{F}}(k) ] \label{QFEwithb4now} \\
&& + \tfrac{1}{4} \mathcal{F}(k) \Big( \mathcal{W}(k) L_{\mathcal{W}}(-k) + \mathcal{W}(-k) L_{\mathcal{W}}(k) \Big)  \notag \, .
\end{eqnarray}
It turns out that we can perform a similar manipulation for $\QWe(k)$ from Eq.~(\ref{QWE_div}), where the divergence is multiplied instead by the propagator structure
\begin{equation}
i \mathcal{W}(k) \Big( \mathcal{F}(k) -   \mathcal{F}^*(k)   \Big) = - 2 \pi \big[ \theta(k^0) + \mathfrak{n}(k) \big] \delta'(k^2+M^2) \, , 
\end{equation}
which we've regulated similar to (\ref{regulatedprops}) above. Again using $x^2 \delta'(x)=0$ this equation implies that
\begin{equation}
(k^2 +M^2)^2 \cdot i \mathcal{W}(k) \Big( \mathcal{F}(k) -   \mathcal{F}^*(k)   \Big) = 0 \label{WW_identity} \, , 
\end{equation}
and so in fact we may more usefully rewrite (\ref{QWE_div}) as
\begin{eqnarray}
\QWe(k) & = & i \mathcal{W}(k) \Big( \mathcal{F}(k) -   \mathcal{F}^*(k)   \Big) \Big( I(k) - \frac{1}{4} \mathrm{Im}[L_{\mathcal{F}}(k)] - a_2 - b_2 k^2  - b_4 (k^2+M^2)^2 \Big) \notag \\
&& - \frac{1}{4} \mathcal{W}(k) \Big( \mathcal{F}(k) +  \mathcal{F}^*(k)   \Big) \mathrm{Re}[ L_{\mathcal{F}}(k) ] \label{QWEwithb4now}  \\
&& + \frac{1}{4}  \mathcal{F}^*(k)   \mathcal{F}(k) L_{\mathcal{W}}(k) + \frac{1}{4} \mathcal{W}(k) \mathcal{W}(k) L_{\mathcal{W}}(-k) \ . \notag
\end{eqnarray}
This shows that we can clearly absorb all divergences now using $a_2$, $b_2$ and $b_4$. Said differently, the divergence proportional to $(k^2+M^2)^2$ in (\ref{QWEwithb4now}) vanishes on account of Eq.~(\ref{WW_identity}).

\subsubsection{Finiteness of the partial self-energies}
\label{app:selfen}

The structure of the two-point loop correction is well-known: the loop integrals are related to the partial self-energies of the (composite) Higgs field \cite{Calzetta:2008iqa}. In fact, one may express each of these as
\begin{equation} \label{matrixSigmaApp}
\left[ \begin{matrix} \QFe(k) & \QWe(-k) \\ \QWe(k)  & \QFeAST(k)  \end{matrix} \right] = \left[ \begin{matrix} \mathcal{F}(k) & \mathcal{W}(-k) \\ \mathcal{W}(k) & \mathcal{F}^*(k) \end{matrix} \right] \left[ \begin{matrix} \Sigma_{\mathcal{F}}(k) & \Sigma_{\mathcal{W}}(-k) \\ \Sigma_{\mathcal{W}}(k) & \Sigma_{\mathcal{F}}^{\ast}(k) \end{matrix} \right] \left[ \begin{matrix} \mathcal{F}(k) & \mathcal{W}(-k) \\ \mathcal{W}(k) & \mathcal{F}^*(k) \end{matrix} \right] \ .
\end{equation}
Multiplying out the matrices and comparing to (\ref{QFEwithb4now}) and (\ref{QWEwithb4now}) we clearly have
\begin{eqnarray}
\mathrm{Re}[ \Sigma_{\mathcal{F}}(k)]  & = & - \tfrac{1}{4} \mathrm{Re}[ L_{\mathcal{F}}(k) ] \\
\mathrm{Im}[ \Sigma_{\mathcal{F}}(k)] & = & I(k) - \tfrac{1}{4} \mathrm{Im}[ L_{\mathcal{F}}(k) ] - a_2 - b_2 k^2 - b_4 (k^2+M^2)^2 \label{SigmaF_app} \\
\Sigma_{\mathcal{W}}(k) & = & \tfrac{1}{4} L_{\mathcal{W}}(k) \, , 
\end{eqnarray}
which are the results quoted in Eq.~(\ref{SigmaF_main}) and (\ref{SigmaW_main}) in the main text. As must be the case, the off-diagonal self-energies are perfectly finite, and only the diagonals need to be renormalised. 

Since all divergences come from the vacuum contributions of the initial state, to render the loop corrections finite all it takes is to consider the vacuum part of Eq.~(\ref{SigmaF_app}), where one finds explicitly:
%
\begin{eqnarray}
\mathrm{Im}[ \Sigma^{\mathrm{vac}}_{\mathcal{F}}(k) ] & : = & \Ivac(k) - \tfrac{1}{4} \mathrm{Im}[ \LFvac(k) ] - a_2 - b_2 k^2 - b_4 (k^2+M^2)^2 \label{Sigmavac_TBR} \\
& = & \frac{(k^2+M^2)^2 - 4 k^2 M^2 + 3 M^4}{32\pi^2(4-D)} - a_2 - b_2 k^2 - b_4 (k^2+M^2)^2 \\
& \ & - \frac{11 (k^2)^2-10 k^2 M^2-6 M^4}{480 \pi ^2} - \frac{M^2 (k^2-M^2)}{8 \pi ^2} \log \left(\frac{M^2}{4 \pi  e^{1-\gamma } \mu ^2}\right) \notag \\
&& - \int_{-1}^{+1} \exd y\; \frac{y^2 (5 y^2+3) (k^2)^2 -12 (2 y^2+1)k^2 M^2 +24 M^4}{256 \pi ^2} \log \bigg|\frac{(1-y^2)k^2 +4 M^2 }{16 \pi  e^{{3}/{5}-\gamma } \mu ^2}\bigg| \notag \, .
\end{eqnarray}
%
As discussed in the main text, $\Sigma_{\mathcal{F}}$ and $\Sigma_{\mathcal{W}}$ are only the partial self-energies since interactions of $\mathscr{D}\chi$ will also contribute to each once they are also integrated out. For this reason we simply cancel the divergences using a modified minimal subtraction ($\overline{\mathrm{MS}}$) scheme so that
\begin{small}
\begin{eqnarray}
\mathrm{Im}[ \Sigma^{\mathrm{vac}}_{\mathcal{F}}(k)] & = & - \frac{M^2 (k^2-M^2)}{8 \pi ^2} \log \left(\frac{M^2}{\mu ^2}\right) \notag \\
&& - \int_{-1}^{+1} \exd y\; \frac{y^2 (5 y^2+3) (k^2)^2 -12 (2 y^2+1)k^2 M^2 +24 M^4}{256 \pi ^2} \log \bigg|\frac{(1-y^2)k^2 +4 M^2 }{\mu ^2}\bigg| \notag \, ,
\end{eqnarray}
\end{small}\ignorespaces
with appropriately chosen $a_{2}$, $b_2$ and $b_4$.

\subsection{Evaluation of non-vacuum contribution}

Let us now examine the non-vacuum contribution from Eq.~(\ref{deltasplit_ints}). These integrals can all be expressed in terms of the distribution function $\mathfrak{n}$ from the splitting of the free propagators in Eq.~(\ref{FWwithn_MAIN}). One has
\begin{equation}
\delta I(k) = - 2 \int \frac{\exd^4 \ell}{(2 \pi)^4} \big( k^2 + \ell^2 \big) \cdot 2 \pi \mathfrak{n}(\ell) \delta(\ell^2 + M^2) \, , \label{deltaI_n}
\end{equation}
and we can also write
\begin{eqnarray}
\delta L_{\mathcal{F}}(k) = 2 L_{\mathcal{F}\mathfrak{n}}(k) +  L_{\mathfrak{nn}}(k) \qquad \mathrm{and} \qquad \delta L_{\mathcal{W}}(k) = 2 L_{\mathcal{W}\mathfrak{n}}(k) +  L_{\mathfrak{nn}}(k) \, , \label{deltaLFW_n}
\end{eqnarray}
with the definitions:
\begin{eqnarray}
L_{\mathcal{F}\mathfrak{n}}(k) & \equiv & \int \frac{\exd^4 \ell}{(2\pi)^4}  \; \Big( k^2 + (k -\ell)^2 + \ell^2 \Big)^2 \cdot 2 \pi \mathfrak{n}(\ell) \delta(\ell^2 + M^2 )  \cdot \frac{- i}{(k - \ell)^2 + M^2 - i \epsilon} \label{LFn_int} \\
L_{\mathcal{W}\mathfrak{n}}(k) & \equiv & \int \frac{\exd^4 \ell}{(2\pi)^4}  \; \Big( k^2 + (k -\ell)^2 + \ell^2 \Big)^2 \cdot 2 \pi \mathfrak{n}(\ell) \delta(\ell^2 + M^2 ) \cdot 2 \pi \theta(k^0 - \ell^0) \delta\big( (k - \ell)^2 + M^2 \big) \qquad \quad \\
L_{\mathfrak{nn}}(k) & \equiv & \int \frac{\exd^4 \ell}{(2\pi)^4}  \; \Big( k^2 + (k -\ell)^2 + \ell^2 \Big)^2 \cdot 2 \pi \mathfrak{n}(\ell) \delta(\ell^2 + M^2 ) \cdot 2 \pi \mathfrak{n}(k - \ell) \delta\big( (k - \ell)^2 + M^2 \big)  \, .
\end{eqnarray}
Without further assumptions about the distribution function $\mathfrak{n}$  these integrals cannot be simplified further.

\subsubsection{Isotropic state}

To illustrate how these integrals behave more transparently, let us specialize to the case of an isotropic state ({\it cf.} Eq.~(\ref{matrixexcitedstate}) with $n(\mathbf{k}) \to n(|\mathbf{k}|)$) such that
\begin{eqnarray}
\mathfrak{n}(k) = \theta(k^0) n(|\mathbf{k}|) + \theta(- k^0) n(|\mathbf{k}|) = n(|\mathbf{k}|) \, , 
\end{eqnarray}
so that the occupation number $n(|\mathbf{k}|)$ is rotationally invariant in $\mathbf{k}$. It is trivial to simplify (\ref{deltaI_n}) to
\begin{equation}
\delta I(k) = \frac{M^2 - k^2}{\pi^2} \int_{M}^\infty \exd \Omega \; \sqrt{\Omega^2 - M^2} \; n\big( \sqrt{\Omega^2 - M^2} \, \big) \, ,
\end{equation}
where we have integrated over $\ell_0$ and all angles and defined $\Omega = \sqrt{ |\boldsymbol{\ell}|^2 + M^2 }$. The integral (\ref{LFn_int}) can be similarly simplified to
\begin{eqnarray}
L_{\mathcal{F}\mathfrak{n}}(k) & = & \frac{ i (-3 k^2 + 4M^2) }{2 \pi^2 } \int_M^\infty \exd \Omega \; \sqrt{\Omega^2 - M^2} \; n\big( \sqrt{ \Omega^2 - M^2 } \, \big)  \\
&  & + \frac{ i (k^2 - 2M^2)^2 }{16 \pi^2 |\mathbf{k}| }\int_M^\infty \exd \Omega \;  n\big( \sqrt{ \Omega^2 - M^2 } \, \big) \bigg[ \; g\left( \frac{k^2/2 + k_0 \Omega }{|\mathbf{k}| \sqrt{\Omega^2 - M^2}} \right) + g\left( \frac{k^2/2 - k_0 \Omega }{|\mathbf{k}| \sqrt{\Omega^2 - M^2}} \right) \bigg] \notag \, , 
\end{eqnarray}
where we have used the definition of the function $g(x)$ from Eq.~(\ref{f_def_int}). The real and imaginary part of this integral then turns out to be 
\begin{eqnarray}
\mathrm{Re}\big[ L_{\mathcal{F}\mathfrak{n}}(k) \big] & = & \frac{ \theta(k^2) \, (k^2 - 2M^2)^2 }{16 \pi |\mathbf{k}| } \int_{\Omega_{+}}^\infty \exd \Omega \; \Big[ n\big( \sqrt{ \Omega^2 - M^2 } + n\big( \sqrt{ (\Omega - |k_0|)^2 - M^2 } \, \big) \, \Big] \\
& & + \frac{ \theta(-k^2 - 4M^2) \, (k^2 - 2M^2)^2 }{8 \pi |\mathbf{k}| } \int_{\Omega_{-}}^{\Omega_{+}} \exd \Omega \;  n\big( \sqrt{ \Omega^2 - M^2 } \, \big) \notag \\
\mathrm{Im}\big[ L_{\mathcal{F}\mathfrak{n}}(k) \big] & = & \frac{ (-3 k^2 + 4M^2) }{2 \pi^2 } \int_M^\infty \exd \Omega \; \sqrt{\Omega^2 - M^2} \; n\big( \sqrt{ \Omega^2 - M^2 } \, \big)  \\
&  & + \frac{ (k^2 - 2M^2)^2 }{16 \pi^2 |\mathbf{k}| }\int_M^\infty \exd \Omega \;  n\big( \sqrt{ \Omega^2 - M^2 } \, \big) \; \log \bigg| \frac{(k^2 - 2 |\mathbf{k}| \sqrt{\Omega^2 - M^2})^2 - 4 k_0^2 \Omega^2 }{ (k^2 + 2 |\mathbf{k}| \sqrt{\Omega^2 - M^2})^2 - 4 k_0^2 \Omega^2  } \bigg| \notag \, .
\end{eqnarray}
Next we notice that the integrals $L_{\mathcal{W}\mathfrak{n}}$ and $L_{\mathfrak{nn}}$ are precisely of the form (\ref{loopLf_def}). It then trivially follows using the result (\ref{loopLf_ans}) that
\begin{eqnarray}
L_{\mathcal{W}\mathfrak{n}}(k) & = & \frac{\theta(k^0) \theta(-k^2 - 4M^2) (k^2 -2M^2)^2}{8 \pi |\mathbf{k}|} \int_{ \Omega_{-} }^{\Omega_{+} } \exd \Omega \; n\big( \sqrt{ \Omega^2 - M^2} \, \big) \\
&&  +\frac{\theta(k^2) (k^2 - 2M^2)^2 }{8\pi |\mathbf{k}|} \int_{\Omega_{+} }^{\infty} \exd \Omega\; \bigg[ \theta(-k^0) n\big( \sqrt{ \Omega^2 - M^2} \, \big) + \theta(k^0) n\big( \sqrt{ (\Omega-|k_0|)^2 - M^2} \, \big) \bigg] \notag \\
L_{\mathfrak{nn}}(k) & = & \frac{\theta(-k^2 - 4 M^2) (k^2 - 2M^2)^2}{8\pi |\mathbf{k}|} \int_{ \Omega_{-} }^{\Omega_{+} } \exd \Omega \; n\big( \sqrt{\Omega^2 - M^2} \big) \, n\big( \sqrt{ (\Omega-|k_0|)^2 - M^2} \big) \\
&&  + \frac{\theta(k^2) (k^2 -2M^2)^2 }{4\pi |\mathbf{k}|} \int_{\Omega_{+} }^{\infty} \exd \Omega\; n\big( \sqrt{\Omega^2 - M^2} \big) \, n\big( \sqrt{ (\Omega-|k_0|)^2 - M^2} \big) \notag \, .
\end{eqnarray}
Putting these together into the formulae (\ref{deltaI_n}) and (\ref{deltaLFW_n}), along with the vacuum contributions gives the results in the main text.

\subsubsection{Thermal state}

We specialize further to a thermal state such that
\begin{equation}
n(|\mathbf{k}|) = \frac{1}{e^{\beta \sqrt{ |\mathbf{k}|^2 + M^2 }} + 1} \, .
\end{equation}
In the high temperature limit, the main contribution to the self-energies is $L_{\mathfrak{nn}}(k)$, since this contains two Bose-Einstein distributions. Using the above formula one finds
\begin{eqnarray}
L_{\mathfrak{nn}}(k) & = &  \frac{\theta(-k^2 - 4 M^2) ( - k^2 + 2 M^2)^2}{ 8 \pi |\mathbf{k}|} \int_{ \Omega_{-} }^{\Omega_{+} } \frac{ \exd \Omega }{[ e^{\beta \Omega } - 1] [ e^{\beta ( | k_0 |  - \Omega )} - 1 ]} \notag \\
&&  + \frac{\theta(k^2) ( - k^2 + 2 M^2)^2 }{4 \pi |\mathbf{k}|} \int_{\Omega_{+} }^{\infty}  \frac{\exd \Omega}{[ e^{\beta \Omega } - 1] [ e^{\beta ( \Omega - |k_0| ) } - 1 ]} \, .
\end{eqnarray}
It turns out that one can write down a closed-form expression for this integral:
\begin{eqnarray}
L_{\mathfrak{nn}}(k) & = &  \frac{\theta(-k^2 - 4 M^2) ( - k^2 + 2 M^2)^2}{ 8 \pi |\mathbf{k}|} \bigg[ \frac{ \log(e^{\beta  \Omega }-1) -\log(e^{\beta  (| k_0| -\Omega )}-1)  -\beta  \Omega }{\beta  \left(e^{\beta  | k_0 | }-1\right)} \bigg] \bigg|_{\Omega_{-}}^{\Omega_{+}} \\
&&  + \frac{\theta(k^2) ( - k^2 + 2 M^2)^2 }{4 \pi |\mathbf{k}|} \bigg[ \frac{\log(e^{\beta  \Omega }-1 )}{\beta (e^{-\beta  | k_0 | }-1)}+\frac{\log(e^{\beta  (\Omega -| k_0 | )}-1)}{\beta (e^{\beta  |k_0 | }-1)}+\Omega \bigg]\bigg|_{\Omega_{+}}^{\infty} \notag \, .
\end{eqnarray}
Of course $\Sigma_{\mathcal{F}}(k) \simeq - \frac{1}{4} L_{\mathfrak{nn}}(k)$ and $\Sigma_{\mathcal{W}}(k) \simeq + \frac{1}{4} L_{\mathfrak{nn}}(k)$ and taking the high temperature limit gives the result Eq.~(\ref{thermalSigmas}).

\subsection{Cancellations in r/a basis}
\label{app:ra_cancel}

We here show how cancellations happen in the influence functional when we switch to the r/a basis. It is convenient to here rewrite the propagators given in (\ref{FWwithn_MAIN}) in the form
\begin{eqnarray}
\mathcal{F}(k) & = & \frac{ - i}{k^2 + M^2} + f(k) \delta(k^2 + M^2)  \qquad \mathrm{with}\ f(k) = 2 \pi \left( \tfrac{1}{2} + \mathfrak{n}(k) \right) \label{Fwithf}\\
\mathcal{W}(k) & = & w(k) \delta(k^2 + M^2)  \hspace{24mm} \mathrm{with}\ w(k) = 2 \pi \left( \theta(k_0) + \mathfrak{n}(k) \right) \label{Wwithw} \, ,
\end{eqnarray}
where $\mathfrak{n}(k) = \theta(k^0) n(\mathbf{k}) + \theta(-k^0) n(-\mathbf{k})$ which is even in 4-momenta (although $n(\mathbf{k})$ need not be even in three-momenta for anisotropic states). Notice that this implies that $f(k) = f(-k)$ however $ w(k) \neq w(-k)$ which we use below.

\subsubsection{Proof that $\Gamma_{\mathcal{U}}+\Gamma_{\mathcal{U}}^{\ast} - 3 \Gamma_{\mathcal{N}} - 3 \Gamma_{\mathcal{N}}^{\ast} \to 0$}

We here provide the basic steps showing how for any excited state
\begin{equation}
\Gamma_{\mathrm{rrr}}(x,y,z) \equiv \Gamma_{\mathcal{U}}(x,y,z) + \Gamma^{\ast}_{\mathcal{U}}(x,y,z) - 3 \Gamma_{\mathcal{N}}(x,y,z) - 3 \Gamma^{\ast}_{\mathcal{N}}(x,y,z)  \, , 
\end{equation}
vanishes when integrated symmetrically over the three points. Using the results (\ref{GammaU_def}) and (\ref{GammaN_def}) one can write:
\begin{eqnarray}
\Gamma_{\mathrm{rrr}}(x,y,z) = \int \frac{\exd^4 k}{(2\pi)^4} \int \frac{\exd^4 p}{(2\pi)^4} \int \frac{\exd^4 q}{(2\pi)^4} \; (2\pi)^4 \delta^{(4)}(k+p+q)  \overline{\Gamma}_{\mathrm{rrr}}(k,p,q) e^{i k \cdot x + i p \cdot y + i q \cdot z} \, , 
\end{eqnarray}
where
\begin{footnotesize}
\begin{eqnarray}
\frac{ \overline{\Gamma}_{\mathrm{rrr}}(k,p,q) }{k^2 + p^2 + q^2} & = & \frac{ \Gamma_{\mathcal{U}}(k,p,q) + \Gamma^{\ast}_{\mathcal{U}}(-k,-p,-q)  - 3\Gamma_{\mathcal{N}}(k,p,q) -3 \Gamma^{\ast}_{\mathcal{N}}(-k,-p,-q) }{k^2 + p^2 + q^2} \\
& = & -2 \left(\frac{f(k) \delta \left(k^2+M^2\right)}{\left(M^2+p^2\right) \left(M^2+q^2\right)}+\frac{f(p) \delta \left(M^2+p^2\right)}{\left(k^2+M^2\right) \left(M^2+q^2\right)}+\frac{\left(f(q)-\frac{3}{2} (w(-q)+w(q))\right) \delta \left(M^2+q^2\right)}{\left(k^2+M^2\right) \left(M^2+p^2\right)}\right) \notag \\
& \ & -3 i \bigg(\frac{f(p) (w(-q)-w(q)) \delta \left(M^2+p^2\right) \delta \left(M^2+q^2\right)}{k^2+M^2}+\frac{f(k) (w(-q)-w(q)) \delta \left(k^2+M^2\right) \delta \left(M^2+q^2\right)}{M^2+p^2} \notag \\
&& \qquad \qquad +\frac{\delta \left(k^2+M^2\right) (w(k) w(p)-w(-k) w(-p)) \delta \left(M^2+p^2\right)}{M^2+q^2}\bigg) \notag \\
&& +  \Big[ 3 f(q) w(-k) w(-p)+3 f(q) w(k) w(p)-3 f(k) f(p) w(-q) -3 f(k) f(p) w(q) \notag\\
&& \qquad +2 f(k) f(p) f(q)-w(-k) w(-p) w(-q)-w(k) w(p) w(q) \Big] \delta \left(k^2+M^2\right) \delta \left(M^2+p^2\right) \delta \left(M^2+q^2\right) \notag
\end{eqnarray}
\end{footnotesize}\ignorespaces
All we have used is evenness of $f(k)$ in the above. Next notice that $w(k)+w(-k) = 2f(k)$, and so we replace everywhere $w(-k)= 2f(k) - w(k)$. This simplifies the above into 
\begin{footnotesize}
\begin{eqnarray}
\frac{ \overline{\Gamma}_{\mathrm{rrr}}(k,p,q) }{k^2 + p^2 + q^2} & = & -2 \left(\frac{f(k) \delta \left(k^2+M^2\right)}{\left(M^2+p^2\right) \left(M^2+q^2\right)}+\frac{f(p) \delta \left(M^2+p^2\right)}{\left(k^2+M^2\right) \left(M^2+q^2\right)}-\frac{2 f(q) \delta \left(M^2+q^2\right)}{\left(k^2+M^2\right) \left(M^2+p^2\right)}\right) \\
& \ & -6 i \bigg[ \frac{ \delta \left(k^2+M^2\right) \delta \left(M^2+p^2\right) (f(p) w(k)+f(k) w(p)-2 f(k) f(p))}{M^2+q^2} \notag \\
&& +\frac{\delta \left(M^2+p^2\right) \delta \left(M^2+q^2\right) (f(p) f(q)-f(p) w(q))}{k^2+M^2}+\frac{\delta \left(k^2+M^2\right) \delta \left(M^2+q^2\right) (f(k) f(q)-f(k) w(q))}{M^2+p^2}\bigg] \notag \\
& \ & - 2 \Big[ f(p) f(q) w(k)+f(k) f(q) w(p)-2 f(k) f(p) w(q) \notag \\
&& \qquad \qquad +f(p) w(k) w(q)-2 f(q) w(k) w(p)+f(k) w(p) w(q)\Big]\delta \left(k^2+M^2\right) \delta \left(M^2+p^2\right) \delta \left(M^2+q^2\right) \notag \, .
\end{eqnarray}
\end{footnotesize}\ignorespaces
If we could interchange $k$, $p$ and $q$ freely then this would vanish, but the position labels in the $e^{i k \cdot x + i p \cdot y + i q \cdot z}$ spoil this symmetry. However, when we recall that this appears integrated under $\int \exd^4 x \int \exd^4 y \int \exd^4 z$ in the influence functional, we can relabel the $x,y,z$ freely and so in fact the quantity in question exactly vanishes:
\begin{eqnarray}
\int \exd^4 x \; \big( \mathscr{D}_{\mathrm{r}}\chi_{\mathrm{r}}(x) \big)^2 \int \exd^4 y\; \big( \mathscr{D}_{\mathrm{r}}\chi_{\mathrm{r}}(y) \big)^2  \int \exd^4 z\;  \big( \mathscr{D}_{\mathrm{r}}\chi_{\mathrm{r}}(z) \big)^2 \; \overline{\Gamma}_{\mathrm{rrr}}(x,y,z) = 0 \, .
\end{eqnarray}
This is related to cancellations that are described in  \cite{Bedaque:1996af, Gelis:1997zv} giving cutting rules at finite temperature.

\subsubsection{Proof that $[\mathcal{Q}_{\mathcal{F}} + \mathcal{Q}_{\mathcal{F}}^{\ast} - \mathcal{Q}_{\mathcal{W}} - \mathcal{Q}^{\ast}_{\mathcal{W}}](x,y) \to 0$}

Let us consider the combination 
\begin{equation}
\mathcal{Q}_{\mathrm{rr}}(x,y)\equiv \int \frac{\exd^4 k}{(2\pi)^4} \overline{\mathcal{Q}}_{\mathrm{rr}}(k) e^{i k \cdot (x-y))} \quad \mathrm{with} \ \ 
\overline{\mathcal{Q}}_{\mathrm{rr}}(k) \equiv \mathcal{Q}_{\mathcal{F}(k)}+ \mathcal{Q}_{\mathcal{F}}^{\ast}(k) - \mathcal{Q}_{\mathcal{W}}(k) - \mathcal{Q}_{\mathcal{F}}(-k) \, ,
\end{equation}
where we have used $\mathcal{Q}_{\mathcal{F}}(-k) = \mathcal{Q}_{\mathcal{F}}(k)$ and $\mathcal{Q}^{\ast}_{\mathcal{W}}(k) = \mathcal{Q}_{\mathcal{W}}(k)$. We show here that $\overline{\mathcal{Q}}_{\mathrm{rr}}(k) = 0$ exactly.
Using the explicit form of the loop corrections one finds
\begin{eqnarray}
\overline{\mathcal{Q}}_{\mathrm{rr}}(k) & = & i \Big( I(k) - \frac{1}{4} \mathrm{Im}[L_{\mathcal{F}(k)}] \Big) \bigg\{  \mathcal{F}(k)^2 - \mathcal{F}^*(k) ^2 - \Big( \mathcal{W}(k) + \mathcal{W}(-k) \Big) \Big( \mathcal{F}(k) -   \mathcal{F}^*(k)   \Big) \bigg\} \notag \\
&& - \frac{1}{4} \mathrm{Re}[L_{\mathcal{F}(k)}] \bigg\{  \mathcal{F}(k)^2 + \mathcal{F}^*(k) ^2 + 2 \mathcal{W}(k) \mathcal{W}(-k) \notag \\
&& \hspace{45mm}+ \Big( \mathcal{W}(k) + \mathcal{W}(-k) \Big) \Big( \mathcal{F}(k) +  \mathcal{F}^*(k)   \Big) \bigg\} \notag \\
&& + \frac{1}{4} \bigg\{ \Big( \mathcal{F}(k) +  \mathcal{F}^*(k)   \Big) \Big( \mathcal{W}(k) L_{\mathcal{W}}(-k) + \mathcal{W}(-k) L_{\mathcal{W}}(k) \Big) \\
&& \hspace{40mm} -  \mathcal{F}^*(k)   \mathcal{F}(k) L_{\mathcal{W}}(k) - \mathcal{W}(k) \mathcal{W}(k) L_{\mathcal{W}}(-k) \notag \\
&& \hspace{40mm} -  \mathcal{F}^*(k)   \mathcal{F}(k) L_{\mathcal{W}}(-k) - \mathcal{W}(-k) \mathcal{W}(-k) L_{\mathcal{W}}(k) \bigg\} \, . \notag
\end{eqnarray}
Let us simplify each of these terms using the decomposition (\ref{Fwithf}) and (\ref{Wwithw}) for the Feynman and Wightman propagators. Recall that $w(- k) = 2 f(k) - w(k)$, and use $f(k) = 2 \pi (\tfrac{1}{2} + \mathfrak{n}(k))$ and $w(k) = 2 \pi (\theta(k^0) + \mathfrak{n}(k))$ to find 
\begin{eqnarray}
\overline{\mathcal{Q}}_{\mathrm{rr}}(k) & = & - \frac{1}{4} \mathrm{Re}[L_{\mathcal{F}(k)}] \bigg\{ - \frac{2}{(k^2 + M^2)^2} - 2 \pi^2 \big[ \delta(k^2 + M^2) \big]^2 \bigg\} \\
&& + \frac{1}{4} \bigg\{ - \Big( L_{\mathcal{W}}(k) + L_{\mathcal{W}}(-k) \Big) \Big( \frac{1}{(k^2 + M^2)^2} + \pi^2 \big[ \delta(k^2 + M^2) \big]^2  \Big) \bigg\} \notag \\
& = & + \frac{1}{4} \bigg( 2 \mathrm{Re}[L_{\mathcal{F}(k)}] - L_{\mathcal{W}}(k) - L_{\mathcal{W}}(-k) \bigg)  \mathcal{F}^{\rm vac}(k) \mathcal{F}^{{\rm vac} *}(k) \ .  
\end{eqnarray}
Notice that all $\mathfrak{n}$ distributions cancel out. Now let us recall the definition of the loops, one finds
\begin{eqnarray}
&& 2 \mathrm{Re}[L_{\mathcal{F}}(k)] - L_{\mathcal{W}}(k) - L_{\mathcal{W}}(-k) \notag \\
&& \quad = \int \frac{\exd^4 \ell}{(2\pi)^4}  \; \Big( k^2 + (k -\ell)^2 + \ell^2 \Big)^2 \bigg[  \mathcal{F}(\ell)   \mathcal{F}(k - \ell )  + \mathcal{F}^*(\ell)   \mathcal{F}^*(k - \ell )  \\
&& \hspace{65mm} - \mathcal{W}(\ell) \mathcal{W}(k - \ell ) - \mathcal{W}(-\ell) \mathcal{W}(- k + \ell ) \bigg] =0  \, .\notag
\end{eqnarray}
This is most easily seen to vanish when the propagators are expressed in terms of advanced and retarded propagators, due to causality.

\bibliographystyle{JHEP}
\bibliography{references.bib}

\end{document}